\documentclass{article}
\topmargin = - 0.5 cm
\usepackage{amsmath}
\usepackage{amssymb}

\newcommand{\epr}{{\sc epr}}
 \newcommand{\beq}{\begin{equation}}
\newcommand{\eeq}{\end{equation}} 
\newcommand{\bea}{\begin{eqnarray}}
\newcommand{\eea}{\end{eqnarray}} \newcommand{\nn}{\nonumber}

 \newcommand{\qm}{quantum
mechanics} 
\newcommand{\ca}{$C^*$-algebra} 
 \newcommand{\rep}{representation}
\newcommand{\irrep}{irreducible representation}
\newcommand{\Hs}{Hilbert space}

   \newcommand{\vna}{von
Neumann algebra}

\newcommand{\id}{\mbox{\rm id}}

 \newcommand{\ovl}{\overline}
 \newcommand{\til}{\tilde}
\newcommand{\raw}{\rightarrow} 
\newcommand{\law}{\leftarrow} 
\newcommand{\hraw}{\hookrightarrow}

\newcommand{\ot}{\otimes} 
 
 \newcommand{\wed}{\wedge}
\renewcommand{\Re}{{\rm Re}\,} \renewcommand{\Im}{{\rm Im}\,}
\newcommand{\rst}{\upharpoonright} 
\newcommand{\x}{\times} 
 \newcommand{\ran}{{\rm ran}}
\newcommand{\Tr}{\mbox{\rm Tr}\,}

 \newcommand{\BH}{\mathcal{B}({\mathcal H})} 
\newcommand{\cin}{C^{\infty}} \newcommand{\cci}{C^{\infty}_c}
\newcommand{\half}{\mbox{\footnotesize $\frac{1}{2}$}}
\newcommand{\third}{\mbox{\footnotesize $\frac{1}{3}$}}

\newcommand{\lho}{\lim_{\hbar\rightarrow 0}}   \newcommand{\qw}{{\mathcal Q}_{\hbar}^W}
\newcommand{\qp}{\CQ_{\hbar}^{\mbox{\tiny pos}}}
\newcommand{\qh}{q_{\hbar}} 
 
 \newcommand{\qb}{{\mathcal
Q}_{\hbar}^{B}}

\newcommand{\inv}{^{-1}} 
\newcommand{\Exp}{{\rm Exp}} 

\newcommand{\er}{\eqref}
\newcommand{\al}{\alpha} \newcommand{\bt}{\beta}
\newcommand{\gm}{\gamma} \newcommand{\Gm}{\Gamma}
\newcommand{\dl}{\delta} \newcommand{\Dl}{\Delta}
\newcommand{\ep}{\epsilon} \newcommand{\varep}{\varepsilon}

\newcommand{\lm}{\lambda} \newcommand{\Lm}{\Lambda}
\newcommand{\rh}{\rho} \newcommand{\sg}{\sigma}
\newcommand{\Sg}{\Sigma} \newcommand{\ta}{\tau} 
\newcommand{\Ph}{\Phi} \newcommand{\phv}{\varphi}
\newcommand{\ch}{\chi} \newcommand{\ps}{\psi} \newcommand{\Ps}{\Psi}
\newcommand{\om}{\omega} \newcommand{\Om}{\Omega}

\newcommand{\GC}{\mathfrak{C}}

\newcommand{\GS}{\mathfrak{S}} \newcommand{\g}{\mathfrak{g}}
\newcommand{\GQ}{\mathfrak{Q}} 
\newcommand{\h}{\mathfrak{h}}

\newcommand{\CA}{{\mathcal A}} \newcommand{\CB}{{\mathcal B}}
\newcommand{\CC}{{\mathcal C}} \newcommand{\CF}{{\mathcal F}}
\newcommand{\CE}{{\mathcal E}}
 \renewcommand{\H}{{\mathcal H}}
 
\newcommand{\CK}{{\mathcal K}}   \newcommand{\CL}{{\mathcal L}}   
 \newcommand{\CM}{{\mathcal M}}
 \newcommand{\CS}{{\mathcal S}}
\newcommand{\CO}{{\mathcal O}} \newcommand{\CP}{{\mathcal P}}
\newcommand{\CQ}{{\mathcal Q}}

\newcommand{\C}{{\mathbb C}} 
 
\newcommand{\N}{{\mathbb N}} \newcommand{\R}{{\mathbb R}}
 \newcommand{\Z}{{\mathbb Z}}
  \makeatletter
\newskip\tempskip \def\endproof{{\parfillskip24\p@ plus\@ne
fil\@@par}\tempskip\prevdepth
\ifdim\lastskip=\z@\tempskip\z@\else\vskip-\lastskip
\ifdim\tempskip>4\p@ \tempskip.5\tempskip \else \tempskip\z@\fi\fi
\nobreak\vskip-\baselineskip\vskip-\tempskip\noindent\hbox
to\hsize{\hfill
$\blacksquare$}\par\vskip\tempskip\vskip\abovedisplayskip\@doendpe}
\makeatother \makeatletter
\newskip\tempskip \def\endiproof{{\parfillskip24\p@ plus\@ne
fil\@@par}\tempskip\prevdepth
\ifdim\lastskip=\z@\tempskip\z@\else\vskip-\lastskip
\ifdim\tempskip>4\p@ \tempskip.5\tempskip \else \tempskip\z@\fi\fi
\nobreak\vskip-\baselineskip\vskip-\tempskip\noindent\hbox
to\hsize{\hfill
$\Box$}\par\vskip\tempskip\vskip\abovedisplayskip\@doendpe}
\makeatother 

\newcommand{\BBH}{\mathbb{H}}

\renewcommand{\qh}{\CQ_{\hbar}}
\renewcommand{\qp}{\CQ^{pre}_{\hbar}}
\newcommand{\gq}{geometric quantization}
\newcommand{\spinc}{\mathrm{Spin}^c}
\def\Dslash{\setbox0=\hbox{$D$}D\hskip-\wd0\hbox to\wd0{\hss\sl/\/\hss}}
\newcommand{\DS}{\Dslash}

\newcommand{\ind}{\mathrm{index}}
\newcommand{\coker}{\mathrm{coker}}

\renewcommand{\lho}{\lim_{\hbar\raw 0}}
\newcommand{\lni}{\lim_{N\raw\infty}}
\newcommand{\up}{\uparrow}
\newcommand{\CD}{\mathcal{D}}
\newcommand{\down}{\downarrow}
\begin{document}
\setlength{\baselineskip}{1\baselineskip}
\thispagestyle{empty}
\title{Between classical and quantum\footnote{To appear in Elsevier's forthcoming  {\it  Handbook of the Philosophy of Science}, Vol.\ 2: {\it Philosophy of Physics} (eds.\ John Earman \&\ Jeremy Butterfield). The author is indebted to  Stephan de Bi\`{e}vre, Jeremy Butterfield, Dennis Dieks, Jim Hartle, Gijs Tuynman, 
Steven Zelditch, and Wojciech Zurek for detailed comments on various drafts of this paper. The final version has greatly benefited from the 7 Pines Meeting on `The Classical-Quantum Borderland' (May, 2005); the author wishes to express his gratitude to Lee Gohlike and the Board of the 7 Pines Meetings for the invitation, and to the other speakers (M. Devoret, J. Hartle, E. Heller, G. `t Hooft, D. Howard, M. Gutzwiller, M. Janssen, A. Leggett, R. Penrose, P. Stamp, and W. Zurek) for sharing their insights with him. 
}}
\author{\textbf{N.P. Landsman} \\ \mbox{} \hfill \\
Radboud Universiteit Nijmegen\\
Institute for Mathematics, Astrophysics, and Particle Physics\\
Toernooiveld 1, 6525 ED NIJMEGEN\\
THE NETHERLANDS\\
\mbox{} \hfill \\
email \texttt{landsman@math.ru.nl}}
\date{\today}
\maketitle
\begin{abstract}
The relationship between classical and quantum theory is of central importance to the philosophy of physics, and any interpretation of quantum mechanics has to clarify it.   Our discussion of this relationship is partly historical and conceptual, but mostly technical and mathematically rigorous, including over 500 references.  For example, we sketch how certain intuitive ideas of the founders of quantum theory have fared in the light of current mathematical knowledge. One such idea that has certainly stood the test of time is Heisenberg's `quantum-theoretical \textit{Umdeutung} (reinterpretation) of classical observables', which lies at the basis of  quantization theory. Similarly, Bohr's correspondence principle (in somewhat revised form) and Schr\"{o}dinger's wave packets (or coherent states)
continue to be of great importance in understanding classical behaviour from \qm. 
On the other hand, no consensus has been reached on the Copenhagen Interpretation, 
but in view of the parodies of it one typically finds in the literature we describe it in detail.

 On the assumption that \qm\ is universal and complete, we discuss three ways in which classical physics has so far been believed to emerge from quantum physics, namely in the limit $\hbar\rightarrow 0$ of small Planck's constant (in a finite system), in the limit $N\raw\infty$ of a large system with $N$ degrees of freedom (at fixed $\hbar$), and through decoherence and consistent histories. The first limit is closely related to modern quantization theory and microlocal analysis, whereas the second involves methods of \ca s and the concepts of superselection sectors and macroscopic observables. In  these limits,
  the classical world does not emerge as a sharply defined objective reality, but rather as an approximate {\it appearance} relative to certain ``classical" states and observables. Decoherence subsequently clarifies the role of such states, in that they are ``einselected", i.e.\ robust against coupling to the environment. Furthermore, the nature of classical observables is elucidated by the fact that they typically define (approximately) consistent sets of histories. 

This combination of ideas and techniques does not quite  resolve the measurement problem, but it does  make the point that classicality results from the {\it elimination} of certain states and observables from quantum theory. 
  Thus the classical world is not created by observation (as  Heisenberg once claimed),  but rather by the lack of it.
  \end{abstract}
  \newpage\tableofcontents \newpage
\begin{quote}
`But the worst thing is that I am quite unable to clarify the transition [of matrix mechanics] 
to the classical theory.' 
 (Heisenberg to Pauli, October 23th, 1925)\footnote{`Aber das Schlimmste ist, da\ss\ ich \"{u}ber den \"{U}bergang in die klassische Theorie nie Klarheit bekommen kann.' See Pauli (1979), p.\ 251.}
\end{quote}
\begin{quote}
`Hendrik Lorentz considered the establishment of the correct relation between the classical and the quantum theory as the most fundamental problem of future research. This problem bothered him as much as it did Planck.' (Mehra \&\ Rechenberg, 2000, p.\ 721)
\end{quote} 
\begin{quote}
`Thus \qm\ occupies a very unusual place among physical theories: it contains classical mechanics as a limiting case, yet at the same time it requires this limiting case for its own formulation.' (Landau \&\ Lifshitz, 1977, p.\ 3)
\end{quote} 
\section{Introduction}\label{S1}
Most modern physicists and philosophers would agree that a decent interpretation of quantum mechanics should fullfil at least two criteria. Firstly, it has to elucidate  the physical meaning of its mathematical formalism and thereby secure the empirical content of the theory. This point (which we address only in a derivative way) was clearly recognized by all the founders of quantum theory.\footnote{The history of quantum theory has been described in a large number of books. The most detailed presentation is in Mehra \&\ Rechenberg (1982--2001), but this multi-volume series has by no means superseded smaller works such as Jammer (1966),  vander Waerden (1967), Hendry (1984), Darrigol (1992), and Beller (1999). Much information may also be found in biographies such as Heisenberg (1969), Pais (1982), Moore (1989), Pais (1991), Cassidy (1992), Heilbron (2000), Enz (2002), etc. See also Pauli (1979). A new project on the history of matrix mechanics led by J\"{u}rgen Renn is on its way.\label{historybooks}} 
Secondly (and this {\it is}  the subject of this paper), it has to explain at least the {\it appearance} of the classical world.\footnote{That these point are quite distinct is shown by the Copenhagen Interpretation, which exclusively addresses the first at utter neglect of the second. Nonetheless, in most other approaches to \qm\ there is substantial overlap between the various mechanisms that are proposed to fullfil the two criteria in question.}
 As shown by our second quotation above, Planck saw the difficulty this poses, and 
as a first contribution he noted that the high-temperature limit of his formula for black-body radiation converged to the classical expression. 
 Although Bohr believed that {\it \qm\ should be interpreted through classical physics}, among the founders of the theory he seems to have been unique in his lack of appreciation of the problem of {\it deriving classical physics from quantum theory}. Nonetheless, through his correspondence principle (which he proposed in order to address the {\it first} problem above rather than the second) Bohr made one of the most profound contributions to the issue.
Heisenberg initially recognized the problem, but quite erroneously came to believe he had solved it in his renowned paper on the uncertainty relations.\footnote{`One can see that the transition from micro- to macro-mechanics is now very easy to understand: classical mechanics is altogether part of \qm.' (Heisenberg to Bohr, 19 March 1927, just before 
the submission on 23 March of Heisenberg (1927). See {\it Bohr's Scientific Correspondence} in the {\it Archives for the History of Quantum Physics}).}
Einstein famously did not believe in the fundamental nature of quantum theory, whereas Schr\"{o}dinger was well aware of the problem from the beginning, later highlighted the issue with his legendary cat, and at various stages in his career made  important technical contributions towards its resolution. Ehrenfest stated the well-known theorem named after him.  Von Neumann saw the difficulty, too, and addressed it by means of his well-known analysis of the measurement procedure in \qm. 

The problem is actually even more acute than the founders of quantum theory foresaw. 
The experimental realization of  Schr\"{o}dinger's cat  is nearer than most physicists would feel comfortable with (Leggett, 2002;  Brezger et al., 2002; Chiorescu et al., 2003; 
Marshall et al., 2003;  Devoret et al., 2004). 
Moreover, awkward superpositions are by no means confined to physics 
 laboratories:  due to its chaotic motion, Saturn's moon Hyperion (which is about the size of New York) has been estimated to spread out all over its orbit within  20 years if treated as an isolated  quantum-mechanical wave packet (Zurek \&\ Paz, 1995). Furthermore, decoherence theorists have made the point  that ``measurement" is not only a procedure carried out by experimental physicists in their labs, but takes place in Nature all the time without any human intervention. On the conceptual side, parties as diverse as  Bohm \&\ Bell and their followers on the one hand  and the quantum cosmologists on the other have argued
that a ``Heisenberg cut" between object and observer cannot possibly lie at the basis of
a fundamental theory of physics.\footnote{Not to speak of the problem, also raised
by quantum cosmologists, of deriving classical space-time from some theory of quantum gravity. This is certainly part of the general program of deriving classical physics from quantum theory, but unfortunately it cannot be discussed in this paper. }
  These and other remarkable  insights of the past few decades have drawn wide attention to the importance of the problem of interpreting \qm, and in particular of explaining classical physics from it. 

We will discuss these ideas in more detail below, and indeed our discussion of the relationship between classical and \qm\ will be partly historical. However, other than that it will be technical and mathematically rigorous. For the problem at hand is so delicate that in this area sloppy mathematics is almost guaranteed to lead to unreliable physics and conceptual confusion (notwithstanding the undeniable success of poor man's math elsewhere in theoretical physics). Except for von Neumann, this was not the attitude of the pioneers of \qm; but while it has to be acknowledged that many of their ideas are still central to the current discussion, these ideas {\it per se} have {\it not} solved the problem. Thus we assume the reader to be familiar with the Hilbert space formalism of \qm,\footnote{
Apart from seasoned classics such as Mackey (1963), Jauch (1968), Prugovecki (1971), Reed \&\ Simon (1972), or Thirring (1981), the reader might consult more recent books such as Gustafson \&\ Sigal (2003) or  Williams (2003). See also Dickson (2005).} 
and for some parts of this paper (notably Section \ref{S6} and parts of Section \ref{S4}) also with the basic theory of \ca s and its applications to quantum theory.\footnote{For physics-oriented introductions to \ca s see Davies (1976), Roberts \&\ Roepstorff (1969), Primas (1983), Thirring (1983), Emch (1984), Strocchi (1985), Sewell (1986), Roberts (1990), Haag (1992), Landsman (1998), Araki (1999), and Sewell (2002).  Authoratitive mathematical texts include Kadison \&\ Ringrose (1983, 1986) and Takesaki (2003). \label{Cstarlit}} In addition, some previous encounter with the conceptual problems of quantum theory would be helpful.\footnote{Trustworthy books include, for example,  
Scheibe (1973), Jammer (1974), van Fraassen (1991), dÕEspagnat (1995), Peres (1995), 
Omn\`{e}s (1994, 1999), Bub (1997), and Mittelstaedt  (2004).\label{QMtexts}}

Which ideas {\it have} solved the problem of explaining the appearance of the classical world from quantum theory? In our opinion, none have, although since the founding days of \qm\ a number of new ideas have been proposed that almost certainly will play a role in the eventual resolution, should it ever be found. These ideas surely include:
\begin{itemize}
\item 
The limit $\hbar\rightarrow 0$ of small Planck's constant (coming of age with the mathematical field of microlocal analysis);
\item The limit $N\raw\infty$ of a large system with $N$ degrees of freedom (studied in a serious only way after the emergence of \ca ic methods);
\item 
Decoherence and consistent histories. 
\end{itemize}
Mathematically, the second limit may be seen as a special case of the first, though the underlying physical situation is of course quite different. In any case, after a detailed analysis our 
conclusion will be that none of these ideas in isolation is capable of explaining the classical world, but that there is some hope that by combining all three of them, one might do so in the future. 

Because of the fact that the subject matter of this review is unfinished business, to date one may adopt a number of internally consistent but mutually incompatible philosophical stances on the relationship between classical and quantum theory. Two extreme ones, which are always useful to keep in mind whether one holds one of them or not, are:
\begin{enumerate}
\item Quantum theory is fundamental and universally valid, and the classical world has 
only ``relative" or ``perspectival" existence.
\item Quantum theory is an approximate and derived theory, possibly false, and the classical world exists absolutely.
\end{enumerate}
An example of a position that our modern understanding of the measurement problem\footnote{See the books cited in footnote \ref{QMtexts}, 
especially Mittelstaedt (2004).}  has rendered internally inconsistent is: 
\begin{quote}
3.  Quantum theory is fundamental and universally valid, and (yet) the classical world exists absolutely.
\end{quote}

In some sense stance 1 originates with Heisenberg (1927), but the modern era started with
Everett (1957).\footnote{\label{MWM} Note, though,  that stance 1 by no means implies the so-called  Many-Worlds Interpretation, which also in our opinion is `simply a meaningless collage of words' (Leggett, 2002).} These days, most decoherence theorists, consistent historians,
and modal interpreters seem to support it. Stance 2 has a long and respectable pedigree unequivocally, including among others Einstein, Schr\"{o}dinger, and Bell. More recent backing has come from  Leggett as well as from ``spontaneous collapse" theorists such as Pearle, Ghirardi, Rimini, Weber, and others.
As we shall see in Section \ref{S3},  Bohr's position eludes classification according to these terms; our three stances being of an ontological nature, he probably would have found each of them unattractive.\footnote{To the extent that it was inconclusive, Bohr's debate with Einstein certainly suffered from the fact that the latter attacked strawman 3 (Landsman, 2006).
The fruitlessness of discussions such as those between Bohm and Copenhagen (Cushing, 1994) or between  Bell (1987, 2001) and Hepp (1972) has the same origin.} 

Of course, one has to specify what the terminology involved means. By quantum theory we mean standard \qm\  including the eigenvector-eigenvalue link.\footnote{Let $A$ be a selfadjoint operator on a \Hs\ $\H$, with associated
projection-valued measure $P(\Delta)$, $\Delta\subset \R$, so that $A=\int dP(\lm)\, \lm$ (see also footnote  \ref{PVM} below). The eigenvector-eigenvalue link states that a state $\Psi$ of the system lies in $P(\Delta)\H$ if and only if $A$ takes some value in $\Dl$ for sure. In particular, if $\Psi$ is an eigenvector of $A$ with eigenvalue $\lm$ (so that $P(\{\lm\})\neq 0$ and $\Psi\in P(\{\lm\})\H$), then $A$ takes the value $\lm$ in the state $\Psi$ with probability one. In general, the probability $p_{\Ps}(\Delta)$ that
in a state $\Psi$ the observable $a$ takes some value in $\Dl$ (``upon measurement") is given by the Born--von Neumann  rule $p_{\Ps}(\Delta)=(\Psi, P(\Delta)\Psi)$.} 
Modal interpretations of \qm\ (Dieks (1989a,b; van Fraassen, 1991; Bub, 1999; Vermaas, 2000; Bene \& Dieks, 2002; Dickson, 2005) deny this link, and lead to positions  close to or identical to stance 1.
 The projection postulate is neither endorsed nor denied when we generically speak of quantum theory.

 It is a bit harder to say what ``the classical world" means. In the present discussion we evidently  can {\it not}  define the classical world as the world that exists independently of observation - as Bohr did, see Subsection \ref{Pcl} - but neither can it be taken to mean the
part of the world that is described by the laws of classical physics full stop; for if stance 1 is correct, then  these laws are only approximately valid, if at all. Thus we simply put it like this: 
\begin{quote}{\it The classical world is what observation shows us to behave - with appropriate accuracy - according to the laws of classical physics}.
\end{quote}
There should be little room for doubt as to what `with appropriate accuracy' means: the existence of the colour grey does not imply the nonexistence of black and white!

We {\it can} define the {\it absolute existence of}  the classical world \`{a} la Bohr as its existence  independently of observers or measuring devices. Compare with Moore's (1939)  proof of the existence of the external world:
\begin{quote} 
How? By holding up my two hands, and saying, as I make a certain gesture with the right hand, `Here is one hand', and adding, as I make a certain gesture with the left, `and here is another'.\end{quote}

Those holding position 1, then, maintain that {\it the classical world exists only
as an appearance relative to a certain specification}, where the specification in question could be an observer (Heisenberg), a certain class of observers and states (as in decoherence theory), or some  coarse-graining of the Universe defined by a particular consistent set of histories, etc. If the notion of an observer  is construed in a sufficiently abstract and general sense, one might also 
formulate stance 1 as claiming that the classical world merely exists  {\it from the perspective of the observer} (or the corresponding class of observables).\footnote{The  terminology ``perspectival" was suggested to the author by Richard Healey.}  For example,
 Schr\"{o}dinger's cat ``paradox" dissolves at once when the appropriate perspective is introduced; cf.\ Subsection \ref{hepps}. 
 
Those holding stance 2, on the other hand,  believe that the classical world exists in an absolute sense (as Moore did). Thus stance 2 is akin to common-sense realism, though the distinction between 1 and 2 is largely independent of the issue of scientific realism.\footnote{See  Landsman (1995) for a more elaborate discussion of 
realism in this context.  Words like ``objective" or ``subjective" are not likely to be helpful in drawing the distinction either: the claim that `my children are the loveliest creatures in the world' is at first glance subjective, but it can trivially be turned into an objective one through the reformulation that `Klaas Landsman finds his  children the loveliest creatures in the world'. Similarly, the proposition that 
(perhaps due to decoherence)  `local observers find that the world is classical' is perfectly objective, although it describes a subjective experience. See also Davidson (2001). }
For  defendants of stance 1 usually still believe in the existence of some observer-independent reality (namely somewhere in the quantum realm), but deny that this reality incorporates the world observed around us.  This justifies a pretty vague specification of such an important notion as the classical world: one of the interesting outcomes of the otherwise futile discussions surrounding the Many Worlds Interpretation has been the insight that  {\it if \qm\ is fundamental, then the notion of a classical world is intrinsically vague and approximate}. Hence it would be self-defeating to be too precise at this point.\footnote{See Wallace (2002, 2003); also cf.\ Butterfield (2002). This point was not lost on Bohr and Heisenberg either; see Scheibe (1973).}

Although stance 1 is considered defensive if not cowardly by adherents of stance 2, it is a highly nontrivial mathematical fact that so far it seems supported by the formalism of \qm. In his derision  of what he called `FAPP' (= For All Practical Purposes) solutions to the 
measurement problem (and more general attempts to explain the appearance of the classical world from quantum theory), Bell (1987, 2001)  and others in his wake mistook a profound epistemological stance for a poor defensive move.\footnote{The insistence on ``precision" in such literature is reminiscent of Planck's long-held belief in the absolute nature of irreversibility (Darrigol, 1992; Heilbron, 2002). It should be mentioned that although Planck's stubbornness by historical accident led him to take the first steps towards quantum theory, he eventually gave it up to side with Boltzmann.} It is, in fact, stance 2 that we would recommend to the cowardly: for 
 proving  or disproving stance 1 seems the real challenge of the entire debate, and we regard the technical content of this paper as a survey of progress towards actually proving it. Indeed, to sum up our conclusions, we claim that there is good evidence that:
\begin{enumerate}
\item Classical physics emerges from quantum theory in the limit $\hbar\rightarrow 0$ or  $N\raw\infty$ {\it provided that the system is in certain ``classical" states and is monitored 
with ``classical" observables only};
\item Decoherence and consistent histories will probably  explain  {\it why} the system happens to be in such states and has to be observed in such a way.
\end{enumerate}
However, even if one fine day this scheme will be made to work, the explanation of the appearance of the classical world from quantum theory will be predicated on an external solution of the notorious `from ``and" to ``or" problem': If \qm\ predicts various possible outcomes with certain probabilities, why does only {\it one} of these appear to us?\footnote{It has to be acknowledged that we owe the insistence on this question to the defendants of stance 2.
See also footnote \ref{MWM}.}

For a more detailed outline of this paper we refer to the table of contents above. Most philosophical discussion will be found in Section \ref{S3} on the Copenhagen interpretation, since whatever its merits, it undeniably set the stage for the entire discussion on the relationship between classical and quantum.\footnote{We do not discuss the classical limit of \qm\ in the philosophical setting of theory reduction and intertheoretic relations;
see, e.g.,  Scheibe (1999) and Batterman (2002).}
 The remainder of the paper will be of an almost purely technical nature. Beyond this point we will  try to avoid controversy, but 
when unavoidable it will be  confined to the Epilogues appended to Sections \ref{S3}-\ref{S6}. 
The final Epilogue (Section \ref{S8}) expresses our deepest thoughts on the subject. 
\section{Early history}\label{S2}
This section is a recapitulation of the opinions and contributions of the founders of \qm\ regarding  the relationship between classical and quantum. More detail may be found in the books cited in footnote  \ref{historybooks} and in specific literature to be cited; for an impressive bibliography see also Gutzwiller (1998). 
The early history of quantum theory is of interest in its own right, concerned as it is with one of the most significant scientific revolutions in history. Although this history is not a main focus of this paper,  it is of special significance for our theme. For the usual and mistaken interpretation of Planck's work 
(i.e.\ the idea that he introduced something like a ``quantum postulate", see Subsection \ref{HC} below) 
appears to have triggered the   belief that quantum theory and Planck's constant are related to a universal discontinuity in Nature. Indeed, this discontinuity is sometimes even felt to  mark the basic difference between classical and quantum physics. This belief is particularly evident in the writings of Bohr, but still resonates even today. 
\subsection{Planck and Einstein}
The relationship between classical physics and quantum theory is so subtle and confusing that historians and physicists cannot even agree about the precise way the classical gave way to the quantum! 
 As Darrigol (2001) puts it: `During the past twenty years, historians [and physicists] have disagreed over the meaning of the quanta which Max Planck introduced in his black-body theory of 1900. The source of this confusion is the publication (\ldots) of Thomas Kuhn's [(1978)] iconoclastic thesis that Planck did not mean his energy quanta to express a quantum discontinuity.' 

As is well known (cf.\ Mehra \&\ Rechenberg, 1982a, etc.), Planck initially derived Wien's law for blackbody radiation in the context  of his (i.e.\ Planck's) program of establishing the absolute nature of irreversibility (competing with  Boltzmann's probabilistic approach, which eventually triumphed). When new high-precision measurements in October 1900 turned out to refute Wien's law, Planck first guessed his  expression 
\beq  E_{\nu}/N_{\nu}=h\nu/(e^{h\nu/kT}-1) \label{Planck}\eeq
for the correct law,  \textit{en passant} introducing two new constants of nature $h$ and $k$,\footnote{Hence Boltzmann's constant $k$ was introduced by Planck, who was the first to write down the formula $S= k\log W$.} and subsequently, on December 14, 1900, presented a theoretical derivation of his law in which he allegedly introduced the idea that the energy of the resonators making up his black body was quantized in units of $\varep_{\nu}=h\nu$ (where $\nu$ is the frequency of a given resonator). This derivation is generally seen as the birth of quantum theory, with the associated date of birth just mentioned. 

However, it is clear by now (Kuhn, 1978; Darrigol, 1992, 2001; Carson, 2000; Brush, 2002) that Planck was at best agnostic about the energy of his resonators, and at worst assigned them a continuous energy spectrum. Technically, in the particular derivation of his empirical law that eventually turned out to lead to the desired result (which relied on Boltzmann's concept of entropy),\footnote{Despite the fact that Planck only converted to Boltzmann's approach to irreversibility around 1914.} Planck had to count the number of ways a given amount of energy $E_{\nu}$ could be distributed over a given number of resonators $N_{\nu}$ at frequency $\nu$. This number is, of course, infinite, hence in order to find a finite answer Planck followed Boltzmann in breaking up $E_{\nu}$ into a large number $A_{\nu}$ of portions of identical size $\varep_{\nu}$, so that $A_{\nu}\varep_{\nu}=E_{\nu}$.\footnote{The number in question is then given by $(N+A-1)!/(N-1)!A!$, dropping the dependence on $\nu$ in the notation.}  Now, as we all know, whereas Boltzmann let
$\varep_{\nu}\raw 0$ at the end of his corresponding calculation for a gas, Planck discovered that his empirical blackbody law emerged if he assumed the relation $\varep_{\nu}=h\nu$.

 However, this postulate did \textit{not} imply that Planck quantized the energy of his resonators. In fact, in his definition of a given distribution he counted the number of resonators with energy \textit{between} say $(k-1)\varep_{\nu}$ and
$k\varep_{\nu}$ (for some $k\in\mathbb{N}$), as Boltzmann did in an analogous way for a gas, rather than the number of resonators \text{with} energy $k\varep_{\nu}$, as most physicists came to interpret his procedure. More generally, there is overwhelming textual evidence that Planck himself by no means believed or implied that he had quantized energy; for one thing, in his Nobel Prize Lecture in 1920 he attributed the correct interpretation of the energy-quanta $\varep_{\nu}$ to Einstein. 
Indeed, the modern understanding of the earliest phase of quantum theory is that it was Einstein rather than Planck who, during the period 1900--1905, clearly realized that Planck's radiation law marked a break with classical physics  (B\"{u}ttner, Renn,  \&\ Schemmel, 2003). This insight, then,  led Einstein to  the quantization of energy. This he did in a  twofold way,  both in connection with Planck's resonators - interpreted by Einstein as harmonic oscillators in the modern way - and, in a closely related move, through his concept of a photon.  Although Planck of course introduced the constant named after him, and as such is the founding {\it father} of the theory  characterized by $\hbar$, it is the introduction of the photon that  made Einstein at least the {\it mother} of quantum theory. Einstein himself may well  have regarded the photon as his
  most revolutionary discovery, for what he wrote about his  pertinent paper is not matched in self-confidence by anything he said about relativity:  
 `Sie handelt \"{u}ber die Strahlung und die energetischen Eigenschaften des Lichtes und ist sehr revolution\"{a}r.'\footnote{`[This paper] is about radiation and the energetic properties of light, and is very revolutionary.'  See also the Preface to Pais (1982).} 

Finally, in the light of the present paper, it deserves to be mentioned that Einstein (1905) and Planck (1906) were the first to comment on the classical limit of quantum theory; see the preamble to Section \ref{S5} below. 
\subsection{Bohr}\label{Bohr1}
 Bohr's brilliant model of the atom reinforced his idea that quantum theory was a theory of quanta.\footnote{Although at the time Bohr followed practically all physicists in their rejection of Einstein's photon, since he believed that during a quantum jump the atom emits  electromagnetic radiation in the form of a spherical wave. His model probably would have gained in consistency by adopting the photon picture of radiation, but in fact Bohr was to be the last prominent opponent of the photon, resisting the idea until 1925. See also Blair Bolles (2004) and footnote \ref{Bohropp} below.\label{BFN1}}
 Since this model simultaneously highlighted the clash between classical and quantum physics {\it and} carried  the germ of a resolution of this conflict through Bohr's equally brilliant correspondence principle, it is worth saying a few words about it here.\footnote{Cf.\ Darrigol (1992) for a detailed treatment; also see  Liboff  (1984) and Steiner (1998).} Bohr's atomic model addressed the radiative instability of Rutherford's solar-system-style atom:\footnote{The solar system provides the popular visualization of Rutherford's atom, but his own picture was more akin to Saturn' rings than to a planet orbiting the Sun.}  according to the electrodynamics of Lorentz, an accelerating electron should radiate, and since the envisaged circular or elliptical motion of an electron around the nucleus is a special case of an accelerated motion, the electron should continuously lose energy and spiral towards the nucleus.\footnote{In addition, any Rutherford style atom with more than one electron is mechanically unstable, since the electrons repel each other, as opposed to planets, which attract each other.}  Bohr countered this instability by three simultaneous moves, each of striking originality:
\begin{enumerate}
\item He introduced a quantization condition that singled out only a discrete number of allowed electronic orbits (which subsequently were to be described using classical mechanics, for example, in Bohr's calculation of the Rydberg constant $R$).
\item He replaced the emission of continuous radiation called for by Lorentz by quantum jumps with unpredictable destinations taking place at unpredictable moments, during which the atom emits light with energy equal to the energy difference of the orbits between which the electron jumps.
\item He prevented the collapse of the atom through such quantum jumps by introducing the notion of ground state, below which no electron could fall.
\end{enumerate}
With these postulates, for which at the time there existed no foundation whatsoever,\footnote{\label{SigalF}What has hitherto been mathematically proved of Bohr's atomic model  is the existence of a ground state (see Griesemer,  Lieb, \&\ Loss, 2001, and references therein for the greatest generality available to date) and the metastability of the excited states of the atom after coupling to the electromagnetic field (cf.\ Bach, Fr\"{o}hlich, \&\  Sigal, 1998,  1999 and Gustafson \&\ Sigal, 2003). The energy spectrum is discrete only if the radiation field is decoupled, leading to the usual computation of the
spectrum of the hydrogen atom first performed by Schr\"{o}dinger and Weyl. See also the end of Subsection \ref{PSL}.}
 Bohr explained the spectrum of the hydrogen atom, including an amazingly accurate calculation of  $R$.
Moreover, he proposed what was destined to be the key guiding principle in the search for quantum mechanics in the coming decade, viz.\ the correspondence principle (cf.\ Darrigol, 1992, {\it passim},  and Mehra \&\ Rechenberg, 1982a, pp.\ 249--257). 

In general, there is no relation between the energy that an electron loses during a particular quantum jump and the energy it would have radiated classically (i.e.\ according to Lorentz) in the orbit it revolves around preceding this jump. Indeed, in the ground state it cannot radiate through quantum jumps  at all, whereas according to classical electrodynamics it should radiate  all the time. However, Bohr saw that in the opposite case of very wide orbits (i.e.\ those having very large principal quantum numbers $n$),
the frequency $\nu=(E_n-E_{n-1})/h$ (with $E_n=-R/n^2$) of the emitted radiation  approximately corresponds to the frequency of the lowest harmonic of the classical theory, applied to electron motion in the initial orbit.\footnote{Similarly,  higher harmonics correspond to quantum jumps
$n\raw n-k$ for $k>1$.} Moreover, the measured intensity of the associated spectral line (which theoretically should be related to the probability of the quantum jump, a quantity out of the  reach of early quantum theory), similarly turned out to be given by classical electrodynamics.
This property, which in simple cases could be verified either by explicit computation or by experiment, became a guiding principle in situations where it could not be verified, and was sometimes even extended to low quantum numbers, especially when the classical theory predicted selection rules. 
  
It should be emphasized that {\it Bohr's correspondence principle was concerned with the properties of radiation, rather than with the mechanical orbits themselves}.\footnote{As such, it remains to be verified in a rigorous way.}
This  is not quite  the same as what is usually called the correspondence principle in the modern literature.\footnote{A typical example of the modern version is:  `Non-relativistic \qm\ was founded on the correspondence principle of Bohr: ``When the Planck constant $\hbar$ can be considered small with respect to the other parameters such as masses and distances, quantum theory approaches classical Newton theory."'  (Robert, 1998, p.\ 44). The reference to Bohr is historically inaccurate!} In fact, although also this modern correspondence principle has a certain range of validity (as we shall see in detail in Section \ref{S5}), Bohr never endorsed anything like that, and is even on record as opposing such a principle:\footnote{Quoted from Miller (1984), p.\ 313.}
\begin{quote}
`The place was Purcell's office where Purcell and others had taken Bohr for a few minutes of rest [during a visit to the Physics Department at Harvard University in 1961]. They were in the midst of a general discussion when Bohr commented: ``People say that classical mechanics is the limit of \qm\ when $h$ goes to zero." Then, Purcell recalled, Bohr shook his finger and walked to the blackboard on which he wrote $e^2/hc$. As he made three strokes under $h$, Bohr turned around and said, ``you see $h$ is in the denominator."'  
\end{quote} 
\subsection{Heisenberg}\label{heis}
Heisenberg's (1925) paper \textit{\"{U}ber die quantentheoretische Umdeutung kinematischer und mechanischer Beziehungen}\footnote{\textit{On the quantum theoretical reinterpretation of kinematical and mechanical relations}.  English translation in  vander Waerden, 1967.} is generally seen as a turning point in the development of \qm. Even A. Pais, no friend of Heisenberg's,\footnote{For example, in Pais (2000), claiming to portray the `genius of science', Heisenberg is conspicously absent.}  conceded that  Heisenberg's paper marked  'one of the great jumps - perhaps the greatest - in the development of twentieth century physics.' What did Heisenberg actually accomplish? This question is particularly interesting from the perspective of our theme.

At the time, atomic physics was in a state of crisis, to which various camps 
responded in  different ways. Bohr's approach might best be described as \textit{damage control}: his quantum theory was a hybrid of classical mechanics adjusted by means of \textit{ad hoc} quantization rules, whilst keeping  electrodynamics classical at all cost.\footnote{\label{Bohropp} 
Continuing footnote \ref{BFN1}, we quote from 
 Mehra \&\ Rechenberg, 1982a, pp 256--257: `Thus, in the early 1920s, Niels Bohr arrived at a definite point of view how to proceed forward in atomic theory. He wanted to make maximum use of what he called the ``more dualistic prescription" (\ldots) In it the atom was regarded as a mechanical system having discrete states and emitting radiation of discrete frequencies, determined (in a nonclassical way) by the energy differences between stationary states; radiation, on the other hand, had to be described by the classical electrodynamic theory.'}    Einstein, who had been  the first physicist to recognize the need to quantize classical electrodynamics, in the light of his triumph with General Relativity  nonetheless dreamt of a classical field theory with singular solutions as the ultimate explanation of quantum phenomena. Born led the radical camp, which included Pauli: he saw the need for an entirely new mechanics replacing classical mechanics,\footnote{It was Born who coined the name \textit{quantum mechanics} even before Heisenberg's paper.} which was to be based on discrete quantities satisfying difference equations.\footnote{This idea had earlier occurred to Kramers.}  This was a leap in the dark, especially because of Pauli's frowning upon the correspondence principle (Hendry, 1984; Beller, 1999). 

It was Heisenberg's genius to {\it interpolate} between Bohr and Born.\footnote{Also literally!  Heisenberg's traveled between Copenhagen and G\"{o}ttingen most of the time.} The meaning of his \textit{Umdeutung}  was to keep the classical equations of motion,\footnote{This crucial aspect of  \textit{Umdeutung} was appreciated at once by Dirac (1926): `In a recent paper Heisenberg puts forward a new theory which suggests that  it is not the equations of classical mechanics that are in any way at fault, but that the mathematical operations by which physical results are deduced from them require modification. (\ldots)  The correspondence between the quantum and classical theories lies not so much in the limiting agreement when $\hbar\raw 0$ as in the fact that the mathematical operations on the two theories obey in many cases the same laws.'}  whilst reinterpreting the mathematical symbols occurring therein as (what were later recognized to be) matrices. Thus his \textit{Umdeutung} $x\mapsto a(n,m)$ was a precursor of what now would be called a quantization map $f\mapsto Q_{\hbar}(f)$, where $f$ is a classical observable, i.e.\ a function on phase space, and $Q_{\hbar}(f)$ is a quantum mechanical observable, in the sense of an operator on a \Hs\ or, more abstractly, an element of some \ca. See Section \ref{S4} below. 
As Heisenberg recognized, this move implies the noncommutativity of the quantum mechanical observables; it is this, rather than something like a ``quantum postulate" (see Subsection \ref{HC} below), that is the defining characteristic of quantum mechanics. Indeed, most later work on quantum physics and practically all considerations on the connection between classical and quantum rely on Heisenberg's idea of \textit{Umdeutung}. This even applies to the mathematical formalism as a whole; see Subsection \ref{vNs}. 

We here use the term ``observable" in a loose way. It is now well recognized (Mehra \&\ Rechenberg, 1982b; Beller, 1999; Camilleri, 2005) that Heisenberg's claim that his formalism could be physically interpreted as the replacement of atomic orbits by observable quantities was a red herring, inspired by his discussions with Pauli.
 In fact, in quantum mechanics \textit{any} mechanical quantity has to be ``reinterpreted", whether or not it is observable. As Heisenberg (1969) recalls, Einstein reprimanded him for the illusion that physics admits an \textit{a priori} notion of an observable, and explained that a theory determines what can be observed. Rethinking the issue of observability then led Heisenberg to his second major contribution to \qm, namely his uncertainty relations. 

These relations were Heisenberg's own answer to the quote opening this paper.  Indeed, matrix mechanics was initially an extremely abstract and formal scheme, which lacked not only any visualization but also the concept of a state (see below). Although these features were initially quite to the liking of Born, Heisenberg, Pauli, and Jordan, the success of Schr\"{o}dinger's work forced them to renege on their radical stance, and look for a semiclassical picture supporting their mathematics; this was a considerable U-turn  (Beller, 1999; Camilleri, 2005).  Heisenberg (1927) found such a picture, claiming that his uncertainty relations provided the `intuitive content of the quantum theoretical kinematics and mechanics' (as his paper was called). His idea was that {\it the classical world emerged from \qm\ through observation}: `The trajectory only comes into existence because we observe it.' \footnote{`Die Bahn entsteht erst dadurch, da\ss\ wir sie beobachten.'}  This idea was to become extremely influential, and could be regarded as the origin of stance 1 in the Introduction. 
\subsection{Schr\"{o}dinger} \label{Ssection}
The history of \qm\ is considerably clarified by the insight that Heisenberg  and Schr\"{o}dinger  did not, as is generally believed, discover two equivalent formulations of the theory, but rather that Heisenberg (1925) identified the mathematical nature of the observables, whereas Schr\"{o}dinger (1926a) found the description of states.\footnote{See also Muller (1997).} Matrix mechanics lacked the notion of a state, but by the same token wave mechanics initially had no observables; it was only in his attempts to relate wave mechanics to matrix mechanics that Schr\"{o}dinger (1926c) introduced the position and momentum operators\footnote{Here $j=1,2,3$. In modern terms, the expressions on the right-hand side are unbounded operators on the \Hs\ $\H=L^2(\R^n)$. See Section \ref{S4} for more details. The expression $x^i$ is a multiplication operator, i.e.\ $(x^j\Psi)(x)=x^j\Psi(x)$, whereas, obviously, $(\partial/\partial x^j \Psi)(x)=
(\partial\Psi/\partial x^j)(x)$.}
\begin{eqnarray}
\qh(q^j) & =& x^j; \nn \\ \qh(p_j) & =& -i\hbar \frac{\partial}{\partial x^j}.\label{SOP}
\end{eqnarray}
This provided a new basis for Schr\"{o}dinger's equation\footnote{Or the corresponding time-independent one, with $E\Psi$ on the right-hand side.}
\beq \left(-\frac{\hbar^2}{2m}\sum_{j=1}^n \frac{\partial^2}{\partial x_j^2} +V(x)\right)\Psi=i\hbar \frac{\partial\Psi}{\partial t},\label{Schreq}\eeq
by interpreting the left-hand side as $H\Psi$, with $H=\qh(h)$ in terms of the classical Hamiltonian $h(p,q)=\sum_j p_j^2/2m +V(q)$. Thus  Schr\"{o}dinger founded the theory of the operators now named after him,\footnote{\label{SOPrefs} See Reed \& Simon (1972, 1975, 1987, 1979), Cycon et al. (1987), Hislop \& Sigal (1996),  Hunziker \&\ Sigal (2000), Simon (2000),  Gustafson \&\ Sigal (2003).
For the mathematical origin of the Schr\"{o}dinger equation also cf.\ Simon (1976).
} and in doing so gave what is still the most important example of Heisenberg's idea of {\it Umdeutung} of classical observables. 

Subsequently, correcting and expanding on certain  ideas of Dirac, Pauli, and  Schr\"{o}dinger, von Neumann (1932) brilliantly glued these two parts together through the concept of a \Hs. He also gave an abstract form of the formulae of Born, Pauli,  Dirac, and Jordan for the transition probabilities, thus completing the mathematical formulation of \qm. 

However, this is not how Schr\"{o}dinger saw his contribution. He intended wave mechanics as a full-fledged classical field theory of reality, rather than merely as one half (namely in modern parlance the state space half) of a probabilistic description of the world that still incorporated the quantum jumps he so detested (Mehra \&\ and Rechenberg,  1987; G\"{o}tsch, 1992; Bitbol \&\ Darrigol, 1992; Bitbol, 1996;  Beller, 1999). 
Particles were supposed to emerge in the form of wave packets, but it was immediately pointed out by Heisenberg, Lorentz,  and others that in realistic situations such wave packets tend to spread in the course of time. This had initially been overlooked by Schr\"{o}dinger (1926b), who had based his intuition on the special case of the harmonic oscillator. On the positive side, in the course of his unsuccessful attempts to derive classical particle mechanics from wave mechanics through the use of wave packets, Schr\"{o}dinger (1926b) gave the first example of what is now called a {\it coherent state}.
Here a quantum wave function $\Psi_z$ is labeled by a `classical' parameter $z$, in such a way that the quantum-mechanical time-evolution $\Psi_z(t)$ is approximately given by
$\Psi_{z(t)}$, where $z(t)$ stands for some associated classical time-evolution; see Subsections \ref{PSQ} and \ref{CEOM} below. This has turned out to be a very important idea in understanding the transition from quantum to classical mechanics. 

Furthermore, in the same paper Schr\"{o}dinger (1926b) proposed the following wave-mechanical version  of Bohr's correspondence principle: classical atomic states should come from superpositions of a very large number (say at least 10,000) of highly excited states (i.e.\ 
energy eigenfunctions with very large quantum numbers). After decades of limited theoretical interest in this idea, interest in wave packets in atomic physics was revived in the late 1980s  due to the development of modern experimental techniques based on lasers (such as pump-probing and phase-modulation). See  Robinett (2004) for a recent technical review, or  Nauenberg,  Stroud, \&\ Yeazell (1994) for an earlier popular account. Roughly speaking, the picture that has emerged is this: a localized wave packet of the said type initially follows a time-evolution with almost classical periodicity, as Schr\"{o}dinger hoped, but subsequently spreads out after a number of orbits. Consequently, during this second phase the probability distribution  approximately fills the classical orbit (though not uniformly). Even more surprisingly, on a much longer time scale there is a phenomenon of {\it wave packet revival}, in which the wave packet recovers its initial localization. Then the whole cycle starts once again, so that one does see periodic behaviour, but not of the expected classical type. Hence even in what naively would be thought of as the thoroughly  classical regime, wave phenomena continue to play a role,  leading to quite unusual and unexpected behaviour. Although a rigorous mathematical description of wave packet revival has not yet been forthcoming, the overall picture (based on both ``theoretical physics" style mathematics and experiments) is clear enough.

It is debatable (and irrelevant) whether the story of wave packets has evolved 
 according to Schr\"{o}dinger's intentions (cf.\ Littlejohn, 1986); what is certain is that his other main idea on the relationship between classical and quantum has been extremely influential. This was, of course,  Schr\"{o}dinger's (1926a) ``derivation" of his wave equation from the Hamilton--Jacobi formalism of classical mechanics. This  gave rise to the WKB approximation and related methods; see Subsection \ref{WKBS}. 

In any case, where Schr\"{o}dinger hoped for a classical interpretation of his wave function, and Heisenberg wanted to have nothing to do with it whatsoever (Beller, 1999), 
Born and Pauli were quick to realize its correct, probabilistic significance.
Thus they deprived the wave function of its naive physical nature, and effectively 
 degraded it to the purely mathematical status of a probability amplitude.
And in doing so, Born and Pauli rendered the connection between \qm\ and classical mechanics almost incomprehensible once again! It was this incomprehensibility that Heisenberg addressed with his uncertainty relations.  
 \subsection{von Neumann}\label{vNs}
Through its creation of the Hilbert space formalism of \qm,  von Neumann's book  (1932) can be seen as a mathematical implementation of Heisenberg's idea of {\it Umdeutung}. Von Neumann in effect proposed the following quantum-theoretical reinterpretations:
\begin{trivlist}
\item Phase space $M$ $\mathbf{\mapsto}$ Hilbert space $\H$;
\item  Classical observable (i.e.\ real-valued measurable function on $M$)  $\mathbf{\mapsto}$ self-adjoint operator on $\H$;
\item Pure state (seen as point in $M$)  $\mathbf{\mapsto}$ unit vector (actually ray)  in $\H$;
\item Mixed state (i.e.\ probability measure on $M$) $\mathbf{\mapsto}$  density matrix on $\H$;
\item Measurable subset of $M$  $\mathbf{\mapsto}$ closed linear subspace  of $\H$;
\item Set complement  $\mathbf{\mapsto}$ orthogonal complement;
\item Union of subsets  $\mathbf{\mapsto}$ closed linear span of subspaces;
\item Intersection of subsets $\mathbf{\mapsto}$ intersection of subspaces;
\item Yes-no question (i.e.\ characteristic function on $M$) $\mathbf{\mapsto}$ projection operator.\footnote{Later on, he of course added the {\it Umdeutung} of a Boolean lattice by a modular lattice, and  the ensuing {\it Umdeutung} of classical logic by quantum logic (Birkhoff \&\ von Neumann,  1936). }
 \end{trivlist}

Here we  assume for simplicity that 
 quantum observables $R$ on a \Hs\ $\H$ are bounded operators, i.e.\ $R\in\BH$. Von Neumann actually {\it derived} his {\it Umdeutung} of classical mixed states as density matrices from his axiomatic characterization of quantum-mechanical states as  linear maps $\Exp: \BH\raw \C$  that satisfy $\Exp(R)\geq 0$ when $R\geq 0$,\footnote{I.e., when $R$ is self-adjoint with positive spectrum, or, equivalently, when $R=S^*S$ for some $S\in \BH$.} $\Exp(1)=1$,\footnote{Where the $1$ in $\Exp(1)$ is the unit operator on $\H$.},  and countable additivity on a commuting set of operators. For he proved that such a map $\Exp$ is necessarily given by a density matrix $\rh$ according to $\Exp(R)=\Tr(\rh R)$.\footnote{This result has been widely misinterpreted (apparently also by von Neumann himself) as a theorem excluding hidden variables in \qm. See Scheibe (1991).  However, Bell's characterization of von Neumann's linearity assumption in the definition of a state as ``silly'' is far off the mark, since it holds  both in classical mechanics and in \qm.  Indeed, von Neumann's theorem {\it does} exclude all hidden variable extensions of \qm\ that are classical in nature, and it is precisely such extensions that many physicists were originally looking for. See R\'{e}dei \&\ St\"{o}ltzner (2001) and Scheibe (2001) 
for  recent discussions of this issue.} 
A unit vector $\Psi\in\H$ defines a pure state in the sense of von Neumann, which we call $\ps$, by $\ps(R)=(\Psi,R\Psi)$ for $R\in\BH$. Similarly, a density matrix $\rh$ on $\H$ defines a (generally mixed) state, called $\rh$ as well, by $\rh(R)=\Tr(\rh R)$.
In modern terminology, a state on $\BH$ as defined by von Neumann would be called a {\it normal} state.
 In the \ca ic formulation of quantum physics (cf.\ footnote  \ref{Cstarlit}), this axiomatization has been maintained until the present day; here $\BH$ is replaced by more general algebras of observables in order to accommodate possible superselection rules (Haag, 1992).  

Beyond his mathematical axiomatization of \qm, which (along with its subsequent extension by the \ca ic formulation)  lies at the basis of all serious efforts to relate classical and \qm, von Neumann contributed to this relationship through his  analysis of the measurement problem.\footnote{Von Neumann (1932) refrained from discussing either the classical limit of \qm\ or (probably) the notion of quantization.
In the latter direction, he declares that `If the quantity $\mathfrak{R}$ has the operator $R$, then the quantity $f(\mathfrak{R})$ has the operator $f(R)$', and that `If the quantities $\mathfrak{R}$, $\mathfrak{S}$, $\cdots$ have the operators $R$, $S$, $\cdots$, then the quantity $\mathfrak{R}+\mathfrak{S}+\cdots$ has the operator $R+S+\cdots$'.
However, despite his legendary clarity and precision, von Neumann is rather vague about the meaning of the transition $\mathfrak{R}\mapsto R$. It is tempting to construe
 $\mathfrak{R}$  as a classical observable whose quantum-mechanical counterpart is $R$, so that the above quotations might be taken as axioms for quantization. However, 
such an interpretation is neither supported by the surrounding text, nor by our current understanding of quantization (cf.\ Section \ref{S4}). For example, a quantization map $\mathfrak{R}\mapsto \qh(\mathfrak{R})$ cannot satisfy $f(\mathfrak{R})\mapsto f(\qh(\mathfrak{R}))$ even for very reasonable functions such as $f(x)=x^2$.}    Since here the apparent clash between classical and quantum physics comes to a head, it is worth summarizing von Neumann's analysis of this problem here. See also Wheeler \&\ Zurek (1983), Busch, Lahti \&\ Mittelstaedt (1991), Auletta (2001) and Mittelstaedt (2004) for general discussions of the measurement problem.

  The essence of the measurement problem is that certain states are never seen in nature, although they are not merely allowed by \qm\ (on the assumption of its universal validity), but are even predicted to arise in typical measurement situations. Consider a system $S$, whose pure states are mathematically described by normalized vectors (more precisely, rays) in a Hilbert space $\H_S$. One wants to measure an observable $\mathcal{O}$, which is mathematically represented by a self-adjoint operator $O$ on $\H_S$. Von Neumann assumes that $O$ has discrete spectrum,
a simplification which does not hide the basic issues in the measurement
problem. Hence $O$ has unit eigenvectors $\Psi_n$ with real eigenvalues $o_n$. 
To measure $\mathcal{O}$, one couples the system to an apparatus $A$ with
Hilbert space $\H_A$ and ``pointer" observable $\mathcal{P}$, represented
by a self-adjoint operator $P$ on $\H_A$, with discrete eigenvalues
$p_n$ and unit eigenvectors $\Phi_n$. The pure states of the total system $S+A$ then correspond to unit vectors in the tensor product $\H_S\otimes \H_A$.  A good (``first kind") measurement is then such that after the measurement, $\Psi_n$ is correlated to $\Phi_n$, that is, for a suitably chosen initial state $I \in{\H}_A$, a state
 $\Psi_n\ot I$ (at $t=0$) almost immediately evolves into $\Psi_n\ot\Phi_n$. This can indeed be achieved by a suitable Hamiltonian. 

The problem, highlighted by Schr\"{o}dinger's cat,  now arises if one selects the initial state of $S$ to be $\sum_n c_n \Psi_n$ (with $\sum |c_n|^2=1$), for then the superposition principle leads to the conclusion that the final state of the coupled system is $\sum_n c_n \Psi_n \ot\Phi_n$. 
 Now, basically all von Neumann said was that if one restricts the final state to the  system $S$, then the resulting density matrix is the mixture $\sum_n |c_n|^2 
 [\Psi_n]$ (where $[\Psi]$ is the orthogonal projection onto a unit vector $\Psi$),\footnote{I.e., $[\Psi]f=(\Psi,f)\Psi$; in Dirac notation one would have $[\Psi]=|\Psi\rangle\langle\Psi|$.} so that, {\it from the perspective of the system alone}, the measurement appears to have caused a transition from the pure state
 $\sum_{n,m} c_n\ovl{c_m} \Psi_n\Psi_m^*$
 to the mixed state 
 $\sum_n |c_n|^2 [\Psi_n]$, in which interference terms $\Psi_n\Psi_m^*$ for $n\neq m$
are absent. Here the operator $\Psi_n\Psi_m^*$ is defined by $\Psi_n\Psi_m^*f=(\Psi_m,f)\Psi_n$; in particular, $\Psi\Psi^*=[\Psi]$.\footnote{In
Dirac notation one would have  $\Psi_n\Psi_m^*=|\Psi_n\rangle\langle\Psi_m|$. }
 Similarly, the apparatus, taken by itself, has evolved from  the pure state $\sum_{n,m} c_n\ovl{c_m} \Phi_n\Phi^*_m$ to the mixed state 
 $\sum_n |c_n|^2  [\Phi_n]$. This is simply a mathematical theorem (granted the possibility of coupling the system to the apparatus in the desired way), rather than a proposal that there exist two different time-evolutions in Nature, viz.\ the unitary propagation according to the Schr\"{o}dinger equation side by side with the above ``collapse" process.

In any case, by itself this move by no means solves the measurement problem.\footnote{Not even  in an ensemble-interpretation of \qm, which was the interpretation von Neumann unfortunately adhered to when he wrote his book.}  
Firstly, in the given circumstances one is not allowed to adopt the ignorance interpretation of mixed states (i.e.\ assume that the system really is in one of the states $\Psi_n$); cf., e.g., Mittelstaedt (2004). Secondly, even if one were allowed to do so, one could restore the problem (i.e.\ the original superposition $\sum_n c_n \Psi_n\ot \Phi_n$) by once again taking the other component of the system into account. 

Von Neumann was well aware of at least this second point, to which he responded by his construction of a {\it chain}: one redefines $S+A$ as the system, and couples it to a new apparatus $B$, etc. This eventually leads to a post-measurement state $\sum_n c_n \Psi_n \ot\Phi_n\ot \ch_n$ (in hopefully self-explanatory notation, assuming the vectors $\ch_n$ form an orthonormal set), whose restriction to $S+A$ is the mixed
state  $\sum_n |c_n|^2 [\Psi_n]\ot [\Phi_n]$.
The restriction of the latter state to $S$ is, once again, $\sum_n |c_n|^2  [\Psi_n]$. This procedure may evidently be iterated; the point of the construction is evidently to pass on superpositions in some given system to arbitrary systems higher up in the chain.
It follows that for the final state of the original system it does not matter where one ``cuts the chain" (that is, which part of  the chain one leaves out of consideration), as long as it is done {\it somewhere}. Von Neumann (1932, in beautiful prose) and others suggested identifying the cutting with the act of observation, but it is preferable and much more general to  simply say that {\it some} end of the chain is omitted in the description. 

The burden of the  measurement problem, then, is to \begin{enumerate}
\item Construct a suitable chain along with an appropriate cut thereof;  it doesn't matter where the cut is made, as long as it is done.
\item Construct a suitable time-evolution accomplishing the measurement.
 \item Justify the  ignorance interpretation of mixed states. 
\end{enumerate}
As we shall see, these problems are addressed, in a conceptually different but mathematically analogous way, in the 
Copenhagen interpretation as well as in the decoherence approach.
(The main conceptual difference will be that the latter aims to solve also the more ambitious problem of explaining the appearance of the classical world, which in the former seems to be taken for granted). 
\bigskip

We conclude this section by saying that despite some brilliant ideas, the founders of \qm\ left  wide open the problem of deriving classical mechanics as a certain regime of their theory. 
 \section{Copenhagen: a reappraisal}\label{S3}\setcounter{equation}{0}
The  so-called ``Copenhagen interpretation" of \qm\ goes back to ideas first discussed and formulated by Bohr, Heisenberg, and Pauli around 1927. 
Against the idea that there has been a ``party line" from the very beginning, 
it has frequently been pointed out that in the late 1920s there were actually sharp differences of opinion between Bohr and Heisenberg on the interpretation of \qm\ and that they never really arrived at a joint doctrine  (Hooker, 1972; Stapp, 1972;  Hendry, 1984;   Beller, 1999;   Howard, 2004; Camilleri, 2005). For example, they never came to agree about the notion of complementarity (see Subsection \ref{compl}). More generally,  Heisenberg usually based  his ideas on the mathematical formalism of quantum theory, whereas Bohr's position was primarily philosophically oriented. Nonetheless, there is a clearly identifiable core of ideas on which they {\it did} agree, and since this core has everything to do with the relationship between classical and quantum, we are going to discuss it in some detail. 

The principal  primary sources are Bohr's Como Lecture, his reply to \epr, and his essay dedicated to Einstein (Bohr, 1927,  1935, 1949).\footnote{ 
These papers were actually written in collaboration with Pauli (after first attempts with Klein), Rosenfeld, and Pais, respectively.}  Historical discussions of the emergence and reception of these papers are given in Bohr (1985, 1996) and in Mehra \&\ Rechenberg (2001).  As a selection of the enormous literature these papers have given rise to, we mention 
among relatively recent works Hooker (1972), Scheibe (1973), Folse (1985), Murdoch (1987), Lahti \&\ Mittelstaedt (1987), Honner (1987), Chevalley (1991, 1999), Faye (1991), Faye \&\ Folse (1994), Held (1994),  Howard (1994),  Beller (1999), Faye (2002), and Saunders (2004).
For Bohr's sparring partners  see Heisenberg (1930, 1942, 1958, 1984a,b, 1985) with associated secondary literature (Heelan, 1965; H\"{o}rz, 1968; Geyer  
et al., 1993; Camilleri, 2005),  and Pauli (1933, 1949, 1979, 1985, 1994), along with Laurikainen (1988) and Enz (2002). 
  
As with Wittgenstein (and many other thinkers), it helps to understand Bohr if one makes a distinction between an ``early" Bohr and a ``later" Bohr.\footnote{Here we side with Held (1994) and Beller (1999) against Howard (1994) and Suanders (2004). See also Pais (2000), p.\ 22: `Bohr's Como Lecture did not bring the house down, however. He himself would later frown on expressions he used there, such as ``disturbing the phenomena by observation". Such language may have contributed to the considerable confusion that for so long has reigned around this subject.'\label{paisnote}}  Despite a good deal of continuity in his thought (see below), the demarcation point is his response  to \epr\  (Bohr, 1935),\footnote{This response is problematic, as is \epr\  itself. Consequently, there exists a considerable exegetical literature on both, marked by the fact that equally competent and well-informed pairs of commentators manage to flatly contradict each other while at the same time both claiming to explain or reconstruct what Bohr ``really" meant. }  and the main shift he  made afterwards lies in his sharp insistence on the indivisible unity of object and observer after 1935, focusing on the concept of a {\it phenomenon}.
  Before \epr, Bohr equally well believed that object and observer were both necessary ingredients of a complete description of quantum theory, but he then thought that although their interaction could never be neglected,  they might at least logically be considered separately. After 1935, Bohr gradually began to claim that object and observer no longer even had  separate identities, together forming a ``phenomenon". Accordingly, also his notion of complementarity changed, increasingly focusing on the idea that the specification of the experimental conditions is crucial for the unambiguous use of (necessarily) classical concepts in quantum theory (Scheibe, 1973; Held, 1994). See also Subsection \ref{compl} below. 
This development culminated in Bohr's eventual denial of the existence of the quantum world: 
\begin{quote}
`There is no quantum world. There is only an abstract quantum-physical description. It is wrong to think that the task of physics is to find out how nature is. Physics concerns what we can say about nature. (\ldots) What is it that we humans depend on? We depend on our words. Our task is to communicate experience and ideas to others. We are suspended in language.' (quoted by Petersen (1963), p.\ 8.)\footnote{See Mermin (2004) 
for a witty discussion of this controversial  quotation.}  \end{quote}
\subsection{The doctrine of classical concepts}\label{Pcl}
Despite this shift, it seems that Bohr stuck to one key thought throughout his  career:
\begin{quote}
 `However far the phenomena transcend the scope of classical physical explanation, the account of all evidence must be expressed in classical terms. (\ldots) The argument is simply that by the word {\it experiment} we refer to a situation where we can tell others what we have done and what we have learned and that, therefore, the account of the experimental arrangements and of the results of the observations must be expressed in unambiguous language with suitable application of the terminology of classical physics.' (Bohr, 1949, p.\ 209).
\end{quote}

This is, in a nutshell, Bohr's {\it doctrine of classical concepts}. 
Although his many drawings and stories may suggest otherwise, Bohr does not quite express the idea here that the goal of physics lies in the description of experiments.\footnote{Which often but misleadingly has been contrasted with
Einstein's  belief that the  goal of physics is rather to describe reality.
See Landsman (2006) for a recent discussion. } In fact, he merely points out the need for ``unambiguous"  communication, which he evidently felt threatened by  \qm.\footnote{Here  ``unambiguous" means ``objective" (Scheibe, 1973; Chevalley, 1991).}
The controversial part of the quote lies in his identification of the means of unambiguous communication with the language of classical physics, involving particles and waves and the like. We will study Bohr's specific argument in favour of this identification shortly, but it has to be said that, like practically all his foundational remarks on quantum mechanics, Bohr presents his reasoning as self-evident, necessary, and not in need of any further analysis (Scheibe, 1973; Beller, 1999).  Nonetheless, young Heisenberg clashed with Bohr on precisely this point, for  Heisenberg felt that the abstract mathematical formalism of quantum theory (rather than Bohr's world of words and pictures) provided those means of unambiguous communication.\footnote{It is hard to disagree with Beller's (1999) conclusion that Bohr was simply not capable of understanding the formalism of post-1925 quantum mechanics, turning his own  need of understanding this theory in terms of words and pictures into a deep philosophical necessity.}

 By classical physics Bohr undoubtedly meant the theories of Newton, Maxwell, and Lorentz, but that is not the main point.\footnote{Otherwise, one should wonder why one shouldn't use the physics of Aristotle and the scholastics for this purpose, which is a much more effective way of communicating  our naive impressions of the world. In contrast, the essence of physics since Newton has been to unmask a reality behind the phenomena. Indeed, Newton himself empasized that his physics was intended for those capable of natural philosophy, in contrast to  \textit{ye vulgar} who believed naive appearances.   The fact that Aristotle's physics is now known to be wrong should not suffice to disqualify its use for Bohr's purposes, since the very same comment may be made about the physics of Newton etc.} For Bohr, the \textit{defining} property of classical physics was the property that it was \textit{objective}, i.e.\ that it could be studied in an observer-independent way:
 \begin{quote}
`All description of experiences so far has been based on the assumption, already inherent in ordinary conventions of language, that it is possible to distinguish sharply between the behaviour of objects and the means of observation. This assumption is not only fully justified by everyday experience, {\it but even constitutes the whole basis of classical physics}'
(Bohr, 1958, p.\ 25; italics added).\footnote{Despite the typical imperative tone of this quotation, Bohr often regarded certain other properties as essential to classical physics, such as  determinism, the combined use of space-time concepts and dynamical conservation laws, and the possibility of pictorial descriptions. However, these properties were in some sense secondary,  as Bohr considered them to be  {\it consequences} of the possibility of isolating an object in classical physics. For example: `The assumption underlying the ideal of causality [is] that the behaviour of the object is uniquely determined, quite independently of whether it is observed or not' (Bohr, 1937), and then again, now negatively: `the renunciation of the ideal of causality [in \qm] is founded logically only on our not being any longer in a position to speak of the autonomous behaviour of a physical object' (Bohr, 1937).
See Scheibe (1973).} \end{quote}
 See also Hooker (1972), Scheibe (1973) and Howard (1994). Heisenberg (1958, p.\ 55)
shared this view:\footnote{As Camilleri (2005, p.\ 161) states: `For Heisenberg,
classical physics is the fullest expression of the ideal of objectivity.'}
 \begin{quote}
 `In classical physics science started from the belief - or should one say from the illusion? - that we could describe the world or at least part of the world without any reference to ourselves. This is actually possible to a large extent. We know that the city of London exists whether we see it or not. It may be said that classical physics is just that idealization in which we can speak about parts of the world without any reference to ourselves. Its success has led to the general idea of an objective description of the world.' \end{quote}
 
 On the basis of his ``quantum postulate" (see Subsection \ref{HC}), Bohr came to believe that, similarly, the 
 \textit{defining} property of quantum physics was precisely the opposite, i.e.\ the necessity of the role of the observer (or apparatus - Bohr did not distinguish between the two and  never assigned a special role to the mind of the observer or endorsed a subjective view of physics). Identifying unambiguous communication with an objective description, in turn claimed to be the essence of classical physics, 
 Bohr concluded that despite itself quantum physics had to be described entirely in terms of classical physics. Thus his doctrine of classical concepts
has an epistemological origin, arising from an analysis of the conditions for human knowledge.\footnote{See, for example, the very {\it title} of Bohr (1958)!} In that sense it may be said to be Kantian in spirit (Hooker, 1972; Murdoch, 1987; Chevalley, 1991, 1999). 

Now,  Bohr himself is on record as saying: `They do it smartly, but what counts is to do it right' (Rosenfeld, p.\ 129).\footnote{`They' refers to \epr.}   
The doctrine of classical concepts is certainly smart, but is it right?
As we have seen, Bohr's argument starts from the claim that  classical physics is objective (or `unambiguous') in being independent of the observer. In fact, nowadays it is widely believed that \qm\ leads to the {\it opposite} conclusion that ``quantum reality" (whatever that may be) is objective (though ``veiled" in the terminology of dÕEspagnat (1995)), while ``classical reality" only comes into existence relative to a certain specification: this is stance 1 discussed in the Introduction.\footnote{ Indeed,  interesting recent attempts to make  Bohr's philosophy of \qm\ precise accommodate the a priori status of classical observables into some version of the modal interpretation; see Dieks (1989b), Bub (1999),  Halvorson \&\ Clifton (1999, 2002), and Dickson (2005). It should give one some confidence in the possibility of world peace that the two most hostile interpretations of \qm, viz.\ Copenhagen and Bohm (Cushing, 1994) have now found a common home in the modal interpretation in the sense of the authors just cited! Whether or not one agrees with Bub's (2004) criticism of the modal interpretation, Bohr's insistence on the necessity of classical concepts is not vindicated by any current version  of it.}  
 Those who disagree with stance 1 cannot use stance 2 (of denying the fundamental nature of quantum theory) at this point either, as that is certainly not what Bohr had in mind.
 Unfortunately, in his most outspoken defence of Bohr, even  Heisenberg (1958, p.\ 55) was unable to find a better argument for Bohr's doctrine than the lame remark that `the use of classical concepts is finally a consequence of the general human way of thinking.'\footnote{And similarly: `We are forced to use the language of classical physics, simply because we have no other language in which to express the results.' (Heisenberg, 1971, p.\ 130). This in spite of the fact that the later Heisenberg thought about this matter very deeply; see, e.g., his (1942), as well as Camilleri (2005). Murdoch (1987, pp.\ 207--210) desperately tries to boost the doctrine of classical concepts into a profound philosophical argument by appealing to Strawson (1959).}

In our opinion, Bohr's {\it motivation} for his doctrine has to be revised in the light of our current understanding of quantum theory; we will do so in Subsection \ref{primas}. In any case, whatever its motivation, 
the doctrine {\it itself} seems worth keeping: apart from the fact that it evidently describes experimental practice, it provides a convincing explanation for the probabilistic nature of \qm\ (cf.\ the next subsection). 
  \subsection{Object and apparatus: the Heisenberg cut}\label{HC}
Describing quantum physics in terms of classical concepts 
 sounds like an impossible and even self-contradictory task (cf.\ Heisenberg, 1958).  For one, it precludes a completely quantum-mechanical description of the world: `However far the phenomena transcend the scope of classical physical explanation, the account of all evidence must be expressed in classical terms.' But at the same time it precludes a purely classical description of the world, for underneath classical physics one has quantum theory.\footnote{This peculiar situation makes it very hard to give a realist account of the Copenhagen interpretation, since quantum reality is denied whereas classical reality is neither fundamental nor real.} The fascination of Bohr's philosophy of \qm\ lies precisely in his brilliant resolution of this apparently paradoxical situation.  
 
 The first step of this  resolution that he and Heisenberg proposed is to divide the system whose description is sought into two parts: one, the object, is to be described quantum-mechanically, whereas the other, the apparatus, is treated \textit{as if it were classical}. 
Despite innumerable claims to the contrary in the literature  (i.e.\ to the effect that Bohr held that a separate realm of Nature was intrinsically classical), there is no doubt that
both Bohr and Heisenberg believed in the fundamental and universal nature of \qm, and 
saw the classical description of the apparatus as a {\it purely epistemological move without any counterpart in ontology},
expressing the fact that a given {\it quantum} system is {\it being used} as a measuring device.\footnote{See especially Scheibe (1973) on Bohr, and Heisenberg (1958). The point in question has also been made by R. Haag (who knew both Bohr and Heisenberg) in most of his talks on \qm\ in the 1990s.
In this respect we disagree with Howard (1994), who claims that according to Bohr a classical description of an apparatus amounts to picking a particular (maximally) abelian subalgebra of its quantum-mechanical algebra of `beables', which choice is dictated by the measurement context. But having a commutative algebra falls far short of a classical description, since in typical examples one obtains only half of the canonical classical degrees of freedom in this way. Finding a classical description of a 
 quantum-mechanical system is a much deeper problem, to which we shall return throughout this paper.} For example: `The construction and the functioning of all apparatus like diaphragms and shutters, serving to define geometry and timing of the experimental arrangements, or photographic plates used for recording the localization of atomic objects, will depend on properties of materials which are themselves essentially determined by the quantum of action' (Bohr, 1948), as well as: 
`We are free to make the cut only within a region where the quantum mechanical description of the process concerned is effectively equivalent with the classical description'
(Bohr, 1935).\footnote{This last point suggests that the cut has something to do with the division between a microscopic and a macroscopic realm in Nature, 
but although this division often facilitates making the cut when it is well defined, this is by no means a matter of principle. Cf.\ Howard (1994).
In particular, all objections to the Copenhagen interpretation to the effect
that the interpretation is ill-defined because the micro-macro distinction is blurred are unfounded. }

The separation between object and apparatus called for here is
usually called the \textit{Heisenberg cut}, and it  plays an absolutely central role in the Copenhagen interpretation of \qm.\footnote{Pauli (1949) went  as far as saying that the Heisenberg cut provides the appropriate generalization modern physics offers of the old Kantian opposition between a knowable object and a knowing subject: 'Auf diese Weise verallgemeinert die moderne Physik die alte Gegen\"{u}berstellung von erkennenden Subjekt auf der einen Seite und des erkannten Objektes auf der anderen Seite zu der Idee des Schnittes zwischen Beobachter oder Beobachtungsmittel und dem beobachten System.'
(`In this way, modern physics generalizes the old opposition between the knowing subject on the one hand and the known object on the other to the idea
of the cut between observer or means of observation and the observed system.')  He then continued  calling the cut a necessary condition for human knowledge: see footnote \ref{PFN2}. } 
The idea, then, is that {\it a quantum-mechanical object is studied exclusively through its influence on an apparatus that is described classically}. Although {\it described} classically, the apparatus {\it is} a quantum system, and is supposed to be  influenced by its {\it quantum-mechanical} coupling to the underlying (quantum) object.

The alleged necessity of including both object and apparatus in the description was initially claimed to be a consequence of the so-called ``quantum postulate". This notion played a key role in Bohr's thinking: his  Como Lecture (Bohr, 1927) was even entitled `The quantum postulate and the recent development of atomic theory'.  There he stated
its contents as follows:  `The essence of quantum theory is the quantum postulate: every atomic process has an essential discreteness -  completely foreign to classical theories - characterized by PlanckÕs quantum of action.'\footnote{Instead of `discreteness', Bohr alternatively used the words  `discontinuity' or `individuality' as well. He rarely omitted amplifications like `essential'.} Even more emphatically, in his reply to \epr\ (Bohr, 1935):
`Indeed the finite interaction between object and measuring agencies conditioned by the very existence of the  quantum of action entails - because of the impossibility of controlling the reaction of the object on the measurement instruments if these are to serve their purpose - the necessity of a final renunication of the classical ideal of causality and a radical revision of our attitude towards the problem of physical reality.'
  Also, Heisenberg's uncertainty relations were originally motivated by the quantum postulate in the above form.
According to Bohr and Heisenberg around 1927, this `essential discreteness' causes an 
`uncontrollable disturbance' of the object by the apparatus during their interaction.
Although the ``quantum postulate" is not supported by the mature mathematical formalism of \qm\ and is basically obsolete, the intuition of Bohr and Heisenberg that a measurement of a quantum-mechanical object causes an `uncontrollable disturbance' of the latter is actually quite right.\footnote{Despite the fact that  Bohr later distanced himself from it; cf.\ Beller (1999) and footnote  \ref{paisnote} above. In a correct analysis, what is disturbed upon coupling to a classical apparatus is the quantum-mechanical state of the object (rather than certain sharp values of  classical observables such as position and momentum, as the early writings of Bohr and Heisenberg suggest). } 

In actual fact, the reason for this disturbance does not lie in the    ``quantum postulate",  but in the phenomenon of entanglement, as further discussed in Subsection  \ref{primas}.
Namely, from the point of view of von Neumann's measurement theory (see Subsection \ref{vNs})
 the Heisenberg cut is just a two-step example of a von Neumann chain, with the special feature that after the quantum-mechanical interaction has taken place, the second link (i.e.\ the apparatus) is {\it described} classically.
The latter feature not only supports  Bohr's philosophical agenda, but, more importantly,
also suffices to guarantee the applicability of the ignorance interpretation of the mixed state that arises after completion of the measurement.\footnote{In a purely quantum-mechanical von Neumann chain the final state of system plus apparatus is pure, but if the
apparatus  is classical, then  the post-measurement state is mixed.} 
 All of von Neumann's analysis of the arbitrariness of the location of the cut applies here, for  one may always extend the definition of the quantum-mechanical object by coupling the original choice to any other purely quantum-mechanical system one likes, and analogously for the classical part. Thus the two-step nature of the Heisenberg cut includes the possibility that the first link or object is in fact a lengthy chain in itself (as long as it is quantum-mechanical), and similarly for the second link (as long as it is classical).\footnote{\label{PFN2} Pauli (1949) once more: 'W\"{a}hrend die {\sc Existenz} eines solchen Schnittes eine notwendige Bedingung menschlicher Erkenntnis ist, fa\ss t sie die {\sc Lage} des Schnittes als bis zu einem gewissen Grade willk\"{u}rlich und als Resultat einer durch Zweckm\"{a}\ss igkeitserw\"{a}gungen mitbestimmten, also teilweise freien Wahl auf.'
(`While the {\sc existence} of such a [Heisenberg] cut is a necessary condition for human knowledge,
its {\sc location} is to some extent arbitrary as a result of a pragmatic and thereby partly 
free choice.')} This arbitrariness, subject to the limitation expressed by the second (1935) Bohr quote in this subsection, was well recognized by Bohr and Heisenberg, and was found at least by Bohr to be of great philosophical importance. 

It is the interaction between object and apparatus that causes  the measurement to `disturb' the former, but it is only and precisely the classical description of the latter that (through the ignorance interpretation of the final state) makes the disturbance `uncontrollable'.\footnote{These points were not clearly separated by Heisenberg (1927) in his paper on the uncertainty relations, but were later clarified by Bohr. See Scheibe (1973).}  
In the Copenhagen interpretation, {\it probabilities arise solely  because we look at the quantum world through classical glasses}.
\begin{quote}
`Just the necessity of accounting for the function of the measuring agencies on classical lines excludes in principle in proper quantum phenomena an accurate control of the reaction of the measuring instruments on the atomic objects.' 
(Bohr, 1956, p.\ 87)
\end{quote}
\begin{quote} `One may call these uncertainties objective, in that they are simply a consequence of the fact that we describe the experiment in terms of classical physics; they do not depend in detail on the observer. One may call them subjective, in that they reflect our incomplete knowledge of the world.'
(Heisenberg, 1958, pp.\ 53--54)
\end{quote}

Thus the picture that arises is this: Although
the quantum-mechanical side of the Heisenberg cut is described by the Schr\"{o}dinger equation (which is deterministic), while the classical side is subject to Newton's laws (which are equally well deterministic),\footnote{But see Earman (1986, 2005).} unpredictability arises because the quantum system serving as an apparatus is approximated by a classical system. 
 The ensuing  probabilities reflect the ignorance arising from the decision (or need) to ignore the  quantum-mechanical degrees of freedom of the apparatus. Hence the probabilistic nature of quantum theory is not intrinsic but extrinsic, and as such is entirely a consequence of 
 the doctrine of classical concepts, which by the same token {\it explains} this nature.

Mathematically, the simplest illustration of this idea is as follows. Take a finite-dimensional Hilbert space $\H=\C^n$ with the ensuing algebra of observables $\CA=M_n(\C)$ (i.e.\ the $n\times n$ matrices). A unit vector $\Ps\in\C^n$ determines a quantum-mechanical state in the usual way. Now describe this quantum system as if it were classical by ignoring all observables except the diagonal matrices. The state then immediately collapses to a probability measure on the set of $n$ points, with probabilities given by the Born rule
$p(i)=|(e_i,\Ps)|^2$, where $(e_i)_{i=1,\ldots,n}$ is the standard basis of $\C^n$. 

Despite the appeal of this entire picture, it is not at all clear that it actually applies! There is no a priori guarantee whatsoever that one may indeed describe a quantum system ``as if it were classical".  Bohr and Heisenberg apparently took this possibility for granted, probably on empirical grounds, blind to  the extremely delicate theoretical nature of their assumption. It is equally astounding that they never reflected in print on the question if and how  the classical worlds of mountains and creeks they loved so much emerges from  a quantum-mechanical world. In our opinion, the main difficulty in making sense of the Copenhagen interpretation is therefore not of a philosophical nature, 
but is a mathematical one. This difficulty is the main topic of this paper, of which Section \ref{S6} is of particular relevance in the present context.
\subsection{Complementarity}\label{compl}
The notion of a Heisenberg cut is subject to a certain arbitrariness even apart from the precise location of the cut within a given chain, for one might in principle construct the chain in various different and incompatible ways. This arbitrariness was analyzed by Bohr in terms of what he called \textit{complementarity}.\footnote{Unfortunately and typically, Bohr 
once again presented complementarity as a necessity of thought rather than as the truly amazing  possible mode of description it really is.} 

Bohr never gave a precise definition of complementarity,\footnote{Perhaps he preferred this approach because he felt a definition could only reveal part of what was supposed to be defined: one of his favourite examples of complementarity was that between definition and observation.} but restricted himself to the analysis of a number of examples.\footnote{We refrain from discussing the complementarity between truth and clarity, science and religion, thoughts and feelings, and objectivity and introspection here, despite the fact that on this basis Bohr's biographer Pais (1997) came to regard his subject as the greatest philosopher since Kant.}
A prominent such example is the complementarity between
a ``causal" \footnote{\label{caudet} Bohr's use the word ``causal"  is quite confusing in view of the fact that in the British empiricist tradition causality is often interpreted in the sense of a space-time description. But Bohr's ``causal" is meant to be {\it complementary} to a space-time description!}  description of a quantum system in which conservation laws hold, and a space-time description that is necessarily statistical in character. Here Bohr's  idea seems to have been that a stationary state (i.e.\ an energy eigenstate)  of an atom is incompatible with an electron moving in its orbit in space and time - see
Subsection \ref{PSL} for a discussion of this issue. Heisenberg (1958), however, took this example of complementarity to mean that 
a system on which no measurement is performed evolves 
deterministically according to the Schr\"{o}dinger equation, whereas  a rapid succession of measurements produces a space-time path whose precise form quantum theory is only able to predict statistically (Camilleri, 2005). In other words, this example reproduces precisely the picture through  which Heisenberg (1927) believed he had established the connection between classical and quantum mechanics; cf.\ Subsection \ref{heis}. 

Bohr's other key example was the complementarity between particles and waves. Here his principal  aim was to make sense of Young's double-slit experiment. The well-known difficulty with a classical visualization of this experiment is that a particle description appears  impossible because a particle has to go through a single slit, ruining the interference pattern gradually built up on the detection screen, whereas a wave description seems incompatible with the point-like localization on the screen once the wave hits it. Thus Bohr suggested that whilst each  of these classical descriptions is incomplete, the union of them is necessary for a complete description of the experiment.

 The deeper epistemological point appears to be that although the {\it completeness} of the quantum-mechanical description of  the microworld systems seems to be endangered by the doctrine of classical concepts, it is actually restored by the inclusion of {\it two} ``complementary" descriptions (i.e.\ of a given quantum system plus a measuring device that is necessarily described classicaly, `if it is to serve its purpose'). Unfortunately, despite this attractive general idea it is unclear to what precise definition of complementarity Bohr's examples should lead. In the first, the complementary notions of determinism and a space-time description are in mutual harmony as far as classical physics is concerned, but are apparently in conflict with each other in \qm. In the second, however, the wave description of some entity contradicts a particle description of the same entity precisely in classical physics, whereas in \qm\ these descriptions somehow coexist.\footnote{On top of this, Bohr mixed these examples in conflicting ways. In discussing bound states of electrons in an atom he jointly made determinism and particles one half of a complementary pair, waves and space-time being the other. In his description of electron-photon scattering he did it the other way round: this time determinism and waves formed one side, particles and space-time the other (cf.\ Beller, 1999).}

Scheibe (1973, p.\ 32) notes a `clear convergence [in the writings of Bohr] towards a preferred expression of a complementarity between phenomena', where a Bohrian {\it phenomenon} is an indivisible union (or ``whole") of a quantum system and a classically described experimental arrangement used to study it; see item \ref{item2list} below. Some of Bohr's early examples of 
complementarity can be brought under this heading, others  cannot (Held, 1994). For many students of Bohr (including the present author), the fog has yet to clear up.\footnote{Even Einstein (1949, p.\ 674) conceded that throughout his debate with Bohr he had never understood the notion of complementarity, `the sharp formulation of which, moreover, 
 I have been unable to achieve despite much effort which I have expended on it.' See Landsman (2006) for the author's view on the Bohr--Einstein debate. }
 Nonetheless, the following mathematical interpretations might assign some meaning to the idea of complementarity in the framework of von Neumann's formalism of \qm.\footnote{This exercise is quite against the spirit of Bohr, who is on record as saying that `von Neumann's approach (\ldots) did not {\it solve} problems but created {\it imaginary difficulties} (Scheibe, 1973, p.\ 11, quoting Feyerabend; italics in original).} 
\begin{enumerate}
\item Heisenberg (1958) {\it  identified complementary pictures  of
a quantum-mechanical system with equivalent mathematical representations thereof}. For example, he thought of the complementarity of $x$ and $p$ as the
existence of what we now call the Schr\"{o}dinger representations of the canonical commutation relations (CCR) 
on $L^2(\R^n)$ and its Fourier transform to momentum space. 
Furthermore, he felt that in quantum field theory particles and waves gave two {\it equivalent} modes of description of quantum theory because of second quantization. 
Thus for Heisenberg complementary pictures are classical because there is an underlying classical variable, with no apparatus in sight, and such pictures are not mutually contradictory   but (unitarily) equivalent. See also Camilleri (2005, p.\ 88), according to whom `Heisenberg never accepted Bohr's complementarity arguments'. 
\item \label{item2list} Pauli (1933) simply stated that {\it two observables are complementary when the corresponding operators fail to commute}.\footnote{ More precisely, one should probably require that the two operators in question generate the ambient algebra of observables, so that complementarity in Pauli's sense is really defined between two commutative subalgebras
of a given algebra of observables (again, provided they jointly generate the latter).}
Consequently, it then follows from Heisenberg's uncertainty relations that complementary observables cannot be measured simultaneously with arbitrary precision. This suggests (but by no means proves) that they should be measured independently, using mutually exclusive experimental arrangements. The latter feature of complementarity was emphasized by Bohr in his later writings.\footnote{We follow Held (1994) and others. Bohr's  earlier writings do not quite conform to Pauli's approach. In  Bohr's discussions of the double-slit experiment particle and wave form a complementary pair, whereas Pauli's complementary observables are position and momentum, which refer to a single side of Bohr's  pair.} This approach makes the notion 
of  complementarity  unambiguous and mathematically precise, and perhaps for this reason
the few physicists who actually use the idea of complementarity in their work tend to follow Pauli and the later Bohr.\footnote{Adopting this point of view, it is tempting to capture the complementarity between position and momentum by means of the following conjecture: \textit{Any normal pure state $\om$ on $\CB(L^2(\R^n))$} (that is, any wave function seen as a state in the sense of \ca s) {\it is determined by the pair $\{\om| L^{\infty}(\R^n), \om| FL^{\infty}(\R^n)F\inv\}$} (in other words, by its restrictions to position and momentum).
Here $L^{\infty}(\R^n)$ is the \vna\ of multiplication operators on $L^2(\R^n)$, i.e.\ 
the \vna\ generated by the position operator, whereas $FL^{\infty}(\R^n)F\inv$ is its Fourier transform, i.e.\  the \vna\ generated by the momentum operator. The idea is that each of its restrictions $\om| L^{\infty}(\R^n)$ and $\om| FL^{\infty}(\R^n)F\inv$ gives a classical picture of $\om$. These restrictions are a measure on $\R^n$ interpreted as position space, and another measure on $\R^n$ interpreted as momentum space. 
Unfortunately, this conjecture is false. The following counterexample was provided by D. Buchholz (private communication): take $\om$ as the state defined by the wave function 
$\Psi(x) \sim \exp(- a x^2 / 2)$ with $\Re (a) > 0$, $\Im(a)\neq 0$, and $|a|^2=1$. Then
$\om$ depends on $\Im(a)$, whereas neither $\om| L^{\infty}(\R^n)$ nor $\om| FL^{\infty}(\R^n)F\inv$ does. 
 There is even a counterexample to the analogous conjecture for the \ca\ of $2\x 2$ matrices, found by H. Halvorson: 
if $A$ is the commutative \ca\ generated by $\sg_x$, and $B$ the one generated by $\sg_y$,
then the two different eigenstates of $\sg_z$ coincide on $A$ and on $B$. One way to improve our conjecture  might be to hope that if, in the Schr\"{o}dinger picture, two states coincide on the two given commutative \vna s for all times, then they must be equal.
But this can only be true for certain ``realistic" time-evolutions, for the trivial Hamiltonian $H=0$ yields the above counterexample. We leave this as a problem for future research. At the time of writing, Halvorson (2004) contains the only sound mathematical interpretation of the complementarity between position and momentum, by relating it to the representation theory of the CCR. He shows that  in any representation where the position operator has eigenstates, there is no momentum operator, and vice versa.}
\item The present author proposes that {\it observables and pure states are complementary}. For in the Schr\"{o}dinger representation of elementary \qm, the former are, roughly speaking,  generated by the position and momentum operators, whereas the latter are given by wave functions. 
Some of Bohr's other examples of complementarity also square with this interpretation (at least if one overlooks the collapse of the wavefunction upon a measurement). Here one captures the idea that both ingredients of a complementary pair are necessary for a complete description, though the alleged mutual contradiction between observables and states is vague. Also, this reading of complementarity relies on a specific representation of the canonical commutation relations. It is not quite clear what one gains with this ideology, but perhaps it deserves to be developed in some more detail. For example, in quantum field theory it is once more the observables that carry the space-time description, especially in the algebraic description of Haag (1992). 
\end{enumerate}
\subsection{Epilogue: entanglement to the rescue?}\label{primas}
Bohr's ``quantum postulate" being obscure and obsolete, it is interesting to consider Howard's (1994)  `reconstruction' of Bohr's philosophy of physics on the basis of entanglement.\footnote{We find little evidence that Bohr himself ever thought along those lines.  With approval we quote Zeh, who, following a statement of the quantum postulate by Bohr similar to the one in Subsection \ref{HC} above, writes: 
`The later revision of these early interpretations of quantum theory (required by the important role of entangled quantum states for much larger systems) seems to have gone unnoticed by many physicists.' (Joos et al., 2003, p.\ 23.) 
See also Howard (1990) for an interesting historical perspective on entanglement, and cf.\ 
Raimond, Brune,  \&\ Haroche (2001) for the experimental situation.
} His case can perhaps be strengthened by an appeal to the analysis Primas (1983) has given of the need for classical concepts in quantum physics.\footnote{See also Amann \&\ Primas (1997) and Primas (1997).}
Primas proposes to define a ``quantum object" as a physical system $\CS$ that  is free from what he calls ``\epr-correlations" with its environment. Here the ``environment" is meant to include apparatus, observer, the rest of the universe if necessary, and what not.  In elementary \qm,  quantum objects in this sense exist only in very special states: if $\H_S$ is the \Hs\ of the system $S$, and $\H_E$ that of the environment $E$, any pure state of the form
$\sum_i c_i \Psi_i\ot\Phi_i$ (with more than one term) by definition correlates $S$ with $E$; the only uncorrelated pure states are those of the form $\Psi\ot\Phi$ for unit vectors $\Psi\in \H_S$, $\Phi\in\H_E$.
The restriction of an \epr-correlated state on $S+E$ to $S$ is mixed, so that the (would-be) quantum object `does not have its own pure state'; equivalently, the restriction of an \epr-correlated state $\om$ to $S$ together with its restriction to $E$ do not jointly determine $\om$. Again in other words, if the state of the total $S+E$ is \epr-correlated, a complete characterization of the state of $S$ requires $E$ (and vice versa). But (against Bohr!) mathematics defeats words: the sharpest characterization of the notion of \epr-correlations can be given in terms of operator algebras, as follows.  
In the spirit of the remainder of the paper we proceed in a rather general and abstract way.\footnote{Though Summers \&\ Werner (1987) give even more general results, where the tensor product $\CA\hat{\ot}\CB$ below is replaced
by an arbitrary \ca\ $\CC$ containing $\CA$ and $\CB$ as $C^*$-subalgebras.
}

Let $\CA$ and $\CB$ be $C^*$-algebras,\footnote{\label{CSQP} Recall that a $C^*$-algebra is a complex algebra $\CA$ that is
complete in a norm $\|\cdot\|$ that satisfies $\| AB\|\,\leq\, \| A\|\,\|
B\|$ for all $A,B\in\CA$, and has an involution $A\raw A^*$ such that
$\| A^*A\|=\| A\|^2$. A basic examples is
$\CA=\BH$, the algebra of all bounded operators on a \Hs\ $\H$, equipped with the usual operator norm and adjoint.   By the Gelfand--Naimark theorem, any \ca\ is isomorphic to a norm-closed self-adjoint subalgebra of $\BH$, for some \Hs\ $\H$. Another key example is  $\CA=C_0(X)$, the space of
all continuous complex-valued functions on a (locally compact Hausdorff) space $X$ that vanish at infinity (in the sense that for every $\varep>0$ there is a {\it compact} subset $K\subset X$ such that $|f(x)|<\varep$ for all $x\notin K$), equipped with the supremum norm
$\| f\|_{\infty}:=\sup_{x\in X} |f(x)|$, and involution given by (pointwise)
complex conjugation. By the Gelfand--Naimark lemma, any commutative \ca\ is isomorphic to $C_0(X)$ for some locally compact Hausdorff space $X$.
 \label{Cstar}}
 with tensor product $\CA\hat{\ot}\CB$.\footnote{\label{tensorproducts}  The tensor product of two (or more) \ca s is not unique, and we here need the so-called {\it projective} tensor product $\CA\hat{\ot}\CB$,  defined as the completion of the algebraic tensor product $\CA\ot\CB$  in the {\it maximal} $C^*$-cross-norm.
The choice of the projective tensor product guarantees that each state on
$\CA\ot\CB$ extends to a state on $\CA\hat{\ot}\CB$ by continuity; conversely,
since $\CA\ot\CB$ is dense in $\CA\hat{\ot}\CB$, each state on the latter is uniquely determined by its values on the former. See Wegge-Olsen (1993), Appendix T, or Takesaki (2003), Vol.\ {\sc i}, Ch.\ {\sc iv}. 
In particular, product states $\rh\ot\sg$ and mixtures 
$\om=\sum_i p_i \rh_i\ot\sg_i$ thereof as considered below are well defined on $\CA\hat{\ot}\CB$. If $\CA\subset \CB(\H_S)$ and $\CB\subset \CB(\H_E)$ are von Neumann algebras, as in the analysis of Raggio (1981, 1988), it is easier (and sufficient) to work with the {\it spatial} tensor product 
$\CA\ovl{\ot}\CB$, defined as the double commutant (or weak completion)
of $\CA\ot\CB$ in $\CB(\H_S\ot \H_E)$. For any {\it normal} state on $\CA\ot\CB$ extends to a normal state on $\CA\ovl{\ot}\CB$ by continuity.} 
Less abstractly, just think of two \Hs s $\H_S$ and $\H_E$ as above, with tensor product
$\H_S\ot\H_E$, and assume that $\CA=\CB(\H_S)$ while $\CB$ is either $\CB(\H_E)$ itself or some
(norm-closed and involutive) commutative subalgebra thereof. The tensor product $\CA\hat{\ot}\CB$ is then a 
(norm-closed and involutive) subalgebra of $\CB(\H_S\ot\H_E)$, the algebra of all bounded operators on $\H_S\ot\H_E$.  

 A {\it product state} on $\CA\hat{\ot}\CB$ is a state of the form
$\om=\rh\ot\sg$, where the states $\rh$ on $\CA$ and $\sg$ on $\CB$ may be either pure or mixed.\footnote{
We use the notion of a state that is usual in the algebraic framework. Hence  a {\it state} on a \ca\ $\CA$  is a linear functional  
$\rh:\CA\raw\C$ that is {\it positive} in that $\rh(A^*A)\geq 0$ for all $A\in\CA$
and {\it normalized} in that $\rh(1)=1$, where $1$ is the unit element of $\CA$.
If $\CA$ is a von Neumann algebra, one has the notion of a
 {\it normal} state, which satisfies an additional continuity condition. If $\CA=\CB(\H)$, then a fundamental theorem of von Neumann states that
each normal state $\rh$ on $\CA$ is given by a density matrix $\hat{\rh}$ on $\H$, so that $\rh(A)=\Tr (\hat{\rh} A)$ for each $A\in\CA$. In particular, a normal pure state on $\CB(\H)$ (seen as a von Neumann algebra) is necessarily of the form $\ps(A)=(\Ps,A\Ps)$ for some unit vector $\Ps\in\H$.}
We say that a state  $\om$ on $\CA\hat{\ot}\CB$
is {\it decomposable} when it is a mixture of product states, i.e.\ when  
 $\om=\sum_i p_i \rh_i\ot\sg_i$, where the coefficients $p_i>0$ satisfy $\sum_i p_i=1$.\footnote{Infinite sums are allowed here. More precisely, $\om$ is decomposable if it is in the $w^*$-closure of the convex hull of the product states  on $\CA\hat{\otimes}\CB$.} A decomposable state $\om$ is pure precisely when it is a product of pure states. This has the important consequence that both its restrictions $\om_{|\CA}$ and $\om_{|\CB}$ to $\CA$ and $\CB$, respectively, are pure as well.\footnote{The restriction $\om_{|\CA}$ of a state $\om$ on $\CA\hat{\otimes}\CB$ to, say, $\CA$ is given by $\om_{|\CA}(A)=\om(A\ot 1)$, where $1$ is the unit element of $\CB$, etc.} On the other hand, a state on $\CA\hat{\otimes}\CB$ may be said to be  {\it \epr-correlated} (Primas, 1983) when it is {\it not} decomposable.  An \epr-correlated {\it pure} state has the property that its
 restriction  to $\CA$ or $\CB$ is {\it mixed}.

Raggio (1981) proved  that {\it 
each normal state on $\CA\hat{\otimes}\CB$  is decomposable 
 if and only if  $\CA$ or $\CB$ is commutative.} In other words, {\it \epr-correlated states exist  if and only if  $\CA$ and $\CB$ are both noncommutative.}\footnote{Raggio (1981) proved this for von Neumann algebras and normal states. His proof was adapted to \ca s by Bacciagaluppi (1993).}
As one might expect, this result is closely related to the  Bell inequalities. Namely,
 the Bell-type (or Clauser--Horne) inequality
\beq \sup\{\om(A_1(B_1+B_2)+A_2(B_1-B_2))\} \leq 2, \label{bell}
\eeq
 where {\it for a fixed state $\om$} the supremum is taken over all self-adjoint operators $A_1,A_2\in \CA$, $B_1,B_2\in\CB$, each of norm $\leq 1$, {\it holds if and only if $\om$ is decomposable} (Baez, 1987; Raggio, 1988). 
Consequently, the inequality \er{bell} can only be violated in some state $\om$ when 
 the algebras $\CA$ and $\CB$ are both noncommutative. If, on the other hand, \er{bell} is satisfied, then one knows that there exists a classical probability space and probability measure
(and hence a ``hidden variables" theory) reproducing the given correlations (Pitowsky, 1989). As stressed by  Bacciagaluppi (1993), such a description does {\it not} require the entire setting to be classical; as we have seen, only one of the algebras $\CA$ and $\CB$ has to be commutative for the  Bell inequalities to hold. 

Where does this leave us with respect to Bohr? If we follow Primas (1983) in 
describing a (quantum) object as a system free from \epr-correlations with its environment, then the mathematical results just reviewed leave us with two possibilities.  Firstly, we may pay lip-service to Bohr in taking the algebra $\CB$ (interpreted as the algebra of observables of the environment in the widest possible sense, as above) to be commutative {\it as a matter of description}. In that case, our object is really an ``object" in {\it any} of its states. 
 But clarly this  is not the only possibility. For even in the case of elementary \qm\ - where $\CA=\CB(\H_S)$ and $\CB=\CB(\H_E)$ -  the system is still an ``object" in the sense of Primas as long as the total state
$\om$ of $S+E$ is decomposable. In general, for pure states this just means that $\om=\psi\ot\phi$, i.e.\ that the total state is a product of pure states. To accomplish this, one has to define the Heisenberg cut in an appropriate way, and subsequently hope that
the given product state remains so under time-evolution (see  
Amann \&\ Primas (1997) and Atmanspacher,  Amann \&\  M\"{u}ller-Herold, 1999, and references therein). This selects certain states on $\CA$ as ``robust" or ``stable", in much the same way as in the decoherence approach. We therefore continue this discussion in Section \ref{S7} (see especially point \ref{CHD} in Subsection \ref{DSS}).
\section{Quantization}\label{S4}\setcounter{equation}{0}
Heisenberg's (1925) idea of {\it Umdeutung} ({\it reinterpretation}) suggests that it is possible to construct a quantum-mechanical description of a physical system whose classical description is known. As we have seen, this possibility was realized by Schr\"{o}dinger (1925c), who found the simplest example \er{SOP} and \er{Schreq} of {\it Umdeutung} in the context of atomic physics. This early example was phenomenally successful, as almost all of atomic and molecular physics is still based on it. 

Quantization theory is an attempt to understand this example, make it mathematically precise, and generalize it to more complicated systems. It has to be stated from the outset that, like the entire classical-quantum interface, the nature of quantization is not yet well understood. This fact is reflected by the existence of a fair number of competing quantization procedures,  the most transparent of which we will review below.\footnote{The path integral approach to quantization is still under development and so far has had no impact on foundational debates, so  we will not discuss it here.
See  Albeverio \&\ H\o egh-Krohn (1976) and Glimm \&\ Jaffe (1987).} Among  the first mathematically serious discussions of quantization are Mackey (1968) and  Souriau (1969); more recent and comprehensive treatments are, for example, Woodhouse (1992), Landsman (1998),  and Ali \&\ Englis (2004).
\subsection{Canonical quantization and systems of imprimitivity}\label{Mackey}
The approach based on \er{SOP} is often called {\it canonical quantization}. Even apart from the issue of mathematical rigour, one can only side with Mackey (1992, p.\ 283), who wrote: `Simple and elegant as this model is, it appears at first sight to be quite arbitrary and ad hoc. It is difficult to understand how anyone could have guessed it and by no means obvious how to modify it to fit a model for space different from $\R^r$.' 

One veil of the mystery of quantization was lifted by von Neumann (1931), who  (following earlier heuristic proposals by Heisenberg, Schr\"{o}dinger, Dirac, and Pauli) recognized that \er{SOP} does not  merely provide {\it a} \rep\ of the canonical commutation relations
\beq [\qh(p_j),\qh(q^k)]=-i\hbar\dl^k_j, \label{ccr}\eeq
but (subject to a regularity condition)\footnote{It is required that the unbounded operators $\qh(p_j)$ and $\qh(q^k)$ integrate to a unitary \rep\ of the $2n+1$-dimensional Heisenberg group $H_n$, i.e.\ the unique connected and simply connected Lie group with $2n+1$-dimensional Lie algebra with generators $X_i,Y_i,Z$ ($i=1,\ldots, n$)  subject to the Lie brackets $[X_i,X_j]=[Y_i,Y_j]=0$,
$[X_i,Y_j]=\dl_{ij}Z$, $[X_i,Z]=[Y_i,Z]=0$. Thus von Neumann's uniqueness theorem for \rep s of the canonical commutation relations is (as he indeed recognized himself) really a uniqueness theorem for unitary \rep s of $H_n$ for which the central element $Z$ is mapped to $-i\hbar\inv 1$, where $\hbar\neq 0$ is a {\it fixed} constant. See, for example, Corwin \&\ Greenleaf (1989) or Landsman (1998).} is {\it the only} such \rep\ that is irreducible (up to unitary equivalence). In particular, the seemingly different formulations of quantum theory by Heisenberg and Schr\"{o}dinger (amended by the inclusion of states and of observables, respectively - cf.\ Section \ref{S2}) simply involved superficially different but unitarily equivalent \rep s of \er{ccr}: the difference between matrices and waves was just one between coordinate systems in \Hs, so to speak.  Moreover, any other conceivable formulation of \qm\ - now simply {\it defined} as a (regular) \Hs\ \rep\ of \er{ccr} - has to be equivalent to the one of Heisenberg and Schr\"{o}dinger.\footnote{This is unrelated to the issue of the  Heisenberg picture versus the Schr\"{o}dinger picture, which is about the time-evolution of observables versus that of states.}

This, then, transfers the quantization problem of a particle moving on $\R^n$ to the canonical commutation relations \er{ccr}. Although a mathematically rigorous theory of these commutation relations (as they stand) exists (J\o rgensen,\&\ Moore, 1984; Schm\"{u}dgen, 1990), they are problematic nonetheless. Firstly, technically speaking the operators involved are unbounded, and in order to represent physical observables they have to be self-adjoint; yet on their respective domains of self-adjointness the commutator on the left-hand side is undefined. Secondly, and more importantly, \er{ccr} relies on the possibility of choosing  global coordinates on $\R^n$, which precludes  a naive generalization to arbitrary configuration spaces. And thirdly, even if one has managed to quantize $p$ and $q$ by finding a \rep\ of \er{ccr}, the problem of quantizing other observables remains - think of the Hamiltonian and the Schr\"{o}dinger equation.  

About 50 years ago, Mackey set himself the task of making good sense of canonical quantization; see Mackey (1968, 1978, 1992) and the brief exposition below for the result. Although the author now regards Mackey's  reformulation of quantization in terms of induced \rep s and systems of imprimitivity merely as a stepping stone towards our current understanding based on deformation theory and groupoids (cf.\ Subsection \ref{DQsection} below), Mackey's approach is (quite rightly) often used in the foundations of physics, and one is well advised to be familiar with it. In any case,  Mackey (1992, p.\ 283 - continuing the previous quotation) claims with some justification that his approach to quantization  `removes much of the mystery.'

Like most approaches to quantization, Mackey assigns momentum and position a quite different role in \qm, despite the fact that in classical mechanics $p$ and $q$ can be interchanged by a canonical transformation:\footnote{Up to a minus sign, that is. This is true globally on $\R^n$ and locally on any symplectic manifold, where local Darboux coordinates
do not distinguish between position and momentum.}
\begin{enumerate}
\item 
The position operators $\qh(q^j)$ are collectively replaced by a single projection-valued measure $P$ on $\R^n$,\footnote{\label{PVM} A projection-valued measure $P$ on a space $\Om$ with Borel structure (i.e.\ equipped with a $\sg$-algebra of measurable sets defined by the topology) with values in a \Hs\ $\H$ is a map $E\mapsto P(E)$ from the Borel subsets
$E\subset\Om$ to the projections on $\H$ that satisfies $P(\emptyset)=0$,
$P(\Om)=1$, $P(E) P(F)=P(F)P(E)=P(E\cap F)$ for all measurable $E,F\subset\Om$, and
$P(\cup_{i=1}^{\infty} E_i)=\sum_{i=1}^{\infty} P(E_i)$ for all countable collections of mutually disjoint $E_i\subset\Om$. 
} which on $L^2(\R^n)$ is given by $P(E)=\ch_E$ as a multiplication operator. Given this $P$, any multiplication operator defined by a (measurable) function $f:\R^n\raw\R$ can be represented as $\int_{\R^n} dP(x)\, f(x)$,
which is defined and self-adjoint on a suitable domain.\footnote{\label{domain} This domain consists of all
$\Psi\in\H$ for which $\int_{\R^n} d(\Psi,P(x)\Psi)\, |f(x)|^2<\infty$.} In particular, the position operators $\qh(q^j)$ can be reconstructed from $P$ by choosing $f(x)=x^j$, i.e.\ 
\beq \qh(q^j)=\int_{\R^n} dP(x)\, x^j.\eeq
\item The momentum operators $\qh(p_j)$ are collectively replaced by a single unitary group \rep\ $U(\R^n)$, defined  on $L^2(\R^n)$ by $$U(y)\Psi(x):=\Psi(x-y).$$ Each $\qh(p_j)$
can be reconstructed from $U$ by means of 
\beq \qh(p_j)\Psi:=i\hbar \lim_{t_j\raw 0} t_j\inv(U(t_j)-1)\Psi,\eeq
 where $U(t_j)$ is $U$ at $x^j=t_j$ and $x^k=0$ for $k\neq j$.\footnote{By Stone's theorem (cf.\ Reed \&\ Simon, 1972), this operator is defined and self-adjoint on the set of all $\Psi\in H$ for which the limit exists.} \end{enumerate}

 Consequently, it entails no loss of generality to work with the pair $(P,U)$ instead of the   pair $(\qh(q^k),\qh(p_j))$. The commutation relations \er{ccr} are now replaced by 
\beq U(x)P(E)U(x)\inv=P(xE), \label{impr}\eeq
where $E$ is a (Borel) subset of $\R^n$ and $xE=\{x\om\mid\om\in E\}$. On the basis of this reformulation, Mackey proposed  the following sweeping generalization of the the canonical commutation relations:\footnote{All groups and spaces are supposed to be locally compact, and actions and \rep s are assumed continuous.} 
\begin{quote}A {\it system of imprimitivity} $(\H,U,P)$ for
a given action of a group $G$  on a space $Q$  consists of a \Hs\ $\H$, a unitary \rep\ $U$ of $G$ on $\H$, and a projection-valued measure $E\mapsto P(E)$ on $Q$ with values in $\H$, such that  \er{impr} holds for all $x\in G$ and all Borel sets $E\subset Q$.
\end{quote}

In physics such a system describes the \qm\ of a particle moving on a configuration space $Q$ on which $G$ acts by symmetry transformations; see Subsection \ref{DQsection} for a more detailed discussion. When everything is smooth,\footnote{I.e.\ $G$ is a Lie group, $Q$ is a manifold, and the $G$-action is smooth.}  each element $X$ of the Lie algebra $\g$ of $G$ defines a generalized momentum operator \beq
\qh(X)=i\hbar dU(X)\label{mom}\eeq on $\H$.\footnote{This operator is defined and self-adjoint on the domain of vectors $\Psi\in \H$ for which $dU(X)\Psi:=\lim_{t\raw 0} t\inv(U(\exp(tX))-1)\Psi$ exists.\label{dUFN}}  These operators satisfy the generalized canonical commutation relations\footnote{As noted before in the context of \er{ccr}, the  commutation relations \er{Gccr},
\er{Gccr2} and \er{Gccr3}  do not hold on the domain of self-adjointness of the operators involved, but on a smaller common core.}
\beq [\qh(X),\qh(Y)] = i\hbar \qh([X,Y]). \label{Gccr}\eeq Furthermore, in terms of the operators\footnote{For the domain of $\qh(f)$ see footnote  \ref{domain}.} 
\beq \qh(f)=\int_Q dP(x)\, f(x), \label{pifP}\eeq where $f$ is a smooth function on $Q$ and $X\in\g$, 
one in addition has 
\beq [\qh(X),\qh(f)] =i\hbar\qh(\xi^Q_X f), \label{Gccr2}\eeq
where $\xi^Q_X$ is the canonical  vector field on $Q$ defined by the $G$-action,\footnote{I.e.\ $\xi^Q_X f(y)=d/dt|_{t=0} [f(\exp(-tX)y)]$.\label{VFFN}}
 and \beq [\qh(f_1),\qh(f_2)]=0. \label{Gccr3}\eeq

Elementary \qm\ on $\R^n$ corresponds to the special case $Q=\R^n$
and $G=\R^n$ with the usual additive group structure. To see this,
we denote  the standard basis  of $\R^3$ (in its guise as the Lie algebra of $\R^3$)  by the name $(p_j)$, and furthermore take $f_1(q)=q^j$, 
$f_2(q)=f(q)=q^k$. Eq.\ \er{Gccr} for $X=p_j$ and $Y=p_k$ then reads $[\qh(p_j),\qh(p_k)]=0$, eq.\ \er{Gccr2} yields the canonical commutation relations \er{ccr}, and \er{Gccr3} states the commutativity of the position operators, i.e.\  $[\qh(q^j),\qh(q^k)]=0$.

In order to incorporate spin, one picks  $G=E(3)=SO(3)\ltimes\R^3$ (i.e.\ the Euclidean motion group), acting on $Q=\R^3$ in the obvious (defining) way.
The Lie algebra of $E(3)$ is $\R^6=\R^3\x\R^3$ as a vector space; we extend the basis $(p_j)$ of the second copy of $\R^3$ (i.e.\ the Lie algebra of $\R^3$) by a basis $(J_i)$ of the first copy of $\R^3$ (in its guise as the Lie algebra of $SO(3)$) , and find that the $\qh(J_i)$ are just the usual angular momentum operators.\footnote{The commutation relations in the previous paragraph are now extended by the familiar relations 
$[\qh(J_i),\qh(J_j)] = i\hbar\ep_{ijk}\qh(J_k)$,
 $ [\qh(J_i),\qh(p_j)] = i\hbar\ep_{ijk}\qh(p_k)$, and 
 $[\qh(J_i),\qh(q^j)] = i\hbar\ep_{ijk}\qh(q^k)$. \label{AMCCR}}

Mackey's generalization of von Neumann's (1931) uniqueness theorem for the \irrep s of the canonical commutation relations \er{ccr} is his {\it  imprimitivity theorem}. This theorem applies to  the special case where $Q=G/H$ for some (closed) subgroup $H\subset G$, and states that  (up to unitary equivalence) there is a bijective correspondence between:
\begin{enumerate}
\item Systems of imprimitivity  $(\H,U,P)$ for the left-translation of $G$ on $G/H$;
\item Unitary \rep s $U_{\ch}$ of $H$.
\end{enumerate}
This correspondence preserves irreducibility.\footnote{Specifically, given  $U_{\ch}$ the triple $(\H^{\ch},U^{\ch},P^{\ch})$ is a system of imprimitivity, where $\H^{\ch}=L^2(G/H)\ot \H_{\ch}$ carries the \rep\ $U^{\ch}(G)$ induced by $U_{\ch}(H)$, and the $P^{\ch}$ act like multiplication operators. Conversely,
if $(\H,U,P)$ is a system of imprimitivity, then there exists a unitary \rep\ $U_{\ch}(H)$ such that the triple $(\H,U,P)$ is unitarily equivalent to the triple $(\H^{\ch},U^{\ch},P^{\ch})$ just described. For example, for $G=E(3)$ and $H=SO(3)$ one has $\ch=j=0,1,2,\ldots$ and 
$\H^j=L^2(\R^3)\ot \H_j$ (where $\H_j=\C^{2j+1}$
carries the given \rep\ $U_j(SO(3))$). \label{MIT}} 

For example, von Neumann's theorem is recovered as a special case of Mackey's by making  the choice $G=\R^3$ and $H=\{e\}$ (so that $Q=\R^3$, as above):
the uniqueness of the (regular) \irrep\ of the canonical commutation relations here  follows from the uniqueness of the \irrep\ of the trivial group.
 A more illustrative example is $G=E(3)$ and $H=SO(3)$ (so that $Q=\R^3$), in which case the \irrep s of the associated system of imprimitivity are classified by spin $j=0,1,\ldots$.\footnote{By the usual arguments (Wigner's theorem), one may replace $SO(3)$ by $SU(2)$, so as to obtain $j=0,1/2,\ldots$.} Mackey saw this as an explanation for the emergence of spin as a purely quantum-mechanical degree of freedom. Although the opinion that spin has no classical analogue  was widely shared also among the pioneers of quantum theory,\footnote{This opinion goes back to Pauli (1925), who talked about a `klassisch nicht beschreibbare Zweideutigkeit in den quantentheoretischen Eigenschaften des Elektrons,' (i.e.\ an `ambivalence in the quantum theoretical properties of the electron that has no classical description') which was later identified as spin by Goudsmit and Uhlenbeck. Probably the first person to draw attention to the classical counterpart of spin was Souriau (1969).
 Another misunderstanding about spin is that its ultimate explanation must be found in relativistic \qm.} it is now  obsolete (see Subsection \ref{DQsection} below). Despite this unfortunate misinterpretation, Mackey's approach to canonical quantization is hard to surpass in power and clarity, and has many interesting applications.\footnote{This begs the question about the `best' possible proof of Mackey's  imprimitivity theorem. Mackey's own proof was rather measure-theoretic in flavour, and did not shed much light on the origin of his result. Probably the shortest proof has been given by \O rsted (1979), 
but the insight brevity gives is still rather limited. Quite to the contrary, truly transparent proofs reduce a mathematical  claim to a tautology. Such proofs, however, tend to require a formidable machinery to make this reduction work; see  Echterhoff et al. (2002) and
Landsman (2005b) for two different approaches to the  imprimitivity theorem in this  style.}  

 We mention one of specific interest to the philosophy of physics, namely the {\it Newton--Wigner position operator} (as analyzed by Wightman, 1962).\footnote{Fleming \&\  Butterfield (2000) give an up-to-date introduction to particle localization in relativistic quantum theory. See also De Bi\`{e}vre (2003).}  
Here the general question is whether a given unitary \rep\ $U$ of $G=E(3)$ on some \Hs\ $\H$ may be extended to a system of imprimitivity with respect to $H=SO(3)$ (and hence
$Q=\R^3$, as above); in that case, $U$ (or rather the associated quantum system) is  said to be {\it localizable} in $\R^3$.  Following Wigner's (1939) suggestion that a relativistic  particle is described by an irreducible \rep\ $U$ of the Poincar\'{e} group $P$, one obtains a \rep\ $U(E(3))$ by restricting
$U(P)$ to the subgroup $E(3)\subset P$.\footnote{Strictly speaking, this hinges on the choice of an inertial frame in Minkowski space, with associated adapted co-ordinates such that the configuration space $\R^3$ in question is given by $x^0=0$.}  It then follows from the previous analysis that the particle described by $U(P)$ is localizable if and only if $U(E(3))$ is induced by some \rep\ of $SO(3)$. This can, of course, be settled, with the result that massive particles of arbitrary spin can be localized in $\R^3$ (the corresponding position operator being precisely the one of Newton and Wigner), whereas
 massless particles may be localized in $\R^3$ if and only if their helicity is less than one. In particular,  the photon (and the graviton) cannot be localized
in $\R^3$ in the stated sense.\footnote{
Seeing photons as quantized light waves with two possible polarizations
transverse to the direction of propagation, this last result is physically
perfectly reasonable. }

To appreciate our later material on both  phase space quantization and deformation quantization, it is helpful to give a \ca ic reformulation of Mackey's approach. Firstly, by the spectral theorem (Reed \&\ Simon, 1972; Pedersen, 1989), a projection-valued measure $E\mapsto P(E)$ on a space $Q$ taking values in a \Hs\ $\H$ is equivalent to a nondegenerate \rep\ $\pi$ of the commutative \ca\ $C_0(Q)$ on $\H$ through the correspondence \er{pifP}.\footnote{A  {\it representation} of a \ca\ $\CA$ on a \Hs\ $\H$ is a linear map $\pi:\CA\raw\BH$ such that $\pi(AB)=\pi(A)\pi(B)$ and
$\pi(A^*)=\pi(A)^*$ for all $A,B\in\CA$. 
Such a  \rep\ is called {\it nondegenerate} when $\pi(A)\Psi =0$ for all $A\in\CA$ implies $\Psi=0$.} Secondly, if $\H$ in addition carries a unitary \rep\ $U$ of $G$, the defining condition \er{impr} of a system of imprimitivity (given a $G$-action on $Q$) is equivalent to the covariance condition \beq U(x)\qh(f)U(x)\inv=\qh(L_xf) \label{Gcov}\eeq
 for all $x\in G$ and $f\in C_0(Q)$, where $L_xf (m)=f(x\inv m)$.  Thus a  system of imprimitivity for a given $G$-action on $Q$ is ``the same" as a covariant nondegenerate \rep\ of $C_0(Q)$. Thirdly, from a $G$-action on $Q$ one can construct a certain \ca\ $C^*(G,Q)$, the so-called {\it transformation group \ca} defined by the action,
which has the property that its nondegenerate \rep s correspond bijectively (and ``naturally") to covariant nondegenerate \rep s of $C_0(Q)$, and therefore to systems of imprimitivity for the given $G$-action (Effros \&\  Hahn, 1967; Pedersen, 1979; Landsman, 1998).  In the \ca ic approach to quantum physics, $C^*(G,Q)$ is
the algebra of observables of a particle moving on $Q$ subject to the symmetries defined by the $G$-action; its inequivalent \irrep s correspond to the possible superselection sectors of the system (Doebner \&\  Tolar, 1975; Majid, 1988, 1990; Landsman, 1990a, 1990b, 1992).\footnote{Another reformulation of Mackey's approach, or rather an extension of it, has been given by Isham (1984). In an attempt to reduce the whole theory 
to a problem in group \rep s, he proposed that the possible quantizations of a particle 
with configuration space $G/H$ are given by the inequivalent \irrep s of a ``canonical group" $G_c=G\ltimes V$, where $V$ is the lowest-dimensional vector space that carries a \rep\ of $G$ under which $G/H$ is an orbit in the dual vector space  $V^*$. All pertinent systems of imprimitivity then indeed correspond to unitary \rep s of $G_c$, but this group has many other \rep s whose physical interpretation is obscure. See also footnote \ref{Isham2}.\label{Isham1}}
\subsection{Phase space quantization and coherent states}\label{PSQ}
In Mackey's approach to quantization, $Q$ is the {\it configuration space} of the system; 
the associated position coordinates {\it commute} (cf.\ \er{Gccr3}).
This is reflected by the correspondence just discussed between  projection-valued measures on $Q$ and \rep s of the {\it commutative} \ca\ $C_0(Q)$. The noncommutativity of observables (and the associated uncertainty relations) typical of \qm\ is incorporated by adding the symmetry group $G$ to the picture and imposing the relations \er{impr} (or, equivalently, \er{Gccr2} or \er{Gcov}). As we have pointed out, this procedure upsets the symmetry between the phase space variables position and momentum in classical mechanics. 

This somewhat unsatisfactory feature of Mackey's approach may be avoided by replacing $Q$  by the {\it phase space} of the system, henceforth called $M$.\footnote{Here the reader may think of the simplest case $M=\R^6$, the space of $p$'s and $q$'s of a particle moving on $\R^3$. More generally, if $Q$ is the configuration space, the associated phase space is the cotangent bundle $M=T^*Q$. Even more general phase spaces, namely arbitrary symplectic manifolds, may be included  in the theory as well. References for what follows include Busch,  Grabowski, \&\ Lahti, 1998, Schroeck, 1996, and Landsman, 1998, 1999a.  } In this approach, noncommutativity is incorporated by a treacherously  tiny modification to Mackey's setup. Namely, the projection-valued measure $E\mapsto P(M)$ on $M$ with which he starts is now replaced by a {\it positive-operator-valued measure} or {\it POVM} on $M$, still taking values in some \Hs\ $\CK$. This is  a map $E\mapsto A(E)$ from the (Borel)
subsets $E$ of $M$ to the collection of {\it positive} bounded operators on $\CK$,\footnote{A bounded operator $A$ on $\CK$ is called positive when $(\Psi,A\Psi)\geq 0$ for all $\Psi\in\CK$.
Consequently, it is self-adjoint with spectrum contained in $\R^+$.}
satisfying $A(\emptyset)=0$, $A(M)=1$, and 
$A(\cup_i E_i)=\sum_i A(E_i)$ for any countable collection of disjoint Borel sets $E_i$.\footnote{Here the infinite sum is taken in the weak operator topology.
 Note that the above conditions force $0\leq A(E)\leq 1$, in the sense that
$0\leq (\Psi, A(E)\Psi)\leq (\Psi,\Psi)$ for all $\Psi\in\CK$.} A POVM that satisfies
$A(E\cap F)  =  A(E)A(F)$ for all (Borel) $E,F\subset M$ is precisely a projection-valued measure, so that a POVM is a generalization of the latter.\footnote{This has given rise to the so-called {\it operational approach} to quantum theory, in which observables are not represented by self-adjoint operators (or, equivalently, by their associated  projection-valued measures), but by POVM's.  The space $M$ on which the POVM is defined is the space of outcomes of the measuring instrument; the POVM
is determined by both $A$ and a calibration procedure for this instrument.  The probability that in a state $\rh$ the outcome of the experiment lies in $E\subset M$ is taken to be $\Tr (\rh A(E))$. See Davies (1976), Holevo (1982), Ludwig (1985),  Schroeck (1996), Busch,  Grabowski, \&\ Lahti (1998),  and De Muynck (2002).} The point, then, is that {\it a given POVM defines a quantization procedure} by the stipulation that a classical observable $f$
(i.e. a measurable function on the phase space $M$, for simplicity assumed bounded)
is quantized by the operator\footnote{The easiest way to define the right-hand side of \er{berezin} is to fix $\Psi\in\CK$ and define a probability measure $p_{\Psi}$ on $M$ by means of $p_{\Psi}(E)=(\Psi,A(E)\Psi)$.
One then {\it defines} $\CQ(f)$ as an operator through its expectation values
$(\Psi, \CQ(f)\Psi)=\int_M dp_{\Psi}(x)\, f(x)$. The expression \er{berezin} generalizes \er{pifP}, and also generalizes the spectral resolution of the operator $f(A)= \int_{\R} dP(\lm) f(\lm)$, where $P$ is the projection-valued measure defined by a self-adjoint operator $A$.  } 
\beq
\CQ(f)=\int_M dA(x) f(x). \label{berezin}
\eeq
Thus the seemingly slight move from projection-valued measures on configuration space to  positive-operator valued measures on phase space gives a wholly new perspective on quantization, actually  reducing this task to the problem of finding such POVM's.\footnote{An important feature of $\CQ$ is that it is {\it positive} in the sense that if $f(x)\geq 0$ for all $x\in M$, then $(\Psi,\CQ(f)\Psi)\geq 0$ for all $\Psi\in\CK$. In other words, $\CQ$ is positive as a map from the \ca\ $C_0(M)$ to the \ca\ $\BH$.\label{BQP}}

 The solution to this problem is greatly facilitated by {\it Naimark's dilation theorem}.\footnote{See, for example, Riesz and  Sz.-Nagy (1990). It is better, however, to see  Naimark's theorem as a special case of Stinesprings's, as explained e.g.\ in Landsman, 1998, and below.}  This states that, given a POVM $E \mapsto A(E)$ on $M$ in a \Hs\ $\CK$, there exists a \Hs\ $\H$
carrying a projection-valued measure $P$ on $M$ and an isometric injection $\CK\hraw\H$,
such that \beq A(E)=[\CK]P(E)[\CK] \label{AEP} \eeq for all $E\subset M$ (where $[\CK]$ is the orthogonal projection from $\H$ onto $\CK$). Combining this with Mackey's imprimitivity theorem yields a powerful  generalization of the latter (Poulsen, 1970; Neumann, 1972;  Scutaru, 1977; Cattaneo, 1979; Castrigiano \&\
Henrichs, 1980).  

First, define a {\it generalized system of imprimitivity} $(\CK,U,A)$ for
a given action of a group $G$  on a space $M$ as a POVM $A$ on $M$ taking values in a \Hs\ $\CK$, along with  a unitary \rep\ $V$ of $G$ on $\CK$ such that
\beq V(x)A(E)V(x)\inv=A(xE) \label{gimpr}\eeq
for all $x\in G$ and $E\subset M$; cf.\ \er{impr}. Now assume  $M=G/H$ (and the associated canonical left-action on $M$). The {\it generalized imprimitivity theorem} states that a generalized system of imprimitivity $(\CK,V,A)$ for this action is necessarily (unitarily equivalent to) a reduction of a system of imprimitivity $(\H, U,P)$ for the same action. In other words, the \Hs\ $\H$  in Naimark's theorem carries a unitary \rep\ $U(G)$ that commutes with the projection $[\CK]$, and the \rep\ $V(G)$ is simply the restriction of $U$ to $\CK$. Furthermore, the POVM $A$ has the form \er{AEP}. The structure of $(\H, U,P)$ is fully described by Mackey's imprimitivity theorem, so that one has a complete classification of generalized  systems of imprimitivity.\footnote{Continuing footnote  \ref{MIT}: $V(G)$ is necessarily a sub\rep\ of some \rep\ $U^{\ch}(G)$ induced by $U_{\ch}(H)$.} One has
\beq \CK=p\H; \:\:\: \H= L^2(M)\ot\H_{\ch}, \label{pH}\eeq
where $L^2$ is defined with respect to a suitable measure on $M=G/H$,\footnote{In the physically relevant case  that $G/H$ is symplectic 
(so that it typically is a coadjoint orbit for $G$) one should take a multiple of the Liouville measure.} the \Hs\ 
$\H_{\ch}$ carries a unitary \rep\ of $H$, and $p$ is a projection in the commutant
of the \rep\ $U^{\ch}(G)$ induced by $U_{\ch}(G)$.\footnote{The explicit form of $U^{\ch}(g)$, $g\in G$, depends on the choice of a cross-section $\sg:G/H\raw G$ of the projection $\pi: G\raw G/H$ (i.e.\ $\pi\circ\sg=\id$). If the measure on $G/H$ defining $L^2(G/H)$ is $G$-invariant, the explicit formula is $U^{\ch}(g)\Psi(x)=U_{\ch}(s(x)\inv g s(g\inv x))\Psi(g\inv x)$.} The quantization \er{berezin} is given by
\beq \CQ(f)=pfp, \label{PSQE}\eeq
where $f$ acts on $L^2(M)\ot\H_{\ch}$ as a multiplication operator, i.e.\ $(f\Psi)(x)=f(x)\Psi(x)$. In particular, one has $P(E)=\ch_E$ (as a multiplication operator) for
 a region $E\subset M$ of phase space, so that $\CQ(\ch_E)=A(E)$. Consequently, the probability that in a state $\rh$ (i.e.\ a density matrix on $\CK$) the system is localized in $E$ is given by  $\Tr (\rh A(E))$. 

In a more natural way than in  Mackey's approach, the covariant POVM quantization method allows one to incorporate space-time symmetries {\it ab initio}  by taking $G$ to be 
 the Galilei group or the Poincar\'{e} group, and choosing $H$ such that $G/H$ is a physical phase space (on which $G$, then, canonically acts).
 See  Ali et al.\ (1995) and  Schroeck (1996).
 
Another powerful method of constructing POVM's on phase space (which in the presence of symmetries overlaps with the preceding one)\footnote{
Suppose there is a vector $\Om\in\CK$ such that
$\int_{G/H}d\mu(x) |(\Om, V(\sg(x))\Om)|^2 <\infty$
with respect to some cross-section $\sg: G/H\raw G$ and a
$G$-invariant measure $\mu$, as well as $V(h)\Om=U_{\ch}(h)\Om$ for all $h\in H$, where $U_{\ch}:H\raw \C$ is {\it one-dimensional}. Then (taking $\hbar=1$)
the vectors $V(\sg(x))\Om$ (suitably normalized) form a family of coherent states on $G/H$ (Ali et al., 1995; Schroeck, 1996;  Ali,  Antoine,  \&\ Gazeau, 2000). For example, the coherent states \er{pqcohst} are of this form for the Heisenberg group.}
 is based on {\it coherent states}.\footnote{\label{CSFNO}See  Klauder \&\  Skagerstam, 1985, 
Perelomov, 1986,   Odzijewicz, 1992, Paul \&\ Uribe, 1995, 1996, Ali et al., 1995,  and   Ali,  Antoine,  \&\ Gazeau, 2000, for  general discussions of coherent states. } The minimal definition of coherent states in a \Hs\ $\H$ for a phase space $M$ is that 
(for some fixed value of Planck's constant $\hbar$, for the moment)
one has an injection\footnote{This injection must be continuous as a map from $M$ to
$\mathbb{P}\H$, the projective \Hs\ of $\H$.}
 $M\hraw \H$, $z\mapsto\Psi_z^{\hbar}$, such that 
\beq \|\Psi_z^{\hbar}\|=1 \label{normcs}
\eeq
 for all $z\in M$, and 
\beq
 c_{\hbar} \int_M  d\mu_L(z)\,|(\Psi_z^{\hbar},\Phi)|^2 =1, \label{qhnorm}
\eeq
for each $\Phi\in\H$ of unit norm (here $\mu_L$ is the Liouville measure on
$M$ and $c_{\hbar}>0$ is a suitable constant).\footnote{Other measures might occur here; see, for example,   Bonechi \&\  De Bi\`{e}vre (2000).}
Condition (\ref{qhnorm}) guarantees that we may define a POVM on $M$ in $\CK$
by\footnote{Recall that $[\Psi]$ is the orthogonal projection onto a unit vector $\Psi$.}
\beq
A(E)= c_{\hbar} \int_E  d\mu_L(z)\,[\Psi_z^{\hbar}]. \label{pov}
\eeq
Eq.\  (\ref{berezin}) then simply reads (inserting the $\hbar$-dependence of $\CQ$ and a suffix $B$ for later use)
\beq
\qh^B(f)= c_{\hbar}\int_{M}  d\mu_L(z)\, f(z) [\Psi^{\hbar}_z].
\label{b2}
\eeq

The time-honoured example,
due to Schr\"{o}dinger (1926b), is $M=\R^{2n}$, $\H=L^2(\R^n)$, and 
\beq
\Psi_{(p,q)}^{\hbar}(x)=(\pi\hbar)^{-n/4}e^{-
ipq/2\hbar}e^{ipx/\hbar}e^{-(x-q)^2/2\hbar}.\label{pqcohst} 
\eeq
Eq.\ (\ref{qhnorm})  then holds with $d\mu_L(p,q)=(2\pi)^{-n}d^npd^nq$ and
$c_{\hbar}=\hbar^{-n}$. One may verify that $\qh^B(p_j)$ and $\qh^B(q^j)$ coincide with Schr\"{o}dinger's operators \er{SOP}.
This example illustrates that coherent states need {\it not} be mutually orthogonal; in fact, in terms of $z=p+iq$
one has for the states in \er{pqcohst} 
\beq
 |(\Psi^{\hbar}_{z},\Psi^{\hbar}_{w})|^2=e^{-|z-w|^2/2\hbar}; \label{tpbt}
\eeq
the significance of this result will emerge later on.

 In the general case, it is an easy matter to verify Naimark's dilation theorem for the POVM \er{pov}:
 changing notation so that the vectors $\Psi_z^{\hbar}$ now lie in $\CK$,
 one finds 
 \beq \H=L^2(M, c_{\hbar}\mu_L), \label{HSPSQ}
 \eeq
 the embedding $W:\CK\hraw\H$ being given by $(W\Phi)(z)=(\Psi^{\hbar}_z,\Phi)$. The projection-valued measure $P$ on $\H$ is just $P(E)=\ch_E$ (as a multiplication operator), and the projection $p$ onto $W\CK$ is given by 
 \beq p\Psi(z)=c_{\hbar} \int_M  d\mu_L(w) (\Psi^{\hbar}_{z},\Psi^{\hbar}_{w})\Psi(w).\eeq
 Consequently, \er{b2} is unitarily equivalent
 to \er{PSQE}, where $f$ acts on $L^2(M)$ as a multiplication operator.\footnote{This leads to a close relationship between coherent states and \Hs s with a reproducing kernel; see Landsman (1998) or Ali, Antoine, \&\ Gazeau (2000).}  
\begin{quote}{\it Thus \er{PSQE} and \er{HSPSQ} (or its possible extension \er{pH}) form the essence of phase space quantization.}\footnote{See also footnote  \ref{Koopman} below.}
\end{quote}
  
 We close this subsection in the same fashion as the previous one, namely by pointing out the \ca ic significance of POVM's. This is extremely easy: whereas
 a projection-valued measure on $M$  in $\H$ is the same as a nondegenerate {\it \rep} of $C_0(M)$ in $\H$, a POVM  on $M$  in a \Hs\ $\CK$ is nothing but a nondegenerate {\it completely positive map} $\phv:C_0(M)\raw \mathcal{B}(\CK)$.\footnote{A map  $\phv:\CA\raw\CB$ between \ca s is called positive when $\phv(A)\geq 0$
 whenever $A\geq 0$; such a map is called {\it completely positive} if for all $n\in \N$ the map $\phv_n: \CA\otimes M_n(\C)\raw  \CB\otimes M_n(\C)$, defined by linear extension of $\phv\otimes\id$ on elementary tensors, is positive (here $M_n(\C)$ is the \ca\ of $n\x n$ complex matrices). When $\CA$ is commutative
 a nondegenerate positive map $\CA\raw \CB$ is automatically completely positive for any $\CB$.}
 Consequently, Naimark's dilation theorem becomes a special case of Stinespring's (1955) theorem: if $\CQ:\CA\raw \CB(\CK)$ is a completely positive map,  there exists a \Hs\ $\H$ carrying a \rep\ $\pi$ of $C_0(M)$
 and an isometric injection $\CK\hraw\H$, such that $\CQ(f)=[\CK]\pi(f)[\CK]$ for all $f\in C_0(M)$. In terms of $\CQ(C_0(M))$, the covariance condition
 \er{gimpr} becomes  $U(x)\CQ(f)U(x)\inv=\CQ(L_xf)$, just like 
 \er{Gcov}. 
 \subsection{Deformation quantization}\label{DQsection}
 So far, we have used the word `quantization' in a heuristic way, basing our account on historical continuity rather than on axiomatic foundations. In this subsection and the next we set the record straight by introducing two alternative ways of looking at quantization in an axiomatic way. We start with the approach that historically came last, but which conceptually is closer to the material just discussed. This is {\it deformation quantization}, originating in the work of Berezin (1974, 1975a, 1975b), Vey (1975),  and  Bayen et al.\ (1977). We here follow the \ca ic approach to deformation quantization proposed by Rieffel (1989a, 1994), since it is not only mathematically transparent and rigorous, but also reasonable close to physical practice.\footnote{See also Landsman (1998) for an extensive discussion of the  \ca ic approach to deformation quantization. In other approaches  to deformation quantization, such as the theory of star products, $\hbar$ is a formal parameter rather than a real number. In particular, the meaning of the limit $\hbar\raw 0$ is obscure.} Due to the mathematical language used, this method of course naturally fits into the general  $C^*$-algebraic approach to quantum physics.
 
The key idea of  deformation quantization is that quantization should be defined through the property of having the correct classical limit. Consequently, Planck's ``constant" $\hbar$ is treated as a variable, so that for each of its values one should have a quantum theory. The key requirement is that this family of quantum theories converges to the underlying classical theory as $\hbar\raw 0$.\footnote{Cf.\ the preamble to Section \ref{S5} for further comments on this limit.}  The mathematical implementation of this idea is quite beautiful, in that the classical algebra of observables is ``glued" to the family of quantum  algebras of observables in such a way that the classical theory  literally forms the boundary of the space containing the pertinent quantum theories (one for each value of $\hbar>0$). Technically, this is done through the concept of a {\it continuous field of
 \ca s}.\footnote{See Dixmier (1977), Fell \&\ Doran (1988), and Kirchberg \&\  Wassermann (1995) for three different approaches to the same concept. Our definition follows the latter;
replacing $I$ by an arbitrary locally compact Hausdorff space one finds the general definition.}  What follows may sound unnecessarily technical, but the last 15 years have indicated that this yields exactly the right definition of quantization.

 Let  $I\subset \R$ be the set in which $\hbar$ takes values; one usually has $I=[0,1]$, but when the phase space is compact, $\hbar$ often takes values in a countable subset of $(0,1]$.\footnote{Cf.\ Landsman (1998) and footnote  \ref{FNPER}, but also see Rieffel (1989a) for the example of the noncommutative torus, where one quantizes a compact phase space for each $\hbar\in(0,1]$.} The same situation occurs in the theory of infinite systems; see Section \ref{S6}.  In any case, {\it $I$ should contain zero as an accumulation point.}  A continuous field of $C^*$-algebras over $I$, then,  consists of a
 \ca\ $\CA$, a collection of \ca s $\{\CA_{\hbar}\}_{{\hbar}\in I}$, and a surjective morphism $\phv_{\hbar}:\CA\raw\CA_{\hbar}$ for each $\hbar\in I$ , such that:
\begin{enumerate}
\item
The function ${\hbar}\mapsto  \|\phv_{\hbar}(A)\|_{\hbar}$ is in $C_0(I)$ for all $A\in\CA$;\footnote{Here $\|\cdot\|_{\hbar}$ is the norm in the \ca\ $\CA_{\hbar}$.}
\item
The norm of any $A\in\CA$ is $\| A\|=\sup_{{\hbar}\in I}\|\phv_{\hbar}(A)\|$;
\item
For any $f\in C_0(I)$ and $A\in\CA$ there is an element $fA\in\CA$ for
which $\phv_{\hbar}(fA)=f({\hbar})\phv_{\hbar}(A)$ for all ${\hbar}\in I$.
\end{enumerate}
 
The idea is that the  family $(\CA_{\hbar})_{\hbar\in I}$ of \ca s is glued together by specifying a topology on the bundle 
$\coprod_{\hbar\in [0,1]}\CA_{\hbar}$ (disjoint union). However, this topology is in fact defined rather indirectly, via the specification of the space of continuous sections of the bundle.\footnote{This is reminiscent of the Gelfand--Naimark theorem for commutative \ca s, which specifies the topology on a locally compact Hausdorff space $X$ via the \ca\ $C_0(X)$. Similarly, in the theory of (locally trivial) vector bundles the Serre--Swan theorem allows one to reconstruct the topology on a vector bundle $E\stackrel{\pi}{\raw} X$ from the space $\Gm_0(E)$ of continuous sections of $E$, seen as a (finitely generated projective) $C_0(X)$-module. See, for example,  Gracia-Bond\'{\i}a,    V\'{a}rilly,  \&\ Figueroa (2001). The third condition in our definition of a continuous field of \ca s makes $\CA$ a $C_0(I)$-module in the precise sense that there exits 
a nondegenerate morphism from $C_0(I)$ to the center of the multiplier of $\CA$. This property may also replace our condition 3.} Namely, a {\it continuous section}
of the field is {\it by definition} an element $\{A_{\hbar}\}_{{\hbar}\in I}$ of $\prod_{{\hbar}\in I}\CA_{\hbar}$ (equivalently, a map $\hbar\mapsto A_{\hbar}$
where $A_{\hbar}\in \CA_{\hbar}$)
 for which there is an $A\in \CA$ such that $A_{\hbar}=\phv_{\hbar}(A)$ for all ${\hbar}\in I$. It follows that
the \ca\ $\CA$ may actually be identified with the space of continuous sections of the field: if we do so, the morphism $\phv_{\hbar}$ is just the evaluation
 map at $\hbar$.\footnote{The structure of $\CA$  as a \ca\ corresponds to the operations of pointwise scalar multiplication,
addition, adjointing, and operator multiplication on sections.}
 
Physically, $\CA_0$ is the commutative algebra of observables of the underlying classical system, and for each $\hbar>0$ the noncommutative \ca\ $\CA_{\hbar}$ is supposed to be the algebra of observables of the corresponding quantum system at value $\hbar$ of Planck's constant. The algebra $\CA_0$, then, is of the form $C_0(M)$, where $M$ is the phase space defining the classical theory.  A phase space has more structure than an arbitrary topological space; it is a manifold on which a Poisson bracket $\{\, ,\,\}$ can be defined. 
For example, on $M=\R^{2n}$ one has the familiar expression
\beq \{f,g\}=\sum_j\frac{\partial f}{\partial p_j}\frac{\partial g}{\partial
q^j}-\frac{\partial f}{\partial q^j}\frac{\partial g}{\partial
p_j}. \label{PBRN}\eeq 

Technically, $M$ is taken to be a {\it Poisson manifold}. This  is a manifold equipped with a Lie bracket $\{\, ,\,\}$ on $\cin(M)$ with the property that for each $f\in\cin(M)$ the map $g\mapsto \{f,g\}$ defines  a derivation of  the commutative algebra structure of $\cin(M)$ given by pointwise multiplication.  Hence this map is given by a vector field $\xi_f$, called
the {\it Hamiltonian vector field} of $f$ (i.e.\ one has $\xi_fg=\{f,g\}$).
 {\it Symplectic manifolds} are special instances of Poisson manifolds, characterized by the property that the  Hamiltonian vector fields exhaust the tangent bundle. A Poisson manifold
is foliated by its {\it symplectic leaves}: a given symplectic leaf $L$ is characterized by the property that at each $x\in L$ the tangent space $T_xL\subset T_xM$ is spanned by the collection of all Hamiltonian vector fields at $x$. Consequently, the flow of any Hamiltonian
vector field on $M$ through a given point lies in its entirety within the symplectic leaf containing that point.  The simplest example of a Poisson manifold is $M=\R^{2n}$ with Poisson bracket \er{PBRN}; this manifold is even symplectic.\footnote{See Marsden \&\ Ratiu (1994) for a mechanics-oriented introduction to Poisson manifolds; also cf.\ Landsman (1998)
or Butterfield (2005) for the basic facts. A classical mathematical paper on Poisson manifolds is Weinstein (1983). \label{PMFN}}

After this preparation, our basic definition is this:\footnote{Here 
$\cci(M)$ stands for the space of smooth functions on $M$ with compact support; this is a norm-dense subalgebra of $\CA_0=C_0(M)$. The question whether the maps $\qh$ can be extended from $\cci(M)$ to $C_0(M)$ has to be answered on a case by case basis. Upon such an extension, if it exists, condition \er{Dirac} will lose its meaning, since the Poisson bracket $\{f,g\}$ is not defined for all $f,g\in C_0(M)$.}

\begin{quote}
{\it A deformation quantization of a phase space $M$ consists of a continuous field of \ca s $(\CA_{\hbar})_{\hbar\in [0,1]}$ (with $\CA_0=C_0(M)$), along with a family of self-adjoint\footnote{I.e. $\qh(\ovl{f})=\qh(f)^*$.}  linear maps $\qh:\cci(M)\raw\CA_{\hbar}$, $\hbar\in(0,1]$, such that:}\end{quote}
\begin{enumerate}
\item For each $f\in\cci(M)$ the map defined by $0\mapsto f$ and
$\hbar\mapsto\qh(f)$ ($\hbar\neq 0$) is a continuous section of the given continuous field;\footnote{Equivalently, one could extend the family $(\qh)_{\hbar\in(0,1]}$ to $\hbar=0$ by $\CQ_0=\id$, and state that $\hbar\mapsto\qh(f)$ is a continuous section. Also, one could replace this family of maps by a single map $\CQ:\cci(M)\raw\CA$ and {\it define}
$\qh=\phv_{\hbar}\circ \CQ:\cci(M)\raw \CA_{\hbar}$.}
\item 
 For all $f,g\in \cci(M)$ one has 
\begin{equation}
\lim_{\hbar\rightarrow 0} 
\left\|\frac{i}{\hbar}[\CQ_{\hbar}(f),\CQ_{\hbar}(g)]-\CQ_{\hbar}(\{f,g\})\right\|_{\hbar} =0. \label{Dirac}
\end{equation}\end{enumerate}

Obvious continuity properties one might like to impose, such as \beq\lim_{\hbar\rightarrow 0} \|\qh(f)\qh(g)-\qh(fg)\|=0,\eeq or \beq\lim_{\hbar\rightarrow 0} \|\qh(f)\|=\| f\|_{\infty},\label{normcont} \eeq
 turn out to be an automatic consequence of the definitions.\footnote{That they are automatic should not distract from the fact that especially \er{normcont} is a beautiful connection between classical and \qm.
 See footnote \ref{CSQP} for the meaning of $\| f\|_{\infty}$.} Condition \er{Dirac}, however, transcends the \ca ic setting, and is the key ingredient in proving (among other things) that the quantum dynamics converges to the classical dynamics;\footnote{This insight is often attributed to Dirac (1930), who was the first to recognize the analogy between the commutator in \qm\ and the Poisson bracket in classical mechanics.}
  see Section \ref{S5}. The map $\qh$ is the quantization map at value $\hbar$ of Planck's constant; we feel it is the most precise formulation of Heisenberg's original {\it Umdeutung} of classical observables known to date. It has the same interpretation as the heuristic symbol $\qh$ used so far: the operator $\qh(f)$ is the quantum-mechanical observable whose classical counterpart is $f$. 
 
 This has turned out to be an fruitful definition of quantization, firstly because most well-understood examples of quantization fit into it (Rieffel, 1994; Landsman, 1998), and secondly because it has suggested various fascinating new ones (Rieffel, 1989a; Natsume\&\  Nest, 1999; ÊNatsume,  Nest, \&\  Ingo, 2003). Restricting ourselves to the former, we note, for example, that \er{b2} with \er{pqcohst} defines a  deformation quantization of the phase space $\R^{2n}$ (with standard Poisson bracket) if one takes $\CA_{\hbar}$ to be the \ca\ of compact operators on the \Hs\ $L^2(\R^n)$. This is called the {\it Berezin quantization} of $\R^{2n}$ (as a phase space);\footnote{ In the literature, Berezin quantization on $\R^{2n}$ is often called anti-Wick quantization (or ordering), whereas on compact complex manifolds it is sometimes called Toeplitz or
Berezin--Toeplitz quantization. Coherent states based on other phase spaces often define deformation quantizations as well; see Landsman, 1998.}
 explicitly, for $\Phi\in L^2(\R^n)$ one has
  \beq \qb(f)\Phi(x)=\int_{\R^{2n}} \frac{d^np
d^nqd^ny}{(2\pi\hbar)^n}\, f(p,q) \ovl{\Psi_{(p,q)}^{\hbar}(y)} \Phi(y) \Psi_{(p,q)}^{\hbar}(x) . \label{qbttsrex} \eeq
This quantization has the distinguishing feature of positivity,\footnote{Cf.\ footnote  \ref{BQP}. As a consequence, \er{qbttsrex} is valid not only for $f\in \cci(\R^{2n})$, but even for all $f\in L^{\infty}(\R^{2n})$, and the extension 
of $\qb$ from $\cci(\R^{2n})$ to $L^{\infty}(\R^{2n})$ is continuous.
} a property not shared by its more famous sister called {\it Weyl quantization}.\footnote{The original reference is Weyl (1931). See, for example, Dubin, Hennings, \&\ Smith (2000) 
and Esposito, Marmo, \&\ Sudarshan (2004)  
for a modern physics-oriented yet mathematically rigorous treatment. See also Rieffel (1994) and Landsman (1998) for a discussion from the perspective of deformation quantization.}
The latter is a deformation quantization of $\R^{2n}$ as well, having the same continuous field of \ca s, but differing from  Berezin quantization in its quantization map
\beq \qw(f)\Phi(x)=\int_{\R^{2n}}
 \frac{d^npd^nq}{(2\pi\hbar)^n}\, e^{ip(x-q)/\hbar}f\left(p,\half(x+q)
 \right)\Phi(q). \label{defweylq} \eeq
 Although it lacks good positivity and hence continuity properties,\footnote{Nonetheless, Weyl quantization may be extended from $\cci(\R^{2n})$ to much larger function spaces using techniques from the theory of distributions (leaving the \Hs\ setting typical of \qm). The classical treatment is in H\"{o}rmander (1979, 1985a).} 
  Weyl quantization enjoys better symmetry properties than Berezin quantization.\footnote{\label{GCQQ} Weyl quantization is covariant under the affine symplectic group $\mathrm{Sp}(n,\R)\ltimes \R^{2n}$, whereas Berezin quantization is merely covariant under its subgroup $\mathrm{O}(2n)\ltimes \R^{2n}$.} Despite these differences, which  illustrate the lack of uniqueness of concrete quantization procedures, Weyl and Berezin quantization both reproduce Schr\"{o}dinger's position and momentum operators \er{SOP}.\footnote{This requires a formal extension of the maps $\qh^W$ and $\qh^B$ to unbounded functions on $M$ like $p_j$ and $q^j$.}  
Furthermore, if $f\in L^1(\R^{2n})$, then $\qb(f)$ and $\qw(f)$ are trace class, with
\beq
\Tr \qb(f)=\Tr \qw(f)=\int_{\R^{2n}}
 \frac{d^npd^nq}{(2\pi\hbar)^n}\, f(p,q).\eeq
 Weyl and  Berezin quantization are related by 
  \beq
  \qb(f)=\qw(e^{\frac{\hbar}{4}\Delta_{2n}}f),
  \eeq
  where $\Delta_{2n}=\sum_{j=1}^n (\partial^2/\partial p_j^2 + \partial^2/\partial (q^j)^2)$, from which it may be shown that Weyl and Berezin  quantization
 are {\it asymptotically equal} in the sense that  for any $f\in \cci(\R^{2n})$ one has 
 \beq \lim_{\hbar\rightarrow 0} \|\qh^B(f)-\qh^W(f)\|=0. \label{WBEQ}\eeq

 Mackey's approach to quantization also finds its natural home in the setting of deformation quantization. Let a Lie group $G$ act on a manifold $Q$,  interpreted as a {\it configuration space}, as in  Subsection \ref{Mackey}.  It turns out that the corresponding classical {\it phase space} is the manifold $\g^*\times Q$, equipped with  the so-called {\it semidirect product  Poisson structure}  (Marsden,  Ra\c tiu \&\  Weinstein, 1984; Krishnaprasad \&\ Marsden,  1987).   Relative to a basis $(T_a)$ of the Lie algebra $\g$ of $G$ with structure constants $C_{ab}^c$ (i.e.\ $[T_a,T_b]=\sum_c C_{ab}^cT_c$), the Poisson bracket in question is given by \beq \{f,g\} =  \sum_a\left(\xi^M_a f \frac{\partial g}{\partial\theta_a} -
\frac{\partial f}{\partial\theta_a}\xi^M_a g\right)-\sum_{a,b,c} C_{ab}^c
\theta_c \frac{\partial f}{\partial\theta_a} \frac{\partial
g}{\partial\theta_b},\eeq
where $\xi_a^M=\xi_{T_a}^M$. To illustrate the meaning of this lengthy expression, we consider a few  special
cases. First, take  $f=X\in\g$ and $g=Y\in\g$  (seen as linear functions on the dual $\g^*$). This yields
\beq \{X,Y\}=-[X,Y]. \label{PB1}\eeq
Subsequently, assume that  $g$ depends on position $q$ alone.
This leads to
\beq \{X,g\}=-\xi^M_X g. \label{PB2} \eeq Finally, assume that
 $f=f_1$ and $g=f_2$ depend on $q$ only; this clearly gives
\beq \{f_1,f_2\}=0. \label{PB3}\eeq   

The two simplest physically relevant examples, already considered at the quantum level in Subsection \ref{Mackey}, are as follows. First,  take $G=\R^n$ (as a  Lie group) and $Q=\R^n$ (as a manifold), with $G$ acting on $Q$ by translation. Eqs.\ \er{PB1} - \er{PB3} then yield the Poisson brackets $\{p_j,p_k\}=0$, $\{p_j,q^k\}=\dl_j^k$, and $\{q^j,q^k\}=0$, showing that in this case $M=\g^*\x Q=\R^{2n}$ is the standard phase space of a particle moving in $\R^n$; cf.\ \er{PBRN}.  Second, the case $G=E(3)$ and $Q=\R^3$ yields a phase space
$M=\R^3\x\R^6$, where $\R^6$ is the phase space of a spinless particle just considered, and $\R^3$ is an additional internal space containing spin as a classical degree of freedom. Indeed, beyond the Poisson brackets on $\R^6$ just described, \er{PB1} - \er{PB3} give rise to the additional  Poisson brackets $\{J_i,J_j\}=\ep_{ijk}J_k$, $\{J_i,p_j\}=\ep_{ijk}p_k$, and $\{J_i,q^j\}=\ep_{ijk}q^k$.\footnote{These are the classical counterparts of the 
  commutation relations for angular momentum  written down in footnote  \ref{AMCCR}.}

The analogy between \er{PB1}, \er{PB2},  \er{PB3} on the one hand, and  \er{Gccr},  \er{Gccr2},  \er{Gccr3}, respectively, on the other, is no accident: the Poisson brackets in question {\it are} the  classical counterpart  of the commutation relations just referred to. This observation is made precise by the fundamental theorem relating Mackey's systems of imprimitivity to deformation quantization (Landsman, 1993, 1998): one can equip the family of \ca s 
\begin{eqnarray}
\CA_0 &=& C_0(\g^*\times Q); \nn\\
\CA_{\hbar} &=& C^*(G,Q),
\end{eqnarray}
  where $C^*(G,Q)$ is the transformation grouo \ca\ defined by the given $G$-action on $Q$ (cf.\ the end of Subsection \ref{Mackey}), with the structure of a continuous field, and one can define quantization maps $\qh:\cci(\g^*\times Q)\raw C^*(G,Q)$ so as to obtain a deformation quantization of the phase space $\g^*\x Q$. It turns out that for special functions of the type $X,Y\in\g$, and $f=f(q)$ just considered, the equality
  \begin{equation}
\frac{i}{\hbar}[\CQ_{\hbar}(f),\CQ_{\hbar}(g)]-\CQ_{\hbar}(\{f,g\}) =0 \label{Diracexact}
\end{equation}
  holds  exactly (and  not merely asymptotically for $\hbar\raw 0$, as required in the fundamental axiom \er{Dirac} for deformation quantization).
  
 This result clarifies the status of Mackey's quantization by systems of imprimitivity. The classical theory underlying the relations \er{impr} is not the usual phase space $T^*Q$ of a structureless particle moving on $Q$, but $M=\g^*\times Q$. For simplicity we restrict ourselves to the transitive case $Q=G/H$ (with canonical left $G$-action). Then $M$ coincides with $T^*Q$ only when $H=\{e\}$ and hence $Q=G$;\footnote{For a Lie group $G$ one has $T^*G\cong \g^*\x G$.} in general, the phase space $\g^*\times (G/H)$ is {\it locally} of the form $T^*(G/H)\x\h^*$ (where $\h^*$ is the dual of the Lie algebra of $H$). The internal degree of freedom described by $\h^*$ is a generalization of classical spin, which, as we have seen, emerges in the case $G=E(3)$ and $H=SO(3)$. All this is merely a special case of a vast class of deformation quantizations described by Lie groupoids; see Bellisard \&\ Vittot (1990),  Landsman (1998, 1999b, 2005b) and Landsman \&\ Ramazan (2001).\footnote{A similar analysis can be applied to Isham's (1984) quantization scheme mentioned in footnote \ref{Isham1}. The unitary \irrep s of the canonical group $G_c$ stand in bijective correspondence with the nondegenerate \rep s of the group \ca\ $C^*(G_c)$ (Pedersen, 1979), which is a deformation quantization of 
 the Poisson manifold $\mathfrak{g}^*_c$ (i.e.\  the dual  of the Lie algebra of $G_c$).
This Poisson manifold contains the coadjoint orbits of $G_c$ as ``irreducible" classical phase spaces, of which only one is the cotangent bundle $T^*(G/H)$ one initially thought one was quantizing (see Landsman (1998) for the classification of the coadjoint orbits of semidirect products). All other orbits are mere lumber that one should avoid. See also Robson (1996). If one is ready for groupoids, there is no need for the canonical group approach. \label{Isham2}}
\subsection{Geometric quantization}\label{GQsection}
Because of its use of abstract \ca s, deformation quantization is a fairly sophisticated and recent technique.  Historically, it was preceded by a more concrete and traditional approach called {\it \gq}.\footnote{\label{GQrefs} Geometric quantization was independently introduced by  Kostant (1970) and Souriau (1969).  Major later treatments on the basis of the original formalism are  Guillemin  \&\ Sternberg (1977), \'{S}niatycki (1980),  Kirillov (1990),  Woodhouse (1992), Puta (1993), Chernoff (1995), Kirillov (2004), and Ali \& Englis (2004). The modern era (based on the use of Dirac operators and $K$-theory) was initiated  by unpublished remarks by Bott in the early 1990s; see Vergne (1994) and Guillemin, Ginzburg \&  Karshon (2002).
 The postmodern (i.e.\ functorial) epoch was launched  in Landsman (2005a).}
 Here the goal is to firstly ``quantize" a phase space $M$ by a concretely given Hilbert space $\H(M)$, and secondly to map the classical observables (i.e.\ the real-valued smooth functions on $M$) into self-adjoint operators on $\H$ (which after all play the role of observables in von Neumann's formalism of \qm).\footnote{In \gq\ phase spaces are always seen as symplectic manifolds (with the sole exception of Vaisman,  1991); the reason why it is unnatural to start with the more general class of Poisson manifolds will become clear in the next subsection.} In  principle, this program should align \gq\ much better with the fundamental role unbounded self-adjoint operators play in \qm\ than deformation quantization, but in practice \gq\ continues to be plagued by  problems.\footnote{\label{GQP1} Apart from rather technical issues concerning the domains and  self-adjointness properties of the operators defined by \gq, the main point is that the various mathematical choices one has to make in the \gq\ procedure
 cannot all be justified by physical arguments, although the physical properties of the theory depend on these choices. (The notion of a polarization is the principal case in point; see also footnote \ref{GQP2} below.) Furthermore, as we shall see, one cannot quantize sufficiently many functions in standard \gq. Our functorial approach to \gq\ in Subsection \ref{EFQ} was partly invented to alleviate these problems.}
  However, it would be wrong to see deformation quantization and \gq\ as {\it competitors}; as we shall see in the next subsection, they are natural {\it allies}, forming ``complementary" parts of a conjectural quantization functor. 

In fact, in our opinion \gq\ is best compared and contrasted with phase space quantization in its concrete formulation of Subsection \ref{PSQ} (i.e.\ before its \ca ic abstraction and subsequent absorption into deformation quantization as indicated in Subsection \ref{DQsection}).\footnote{See also Tuynman (1987).} For \gq\ equally well starts with the \Hs\ $L^2(M)$,\footnote{Defined with respect to the Liouville measure times a suitable factor $c_{\hbar}$, as in \er{qhnorm} etc.;  in \gq\ this factor is not very important, as it is unusual to study the limit $\hbar\raw 0$. For $M=\R^{2n}$ the measure on $M$ with respect to which $L^2(M)$ is defined is $d^npd^nq/(2\pi\hbar)^n$.
} and subsequently attempts to construct $\H(M)$ from it, though typically in a different way from \er{pH}. 

Before doing so, however, the \gq\ procedure first tries to define a linear map $\qp$ from $\cin(M)$ to the class of (generally unbounded) operators on $L^2(M)$ that formally satisfies 
 \begin{equation}
\frac{i}{\hbar}[\qp(f),\qp(g)]-\qp(\{f,g\}) =0, \label{Diracexact2}\eeq
i.e.\ 
\er{Diracexact} with $\CQ=\qp$, as well as the nondegeneracy property \beq
\qp(\ch_M)=1, \label{ndGQ}\eeq where $\ch_M$ is the function on $M$ that is identically equal to 1, and the 1 on the right-hand side is the unit operator on $L^2(M)$. Such a  map is called {\it prequantization}.\footnote{The idea of prequantization predates \gq; see van Hove (1951)
and Segal (1960).} 
For $M=\R^{2n}$ (equipped with its standard Poisson bracket 
\er{PBRN}), a prequantization map is given (on $\Phi\in L^2(M)$) by
\beq \qp(f)\Phi=-i\hbar \{f,\Phi\} +\left(f-\sum_j p_j \frac{\partial f}{\partial p_j}\right)\Phi.
 \label{PQRN}\eeq 
This expression is initially defined for $\Phi\in\cci(M)\subset L^2(M)$, on which domain  $\qp(f)$ is symmetric when $f$ is real-valued;\footnote{An operator $A$ defined on a dense subspace $\mathcal{D}\subset\H$ of a \Hs\ $\H$ is called {\it symmetric} when $(A\Psi,\Phi)=(\Psi,A\Phi)$ for all $\Psi,\Phi\in\mathcal{D}$.} 
 note that the operator in question is unbounded even when $f$ is bounded.\footnote{As mentioned, self-adjointness is a problem in \gq; we will not address this issue here. Berezin quantization is much better behaved than \gq\ in this respect, since it maps bounded functions into bounded operators.} This looks complicated; the simpler expression $\qh(f)\Phi=-i\hbar \{f,\Phi\}$, however, would satisfy \er{Diracexact} but not \er{ndGQ}, and the goal of the second term in \er{PQRN} is to satisfy the latter condition while preserving the former.\footnote{One may criticize the \gq\ procedure for emphasizing \er{Diracexact2} against its equally natural counterpart $\CQ(fg)=\CQ(f)\CQ(g)$, which fails to be satisified by $\qp$ (and indeed by any known quantization procedure, except the silly $\CQ(f)=f$ (as a multiplication operator on $L^2(M)$).} 
For example, one has
\begin{eqnarray}
\qp(q^k) & =& q^k+i\hbar \frac{\partial }{\partial p_k} \nn \\ 
\qp(p_j) & =& -i\hbar \frac{\partial}{\partial q^j}.\label{SOP2}
\end{eqnarray}

For general  phase spaces $M$ one may construct a map $\qp$ that satisfies \er{Diracexact2} and \er{ndGQ} when $M$ is ``prequantizable"; a full explanation of this notion requires some differential geometry.\footnote{\label{GQF}A symplectic manifold $(M,\om)$ is called {\it prequantizable} at some fixed value of $\hbar$ when it admits a complex line bundle $L\raw M$ (called the {\it prequantization line bundle}) with connection $\nabla$
such that $F=-i\om/\hbar$ (where $F$ is the curvature of the connection, defined by
$F(X,Y)=[\nabla_X,\nabla_Y]-\nabla_{[X,Y]}$).  This is the case iff $[\om]/2\pi\hbar\in H^2(M,\Z)$, where $[\om]$ is the  de Rham cohomology class of the symplectic form. 
If so, prequantization is defined by the formula $\qp(f)=-i\hbar\nabla_{\xi_f} +f$, where
$\xi_f$ is the Hamiltonian vector field of $f$ (see Subsection \ref{DQsection}). This expression is defined and symmetric on the space $\cci(M,L)\subset L^2(M)$ of compactly supported smooth sections of $L$, and is easily checked to satisfy \er{Diracexact2} and \er{ndGQ}.
To obtain \er{PQRN} as a special case, note that for $M=\R^{2n}$ with the canonical symplectic form $\om =\sum_k dp_k\wedge dq^k$ one has $[\om]=0$, so that $L$ is the trivial bundle $L= \R^{2n}\x\C$. The connection $\nabla=d+A$ with
$A= -\frac{i}{\hbar}\sum_k p_kdq^k$ satisfies $F=-i\om/\hbar$, and this eventually yields \er{PQRN}.} Assuming this to be the case, then
for one thing prequantization is a very effective tool in constructing unitary group \rep s 
of the kind that are interesting for physics. Namely, suppose a Lie group $G$ acts on the phase space $M$ in ``canonical" fashion. This means that there exists a map $\mu:M\raw\g^*$, called the {\it momentum map}, such that $\xi_{\mu_X}=\xi^M_X$ for each $X\in\g$,\footnote{ Here $\mu_X\in\cin(M)$ is defined by $\mu_X(x)=\langle \mu(x),X\rangle$, and $\xi^M_X$ is the  vector field on $M$ defined by the $G$-action (cf.\  footnote  \ref{VFFN}). Hence this condition means that $\{\mu_X,f\}(y)=d/dt_{|t=0} [f(\exp(-tX)y)]$ for all $f\in\cin(M)$ and all $y\in M$.} and in addition
$\{\mu_X,\mu_Y\}=\mu_{[X,Y]}$. See Abraham \& Marsden (1985), Marsden \&\ Ratiu (1994), Landsman (1998), Butterfield (2005), etc. 
On then obtains a \rep\ $\pi$ of the Lie algebra $\g$ of $G$ by skew-symmetric unbounded operators on $L^2(M)$ through
\beq \pi(X)=-i\hbar \qp(\mu_X), \label{LArep}\eeq
which often exponentiates to a unitary \rep\ of $G$.\footnote{An operator $A$ defined on a dense subspace $\mathcal{D}\subset\H$ of a \Hs\ $\H$ is called {\it skew-symmetric} when $(A\Psi,\Phi)=-(\Psi,A\Phi)$ for all $\Psi,\Phi\in\mathcal{D}$. If one has a unitary \rep\ $U$ of a Lie group $G$ on $\H$, then the derived \rep\ $dU$ of the Lie algebra $\g$ (see footnote  \ref{dUFN}) consists of skew-symmetric operators, making one hopeful that a given \rep\ of $\g$  by skew-symmetric operators can be integrated (or exponentiated) to a unitary \rep\ of $G$. See  Barut \&\  Ra\c{c}ka (1977) or J\o rgensen \&\ Moore (1984) and references therein.} 

As the name suggests, prequantization is not yet quantization. For example, the prequantization of $M=\R^{2n}$ does not reproduce Schr\"{o}dinger's wave mechanics: the operators \er{SOP2} are not unitarily equivalent to \er{SOP}. In fact, as a carrier of the \rep\ \er{SOP2} of the canonical commutation relations \er{ccr},
the \Hs\ $L^2(\R^{2n})$ contains $L^2(\R^n)$ (carrying the \rep\ \er{SOP}) with infinite multiplicity (Ali \&\ Emch, 1986). This situation is often expressed by the statement that ``prequantization is reducible" or that the prequantization \Hs\ $L^2(M)$ is `too large', but both claims are misleading: $L^2(M)$ is actually {\it ir}reducible under the action of $\qp(\cin(M))$ (Tuynman, 1998), and saying that for example $L^2(\R^n)$ is ``larger" than $L^2(\R^n)$ is unmathematical  in view of the unitary isomorphism of these \Hs s.
What is true is that in typical examples  $L^2(M)$ is generically reducible under the action of some Lie algebra where one would like it to be irreducible.  This applies, for example, to \er{SOP}, which defines a \rep\ of the Lie algebra of the Heisenberg group. More generally,  in the case where  a phase space $M$ carries a transitive action of a Lie group $G$, so that one would expect the quantization of this $G$-action by unitary operators on a \Hs\ to be irreducible, $L^2(M)$ is typically highly reducible under the \rep\ \er{LArep} of $\g$.\footnote{This can be made precise in the context of the so-called orbit method, cf.\ 
the books cited in footnote  \ref{GQrefs}.}

Phase space quantization encounters this problem as well. Instead of  the complicated expression \er{PQRN}, through \er{berezin} it simply ``phase space prequantizes"
 $f\in\cin(M)$ on $L^2(M)$ by $f$ as a multiplication operator.\footnote{For unbounded $f$ this operator is defined on the set of all $\Phi\in L^2(M)$ for which $f\Phi\in L^2(M)$.}  Under this action of $\cin(M)$ the \Hs\ $L^2(M)$ is of course highly reducible.\footnote{\label{Koopman} Namely, each (measurable) subset $E\subset M$ defines a projection $\ch_E$,  and  $\ch_E L^2(M)$ is stable under all multiplication operators $f$. One could actually decide not to be bothered by this problem and stop here, but then one is simply doing classical mechanics in a \Hs\ setting (Koopman, 1931). This formalism  even turns out to be quite useful for ergodic theory (Reed \&\ Simon, 1972).} 
The identification of an appropriate subspace 
 \beq \H(M)=pL^2(M) \label{pl2}\eeq
 of $L^2(M)$ (where $p$ is a projection) as the \Hs\  carrying the ``quantization" of $M$ (or rather of $\cin(M)$) may be seen as a solution to this reducibility problem,
 for  if the procedure is successful, the projection $p$ is chosen such that $pL^2(M)$ is irreducible under $p\cin(M)p$. Moreover, in this way practically any function on $M$ can be quantized, albeit at the expense of  \er{Diracexact} (which, as we have seen, gets replaced by its asymptotic version \er{Dirac}). See Subsection \ref{SE} for a discussion of 
reducibility versus irreducibility of \rep s of algebras of observables in classical and quantum theory. 
 
 We restrict our treatment of geometric quantization to situations where  it adopts the same strategy as above, in  assuming that  the final \Hs\ has the form \er{pl2}  as well.\footnote{\label{GQP2} Geometric quantization 
has traditionally been based on the notion of a polarization (cf.\ the references
in footnote \ref{GQrefs}).  This device produces a final \Hs\ $\H(M)$ which may not be a subspace of $L^2(M)$, except in the so-called (anti-) holomorphic case.}  But it crucially
differs from phase space quantization in that its first step is  \er{PQRN} (or its generalization to more general phase spaces) rather than just having 
$f\Phi$ on the right-hand side.\footnote{It  also differs from phase space quantization in the ideology that the projection $p$ ought to be constructed  solely from the geometry of $M$: hence the name `\gq'.} Moreover, in \gq\ one merely quantizes a {\it subspace} of the set $\cin(M)$ of classical observables, consisting of those functions that satisfy
\beq [\qp(f),p]=0. \label{GQf}\eeq
If a function $f\in\cin(M)$ satisfies this condition, then  one defines the ``\gq" of $f$ as 
\beq \qh^G(f)=\qp(f)\rst \H(M).\label{QGQ}\eeq
This is well defined, since because of \er{GQf} the operator
$\qp(f)$ now maps $pL^2(M)$ onto itself. Hence \er{Diracexact} holds 
 for $\qh=\qh^G$  because of \er{Diracexact2}; in \gq\ one simply refuses to quantize functions for which  \er{Diracexact} is {\it not} valid.
 
Despite some impressive initial triumphs,\footnote{Such as the orbit method for nilpotent groups and the newly understood Borel--Weil method for compact groups, cf.\ Kirillov (2004) and most other books cited in footnote  \ref{GQrefs}.}  there is no general method that accomplishes the goals of \gq\  with guaranteed success. Therefore, \gq\ has remained something like a hacker's tool, whose applicability largely depends on the creativity of the user.

In any case, our familiar example $M=\R^{2n}$ is well understood, and we illustrate the general spirit of the method in its setting, simplifying further by taking $n=1$.  It is convenient to replace the canonical coordinates $(p,q)$ on $M$ by $z=p+iq$ and $\ovl{z}=p-iq$, and the mathematical toolkit of \gq\ makes it very natural to look at the space of solutions within $L^2(\R^{2})$ of the equations\footnote{Using the formalism explained in footnote  \ref{GQF}, we replace the 1-form $A= -\frac{i}{\hbar}\sum_k p_kdq^k$ defining the connection $\nabla=d+A$ by the gauge-equivalent form $A=\frac{i}{2\hbar}(\sum_k q^kdp_k -\sum_k p_kdq^k)=-\frac{i}{\hbar} \sum_k p_kdq^k+\frac{i}{2\hbar}d (\sum_k p_kq^k)$, which has the same curvature. In terms of this new $A$, which in complex coordinates reads
$A=\sum_k(z_kd\ovl{z}_k-\ovl{z}_kdz_k)/4\hbar$,
eq.\ \er{holz} is just $\nabla_{\partial/\partial\ovl{z}}\Phi=0$. This is an example of the so-called holomorphic polarization in the formalism of \gq.}
\beq
\left(\frac{\partial}{\partial\ovl{z}}+\frac{z}{4\hbar} \right)\Phi(z,\ovl{z})=0.\label{holz}
\eeq
 The general solution of these equations that lies in $L^2(\R^{2})=L^2(\C)$ is
\beq \Phi(z,\ovl{z})=e^{-|z|^2/4\hbar}f(z), \eeq
where $f$ is a holomorphic function such that 
\beq \int_{\C} \frac{dz d
 \ovl{z}}{2\pi \hbar i}e^{-|z|^2/2\hbar}|f(z)|^2<\infty. \label{BFC}\eeq
 The projection $p$, then, is the projection onto the closed subspace of  $L^2(\C)$ consisting of these solutions.\footnote{\label{BFFN}
 The collection of all holomorphic functions on $\C$ satisfying \er{BFC} is a \Hs\  with respect to the inner product $(f,g)=(2\pi \hbar i)^{-1} \int_{\C} dz d\ovl{z}\, \exp(-|z|^2/2\hbar) \ovl{f(z)}g(z)$,
 called the {\it Bargmann--Fock space} $\H_{BF}$. This space may be embedded in $L^2(\C)$ by $f(z) \mapsto \exp(-|z|^2/2\hbar)f(z)$, and the image
 of this embedding is of course just $pL^2(\C)$.}
 The \Hs\  $pL^2(\C)$ is unitarily equivalent to $L^2(\R)$ in a natural way (i.e.\ without the choice of a basis). The condition \er{GQf} boils down to $\partial^2 f(z,\ovl{z})/\partial \ovl{z}_i \partial \ovl{z}_j=0$; in particular, the coordinate functions $q$ and $p$ are quantizable. Transforming to $L^2(\R)$, one finds that the operators $\qh^G(q)$ and $\qh^G(p)$ coincide with Schr\"{o}dinger's expressions \er{SOP}. In particular, the Heisenberg group $H_1$, which acts with infinite multiplicity on $L^2(\C)$, acts irreducibly on $pL^2(\C)$. 
\subsection{Epilogue: functoriality of quantization}\label{EFQ}
A very important aspect of quantization is its interplay with symmetries and constraints. 
Indeed, the fundamental theories describing Nature (viz.\ electrodynamics,  Yang--Mills theory, general relativity, and possibly also string theory)  are a priori formulated as constrained systems. The classical side of constraints and reduction is well understood,\footnote{See  Gotay,
Nester, \&\  Hinds (1978),  Binz,
 \'{S}niatycki and  Fischer (1988),  Marsden (1992), Marsden \&  Ratiu (1994), 
  Landsman (1998), Butterfield (2005), and
Belot (2005).} a large class of important examples being codified by the procedure of symplectic reduction. A special case of this is {\it Marsden--Weinstein reduction}: if a Lie group $G$ acts on a phase space $M$ in canonical fashion with momentum map $\mu:M\raw\g^*$ (cf.\ Subsection \ref{GQsection}), one may form another phase space $M/\hspace{-1mm}/G=\mu\inv(0)/G$.\footnote{Technically, $M$ has to be a symplectic manifold, and if $G$ acts properly and freely on $\mu\inv(0)$, then $M/\hspace{-1mm}/G$ is again a symplectic manifold.} Physically,
in the case where $G$ is a gauge group and $M$ is the unconstrained phase space, $\mu\inv(0)$ is the constraint hypersurface (i.e.\ the subspace of $M$ on which the constraints defined by the gauge symmetry hold), and $M/\hspace{-1mm}/G$ is the true phase space of the system that only contains physical degrees of freedom. 

Unfortunately, the correct way of dealing with constrained quantum systems remains a source of speculation and controversy:\footnote{Cf.\ Dirac (1964), Sundermeyer (1982), Gotay (1986), Duval et al. (1991), Govaerts (1991), Henneaux \&\ Teitelboim (1992), and Landsman (1998) for various perspectives on the quantization of constrained systems.} practically all rigorous results on quantization (like the ones discussed in the preceding subsections) concern unconstrained systems. Accordingly, one would like to quantize a constrained system by reducing the problem to the unconstrained case. This could be done provided the following scenario applies. 
One first quantizes the unconstrained phase space $M$ (supposedly the easiest part of the problem), ans subsequently imposes a quantum version of symplectic reduction. Finally, one proves by abstract means that the quantum theory thus constructed is equal to the theory defined by first reducing at the classical level and then quantizing the constrained classical phase space (usually an impossible task to perform in practice). 

Tragically, sufficiently powerful theorems stating that ``quantization commutes with reduction" in this sense remain elusive.\footnote{\label{GSC} The so-called Guillemin--Sternberg conjecture (Guillemin \&\ Sternberg, 1982) - now a theorem
(Meinrenken, 1998, Meinrenken \&\ Sjamaar, 1999) - merely deals with the case of Marsden--Weinstein reduction where $G$ and $M$ are compact. Mathematically impressive as the  ``quantization commutes with reduction" theorem already is here, it is a far call from gauge theories, where the groups and spaces are not only noncompact but even infinite-dimensional.}
So far, this has blocked, for example, a rigorous quantization of Yang--Mills theory in dimension 4; this is one of the Millenium Problems of the Clay Mathematical Institute, rewarded with 1 Million dollars.\footnote{See \texttt{http://www.claymath.org/millennium/}}

On a more spiritual note, the mathematician 
E. Nelson famously said that `First quantization is a mystery, but second quantization is a functor.' The functoriality of `second' quantization
(a construction involving Fock spaces, see Reed \&\ Simon, 1975) being an almost trivial matter, the deep mathematical and conceptual problem lies in the possible functoriality of `first' quantization, which  simply means quantization in the sense we have been discussing so far. This was initially  taken to mean that canonical transformations $\al$ of the phase space $M$ should be `quantized' by unitary operators $U(\al)$ on $\H(M)$, in such a way 
$U(\al)\qh(f)U(\al)\inv=\CQ(L_{\al}f)$ (cf.\ \er{Gcov}). This is possible only in special circumstances, e.g., when $M=\R^{2n}$ and $\al$ is a linear symplectic map, and more generally when $M=G/H$ is homogeneous and $\al\in G$ (see the end of Subsection \ref{PSQ}).\footnote{ Canonical transformations {\it can} be quantized in approximate sense that becomes precise as $\hbar\raw 0$ by means of so-called Fourier integral operators; see
H\"{o}rmander (1971, 1985b) and Duistermaat (1996).}
Consequently, the functoriality of quantization is widely taken to be a dead end.\footnote{See   Groenewold (1946), van Hove (1951),  Gotay, Grundling, \&\ Tuynman (1996), and Gotay (1999).} 

However, all no-go theorems establishing this conclusion start from wrong and naive categories, both on the classical and on the quantum side.\footnote{Typically, one takes 
the classical category to consist of symplectic manifolds as objects and symplectomorphisms as arrows, and the quantum category to have \ca s as objects and automorphisms as arrows.}  It appears very likely that one may indeed  make quantization functorial by a more sophisticated choice of categories, with the additional bonus that deformation quantization and \gq\ become unified: the former is the object part of the quantization functor, whereas the latter (suitably reinterpreted) is the arrow part. Amazingly, on this formulation the statement that  `quantization commutes with reduction' becomes a special case of the functoriality of quantization (Landsman, 2002, 2005a).

To explain the main idea, we return to the \gq\ of $M=\R^2\cong\C$ 
explained in the preceding subsection. The identification of $pL^2(\C)$\footnote{Or the Bargmann--Fock space $\H_{BF}$, see footnote  \ref{BFFN}.}  as the correct \Hs\ of the problem may be understood in a completely different way, which paves the way for the powerful reformulation of the \gq\ program that will  eventually define the quantization functor. Namely, $\C$  supports a certain linear first-order differential operator $\DS$ that is entirely defined by its geometry as a phase space, called the {\it Dirac operator}.\footnote{\label{DSFN} Specifically, this is the so-called $\spinc$ Dirac operator defined by the complex structure of $\C$, {\it coupled to the prequantization line bundle}. See   Guillemin,  Ginzburg, \&  Karshon (2002).}
This operator is given by\footnote{Relative to the Dirac matrices $\gamma^1=\left( \begin{array}{cc} 0 & i \\ i & 0\end{array}
\right)$ and $\gamma^2=\left( \begin{array}{cc} 0 & -1 \\ 1 & 0\end{array}
\right)$.}
\beq \DS=2\left( \begin{array}{cc}
0&  -\frac{\partial}{\partial z}+\frac{\ovl{z}}{4\hbar}\\ \frac{\partial}{\partial\ovl{z}} +\frac{z}{4\hbar} & 0
\end{array}
\right),\eeq 
acting on $L^2(\C)\ot\C^2$ (as a suitably defined unbounded operator).
This operator
has the generic form $$\DS=\left( \begin{array}{cc} 0&\DS_-\\ \DS_+& 0\end{array}
\right).$$ The {\it index} of  such an  operator  is given by
\beq \mathrm{index}(\DS)=[\ker(\DS_+)]-[\ker(\DS_-)], \label{index}\eeq
where $[\ker(\DS_{\pm})]$ stand for the (unitary) isomorphism class of $\ker(\DS_{\pm})$ {\it seen as a \rep\ space of a suitable algebra of operators}.\footnote{\label{indexFN} 
The left-hand side of \er{index} should really be written as $\ind(\DS_+)$,
since $\coker(\DS_+)=\ker(\DS_+^*)$ and $\DS_+^*=\DS_-$, but since the index is naturally associated to $\DS$ as a whole, we abuse notation in writing $\ind(\DS)$ for $\ind(\DS_+)$. 
The usual index of a linear map $L:V\raw W$ between finite-dimensional vector spaces is defined as
$\ind(L)=\dim(\ker(L))-\dim(\coker(L))$, where $\coker(L)=W/\ran(L)$. Elementary linear algebra yields $\ind(L)=\dim(V)-\dim(W)$. This is surprising because it is independent of $L$, whereas $\dim(\ker(L))$ and $\dim(\coker(L))$ quite sensitively depend on it. For, example,
take $V=W$ and $L=\varepsilon\cdot 1$. If $\varepsilon\neq 0$ then $\dim(\mathrm{ker}(\varepsilon\cdot 1))=\dim(\coker(\varepsilon\cdot 1))=0$,
whereas for $\varepsilon=0$ one has $\dim(\ker(0))=\dim(\coker(0))=\dim(V)$!
Similarly, the usual definiton of \gq\ through \er{holz} etc.\ is unstable against perturbations of the underlying symplectic structure, whereas the refined definition through \er{index} is not. To pass to the latter from the above notion of an index, we first write
$\ind(L)=[\ker(L)]-[\coker(L)]$, where $[X]$ is the isomorphism class of a linear space $X$ as a $\C$-module. This expression is an element of $K_0(\C)$, and we recover the earlier index
through the realization that the class $[X]$ is entirely determined by $\dim(X)$, along with  and the corresponding isomorphism $K_0(\C)\cong\Z$. 
When a more complicated finite-dimensional \ca\ $\CA$ acts on $V$ and $W$ with the property that  $\ker(L)$ and $\coker(L)$ are stable under the $\CA$-action, one may define $[\ker(L)]-[\coker(L)]$ and hence $\ind(L)$ as an element of the so-called \ca ic K-theory group $K_0(\CA)$. Under certain technical conditions, this notion of an index may be generalized to infinite-dimensional \Hs s and \ca s; see Baum, Connes \&\ Higson (1994) and Blackadar (1998). 
The $K$-theoretic index is best understood when $\CA=C^*(G)$ is the group \ca\ of some locally compact group $G$. In the example $M=\R^2$ one might take $G$ to be the Heisenberg group $H_1$, so that $\ind(\DS)\in K_0(C^*(H_1))$.}  In the case at hand, one has
$\ker(\DS_+)=pL^2(\C)$ (cf.\ \er{holz} etc.) and $\ker(\DS_-)=0$, \footnote{Since $(-\frac{\partial}{\partial z}+\frac{\ovl{z}}{4\hbar})\Phi=0$ implies $\Phi(z,\ovl{z})=\exp(|z^2|/4\hbar)f(\ovl{z})$, which lies in $L^2(\C)$ iff $f=0$.} where we  regard $\ker(\DS_+)$ as a \rep\ space of the Heisenberg group $H_1$. Consequently, the \gq\ of the phase space $\C$ is given 
{\it modulo unitary equivalence}  by  $\mathrm{index}(\DS)$, seen as a 
``formal difference" of \rep\ spaces of $H_1$. 

This procedure may be generalized to arbitrary phase spaces $M$,
where $\DS$ is a certain operator naturally defined by the phase space geometry of $M$ and the demands of quantization.\footnote{ Any symplectic manifold carries an almost complex structure compatible with the symplectic form, leading to a $\spinc$ Dirac operator 
as described in footnote  \ref{DSFN}.  See, again,   Guillemin,  Ginzburg, \&  Karshon (2002). If $M=G/H$, or, more generally, if $M$ carries a canonical action of a Lie group $G$ with compact quotient $M/G$, then $\mathrm{index}(\DS)$ defines an element of $K_0(C^*(G))$. 
See footnote  \ref{indexFN}. In complete generality, $\mathrm{index}(\DS)$ ought to be an element of $K_0(\CA)$, where $\CA$ is the \ca\ of observables of the quantum system.} 
This has turned out to be the most promising formulation of \gq\ - at some cost.\footnote{On the benefit side, the invariance of the index under continuous deformations of $\DS$ seems to obviate the ambiguity of traditional quantization procedures with respect to different `operator orderings' not prescribed by the classical theory.}
 For the original goal of quantizing a phase space by a \Hs\ has now been replaced by a much more abstract procedure, in which the result of quantization is a formal difference of certain isomorphism classes of \rep\ spaces of the quantum algebra of observables. To illustrate the  degree of abstraction involved here, suppose  we ignore the action of the observables
 (such as position and momentum in the example just considered). In that case the isomorphism class $[\H]$ of a \Hs\ $\H$ is entirely characterized by its dimension $\dim(\H)$, so that (in case that $\ker(\DS_-)\neq 0$) quantization (in the guise of $\mathrm{index}(\DS)$) can even  be a negative number! Have we gone mad?

Not quite. The above picture of \gq\ is indeed quite irrelevant to physics, {\it unless it is supplemented by deformation quantization}. It is convenient to work at some fixed value of $\hbar$ in this context, so that deformation quantization merely associates some \ca\ $\CA(P)$ to a given phase space $P$.\footnote{Here $P$ is not necessarily symplectic; it may be a Poisson manifold, and to keep Poisson and symplectic manifolds apart we denote the former by $P$ from now on, preserving the notation $M$ for the latter.} Looking for 
a categorical interpretation of quantization, it is therefore natural to assume that the objects of the classical category $\GC$ are phase spaces $P$,\footnote{Strictly speaking, to be an object in this category a Poisson manifold $P$ must be {\it integrable}; see Landsman (2001).} whereas the objects of the quantum category $\GQ$ are \ca s.\footnote{For technical reasons involving $K$-theory these have to be separable.} The object part of the hypothetical quantization functor is to be deformation quantization, symbolically written as $P\mapsto\CQ(P)$.

Everything then fits together if \gq\ is reinterpreted as the arrow part of the conjectural  quantization functor. To accomplish this, the arrows in the classical category $\GC$ should not be taken to be maps between phase spaces, but {\it symplectic bimodules} $P_1\law M\raw P_2$.\footnote{Here $M$ is a symplectic manifold and $P_1$ and $P_2$ are integrable Poisson manifolds; the map $M\raw P_2$ is anti-Poisson, whereas the map $P_1\law M$ is Poisson. Such bimodules (often called {\it dual pairs}) were introduced by Karasev (1989) and Weinstein (1983). In order to occur as arrows in $\GC$, symplectic bimodules have to satisfy a number of regularity conditions (Landsman, 2001).} More precisely, the arrows in $\GC$ are  suitable isomorphism classes of such bimodules.\footnote{This is necessary in order to make arrow composition associative; this is given by a generalization of the symplectic reduction procedure.}
Similarly, the arrows in the quantum category $\GQ$ are not morphisms of \ca s, as might naively be expected, but certain isomorphism classes of bimodules for \ca s, equipped with the additional structure of a generalized Dirac operator.\footnote{The category $\GQ$ is nothing but the category KK introduced by Kasparov, whose objects are  separable \ca s, and whose  arrows are the so-called Kasparov group $KK(A,B)$, composed with Kasparov's product $KK(A,B)\x KK(B,C)\raw KK(A,C)$. See Higson (1990) and Blackadar (1998).} 

Having already defined the object part of the quantization map $\CQ:\GC\raw\GQ$ as deformation quantization, we now propose that the arrow part is \gq, in the sense of a suitable generalization of \er{index}; see Landsman (2005a) for details. We then conjecture that $\CQ$ is a functor; in the cases where this can and has been checked, the functoriality of $\CQ$ is precisely the statement that quantization commutes with reduction.\footnote{A canonical $G$-action on a symplectic manifold $M$ with momentum map $\mu:M\raw\g^*$ gives rise to a dual pair $pt\law M\raw \g^*$, which in $\GC$
is interpreted as an arrow from the space $pt$ with one point to $\g^*$. The composition of this arrow with the arrow $\g^*\hookleftarrow 0\raw pt$ from $\g^*$ to $pt$
is $pt\law M/\hspace{-1mm}/G\raw pt$. If $G$ is connected, functoriality of quantization on these two pairs is equivalent to the Guillemin--Sternberg conjecture (cf.\ footnote  \ref{GSC}); see Landsman (2005a).}

Thus  Heisenberg's idea of {\it Umdeutung}  finds it ultimate realization in  the quantization functor.
\section{The limit $\hbar\raw 0$}\label{S5}\setcounter{equation}{0}
It was recognized at an early stage  that the limit $\hbar\raw 0$ of Planck's constant going to zero should play a role in the explanation of the classical world from quantum theory. Strictly speaking, $\hbar$ is a dimensionful  {\it constant}, but in practice one studies the semiclassical regime of a given quantum theory by forming a dimensionless combination of $\hbar$ and other parameters; this combination then re-enters the theory as if it were a dimensionless version of  $\hbar$ that can indeed be varied. The oldest example is Planck's radiation formula \er{Planck}, with temperature $T$ as the pertinent variable. Indeed, the observation of Einstein (1905) and Planck (1906)
that in the limit $\hbar\nu/kT\raw 0$ this formula converges to the classical equipartition law $E_{\nu}/N_{\nu}=kT$ may well be the first use of the $\hbar\raw 0$ limit of quantum theory.\footnote{Here Einstein (1905) put  
$\hbar\nu/kT\raw 0$ by letting $\nu\raw 0$ at fixed $T$ and $\hbar$, whereas
Planck (1906) took $T\raw \infty$ at fixed $\nu$ and  $\hbar$.}

Another example is the Schr\"{o}dinger equation \er{Schreq} with Hamiltonian
$H=-\frac{\hbar^2}{2m}\Delta_x +V(x)$, where $m$ is the mass of the pertinent particle. Here one may pass to dimensionless parameters by introducing 
an energy scale  $\epsilon$ typical of $H$, like $\epsilon=\sup_x |V(x)|$,  as well as a typical  length scale  $\lm$, such as $\lm=\epsilon/\sup_x |\nabla V(x)|$  (if these quantities are finite).
 In terms of the dimensionless variable $\til{x}=x/\lm$, the rescaled Hamiltonian $\til{H}=H/\epsilon$ is then dimensionless  and equal to
   $\til{H}=-\til{\hbar}^2\Delta_{\til{x}} +\til{V}(\til{x})$, where
$\til{\hbar}=\hbar/\lm\sqrt{2m\epsilon}$ and $\til{V}(\til{x})=V(\lm\til{x})/\epsilon$.
Here $\til{\hbar}$ is dimensionless, and one might study the regime where it is small (Gustafson \&\ Sigal, 2003).
Our last example will occur in the theory of large quantum systems, treated in the next Section.  In what follows, whenever it is considered variable $\hbar$ will denote such a dimensionless version of  Planck's constant. 
 
Although, as we will argue, the limit $\hbar\raw 0$ cannot  by itself explain the classical world, it does give rise to a number of truly pleasing mathematical results. These, in turn, render  almost inescapable the conclusion that the limit in question is indeed a relevant one for the recovery of classical physics from quantum theory. Thus the present section is
meant to be  a catalogue of those pleasantries that might be of direct interest to researchers in the foundations of quantum theory.

There is another, more technical use of the $\hbar\raw 0$ limit, which is to perform computations in \qm\ by approximating the time-evolution of states and observables in terms of associated classical objects. This endeavour is known as {\it semiclassical analysis}. Mathematically, this use of the $\hbar\raw 0$ limit is closely related to the goal of recovering classical mechanics from \qm, but conceptually the matter is quite different. We will attempt to bring the pertinent differences out in what follows. 
\subsection{Coherent states revisited}\label{CSR}
As Schr\"{o}dinger (1926b) foresaw, coherent states play an important role in the 
limit $\hbar\raw 0$.  We recall from Subsection \ref{PSQ} that {\it for some fixed value $\hbar$ of Planck's constant} coherent states in a \Hs\ $\H$ for a phase space $M$ are defined by an injection $M\hraw \H$, $z\mapsto\Psi_z^{\hbar}$, such that \er{normcs} and \er{qhnorm} hold. 
In what follows, we shall say that $\Psi_z^{\hbar}$ is {\it centered at $z\in M$}, a terminology justified by the key example \er{pqcohst}. 

To be relevant to the classical limit, coherent states must satisfy an additional property concerning their dependence on $\hbar$, which
also largely clarifies their nature (Landsman, 1998). Namely, we require that for each $f\in C_c(M)$ and each $z\in M$ the following  function 
from the set $I$ in which $\hbar$ takes values (i.e.\ usually $I=[0,1]$, but in any case containing zero as an accumulation point) to $\C$ is continuous:
\begin{eqnarray}
\hbar &\mapsto&  c_{\hbar} \int_M  d\mu_L(w)\,|(\Psi_w^{\hbar},\Psi_z^{\hbar})|^2 f(w)\:\:\:(\hbar>0);\label{chca} \\
0 &\mapsto& f(z).  \label{chcb}
\end{eqnarray}
 In view of \er{b2},  the right-hand side of \er{chcb} is the same as 
$(\Psi_z^{\hbar},\qh^B(f)\Psi_z^{\hbar})$. In particular, this continuity  condition implies
\beq \lho (\Psi_z^{\hbar},\qh^B(f)\Psi_z^{\hbar})=f(z). \label{hnulB}
\eeq
This means that the classical limit of the quantum-mechanical expectation value of the phase space quantization \er{b2} of the classical observable $f$ in a coherent state centered at $z\in M$ is precisely the classical expectation value of $f$ in the state $z$. This interpretation rests on the identification of classical states with probability measures on phase space $M$, under which points of $M$ in the guise of Dirac measures (i.e.\ delta functions) are pure states.
 Furthermore, it can be shown (cf.\ Landsman, 1998) that the continuity of all functions \er{chca} - \er{chcb}  implies the property
\beq \lho |(\Psi_w^{\hbar},\Psi_z^{\hbar})|^2 =\dl_{wz}, \label{dlxy}\eeq
where $\dl_{wz}$ is the ordinary Kronecker delta (i.e.\ $\dl_{wz}=0$ whenever $w\neq z$ and
$\dl_{zz}=1$ for all $z\in M$). This has a natural physical interpretation as well: the classical limit of the quantum-mechanical transition probability between two coherent states centered at $w,z\in M$ is equal to the classical (and trivial) transition probability between $w$ and $z$. In other words, when $\hbar$ becomes small, coherent states at different values of $w$ and $z$ become increasingly orthogonal to each other.\footnote{See
Mielnik (1968),  Cantoni (1975), Beltrametti \&\ Cassinelli (1984), 
 Landsman (1998), and Subsection \ref{SE} below for the general meaning of the concept of a transition probability.} This has the interesting consequence that 
 \beq \lho (\Psi_w^{\hbar},\qh^B(f)\Psi_z^{\hbar})=0 \:\:\: (w\neq z). \label{hnulB2}
\eeq
for all $f\in C_c(M)$. In particular, the following phenomenon  of the Schr\"{o}dinger cat type occurs in the classical limit:
if $w\neq z$ and one has continuous functions $\hbar\mapsto c_w^{\hbar}\in\C$ and $\hbar\mapsto c_z^{\hbar}\in\C$ on $\hbar\in[0,1]$
such that \beq\Psi^{\hbar}_{w,z}=c_w^{\hbar}\Psi_w^{\hbar}+c_z^{\hbar}\Psi_z^{\hbar}\eeq is a unit vector
for $\hbar\geq 0$ and also $|c_w^0|^2+|c_z^0|^2=1$, then
\beq \lho \left(\Psi^{\hbar}_{w,z},\qh^B(f)\Psi^{\hbar}_{w,z}\right)
=|c_w^0|^2f(w)+|c_z^0|^2f(z). \label{Cat}\eeq
Hence the family of (typically) pure states $\psi^{\hbar}_{w,z}$ 
(on the \ca s $\CA_{\hbar}$ in which the map $\qh^B$ takes values)\footnote{For example, for $M=\R^{2n}$ each   $\CA_{\hbar}$ is equal to the \ca\ of compact 
operators on $L^2(\R^n)$, on which each vector state is certainly pure.}
defined by the vectors $\Psi^{\hbar}_{w,z}$ in some sense converges to the mixed state 
on $C_0(M)$ defined by the right-hand side of \er{Cat}. This is  made precise at the end of this subsection. 

 It goes without saying that Schr\"{o}dinger's coherent states \er{pqcohst} satisfy our axioms; one may also verify \er{dlxy} immediately  from \er{tpbt}. Consequently, by  \er{WBEQ} one has  the same property \er{hnulB} for Weyl quantization (as long as $f\in\CS(\R^{2n})$),\footnote{Here $\CS(\R^{2n})$ is the usual Schwartz space of smooth  test functions with rapid decay at infinity.} that is,
\beq \lho (\Psi_z^{\hbar},\qh^W(f),\Psi_z^{\hbar})=f(z).\label{hnulW}
\eeq
Similarly, \er{hnulB2} holds for $\qw$ as well.  

In addition, many constructions referred to as coherent states in the literature (cf.\ the references in footnote   \ref{CSFNO}) satisfy \er{normcs}, \er{qhnorm}, and  \er{dlxy}; see Landsman (1998).\footnote{For example, coherent states of the type 
introduced by Perelomov (1986) fit into our setting as follows (Simon, 1980). Let $G$ be a
compact connected Lie group, and $\CO_{\lm}$ an integral coadjoint
orbit, corresponding to a highest weight
$\lm$. (One may think here of $G=SU(2)$ and $\lm=0,1/2,1,\ldots$.)
Note that $\CO_{\lm}\cong G/T$, where $T$ is the maximal torus in $G$ with respect to
which weights are defined. Let $\H^{\mbox{\tiny hw}}_{\lm}$ be the carrier
space of the \irrep\ $U_{\lm}(G)$ with highest weight
$\lm$, containing the highest weight vector $\Om_{\lm}$. (For $G=SU(2)$ one has $\H^{\mbox{\tiny hw}}_j=\C^{2j+1}$, the well-known \Hs\ of spin $j$, in which
$\Om_j$  is the vector with spin $j$ in the $z$-direction.) 
 For $\hbar=1/k$, $k\in\N$, define
$\H_{\hbar}:=\H^{\mbox{\tiny hw}}_{\lm/\hbar}$.
 Choosing a section $\sg:\CO_{\lm}\raw G$ of the projection $G\raw G/T$, one then obtains
coherent states $x\mapsto U_{\lm/\hbar}(\sg(x))\Om_{\lm/\hbar}$ with respect to
the Liouville measure on $\CO_{\lm}$ and $c_{\hbar}=\dim(\H^{\mbox{\tiny hw}}_{\lm/\hbar})$. 
These states are obviously not defined for all values of $\hbar$ in $(0,1]$, but only for the discrete set $1/\N$.\label{FNPER}}
 The general picture that emerges is that a coherent state centered at $z\in M$ is the {\it Umdeutung}  of $z$ (seen as a classical pure state, as explained above) as a quantum-mechanical pure state.\footnote{This idea is also confirmed by the fact that at least Schr\"{o}dinger's coherent states are states of minimal uncertainty; cf.\ the references in footnote   \ref{CSFNO}.} 

Despite their wide applicability (and some would say beauty), one has to look beyond coherent states for a complete picture of the $\hbar\raw 0$ limit of \qm. The appropriate generalization is 
the concept of  a {\it continuous field of states}.\footnote{The use of this concept in various mathematical approaches to quantization is basically folklore. 
 For the \ca ic setting see Emch (1984), Rieffel (1989b), Werner (1995), Blanchard (1996), Landsman (1998), and Nagy (2000).}
 This is defined relative to a given deformation quantization of a phase space $M$; cf.\  Subsection \ref{DQsection}. If one now has a state $\om_{\hbar}$ on $\CA_{\hbar}$
for each $\hbar\in[0,1]$ (or, more generally, for a discrete subset of $[0,1]$ containing 0 as an accumulation point),
one may call the ensuing family of states a {\it continuous field} whenever the function $\hbar\mapsto \om_{\hbar}(\qh(f))$ is continuous on $[0,1]$ for each $f\in\cci(M)$; this notion is actually intrinsically defined by the continuous field of \ca s, and is therefore independent of the quantization maps $\qh$.  In particular, one has
\beq
\lho \om_{\hbar}(\qh(f))=\om_0(f). \label{CFS0}
\eeq 
Eq. \er{hnulB} (or \er{hnulW}) shows that coherent states are indeed examples of continuous fields of states, with the additional property that each $\om_{\hbar}$ is pure. As an example where all states $ \om_{\hbar}$ are mixed, we mention the convergence of quantum-mechanical partition functions to their classical counterparts  in statistical mechanics along these lines; see Lieb (1973), Simon (1980), Duffield (1990), and Nourrigat \&\ Royer (2004). Finally, one encounters the surprising phenomenon that pure quantum states may coverge to mixed classical ones. The first example of this has just been exhibited in \er{Cat}; 
other cases in point are energy eigenstates and WKB states (see Subsections \ref{PSL}, \ref{WKBS}, and \ref{QC}  below).
\subsection{Convergence of quantum dynamics to classical motion}\label{CEOM}
Nonrelativistic \qm\ is based on the Schr\"{o}dinger equation \er{Schreq}, which more generally reads
\beq H\Psi(t)=i\hbar \frac{\partial\Psi}{\partial t}.\label{Schreq2}\eeq
The formal solution  with initial value $\Ps(0)=\Ps$ is 
\beq \Psi(t)=e^{-\frac{it}{\hbar} H}\Ps. \label{SSEOM}\eeq
Here we have assumed that $H$ is a given self-adjoint operator on the \Hs\ $\H$ of the system, so that this solution indeed exists and evolves unitarily by Stone's theorem; cf.\ Reed \&\ Simon (1972) and Simon (1976).  Equivalently, one may transfer the time-evolution from states (Schr\"{o}dinger picture) to operators (Heisenberg picture) by putting
\beq A(t)=e^{\frac{it}{\hbar} H}Ae^{-\frac{it}{\hbar} H}.\label{HSEOM}\eeq

We here restrict ourselves to particle motion in $\R^n$, so that $\H=L^2(\R^n)$.\footnote{See  Hunziker \&\ Sigal  (2000) for a recent survey of  $N$-body 
 Schr\"{o}dinger operators.}
 In that case, $H$ is typically given by a formal  expression like \er{Schreq} (on some specific domain).\footnote{
 One then has to prove self-adjointness (or the lack of it) on a larger domain on which the operator is closed; see the literature cited in footnote  \ref{SOPrefs}.} Now, the first thing that comes to mind is {\it Ehrenfest's  Theorem} (1927), which states that for any (unit) vector $\Ps\in  L^2(\R^n)$ in the domain of $\qh(q^j)=x^j$ and $\partial V(x)/\partial x^j$ one has  \beq m\frac{d^2}{dt^2} \langle x^j\rangle(t)=-\left\langle  \frac{\partial V(x)}{\partial x^j} \right\rangle(t),
\label{EhrThm}
\eeq 
with the notation
\begin{eqnarray} \langle x^j\rangle(t) & =& (\Ps(t),x^j\Ps(t)); \nn \\
 \left\langle  \frac{\partial V(x)}{\partial x^j} \right\rangle(t) & =&  \left(\Ps(t), \frac{\partial V(x)}{\partial x^j}\Ps(t)\right).\label{EhrThm2}
\end{eqnarray}
This looks like Newton's second law for the expectation value of $x$ in the state $\ps$, with the tiny but crucial difference that Newton would have liked to see $(\partial V/\partial x^j)(\langle x\rangle(t))$ on the right-hand side of \er{EhrThm}. Furthermore, even apart from this point Ehrenfest's  Theorem by no means suffices to have classical behaviour, since it gives no guarantee whatsoever that $\langle x\rangle(t)$ behaves like a point particle. Much of what follows can be seen as an attempt to sharpen Ehrenfest's  Theorem to the effect that it {\it does} indeed yield  appropriate classical equations of motion for the expectation values of suitable operators. 

 We assume that the quantum Hamiltonian has the more general form 
\beq H=h(\qh(p_j),\qh(q^j)), \label{Hh}
\eeq where $h$ is the classical Hamiltonian (i.e.\ a function defined on classical phase space $\R^{2n}$) and 
$\qh(p_j)$ and $\qh(q^j)$ are the operators given in \er{SOP}. Whenever this expression is ambiguous (as in cases like $h(p,q)=pq$), one has to assume a specific quantization prescription such as Weyl quantization $\qw$ (cf.\ \er{defweylq}), so that formally one has 
\beq H=\qh^W(h). \label{HWeyl}\eeq
In fact, in the literature to be cited an even larger class of quantum Hamiltonians is treated by the methods explained here.
The  quantum Hamiltonian $H$ carries an explicit (and rather singular) $\hbar$-dependence, and for $\hbar\raw 0$ one then expects \er{SSEOM} or \er{HSEOM} to be related in one way or another to the flow of the classical Hamiltonian $h$. This relationship was already foreseen by Schr\"{o}dinger (1926a), and was formalized almost immediately after the birth of \qm\ by the well-known WKB approximation (cf.\ Landau \&\ Lifshitz (1977) and Subsection \ref{WKBS} below). A mathematically rigorous understanding of this and analogous approximation methods only emerged much later,
when a technique called {\it microlocal analysis} was adapted from its original setting of partial differential equations (H\"{o}rmander, 1965; Kohn \&\ Nirenberg, 1965; Duistermaat, 1974, 1996; Guillemin \&\ Sternberg, 1977; Howe, 1980; H\"{o}rmander, 1979, 1985a, 1985b; 
Grigis \&\ Sj\"ostrand, 1994) to the study of the $\hbar\raw 0$ limit of \qm. This adaptation (often called {\it semiclassical analysis}) and its results have now been explained in various reviews written by the main players, notably Robert (1987, 1998), Helffer (1988),  Paul \&\ Uribe (1995), Colin de Verdi\`ere (1998), Ivrii (1998), Dimassi \&\  Sj\"ostrand (1999), and  Martinez (2002) (see also the papers in Robert (1992)). More specific references will be given below.\footnote{For the heuristic theory of semiclassical asymptotics Landau \&\ Lifshitz (1977) is a goldmine.}

As mentioned before, the relationship between $H$ and $h$ provided by semiclassical analysis is double-edged. On the one hand, one obtains approximate solutions of \er{SSEOM} or \er{HSEOM},  or approximate energy eigenvalues and energy eigenfunctions (sometimes called quasi-modes)  for small values of $\hbar$ in terms of classical data. This is how the results are usually presented; one computes specific properties of quantum theory in a certain regime  in terms of an underlying  classical theory. 
On the other hand, however, with some effort the very same results can often be reinterpreted as  a partial explanation of the emergence of classical dynamics from \qm. It is the latter aspect of  semiclassical analysis, somewhat understated in the literature, that is of interest to us.  In this and the next three subsections we restrict ourselves to the simplest type of results, which nonetheless provide a good flavour of what can be achieved and understood by these means. By the same token, we just work with the usual flat phase space $M=\R^{2n}$ as before. 

 The simplest of all results relating classical and quantum dynamics
is this:\footnote{More generally, Egorov's Theorem states that for a large class of Hamiltonians one has $\qw(f)(t) =\qw(f_t)+O(\hbar)$. See, e.g., Robert (1987), Dimassi \&\  Sj\"ostrand (1999), and Martinez (2002).}
\begin{quote}
{\it If the classical Hamiltonian $h(p,q)$ is at most quadratic in $p$ and $q$, and the Hamiltonian in \er{HSEOM} is given by \er{HWeyl},
then} \end{quote}
\beq \qw(f)(t) =\qw(f_t). \label{QCC}
\eeq
 Here $f_t$ is the solution
of the classical equation of motion $df_t/dt=\{h,f_t\}$; equivalently, one may write
\beq f_t(p,q)=f(p(t),q(t)), \label{ft}\eeq
where $t\mapsto (p(t),q(t))$ is the classical Hamiltonian flow of $h$
with initial condition $(p(0),q(0))=(p,q)$. 
This holds for all decent $f$, e.g., $f\in\CS(\R^{2n})$. 

This result explains quantum in terms of classical, but the converse may be achieved by combining \er{QCC} with \er{CFS0}. This yields
\beq 
\lho \om_{\hbar}(\qh(f)(t))=\om_0(f_t) \label{CFS1}
\eeq 
for any continuous field of states $(\om_{\hbar})$. In particular, for Schr\"odinger's coherent states \er{pqcohst} one obtains
\beq
\lho \left(\Psi_{(p,q)}^{\hbar}, \qh(f)(t) \Psi_{(p,q)}^{\hbar}\right)=f_t(p,q). \label{hepp}
\eeq
Now, whereas \er{QCC} merely reflects the good symmetry properties of Weyl quantization,\footnote{Eq.\  \er{QCC} is equivalent to the covariance of Weyl quantization under the affine symplectic group; cf.\ footnote  \ref{GCQQ}.} 
(and is false for $\qh^B$), eq.\ \er{hepp} is actually valid for a large class of realistic Hamiltonians and for any deformation quantization map $\qh$ that is asymptotically equal to $\qw$ (cf.\ \er{WBEQ}). A result of this type was first established by Hepp (1974); further work in this direction includes Yajima (1979), Hogreve, Potthoff \&\ Schrader (1983),  Wang (1986), Robinson (1988a, 1988b), Combescure (1992),  Arai (1995),  Combescure \&\ Robert (1997), Robert (1998), and Landsman (1998).

Impressive results are available also in the Schr\"{o}dinger picture.  The counterpart of \er{QCC} is that for any suitably smooth classical Hamiltonian $h$ (even a time-dependent one) that is at most quadratic in the canonical coordinates $p$ and $q$ on phase space $\R^{2n}$ one may construct generalized coherent states $\Psi^{\hbar}_{(p,q,C)}$, labeled by a set $C$ of classical parameters  dictated by the form of $h$,  such that
\beq
e^{-\frac{it}{\hbar} \qw(h)}\Ps^{\hbar}_{(p,q,C)}=e^{iS(t)/\hbar}\Ps^{\hbar}_{(p(t),q(t),C(t))}. \label{Hag}
\eeq
Here $S(t)$ is the action associated with  the classical trajectory $(p(t),q(t))$  determined by $h$, and $C(t)$ is a solution of a certain system of differential equations that has a classical interpretation as well (Hagedorn, 1998).  Schr\"{o}dinger's coherent states \er{pqcohst} are a special 
case for  the standard harmonic oscillator Hamiltonian. For more general Hamiltonians one then has an asymptotic result  (Hagedorn \&\ Joye, 1999, 2000)\footnote{See also 
 Paul \&\ Uribe (1995, 1996) as well as 
the references listed after \er{hepp}  for analogous statements.}
\beq \lho \left\| e^{-\frac{it}{\hbar} \qw(h)}\Ps^{\hbar}_{(p,q,C)}-e^{iS(t)/\hbar}\Ps^{\hbar}_{(p(t),q(t),C(t))}\right\| =0. \label{HY}\eeq

Once again, at first sight such results merely
contribute to the understanding of quantum dynamics in terms of classical motion. As mentioned, they may be converted into statements on the emergence of classical motion from \qm\ by taking expectation values of suitable $\hbar$-dependent obervables of the type $\qw(f)$.   

For finite $\hbar$, the second term in \er{HY} is a good approximation to the first
- the error even being  as small as $\CO(\exp(-\gm/\hbar))$ for some $\gm>0$ as $\hbar\raw 0$ -
whenever $t$ is smaller than the so-called {\it Ehrenfest time} 
\beq T_E=\lm^{\inv}\log(\hbar\inv), \label{EhrTime} \eeq
where $\lm$ is a typlical inverse time scale of the Hamiltonian (e.g., for chaotic systems it is the largest  Lyapunov exponent).\footnote{Recall that throughout this section we assume that $\hbar$ has been made dimensionless through an appropriate rescaling.}   
This is  the typical time scale on which semiclassical approximations to 
wave packet solutions of the time-dependent Schr\"{o}dinger equation with a general Hamiltonian tend to be valid (Ehrenfest, 1927;  Berry et al., 1979; Zaslavsky, 1981; Combescure \&\ Robert, 1997; Bambusi,   Graffi,  \&\  Paul, 1999; Hagedorn \&\ Joye, 2000).\footnote{
One should distinguish here between two distinct approximation methods to 
 the time-dependent Schr\"{o}dinger equation. Firstly, one has 
the semiclassical propagation of a quantum-mechanical wave packet, i.e.\ its propagation as computed from the time-dependence of the parameters on which it depends {\it according to  the underlying classical equations of motion}. It is shown in the references just cited that
this approximates the full quantum-mechanical propagation of the wave packet well until 
$t\sim T_E$. Secondly, one has the time-dependent WKB approximation (for integrable systems) and its generalization to chaotic systems (which typically involve tens of thousands of terms instead of a single one). This second approximation is valid on a much longer time scale, typically $t\sim\hbar^{-1/2}$ (O'Connor,  Tomsovic,  \&\  Heller, 1992;
Heller \&\ Tomsovic, 1993;  Tomsovic,  \&\  Heller, 1993, 2002; Vanicek \&\ Heller, 2003). Adding to the confusion,
Ballentine has claimed over the years that even the semiclassical propagation of a  wave packet approximates its quantum-mechanical propagation for times
much longer than the Ehrenfest time, typically $t\sim\hbar^{-1/2}$ (Ballentine, Yang, \&\ Zibin, 1994; Ballentine, 2002, 2003). This claim is based on the criterion that the quantum and classical (i.e.\ Liouville) probabilities are approximately equal on such time scales, but the validity of this criterion hinges on the ``statistical" or ``ensemble" interpretation of \qm.
According to this interpretation, a pure state provides a description of certain statistical properties of an ensemble of similarly prepared systems, but need not provide a complete description of an individual system. See Ballentine (1970, 1986). Though once defended by von Neumann, Einstein and Popper, this interpretation  has now completely fallen out of fashion.} For example, Ehrenfest (1927) himself
estimated that for a mass of 1 gram a wave packet would double its width only in about $10^{13}$ years under free motion. However, Zurek and Paz (1995) have estimated the Ehrenfest time  for  Saturn's moon Hyperion to be of the order of  20 years! This obviously poses a serious problem for the program of deriving (the appearance of) classical behaviour from \qm, which affects all interpretations of this theory.

Finally, we have not discussed the important problem of combining the limit $t\raw\infty$ 
with the limit $\hbar\raw 0$; this should be done in such a way that $T_E$ is kept fixed.
This double limit is of particular importance for quantum chaos; see Robert (1998) and most of the literature cited in  Subsection \ref{QC}. 
\subsection{Wigner functions} \label{WFSEC}
The $\hbar\raw 0$ limit of \qm\ is often discussed in terms of the so-called {\it Wigner function}, introduced by Wigner (1932).\footnote{The original context was quantum statistical mechanics; one may write down \er{WF} for mixed states as well. See Hillery et al. (1984) for a survey.}  Each unit vector (i.e. wave function) $\Psi\in L^2(\R^n)$ defines such a function
$W^{\hbar}_{\Psi}$  on classical phase space $M=\R^{2n}$ by demanding that for each $f\in\CS(\R^{2n})$ one has
\beq \left(\Ps,\qw(f)\Ps\right)=\int_{\R^{2n}} \frac{d^np
d^nq}{(2\pi)^n} \, W^{\hbar}_{\Psi}(p,q) f(p,q). \label{WF} \eeq
The existence of such a function may be proved by writing it down explicitly as
\beq
W^{\hbar}_{\Ps}(p,q)=\int_{\R^n}
d^nv\,e^{ipv}\ovl{\Ps(q+\half\hbar v)} \Ps(q-\half\hbar v).\label{wiganew}
\eeq
In other words, the quantum-mechanical expectation value of the Weyl quantization of the classical observable $f$ in a quantum state $\Ps$ formally equals  the classical expectation value of $f$ with respect to the  distribution $W_{\Ps}$.
 However, the latter may not be regarded as a probability distribution because it is not necessarily positive definite.\footnote{Indeed, it may not even be in $L^1(\R^{2n})$, so that its total mass is not necessarily defined, let alone equal to 1. Conditions for the positivity of Wigner functions defined by pure states are given by Hudson (1974); see Br\"ocker \&\ Werner (1995) for the case of mixed states.} Despite this drawback, the Wigner function possesses some attractive properties.  For example,  one has 
\beq \qw(W^{\hbar}_{\Ps})=\hbar^{-n}[\Psi]. \eeq
 This somewhat perverse result means that if the Wigner function defined by $\Ps$ is seen as a classical observable (despite its manifest $\hbar$-dependence!), then its Weyl quantization is precisely ($\hbar^{-n}$ times) the projection operator onto $\Ps$.\footnote{In other words, $W_{\Psi}$ is the Weyl symbol of the projection operator $[\Ps]$.}  Furthermore, one may derive the following formula for the transition probability:\footnote{This formula is well defined since $\Ps\in\ L^2(\R^n)$ implies $W^{\hbar}_{\Psi}\in\ L^2(\R^{2n})$.}
\beq
|(\Phi,\Psi)|^2=\hbar^n \int_{\R^{2n}} \frac{d^np
d^nq}{(2\pi)^n} \, W^{\hbar}_{\Psi}(p,q) W^{\hbar}_{\Phi}(p,q). \label{WFTP} \eeq
This expression has immediate intuitive appeal, since the integrand on the right-hand side is supported by the area in phase space where the two Wigner functions overlap, which is well in tune with the idea of a transition probability.

The potential lack of positivity of a Wigner function  may be remedied by noting that Berezin's deformation quantization scheme   (see \er{qbttsrex}) analogously defines functions $B^{\hbar}_{\Psi}$ on phase space by means of 
 \beq \left(\Ps,\qb(f)\Ps\right)=\int_{\R^{2n}} \frac{d^np
d^nq}{(2\pi)^n} \, B^{\hbar}_{\Psi}(p,q) f(p,q). \label{BF} \eeq
Formally,  \er{qbttsrex} and \er{BF} immediately  yield
\beq
B^{\hbar}_{\Psi}(p,q)=|(\Psi_{(p,q)}^{\hbar},\Psi)|^2 \label{BFexp} \eeq
in terms of Schr\"odinger's coherent states \er{pqcohst}. This expression is manifestly positive definite.
The existence of $B^{\hbar}_{\Psi}$ may be proved rigorously by recalling that the Berezin quantization map $f\mapsto\qb(f)$ is {\it positive} from $C_0(\R^{2n})$ to $\CB(L^2(\R^n))$. This implies that for each (unit) vector $\Ps\in L^2(\R^n)$ the map $f\mapsto (\Ps,\qb(f)\Ps)$ is  positive from $C_c(\R^{2n})$ to $\C$, so that (by the Riesz theorem of measure theory) there must be a measure $\mu_{\Psi}$ on $\R^{2n}$ such that $(\Ps,\qb(f)\Ps)=\int d\mu_{\Ps}\, f$. This measure, then, is precisely given by $d\mu_{\Ps}(p,q)=(2\pi)^{-n} d^npd^nq\,  B^{\hbar}_{\Psi}(p,q)$. If $(\Ps,\Ps)=1$, then $\mu_{\Ps}$ is a probability measure.
Accordingly, despite its $\hbar$-dependence, $B^{\hbar}_{\Ps}$ defines a bona fide classical probability distribution on phase space, in terms of which one might attempt to visualize \qm\ to some extent.

For finite values of $\hbar$, the Wigner and Berezin distribution functions are different, because the quantization maps $\qw$ and $\qb$ are. The connection between  $B^{\hbar}_{\Psi}$ and 
 $W^{\hbar}_{\Psi}$ is easily computed to be
 \beq
 B^{\hbar}_{\Psi}=W^{\hbar}_{\Psi}*g^{\hbar},\label{CBW} \eeq
 where $g^{\hbar}$ is the Gaussian function 
\beq g^{\hbar}(p,q)=(2/\hbar)^n \exp(-(p^2+q^2)/\hbar).\eeq
 This is how physicists look at the Berezin function,\footnote{
The `Berezin' functions $B^{\hbar}_{\Psi}$ were introduced by Husimi (1940) from a different point of view, and are therefore actually called {\it Husimi functions} by physicists.} viz.\  as a Wigner function smeared with a Gaussian so as to become positive.  But since $g^{\hbar}$ converges to a Dirac delta function as $\hbar\raw 0$ (with respect to the measure $(2\pi)^{-n} d^npd^nq$ in the sense of distributions), it is clear from \er{CBW} that
as distributions one has\footnote{\label{BWD} Eq.\ \er{BWAE} should be interpreted as a limit of the distribution on $\mathcal{D}(\R^{2n})$ or $\mathcal{S}(\R^{2n})$ defined by 
$B^{\hbar}_{\Psi}-W^{\hbar}_{\Psi}$. Both  functions are continuous for $\hbar>0$, but lose this property in the limit $\hbar\raw 0$, generally converging to distributions.}
 \beq
 \lho \left(B^{\hbar}_{\Psi}-W^{\hbar}_{\Psi}\right)=0. \label{BWAE}\eeq  
See also \er{WBEQ}. Hence in the study of the limit $\hbar\raw 0$ there is little advantage in the use of Wigner functions; quite to the contrary,  in limiting procedures their generic lack of positivity  makes them more difficult to handle than Berezin functions.\footnote{ See, however, Robinett (1993) and Arai (1995). It should be mentioned that \er{BWAE} expresses the asymptotic equivalence of  Wigner  and Berezin functions as distributions on $\hbar$-independent test functions. Even in the limit $\hbar\raw 0$ one is sometimes interested in studying $O(\hbar)$ phenomena, in which case one should make a choice.}  
 For example, one would like
to write the asymptotic behaviour \er{hnulW} of coherent states in the form $\lho W^{\hbar}_{\Ps^{\hbar}_z}=\dl_z$. 
Although this is indeed true in the sense of  distributions,  the
  corresponding limit  
\beq \lho B^{\hbar}_{\Ps^{\hbar}_z}=\dl_z, \label{lhoB} \eeq
 exists in the sense of  (probability) measures, and is therefore defined on a much larges class of test functions.\footnote{Namely those in $C_0(\R^{2n})$ rather than in  $\mathcal{D}(\R^{2n})$  or $\mathcal{S}(\R^{2n})$.}
Here and in what follows, we abuse notation: if $\mu^0$ is some  probability measure on $\R^{2n}$
 and $(\Psi^{\hbar})$ is a sequence of unit vectors in $L^2(\R^n)$  indexed by $\hbar$ (and perhaps other labels), then  $B^{\hbar}_{\Ps^{\hbar}}\raw 
 \mu^0$ for $\hbar\raw 0$  by definition means that for any $f\in\cci(\R^{2n})$ one has\footnote{Since $\qb$ may be extended from $\cci(\R^{2n})$ to $L^{\infty}(\R^{2n})$, one may omit the stipulation that $\mu^0$ be a {\it probability} measure in this definition if one requires convergence for all $f\in L^{\infty}(\R^{2n})$, or just for all $f$  in the unitization of the \ca\ $C_0(\R^{2n})$.}
 \beq
\lho  \left(\Psi^{\hbar},\qb(f)\Psi^{\hbar}\right)= \int_{\R^{2n}} d\mu^0\, f.\eeq
\subsection{The classical limit of energy eigenstates} \label{PSL}
Having dealt with coherent states in \er{lhoB}, in this subsection we discuss the much more difficult 
problem of computing the limit measure $\mu^0$ for eigenstates of the quantum Hamiltonian $H$. Thus we assume that  $H$ has eigenvalues $E^{\hbar}_n$ labeled by $n\in\N$ (defined with or without 0 according to convenience), and also depending on $\hbar$ because of the explicit dependence of $H$ on this parameter. The associated 
eigenstates $\Ps^{\hbar}_{\mathsf{n}}$ then by definition satisfy \beq H\Ps^{\hbar}_{\mathsf{n}}=E^{\hbar}_n\Ps^{\hbar}_{\mathsf{n}}.\eeq
 Here we incorporate the possibility that the eigenvalue   $E^{\hbar}_n$ is degenerate, so that the label $\mathsf{n}$ extends $n$. For example, for the one-dimensional harmonic oscillator one has $E^{\hbar}_{n}=\hbar\omega (n+\half)$ ($n=0,1,2,\ldots$) without multiplicity, but for the hydrogen atom the Bohrian eigenvalues $E^{\hbar}_n=-m_e e^4/2\hbar^2 n^2$ (where $m_e$ is the mass of the electron and $e$ is its charge) are degenerate, with the well-known eigenfunctions $\Psi^{\hbar}_{(n,l,m)}$ (Landau \&\ Lifshitz, 1977). Hence in this case one has $\mathsf{n}=(n,l,m)$ with $n=1,2,3,\ldots$, subject to $l=0, 1, \ldots, n-1$, and $m=-l, \ldots, l$. 

In any case, it makes sense to let ${\mathsf{n}}\raw\infty$; this certainly means $n\raw\infty$, and may in addition involve sending the other labels in $\mathsf{n}$ to infinity  (subject to the appropriate restrictions on ${\mathsf{n}}\raw\infty$, as above). One then expects classical behaviour \`a la  Bohr if one simultaneously lets $\hbar\raw 0$  whilst $E^{\hbar}_n\raw E^0$ converges to some `classical' value $E^0$. 
Depending on how one lets the possible other labels behave in this limit, this may also involve similar asymptotic conditions on the eigenvalues of operators commuting with $H$ - see below for details in the integrable case.
We denote the collection of such eigenvalues (including $E^{\hbar}_n$) by $\mathsf{E}^{\hbar}_{\mathsf{n}}$. (Hence in the case where the energy levels $E^{\hbar}_n$ are nondegenerate, the label $\mathsf{E}$ is just $E$.)
 In general, we denote the collective limit of the eigenvalues $\mathsf{E}^{\hbar}_{\mathsf{n}}$ as $\hbar\raw 0$ and ${\mathsf{n}}\raw\infty$ by $\mathsf{E}^0$.

 For example, for the hydrogen atom one has the additional operators $J^2$ of total angular momentum as well as the operator $J_3$ of angular momentum in the $z$-direction. The eigenfunction $\Psi^{\hbar}_{(n,l,m)}$ of $H$ with eigenvalue $E^{\hbar}_n$ is in addition an eigenfunction of $J^2$ with eigenvalue  $j_{\hbar}^2=\hbar^2l(l+1)$ and of  $J_3$ with eigenvalue $j^{\hbar}_3=\hbar m$. Along with $n\raw\infty$ and $\hbar\raw 0$, one may then send $l\raw\infty$ and $m\raw\pm\infty$ in such a way that $j_{\hbar}^2$ and $j^{\hbar}_3$ approach specific constants.
 
The object of interest, then, is the measure on phase space obtained as the limit of the Berezin functions
\er{BFexp}, i.e.\ 
\beq \mu_{\mathsf{E}}^0=\lim_{\hbar\raw 0,\mathsf{n}\raw\infty} B^{\hbar}_{\Ps^{\hbar}_{\mathsf{n}}}. \label{measure}
\eeq
Although the pioneers of \qm\ were undoubtedly interested in quantities like this, it was only in the 1970s that rigorous results were obtained. Two cases are well understood:   in this subsection we discuss the {\it integrable} case, leaving chaotic and more generally {\it ergodic} motion to Subsection \ref{QC}.

In the physics literature,  it was argued that for an integrable system the limiting measure $\mu_{\mathsf{E}}^0$ is concentrated (in the form of a $\dl$-function) on the invariant torus associated to $\mathsf{E}^0$ (Berry, 1977a).\footnote{This conclusion was, in fact, reached from the Wigner function formalism. 
See  Ozorio de Almeida (1988) for a review of work of Berry and his collaborators on this subject.} 
Independently, mathematicians began to study a quantity very similar to $\mu_{\mathsf{E}}^0$, defined by limiting sequences of eigenfunctions of the Laplacian on a Riemannian manifold $M$. Here the underlying classical flow is Hamiltonian as well, the corresponding trajectories being  the geodesics of the given metric (see, for example, Klingenberg (1982), Abraham \&\ Marsden (1985), Katok \&\ Hasselblatt (1995), or Landsman (1998)).\footnote{
 The simplest examples of integrable geodesic motion are $n$-tori, where the geodesics are projections of lines, and the sphere, where   the geodesics are great circles (Katok \&\ Hasselblatt, 1995).}
 The ensuing picture largely confirms the folklore of the physicists: 
\begin{quote}
 {\it In the integrable case the limit measure   $\mu_{\mathsf{E}}^0$ is
concentrated on invariant tori}. 
\end{quote}
See  Charbonnel (1986, 1988), Zelditch (1990, 1996a), Toth (1996, 1999),  Nadirashvili, Toth, \&\ Yakobson (2001), and Toth \&\ Zelditch (2002, 2003a, 2003b).\footnote{These papers consider the limit $n\raw\infty$ without $\hbar\raw 0$; in fact, a physicist would say that they put $\hbar =1$. In that case $E_n \raw\infty$; in this procedure the physicists' microscopic $E\sim \CO(\hbar)$ and macroscopic $E\sim \CO(1)$ regimes correspond to $E\sim \CO(1)$ and $E\raw\infty$, respectively.}
Finally, as part of the transformation of microlocal analysis to semiclassical analysis (cf.\ Subsection \ref{CEOM}), these results were adapted to \qm\ 
(Paul \&\ Uribe, 1995, 1996). 

Let us now give some details for  integrable systems (of Liouville type); these include the hydrogen atom as a special case. Integrable systems are defined by the property that
on a $2p$-dimensional phase space $M$ one has $p$ independent\footnote{I.e.
$df_1\wed \cdots\wed df_p\neq 0$ everywhere. At this point we write $2p$ instead of $2n$ for the dimension of phase space in order to avoid notational confusion.} classical observables $(f_1=h,f_2,\ldots, f_p)$ whose mutual Poisson brackets all vanish (Arnold, 1989).   One then hopes  that an appropriate quantization scheme $\qh$ exists under which the corresponding quantum observables $(\qh(f_1)=H,\qh(f_2), \ldots, \qh(f_p))$ are all self-adjoint and mutually commute (on a common core).\footnote{There is no general theory of quantum integrable systems. Olshanetsky \&\ Perelomov (1981, 1983) form a good starting point.} This is indeed the case for the hydrogen atom, where
$(f_1,f_2,f_3)$ may be taken to be $(h,j^2,j_3)$ (where $j^2$ is the total angular momentum and $j_3$ is its $z$-component),\footnote{In fact, if $\mu$ is the 
momentum map for the standard $SO(3)$-action on $\R^3$, then 
$j^2=\sum_{k=1}^3 \mu_k^2$ and $j_3=\mu_3$.}  $H$ is given by \er{HWeyl}, $J^2=\qw(j^2)$, and $J_3=\qw(j_3)$. In general, the energy eigenfunctions
 $\Ps^{\hbar}_{\mathsf{n}}$ will be joint eigenfunctions of the operators
 $(\qh(f_1), \ldots, \qh(f_p))$, so that $\mathsf{E}^{\hbar}_{\mathsf{n}}=(E^{\hbar}_{n_1},\ldots, E^{\hbar}_{n_p})$, with $\qh(f_k)\Ps^{\hbar}_{\mathsf{n}}=E^{\hbar}_{n_k}\Ps^{\hbar}_{\mathsf{n}}$.
 We assume that the submanifolds $\cap_{k=1}^p f_k\inv(x_k)$ are compact and connected for each $x\in\R^p$, so that they are tori by the Liouville--Arnold Theorem (Abraham \&\ Marsden, 1985, Arnold, 1989). 

Letting $\hbar\raw 0$ and $\mathsf{n}\raw\infty$ so that $E^{\hbar}_{n_k}\raw E_k^0$ for some point $E^0=(E^0_1,\ldots, E^0_p)\in\R^p$, it  follows that the limiting measure $\mu_{\mathsf{E}}^0$ as defined in \er{measure} is concentrated on the invariant torus  $\cap_{k=1}^p f_k\inv(E^0_k)$. This torus is generically $p$-dimensional, but for singular points
 $E^0$ it may be of lower dimension.
In particular, in the exceptional circumstance where the invariant torus is one-dimensional, 
 $\mu_{\mathsf{E}}^0$ is concentrated on a classical orbit. Of course, for $p=1$
 (where any Hamiltonian system is integrable) this singular case is generic. Just think of the foliation of $\R^2$ by the ellipses that form the closed orbits 
 of the harmonic oscillator motion.\footnote{\label{Zelditch} It may be enlightening to consider
 geodesic motion on the sphere; this example may be seen as the hydrogen atom without the radial degree of freedom (so that the degeneracy in question occurs in the hydrogen atom as well). If one sends $l\raw\infty$ and
 $m\raw\infty$ in the spherical harmonics $Y^m_l$ (which are eigenfunctions of the Laplacian on the sphere) in such a way that $\lim m/l=\cos\phv$, then
 the invariant tori are generically two-dimensional, and occur when $\cos\phv\neq\pm 1$; an invariant  torus labeled by such a value of $\phv\neq 0,\pi$ comprises all great circles (regarded as part of phase space by adding  to each point of the geodesic a velocity of unit length and direction tangent to the geodesic) whose angle with the $z$-axis is $\phv$ (more precisely, the angle in question is the one between the normal of the plane through the given great circle and the $z$-axis). For $\cos\phv=\pm 1$ (i.e.\ $m=\pm l$), however, there is only one great circle with 
 $\phv=0$  namely the equator (the case $\phv=\pi$ corresponds to the same equator traversed in the opposite direction). Hence in this case  the invariant torus is one-dimensional. The reader may be surprised that the invariant tori explicitly depend on the choice of variables, but this feature is typical of so-called degenerate systems; see Arnold (1989), \S 51.}

What remains, then,  of Bohr's picture of the hydrogen atom in this light?\footnote{We ignore coupling to the electromagnetic field here; see footnote \ref{SigalF}.} Quite a lot, in fact, confirming his remarkable physical intuition. 
The energy levels Bohr calculated are those given by the Schr\"{o}dinger equation, and hence remain correct in mature \qm. His orbits make literal sense only in 
the ``correspondence principle" limit $\hbar\raw 0$, $n\raw\infty$, where, however, the situation is even better than one might expect for integrable systems: because of the high degree of symmetry of the Kepler problem (Guillemin \&\ Sternberg, 1990),  one may construct energy eigenfunctions whose limit measure $\mu^0$ concentrates on any desired classical orbit (Nauenberg, 1989).\footnote{Continuing footnote \ref{Zelditch}, for a given principal quantum number $n$ one forms the eigenfunction $\Ps^{\hbar}_{(n,n-1,n-1)}$ by multiplying the spherical harmonic $Y^{n-1}_{n-1}$ with the appropriate radial wave function. The limiting measure \er{measure} as $n\raw\infty$ and $\hbar\raw 0$ is then concentrated on an orbit (rather than on an invariant torus). Now, beyond what it possible for general integrable systems, one may use the $SO(4)$ symmetry of the Kepler problem and the construction in footnote \ref{FNPER} for the group-theoretic coherent states of Perelomov (1986) to find
 the desired eigenfunctions. See also De Bi\`evre (1992) and  De Bi\`evre et al. (1993).}
In order to recover a travelling wave packet, one has to form wave packets from a very large number of energy eigenstates with very high quantum numbers, as explained in Subsection \ref{Ssection}. For finite $n$ and $\hbar$ Bohr's orbits seem to have no meaning, as already recognized by Heisenberg (1969)   in his pathfinder days!\footnote{The later Bohr also conceded this through his idea that causal descriptions are complementary to space-time pictures; see Subsection \ref{compl}.}   
\subsection{The  WKB approximation} \label{WKBS}
One might have expected a section on the $\hbar\raw 0$ limit of \qm\ to be centered around the WKB approximation, as practically all textbooks base their discussion of the classical limit on this notion. Although the scope of this method is actually rather limited, it is indeed worth saying a few words about it. For simplicity we restrict ourselves to the time-independent case.\footnote{Cf.\ Robert (1998) and references therein for the time-dependent case.} In its original formulation, the time-independent WKB method involves an attempt to approximate solutions of the time-independent Schr\"{o}dinger equation $H\Ps=E\Ps$ by  wave functions of the type
\beq \Ps(x)=a_{\hbar}(x)e^{\frac{i}{\hbar} S(x)}, \label{WKB}
\eeq
 where $a_{\hbar}$ admits an expansion in $\hbar$ as a power series.
 Assuming the Hamiltonian $H$ is of the form \er{Hh}, 
 plugging the Ansatz \er{WKB} into the Schr\"{o}dinger equation, and expanding in $\hbar$, yields in  lowest order the classical (time-independent) Hamilton--Jacobi equation
 \beq
 h\left( \frac{\partial S}{\partial x},x\right)=E, \label{HJE}
 \eeq
supplemented by the so-called (homogeneous) transport equation\footnote{
Only stated here for a classical Hamiltonian  $h(p,q)=p^2/2m + V(q)$. Higher-order terms in $\hbar$ yield further, inhomogeneous transport equations for the expansion coefficients $a_j(x)$  in $a_{\hbar}=\sum_j a_j \hbar^j$. These can be solved in a recursive way, starting with \er{TPE}.} 
\beq 
\left(\half\Delta S +  \sum_k\frac{\partial S}{\partial x^k}\frac{\partial }{\partial x^k}\right) a_0=0.\label{TPE}\eeq
 In particular, $E$ should be a classically allowed value of the energy.
Even when it applies (see below), in most cases of interest 
 the Ansatz \er{WKB} is only valid locally (in $x$), leading to problems with caustics. These problems turn out to be an artefact of the use of the coordinate representation that lies behind the choice of the \Hs\ $\H=L^2(\R^n)$,
 and can be avoided (Maslov \&\  Fedoriuk, 1981):  the WKB method really comes to its own in a geometric reformulation in terms of symplectic geometry. See   Arnold (1989), Bates \&\ Weinstein (1995), and  Dimassi \&\  Sj\"ostrand (1999) for  (nicely complementary) introductory treatments, and Guillemin \&\ Sternberg (1977), H\"{o}rmander (1985a, 1985b), and Duistermaat (1974, 1996) for advanced accounts. 
  
  The basic observation leading to this reformulation is that in the rare cases that $S$ is defined globally as a  smooth function  on the configuration space $\R^{n}$, it  defines a submanifold $\CL$ of the phase space $M=\R^{2n}$ by $\CL=\{(p=dS(x),q=x), x\in\R^n\}$. This submanifold is {\it Lagrangian} in having two defining properties: firstly, 
$\CL$ is $n$-dimensional, and secondly,  the restriction of the symplectic form (i.e.\  $\sum_k dp_k\wedge dq^k$) to $\CL$ vanishes. The Hamilton--Jacobi equation \er{HJE} then guarantees that the Lagrangian  submanifold $\CL\subset M$ is contained in the surface $\Sigma_E=h\inv(E)$ of constant energy $E$ in $M$. Consequently, any solution of the Hamiltonian equations of motion that starts in $\CL$ remains in $\CL$.
 
 In general, then, the starting point of the WKB approximation is a Lagrangian submanifold $\CL\subset \Sigma_E\subset M$, rather than some function $S$ that defines it locally.
By a certain adaptation of the geometric quantization procedure, one may, under suitable conditions, associate a unit vector $\Ps_{\CL}$ in a suitable \Hs\ to $\CL$, which for small $\hbar$ happens to be a good approximation to an eigenfunction of $H$ at eigenvalue $E$. This strategy is successful in the integrable case, where 
the nondegenerate tori (i.e. those of maximal dimension $n$) provide such Lagrangian submanifolds of $M$; the associated  unit vector   $\Ps_{\CL}$
then turns out to be well defined precisely when $\CL$ satisfies (generalized) Bohr--Sommerfeld quantization conditions. In fact, this is how the measures
 $\mu^0_{\mathsf{E}}$ in \er{measure} are generally computed in the integrable case. 

 If the underlying classical system is not integrable, it may still be close enough to  integrability for invariant tori to be defined. Such systems are called quasi-integrable or perturbations of integrable systems, and are described by the Kolmogorov--Arnold--Moser (KAM) theory;  see  Gallavotti (1983), Abraham \&\ Marsden (1985), Ozorio de Almeida (1988), Arnold (1989),   Lazutkin (1993), Gallavotti,  Bonetto \&\  Gentile (2004), and many other books. In such systems the  WKB method continues to provide approximations to the energy eigenstates  relevant to the surviving invariant tori (Colin de Verdi\`ere, 1977;  Lazutkin, 1993; Popov, 2000), but already loses some of its appeal. 
 
 In general systems, notably chaotic ones, the WKB method is almost useless. Indeed, the following theorem of Werner (1995)
shows that the measure $\mu^0_{\mathsf{E}}$ defined by a  WKB function \er{WKB} is concentrated on the Lagrangian submanifold $\CL$ defined by $S$:
\begin{quote} {\it Let $a_{\hbar}$ be in $L^2(\R^n)$ for each $\hbar>0$ with pointwise limit $a_0=\lim_{\hbar\raw 0}a_{\hbar}$
also in $L^2(\R^n)$,\footnote{This assumption is not made in Werner (1995), who directly assumes that $\Ps=a_0 \exp (iS/\hbar)$ in \er{WKB}.} and suppose that $S$ is almost everywhere differentiable. Then for each $f\in\cci(\R^{2n})$:}
\end{quote}
\beq \lho \left(a_{\hbar}e^{\frac{i}{\hbar}S},\qb(f)a_{\hbar}e^{\frac{i}{\hbar}S}\right)=
\int_{\R^n} d^nx\, |a_0(x)|^2 f\left( \frac{\partial S}{\partial x},x\right).\label{werner}
\eeq
  As we shall see shortly, this behaviour is impossible for ergodic systems, and this is enough to seal the fate of WKB for chaotic systems in general (except perhaps as a hacker's tool).
  
  Note, however, that for a given energy level $E$ the discussion so far has been  concerned with properties of the classical trajectories on $\Sigma_E$ (where they are constrained to remain by conservation of energy). Now, it belongs to the essence of \qm\ that other parts of phase space than   $\Sigma_E$ might be relevant to the spectral properties of $H$ as well. For example, for a classical Hamiltonian of the simple form $h(p,q)=p^2/2m + V(q)$, this concerns the so-called {\it classically forbidden area} $\{q\in\R^n\mid V(q)>E\}$ (and any value of $p$). Here the classical motion can have no properties like integrability or ergodicity, because it does not exist. Nonetheless, and perhaps counterintuitively, it is precisely here that a slight adaptation of the WKB method 
tends to be most effective. For $q=x$ in the classically forbidden area, the Ansatz \er{WKB} should be replaced by
 \beq \Ps(x)=a_{\hbar}(x)e^{-\frac{S(x)}{\hbar}}, \label{WKB2}
\eeq
 where this time $S$ obeys  the  Hamilton--Jacobi equation `for imaginary time', \footnote{This terminology comes from the Lagrangian formalism, where the classical action $S=\int dt\, L(t)$ is replaced by $iS$ through the substitution
 $t=-i\ta$ with $\ta\in\R$.}
 i.e.
 \beq
 h\left( i\frac{\partial S}{\partial x},x\right)=E, \label{HJE2}
 \eeq
and the transport equation \er{TPE} is unchanged. 
For example, it follows that in one dimension (with a Hamiltonian of the type \er{Schreq}) the WKB function \er{WKB2} assumes the form
\beq
\Ps(x) \sim e^{-\frac{\sqrt{2m}}{\hbar}\int^{|x|} dy\,\sqrt{V(y)-E}}\eeq
in the  forbidden region, which explains both the  tunnel effect in \qm\
(i.e. the propagation of the wave function into the forbidden region) {\it and} the fact that this effect disappears in the limit $\hbar\raw 0$. However, even here the use of WKB methods has  now largely been superseded by techniques developed by Agmon (1982); see, for example,  Hislop \&\ Sigal (1996) and   Dimassi \&\  Sj\"ostrand (1999) for reviews. 
 \subsection{Epilogue: quantum chaos}\label{QC}
Chaos in classical mechanics was probably known to Newton and was famously highlighted by Poincar\'e (1892--1899),\footnote{See also Diacu \&\ Holmes (1996) and Barrow-Green (1997) for historical background.} 
but its relevance for (and potential threat to) quantum theory was apparently first recognized by  Einstein (1917) in a paper that was `completely ignored for 40 years' (Gutzwiller, 1992).\footnote{It was the study of the very same Helium atom that led Heisenberg to  believe that a fundamentally new `quantum' mechanics was needed to replace the inadequate old quantum theory of Bohr and Sommerfeld. See  Mehra \&\ and Rechenberg (1982b) and Cassidy (1992). Another microscopic example of a chaotic system is the hydrogen atom in an external magnetic field.} Currently, the study of quantum chaos is one of the most thriving businesses in all of physics, as exemplified by   innumerable  conference proceedings and monographs on the subject, ranging from the classic by Gutzwiller (1990) to the online {\it opus magnum} by Cvitanovic et al.\ (2005).\footnote{Other respectable books include, for example, Guhr, M\"uller-Groeling \&\ Weidenm\"uller (1998), 
Haake (2001) and Reichl (2004).} Nonetheless, the subject is still not completely understood, and provides a fascinating testing ground for the interplay between classical and \qm. 

One should distinguish between various different goals in the field of quantum chaos.   The  majority of papers and books on quantum chaos is concerned with the semiclassical analysis of some concretely given  quantum system having a chaotic system as its classical limit. This means that one tries to approximate (for small $\hbar$)  
a suitable quantum-mechanical expression in terms of data associated with the underlying classical motion. Michael Berry even described this goal as the ``Holy Grail" of quantum chaos. 
The methods described in Subsection \ref{CEOM} contribute to this goal, but are largely
 independent of the nature of the dynamics. In this subsection we therefore concentrate on techniques and results specific to chaotic motion. 

Historically, the first new tool in semiclassical approximation theory that specifically applied to chaotic systems was the so-called  {\it Gutzwiller trace formula}.\footnote{This attribution is based on  Gutzwiller (1971). A  similar result was independently derived by  Balian \&\ Bloch (1972, 1974). See also Gutzwiller (1990)  and Brack \&\ Bhaduri (2003) for  mathematically heuristic but otherwise excellent accounts of semiclassical physics based on the trace formula.
Mathematically rigorous discussions and proofs may be found in
 Colin de Verdi\`{e}re (1973),  Duistermaat \&\ Guillemin (1975),  Guillemin \&\ Uribe (1989), Paul \&\ Uribe (1995), and Combescure,  Ralston, \&\ Robert (1999).
} Roughly speaking, this formula approximates  the eigenvalues of the quantum Hamiltonian in terms of the periodic (i.e.\ closed) orbits of the underlying classical Hamiltonian.\footnote{Such orbits are dense but of Liouville measure zero in chaotic classical systems. Their crucial role was first recognized  by Poincar\'e (1892--1899).} 
The Gutzwiller trace formula does not start from the wave function (as the WKB approximation does), but from the {\it propagator} $K(x,y,t)$. Physicists write this as
$K(x,y,t)=\langle x|\exp(-itH/\hbar)|y\rangle$, whereas mathematicians see it as 
the Green's function  in the formula
\beq e^{-\frac{it}{\hbar} H}\Ps(x)=\int d^n y\, K(x,y,t)\Ps(y), \eeq
where $\Ps\in L^2(\R^n)$. 
Its (distributional) Laplace transform
\beq 
G(x,y,E)=\frac{1}{i\hbar}\int_0^{\infty} dt\, K(x,y,t)e^{\frac{itE}{\hbar} } \eeq
contains information about both the spectrum and the eigenfunctions; for 
if the former is discrete, one has 
\beq G(x,y,E)=\sum_j \frac{\Ps_j(x)\ovl{\Ps_j(y)}}{E-E_j}.\eeq
It is possible to approximate $K$ or $G$ itself by an expression of the type
\beq
 K(x,y,t)\sim (2\pi i\hbar)^{-n/2}\sum_P \sqrt{|\det V_P|}e^{\frac{i}{\hbar}S_P(x,y,t)-\half i\pi \mu_P}, \eeq
where the sum is over {\it all} classical paths $P$ from $y$ to $x$ in time $t$ (i.e.\ paths that solve the classical equations of motion). Such a path has an associated action $S_P$, Maslov index $\mu_P$, and Van Vleck (1928) determinant $\det V_P$ (Arnold, 1989). For chaotic systems one typically has to include tens of thousands of paths in the sum, but if one does so the ensuing approximation turns out to be remarkably successful (Heller \&\ Tomsovic, 1993; Tomsovic \&\ Heller, 1993). The Gutzwiller trace formula is a semiclassical approximation to
\beq
g(E)=\int d^n x\, G(x,x,E)=\sum_j \frac{1}{E-E_j},
\eeq
for a quantum Hamiltonian with discrete spectrum and underlying classical Hamiltonian having chaotic motion. It has the form 
\beq g(E)\sim g_0(E)+ \frac{1}{i\hbar} \sum_P \sum_{k=1}^{\infty} \frac{T_P}{2\sinh(k\ch_P/2)}
e^{\frac{ik}{\hbar}S_P(E)-\half i\pi \mu_P},\label{GTF}\eeq
where $g_0$ is a smooth function giving the mean density of states. 
 This time, the sum is over all (prime)  {\it periodic} paths $P$ of the classical Hamiltonian at energy $E$ with associated action $S_P(E)=\oint pdq$ (where the momentum $p$ is determined by $P$, given $E$), period $T_P$,  and stability exponent $\ch_P$ (this is a measure of how rapidly neighbouring trajectories drift away from $P$). 
Since the frustration expressed by Einstein (1917), this was the first indication that semiclassical approximations had some bearing on chaotic systems. 

 Another important development concerning energy levels was the formulation of  two key conjectures:\footnote{Strictly speaking, both conjectures are wrong;   for example, the  harmonic oscillator yields a counterexamples to the first one. See Zelditch (1996a) for further information.   Nonetheless,  the conjectures  are believed to be true in a deeper sense.}
\begin{itemize}
\item If the classical dynamics defined by the classical Hamiltonian $h$ is integrable, then the spectrum of $H$ is ``uncorrelated" or ``random" (Berry \&\ Tabor, 1977).
\item  If the classical dynamics defined by $h$ is chaotic, then the spectrum of $H$ is ``correlated" or ``regular" (Bohigas, Giannoni, \&\ Schmit, 1984).
\end{itemize}
The notions of correlation and randomness used here can be made precise using notions 
like the distribution of level spacings and the pair correlation function of eigenvalues; see Zelditch (1996a) and  De Bi\`evre  (2001)
 for introductory treatments, and most of the literature cited in this subsection for further details.\footnote{This aspect of quantum chaos has applications to number theory and might even lead to a proof of the Riemann hypothesis; see, for example, Sarnak (1999),  Berry \&\ Keating (1999),  and many other recent papers.
Another relevant connection, related to the one just mentioned, is between energy levels  and random matrices; see especially Guhr, M\"uller-Groeling \&\ Weidenm\"uller (1998). For the plain relevance of all this to practical physics see  Mirlin (2000).}

We now consider energy eigenfunctions instead of eigenvalues, and return to the limit measure \er{measure}. In the non (quasi-) integrable case, the key result is that 
\begin{quote}
{\it for ergodic classical motion,\footnote{Ergodicity is the weakest property that any chaotic dynamical system possesses. See  Katok \&\ Hasselblatt (1995), Emch \&\ Liu (2002), Gallavotti,  Bonetto \&\  Gentile (2004), and countless other books.} the limit measure
 $\mu^0_{\mathsf{E}}$ coincides with the (normalized) Liouville measure induced on the constant energy surface $\Sigma_E\equiv h\inv(E)$.}\footnote{The unnormalized Liouville measure $\mu^u_E$ on $\Sigma_E$ is defined by $\mu^u_E(B)=\int_B dS_E(x)\, (\|dh(x)\|)\inv$, where $dS_E$ is the surface element on $\Sigma_E$ and $B\subset \Sigma_E$ is a Borel set. If $\Sigma_E$ is compact, the normalized
Liouville measure $\mu_E$ on $\Sigma_E$ is given by $\mu_E(B)=\mu^u_E(B)/\mu^u_E(\Sigma_E)$. It is a probability measure on $\Sigma_E$, reflecting the fact that the eigenvectors $\Ps^{\hbar}_{\mathsf{n}}$ are normalized to unit length so as to define quantum-mechanical states.}
\end{quote}
 This result was first suggested in the mathematical literature for ergodic geodetic motion on compact hyperbolic Riemannian manifolds (Snirelman, 1974), where it was subsequently proved with increasing generality  (Colin de Verdi\`ere, 1985; Zelditch, 1987).\footnote{In the Riemannian case with $\hbar=1$  the cosphere bundle $S^*Q$ (i.e.\ the subbundle of the cotangent bundle $T^*Q$ consisting of one-forms of unit length) plays the role of $\Sigma_E$. Low-dimensional  examples of ergodic  geodesic motion are provided by compact hyperbolic spaces. Also cf.\  Zelditch (1992a) for the physically important case of a particle moving in an external gauge field.
See also the appendix to Lazutkin (1993) by A.I. Shnirelman, and  Nadirashvili, Toth, \&\ Yakobson (2001) for  reviews.} For certain other ergodic systems this property was proved by Zelditch (1991), G\'erard \&\ Leichtnam (1993), Zelditch \&\  Zworski (1996), and others;  to the best of our knowledge a completely general proof remains to be given. 

An analogous  version for Schr\"odinger operators on $\R^n$ was independently stated in the physics literature (Berry, 1977b,  Voros, 1979), and was eventually proved under certain assumptions on the potential by Helffer, Martinez \&\ Robert (1987), Charbonnel (1992), and Paul \&\ Uribe (1995).  Under suitable assumptions  one therefore has
\beq \lim_{\hbar\raw 0,\mathsf{n}\raw\infty} \left(\Ps^{\hbar}_{\mathsf{n}},\qb(f)\Ps^{\hbar}_{\mathsf{n}}\right)=
\int_{\Sigma_E} d\mu_E\, f \label{HMR}\eeq
 for any $f\in\cci(\R^{2n})$, where again $\mu_E$ is the (normalized) Liouville measure on
$\Sigma_E\subset \R^{2n}$ (assuming this space to be compact). In particular, in the ergodic case $\mu_{\mathsf{E}}^0$ only depends on $E^0$ 
and is the same for (almost) every  sequence of energy eigenfunctions $(\Ps^{\hbar}_{\mathsf{n}})$ as long as $E^{\hbar}_n\raw E^0$.\footnote{\label{scarss}   The result is not necessarily valid for all sequences $(\Ps^{\hbar}_{\mathsf{n}})$ with the given limiting behaviour, but only for
`almost all' such sequences (technically, for a class of sequences of density 1).
See, for example,  De Bi\`evre  (2001) for a simple explanation of this.} 
Thus the support of the limiting measure is uniformly spread out over the 
largest part of phase space that is dynamically possible.

The result that for ergodic classical motion $\mu_{\mathsf{E}}^0$ is the Liouville measure on  $\Sigma_E$ under the stated condition leaves room for  the phenomenon of `scars', according to which in chaotic systems the limiting measure  is sometimes concentrated on periodic classical orbits. 
This terminology is used in two somewhat different   ways in the literature. 
`Strong' scars survive in the limit $\hbar\raw 0$ and  concentrate on stable closed orbits;\footnote{An orbit $\gm\subset M$ is called {\it stable} when
for each neighbourhood $U$ of $\gm$ there is neighbourhood $V\subset U$ of $\gm$ such that $z(t)\in U$ for all $z\in V$ and all $t$.}
 they  may come from `exceptional' sequences of eigenfunctions.\footnote{Cf.\ footnote  \ref{scarss}.}
These are mainly considered in the mathematical literature; cf.\  Nadirashvili, Toth, \&\ Yakobson (2001) and references therein.

 In the physics literature, on the other hand, the notion of a scar usually refers to an anomalous concentration of the functions $B^{\hbar}_{\Ps^{\hbar}_{\mathsf{n}}}$ (cf.\ \er{BFexp}) near {\it un}stable closed orbits
for {\it finite} values of $\hbar$; see Heller \&\ Tomsovic (1993), ÊTomsovic \&\ Heller (1993),
Kaplan  \&\ Heller (1998a,b), and Kaplan (1999) for surveys. Such scars  turn out to be crucial in attempts to explain the energy spectrum of the associated quantum system. The reason why such scars do not survive  the (double) limit in \er{measure} is that this limit is defined with respect to $\hbar$-independent smooth test functions. Physically, this means that one averages over more and more De Broglie wavelengths as $\hbar\raw 0$, eventually losing information about the single wavelength scale (Kaplan, 1999). Hence to pick them up in a mathematically sound way,  one should redefine \er{measure} as a  pointwise limit (Duclos \&\ Hogreve, 1993, Paul \&\ Uribe, 1996, 1998). In any case, there is no contradiction between the mathematical results cited and what physicists have found.

 Another  goal of quantum chaos is the identification of chaotic phenomena within a given quantum-mechanical model. Here the slight complication arises that one cannot simply copy the classical definition of chaos in terms of diverging trajectories in phase space, since (by unitarity of time-evolution) in \qm\ $\|\Ps(t)-\Phi(t)\|$ is constant in time $t$ for solutions of the Schr\"odinger equation. However, this just indicates that should intrinsic quantum chaos exist, it has to be defined differently from classical chaos.\footnote{As pointed out by Belot \&\ Earman (1997), the Koopman formulation of classical mechanics (cf.\ footnote  \ref{Koopman}) excludes  classical chaos 
if this is formulated in terms of trajectories in \Hs.  The transition from classical to quantum notions of chaos can be smoothened by first  reformulating the classical definition of chaos
(normally put in terms of properties of trajectories in phase space).} This 
has now been largely accomplished in the algebraic formulation of quantum theory (Benatti, 1993; Emch et al., 1994;, Zelditch, 1996b,c;  Belot \&\ Earman, 1997; Alicki \&\ Fannes, 2001; Narnhofer, 2001). The most significant recent development in this direction in the ``heuristic" literature has been the study of the quantity 
\beq
M(t)=|(e^{-\frac{it}{\hbar} (H+\Sigma)}\Ps,e^{-\frac{it}{\hbar} H}\Ps)|^2,
\eeq
where $\Ps$ is a coherent state (or Gaussian wave packet), and $\Sg$ is some perturbation of the Hamiltonian $H$ (Peres, 1984). In what is generally regarded as a breakthrough in the field, Jalabert \&\   Pastawski (2001) discovered that in a certain regime $M(t)$ is independent of the detailed form of $\Sg$ and decays as $\sim \exp(-\lm t)$, where $\lm$ is the (largest) Lyapunov exponent of the underlying classical system. See Cucchietti (2004) for a detailed account and further development. 

In any case, the  possibility that classical chaos appears in the $\hbar\raw 0$  limit of \qm\ is by no means predicated on the existence of intrinsic quantum chaos in the above sense.\footnote{Arguments by  Ford (1988) and others to the effect that \qm\ is wrong because it cannot give rise to chaos in its classical limit have to be discarded for the reasons given here. See also Belot \&\ Earman (1997).
 In fact,  using the same  argument, such authors could simultaneously 
have `proved' the {\it opposite} statement  that  any classical dynamics that  arises as the classical limit of a quantum theory
with non-degenerate spectrum must be ergodic. For the naive definition of quantum ergodic flow clearly is that quantum time-evolution sweeps out all states at some energy $E$; but for non-degenerate spectra this is a tautology by definition of an eigenfunction!}
For  even in the unlikely case that quantum dynamics would turn out to be intrinsically non-chaotic,  its classical limit is sufficiently singular to admit  kinds of classical motion without a qualitative counterpart in quantum theory.  This possibility is not only confirmed by most of the literature on quantum chaos
(little of which makes any use of notions of intrinsic quantum chaotic motion), but even more so by the possibility of {\it incomplete} motion.
This is a type of dynamics in which the flow of the Hamiltonian vector field is only defined until a certain time $t_f<\infty$ (or from an initial time $t_i>-\infty$), which means that the equations of motion have no solution for 
$t>t_f$ (or $t<t_i$).\footnote{\label{crunch} The simplest examples are incomplete Riemannian manifolds $Q$ with geodesic flow; within this class, the case $Q=(0,1)$ with flat metric is hard to match in simplicity. Clearly, 
the particle reaches one of the two boundary points in finite time, and does not know what to do (or even whether its exists) afterwards. Other examples come from potentials $V$ on $Q=\R^n$ with the property that the classical dynamics is incomplete; see Reed \&\ Simon (1975) and  Gallavotti (1983). On a somewhat different note, the Universe itself has incomplete dynamics because of the Big Bang and possible Big Crunch.} The point, then, is that unitary quantum dynamics, though intrinsically complete,   may very well have incomplete motion as its classical limit.\footnote{\label{crunch2}
The quantization of the Universe is unknown at present, but geodesic motion on Riemannian manifolds, complete or not, is quantized by $H=-\frac{\hbar^2}{2m}\Delta$ (perhaps with an additonal term proportional to the Ricci scalar $R$, see Landsman (1998)), where $\Delta$ is the Laplacian, and quantization on $Q=\R^n$ is given by the Schr\"{o}dinger equation \er{Schreq}, whether or not the classical dynamics is complete. In these two cases, and probably more generally, the incompleteness of the classical motion is often (but not always) reflected by the lack of essential self-adjointness of the quantum Hamiltonian on its natural initial domain $\cci(Q)$. For example, if $Q$ is complete as a Riemannian manifold, then 
$\Delta$ is essentially self-adjoint on $\cci(Q)$ (Chernoff, 1973, Strichartz, 1983), and if $Q$ is incomplete then the Laplacian usually fails to be essentially self-adjoint on this domain (but see  Horowitz \&\ Marolf (1995) for counterexamples). One may refer to the latter property as quantum-mechanical incompleteness (Reed \&\ Simon, 1975), although a Hamiltonian
that fails to be essentially self-adjoint on $\cci(Q)$ can often be extended 
(necessarily in a non-unique way) to a self-adjoint operator by a choice of boundary conditions (possibly at infinity). By Stone's theorem,   the quantum dynamics  defined by each self-adjoint extension is unitary  (and therefore defined for all times).
Similarly, although no general statement can be made relating (in)complete classical motion in a potential to (lack of) essential selfadjointness of the corresponding Schr\"odinger operator, it is usually the case that completeness implies essential selfadjointness, and vice versa. See Reed \&\ Simon (1975),  Appendix to \S X.1, where the reader may also find  examples of  classically incomplete but quantum-mechanically complete motion, and vice versa.  Now, here is the central point for the present discussion: as probably first noted by Hepp (1974), {\it different self-adjoint extensions have the same classical limit} (in the sense of \er{hepp} or similar criteria), namely the given {\it incomplete} classical dynamics. This proves that complete quantum dynamics  can have incomplete motion as its classical limit. However, much remains to be understood in this area. See also Earman (2005, 2006).}
 \section{The limit $N\raw\infty$}\label{S6}\setcounter{equation}{0}
In this section we show to what extent  classical physics may approximately emerge from quantum theory when the size of a system becomes large. Strictly classical behaviour would be an idealization reserved for the limit where this size  is infinite, which we symbolically denote by ``$\lim N\raw\infty$".
As we shall see, mathematically speaking this limit is a special case of the limit $\hbar\raw 0$ discussed in the previous chapter. What is more,  we shall show that  formally the limit $N\raw\infty$ even  falls under the heading of continuous fields of \ca s and deformation quantization (see Subsection \ref{DQsection}.) Thus the `philosophical' nature of the idealization involved in assuming that a system is infinite is much the same as that of assuming $\hbar\raw 0$ in a quantum system of given (finite) size; in particular, the introductory comments in Section \ref{S1} apply here as well.

An analogous  discussion pertains to the derivation of thermodynamics from statistical mechanics (Emch \&\ Liu, 2002; Batterman, 2005). For example, {\it in theory} phase transitions only occur in infinite systems, but {\it in practice} one sees them every day. Thus it appears to be valid
to approximate a pot of $10^{23}$ boiling water molecules by an infinite number of such molecules.  The basic point is that the distinction between microscopic and macroscopic regimes is unsharp unless one admits infinite systems as an idealization, so that one can simply say that microscopic systems are finite, whereas macroscopic systems are infinite. This procedure is eventually justified by the results it produces.

Similarly, in the  context of quantum theory classical behaviour is simply not found in finite systems (when $\hbar>0$ is fixed), whereas, as we shall see, it {\it is} found in infinite ones. Given the observed classical nature of the macroscopic world,\footnote{With the well-known mesoscopic exceptions (Leggett, 2002;  Brezger et al., 2002; Chiorescu et al., 2003; 
Marshall et al., 2003;  Devoret et al., 2004). }  at the end of the day one concludes that the idealization in question is apparently a valid one.  
One should not be confused by the fact that the error in the number of particles this approximation involves (viz.\ $\infty-10^{23}=\infty$) is considerably larger than the  number of particles in the actual system.  If all of the $10^{23}$ particles in question were {\it individually} tracked  down, the approximation is indeed a worthless ones, but the point is rather that the limit  $N\raw\infty$ is valid whenever {\it averaging} over  $N=10^{23}$ particles is well approximated by averaging over an arbitrarily larger number $N$ (which, then, one might as well let go to infinity). Below we shall give a precise version of this argument. 

Despite our opening comments above,  the quantum theory of infinite systems has features of its own that deserve a separate section. Our treatment is 
complementary to texts such as Thirring (1983), Strocchi (1985),  Bratteli \&\  Robinson (1987), Haag (1992), Araki (1999), and  Sewell (1986, 2002), which should be consulted for further information on infinite quantum systems. 
The theory in Subsections \ref{MO} and \ref{PSD} is a  reformulation in terms of continuous field of \ca s and  deformation quantization of the more elementary parts of a remarkable series of papers on so-called quantum mean-field systems by Raggio \&\  Werner (1989, 1991),  Duffield \&\  Werner (1992a,b,c), and  Duffield, Roos, \&\  Werner (1992). These models have their origin in the treatment of the BCS theory of superconductivity due to Bogoliubov (1958) and Haag (1962), with important further contributions by Thirring \&\ Wehrl (1967), Thirring (1968), Hepp (1972), Hepp \&\ Lieb (1973), Rieckers (1984), Morchio \&\ Strocchi (1987), Duffner \&\ Rieckers (1988), Bona (1988, 1989, 2000), Unnerstall (1990a, 1990b),  Bagarello \&\ Morchio (1992), Sewell (2002), and others.
 \subsection{Macroscopic observables}\label{MO}
The large quantum systems we are going to study consist of $N$  copies of a single quantum system with unital algebra of observables $\CA_1$. Almost all features already emerge in the simplest example $\CA_1=M_2(\C)$  (i.e.\ the complex $2\x 2$ matrices), so there is nothing wrong with having this case in mind as abstraction increases.\footnote{In the opposite direction of greater generality, it is worth noting that the setting below actually incorporates quantum systems defined on general lattices in $\R^n$ (such as $\mathbb{Z}^n$). For one could relabel things so as to make $\CA_{1/N}$ below
the algebra of observables of all lattice points $\Lm$ contained in, say, a sphere of radius $N$. The limit $N\raw\infty$ then corresponds to the limit $\Lm\raw\mathbb{Z}^n$.} 
The aim of what follows is to describe in what precise sense macroscopic observables (i.e.\ those obtained by averaging over an infinite number of sites) are ``classical".

 From the single
\ca\ $\CA_1$, we construct a continuous field of \ca s $\CA^{\mathrm (c)}$ over 
\beq I=0\cup 1/\N=\{0,\ldots, 1/N,\ldots, \third,\half,1\}\subset [0,1], \label{interval}\eeq
as follows. 
 We put
\begin{eqnarray}
\CA_0^{\mathrm (c)}&=& C(\CS(\CA_1));\nn \\
\CA_{1/N}^{\mathrm (c)}&=& \CA_1^N, \label{fibers}
\end{eqnarray}
where $\CS(\CA_1)$ is the state space of $\CA_1$ (equipped with the weak$\mbox{}^*$-topology)\footnote{In this topology one has $\om_{\lm}\raw\om$ when $\om_{\lm}(A)\raw\om(A)$ for each $A\in\CA_1$.}
 and $\CA_1^N=\hat{\ot}^N \CA_1$ is
 the (spatial) tensor product of $N$ copies of $\CA_1$.\footnote{When $\CA_1$ is finite-dimensional the tensor product is unique. In general, one needs the  {\it projective} tensor product at this point. See footnote \ref{tensorproducts}.
The point is the same here: any tensor product state
$\om_1\ot\cdots\ot\om_N$ on $\ot^N\CA_1$ - defined on elementary tensors by $\om_1\ot\cdots \ot\om_N(A_1\ot\cdots\ot A_N)=\om_1(A_1)\cdots \om_N(A_N)$ - extends to a state on $\hat{\ot}^N \CA_1$ by continuity.
}  This explains the suffix $c$ in $\CA^{\mathrm (c)}$: 
it refers to the fact that the limit algebra $\CA_0^{\mathrm (c)}$ is {\it c}lassical or
 {\it c}ommutative.

 For example, take 
$\CA_1=M_2(\C)$. Each state is given by a density matrix, which is of the form  
\begin{equation}
\rh(x,y,z)=\half   \left(\begin{array}{cc} 1+z & x-iy \\ x+iy & 1-z \end{array}\right),
\label{gens2} \end{equation}
for some $(x,y,z)\in\R^3$ satisfying $x^2+y^2+z^2\leq 1$. Hence $\CS(M_2(\C))$ is isomorphic (as a compact convex set) to the three-ball $B^3$ in $\R^3$. The pure states are precisely the points on the boundary,\footnote{\label{EBfn} The {\it extreme boundary} $\partial_e K$ of a convex set $K$ consists of all $\om\in K$ for which $\om=p\rh+(1-p)\sg$ for some
$p\in (0,1)$ and $\rh,\sg\in K$ implies $\rh=\sg=\om$. If $K=\CS(\CA)$ is the state space of a \ca\ $\CA$, the extreme boundary consists of the pure states on $\CA$ (the remainder of 
$\CS(\CA)$ consisting of mixed states). If $K$ is embedded in a vector space, the extreme boundary $\partial_e K$ may or may not coincide with the geometric boundary $\partial K$ of $K$. In the case $K=B^3\subset \R^3$ it does, but for an equilateral triangle in $\R^2$ 
it does not, since $\partial_e K$ merely consists of the corners of the triangle whereas the geometric boundary includes the sides as well.}
 i.e.\ the density matrices for which $x^2+y^2+z^2=1$ (for these and these alone define one-dimensional projections).\footnote{\label{SSlemma}Eq.\ \er{gens2} has the form $\rh(x,y,z)=\half(x\sg_x+y\sg_y+z\sg_z)$, where the $\sg_i$ are the Pauli matrices.
This yields an isomorphism between $\R^3$ and the Lie algebra of $SO(3)$ in its spin-$\half$ \rep\ $\mathcal{D}_{1/2}$ on $\C^2$. This isomorphism intertwines the defining action of $SO(3)$ on $\R^3$ with its adjoint action on $M_2(\C)$. I.e., for any rotation $R$ one has $\rh(R\mathbf{x})=\mathcal{D}_{1/2}(R)\rh(\mathbf{x})\mathcal{D}_{1/2}(R)\inv$. This will be used later on (see Subsection \ref{PSD}).
}

 In order to define the continuous sections of the field, we introduce the {\it symmetrization maps} $j_{NM}:  \CA^M_1\raw  \CA^N_1$, defined by
\beq
j_{NM}(A_M)=S_N(A_M\ot 1\ot\cdots \ot 1), \label{symmaps}
\eeq 
where one has $N-M$ copies of the unit $1\in\CA_1$ so as to obtain an element of $\CA_1^N$. The symmetrization operator $S_N: \CA^N_1\raw \CA^N_1$ is given by (linear and continuous) extension of \beq
S_N(B_1\ot\cdots \ot B_N)=\frac{1}{N!}\sum_{\sg\in \GS_N} B_{\sg(1)}\ot\cdots\ot B_{\sg(N)}, \label{landc}
\eeq
where $\GS_N$ is the permutation group (i.e.\ symmetric group) on $N$ elements and $B_i\in\CA_1$ for all $i=1,\ldots,N$. For example, $j_{N1}:\CA_1\raw\CA_1^N$ is given by
\beq
j_{N1}(B)=
\ovl{B}^{(N)}=\frac{1}{N}\sum_{k=1}^N 1\ot\cdots\ot B_{(k)}\ot 1\cdots \ot1,\eeq
where $B_{(k)}$ is $B$ seen as an element of the $k$'th copy of $\CA_1$ in $\CA_1^N$. As our notation $\ovl{B}^{(N)}$ indicates, this is just the `average' of $B$ over all copies of $\CA_1$. More generally, in forming $j_{NM}(A_M)$ an
operator $A_M\in\CA_1^M$ that involves $M$ sites is averaged over $N\geq M$ sites. When $N\raw\infty$ this means that one forms a {\it macroscopic} average of an $M$-particle operator.

We say that a sequence $A=(A_1,A_2,\cdots)$ with $A_N\in\CA_1^N$
is {\it symmetric} when  
\beq A_N=j_{NM}(A_M) \label{ass}
\eeq  for some fixed $M$ and all $N\geq M$.  In other words, the tail of a symmetric sequence entirely consists of `averaged' or `intensive' observables, which become macroscopic in the limit $N\raw\infty$.
Such sequences have the important property that they commute in this limit; more precisely, if $A$ and $A'$ are symmetric sequences, then
\beq
\lni \| A_NA_N'-A_N'A_N\|=0. \label{aprc}
\eeq
As an enlightening special case we take $A_N=j_{N1}(B)$ and $A_N'=j_{N1}(C)$ with $B,C\in\CA_1$. One  immediately obtains from the relation $[B_{(k)},C_{(l)}]=0$ for $k\neq l$
 that
\beq \left[\ovl{B}^{(N)},\ovl{C}^{(N)}\right]=\frac{1}{N}\ovl{\left[B,C\right]}^{(N)}.
\label{av0}
\eeq
For example, if $\CA_1=M_2(\C)$ and if for $B$ and $C$ one takes the spin-$\half$ operators
$S_j=\frac{\hbar}{2}\sg_j$ for $j=1,2,3$ (where $\sg_j$ are the Pauli matrices), then
\beq 
\left[\ovl{S}_j^{(N)},\ovl{S}_k^{(N)}\right]=i\frac{\hbar}{N}\epsilon_{jkl} \ovl{S}_l^{(N)}.
\eeq
This shows that averaging one-particle operators leads to commutation relations formally like those of the one-particle operators in question, but with Planck's constant $\hbar$ replaced by a variable $\hbar/N$. For constant $\hbar=1$ this leads to the  interval \er{interval}  over which our continuous
field of \ca s is defined; for any other constant value of $\hbar$ the field would be defined over $I=0\cup \hbar/\N$, which of course merely changes the labeling of the \ca s in question.

We return to the general case, and denote a section of the field with fibers \er{fibers} by a sequence $A=(A_0,A_1,A_2,\cdots)$, with $A_0\in\CA_0^{\mathrm (c)}$ and $A_N\in\CA_1^N$ as before (i.e.\ the corresponding section is $0\mapsto A_0$ and $1/N\mapsto A_N$).
 We then complete the definition of our continuous field by declaring that 
a sequence $A$ defines a  {\it continuous} section iff:
\begin{itemize}
\item $(A_1,A_2,\cdots)$  is {\it approximately symmetric}, in the sense that for any $\varep>0$ there is an $N_{\varep}$ and a symmetric sequence $A'$ such that 
$\|A_N-A_N'\|< \varep$ for all $N\geq N_{\varep}$;\footnote{A symmetric sequence is evidently approximately symmetric.}
\item $A_0(\om)=\lni\om^N(A_N)$, where $\om\in\CS(\CA_1)$ and $\om^N\in \CS(\CA_1^N)$ is the tensor product of $N$ copies of $\om$, defined by (linear and continuous) extension of 
\beq \om^N(B_1\ot\cdots\ot B_N)=\om(B_1)\cdots\om(B_N).\label{omN}\eeq
 This limit exists by definition of an approximately symmetric sequence.\footnote{If  $(A_1,A_2,\cdots)$  is symmetric  with \er{ass}, one has $\om^N(A_N)=\om^M(A_M)$
for $N>M$, so that the tail of the sequence $(\om^N(A_N))$ is even independent of $N$. In the approximately symmetric case one easily proves that  $(\om^N(A_N))$ is a Cauchy sequence.}
\end{itemize}
  It is not difficult to prove that this choice of continuous sections indeed defines a continuous field of \ca s  over $I=0\cup 1/\N$ with fibers \er{fibers}.
The main point is that
\beq
\lni \| A_N\|=\|A_0\|  \label{normeq}
\eeq
whenever $(A_0,A_1,A_2,\cdots)$ satisfies the two conditions above.\footnote{\label{Landsmanfootnote}Given \er{normeq}, the claim follows from Prop.\ II.1.2.3 in Landsman (1998) and the fact that the set of functions $A_0$ on $\CS(\CA_1)$
arising in the said way are dense in $C(\CS(\CA_1))$ (equipped with the supremum-norm). This follows from the Stone--Weierstrass theorem, from which one infers that the functions in question even  exhaust $\CS(\CA_1)$.
}
 This is easy to show for symmetric sequences,\footnote{
Assume \er{ass}, so that $\|A_N\|=\|j_{NN}(A_N)\|$  for $N\geq M$. By the $C^*$-axiom $\|A^*A\|=\|A^2\|$ it suffices to prove \er{normeq} for $A_0^*=A_0$, which implies $A_M^*=A_M$ and hence $A_N^*=A_N$ for all $N\geq M$. One then has $\| A_N\|=\sup\{|\rh(A_N)|, \rh\in\CS(\CA_1^N)\}$. 
Because of the special form of $A_N$ one may replace the supremum over 
the set $\CS(\CA_1^N)$ of all states on $\CA_1^N$ by the supremum over the set $\CS^p(\CA_1^N)$ of all permutation invariant states, which in turn
may be replaced by the supremum over the extreme boundary $\partial \CS^p(\CA_1^N)$ of $\CS^p(\CA_1^N)$. It is well known (St\o rmer, 1969; see also Subsection \ref{QLO})
that the latter consists of all states of the form $\rh=\om^N$, so that 
$\| A_N\|=\sup\{|\om^N(A_N)|, \om\in\CS(\CA_1)\}$. This is actually equal
to $\| A_M\|=\sup\{|\om^M(A_M)|\}$.  Now
the norm in $\CA_0^{\mathrm (c)}$ is $\|A_0\|=\sup\{|A_0(\om)|, \om\in\CS(\CA_1)\}$, and
by definition of $A_0$ one has $A_0(\om)=\om^M(A_M)$. Hence
 \er{normeq} follows.} and follows from this for approximately symmetric ones.

Consistent with \er{aprc}, we conclude that in the limit $N\raw\infty$ the macroscopic observables organize themselves in a commutative \ca\ isomorphic to $C(\CS(\CA_1))$. 
\subsection{Quasilocal observables}\label{QLO}
In the \ca ic approach to quantum theory, infinite systems are usually 
described by means of inductive limit \ca s and the associated quasilocal 
observables (Thirring, 1983; Strocchi, 1985;  Bratteli \&\  Robinson, 1981, 1987; Haag, 1992; 
Araki, 1999; Sewell, 1986, 2002). To arrive at these notions in 
 the case at hand, we proceed as follows (Duffield \&\ Werner, 1992c).
 
   A sequence $A=(A_1,A_2,\cdots)$ (where $A_N\in\CA_1^N$, as before)  is called {\it local} when for some fixed $M$ and all $N\geq M$ one has $A_N=A_M\ot 1\ot\cdots\ot 1$
 (where one has $N-M$ copies of the unit $1\in\CA_1$); cf.\ \er{symmaps}. A sequence is said to be {\it quasilocal} when  for any $\varep>0$ there is an $N_{\varep}$
and a local sequence $A'$ such that 
$\|A_N-A_N'\|< \varep$ for all $N\geq N_{\varep}$.  On this basis, we define 
the {\it inductive limit} \ca\ \beq \ovl{\cup_{N\in\N} \CA_1^N}\label{ILCA} \eeq
of the  family of \ca s  $(\CA_1^N)$ with respect to the inclusion maps $\CA_1^N\hookrightarrow\CA_1^{N+1}$ given by $A_N\mapsto A_N\ot 1$.
As a set,  \er{ILCA} consists of all equivalence classes $[A]\equiv A_0$ of quasilocal sequences $A$ under the equivalence relation $A\sim B$ when $\lni\|A_N-B_N\|=0$. The norm on $\ovl{\cup_{N\in\N} \CA_1^N}$ is 
\beq \| A_0\|=\lni \|A_N\|, \label{normput}\eeq
and the rest of the \ca ic structure is inherited from the quasilocal sequences
in the obvious way (e.g., $A_0^*=[A^*]$ with $A^*=(A_1^*,A_2^*,\cdots)$, etc.).
As the notation suggests, each $\CA_1^N$ is contained in  $\ovl{\cup_{N\in\N} \CA_1^N}$ as a $C^*$-subalgebra by identifying $A_N\in \CA_1^N$ with
the local (and hence quasilocal) sequence $A=(0,\cdots,0, A_N\ot 1,A_N\ot 1\ot1, \cdots)$, and forming its equivalence class $A_0$ in $\ovl{\cup_{N\in\N} \CA_1^N}$ as just explained.\footnote{Of course, the entries  $A_1,\cdots A_{N-1}$, which have been put to zero, are arbitrary.} The assumption underlying the common idea that \er{ILCA} is ``the" algebra of observables of the infinite system under study is that by locality or some other human limitation the infinite tail of the system is not accessible, so that the observables must be arbitrarily close (i.e.\ in norm) to operators of the form $A_N\ot 1\ot1, \cdots$ for some {\it finite} $N$.

This leads us to a second continuous field of \ca s $\CA^{\mathrm (q)}$ over $0\cup 1/\N$, with fibers
 \begin{eqnarray}
\CA^{\mathrm (q)}_0&=& \ovl{\cup_{N\in\N} \CA_1^N};\nn \\
\CA^{\mathrm (q)}_{1/N}&=& \CA_1^N. \label{fibers2}
\end{eqnarray}
Thus the suffix $q$ reminds one of that fact that the limit algebra $\CA^{\mathrm (q)}_0$ consists of {\it q}uasilocal or {\it q}uantum-mechanical observables.
We equip the collection of \ca s \er{fibers2} with the structure of a
 continuous field of \ca s $\CA^{\mathrm (q)}$ over $0\cup 1/\N$  by declaring that the continuous sections are of the form $(A_0,A_1,A_2,\cdots)$
 where $(A_1,A_2,\cdots)$ is quasilocal and $A_0$ is defined by this quasilocal sequence as just explained.\footnote{The fact that this defines a continuous field follows from \er{normput} and Prop.\ II.1.2.3 in Landsman (1998); cf.\ footnote  \ref{Landsmanfootnote}.}
For $N<\infty$ this field has the same fibers 
\beq \CA^{\mathrm (q)}_{1/N}=\CA_{1/N}^{\mathrm (c)}= \CA_1^N\eeq
 as the continuous field $\CA$ of the previous subsection, but the fiber $\CA^{\mathrm (q)}_0$ is completely different from $\CA_0^{\mathrm (c)}$. In particular, if $\CA_1$ is noncommutative then so is $\CA^{\mathrm (q)}_0$, for it contains all $\CA_1^N$. 

The relationship between the continuous fields of \ca s $\CA^{\mathrm (q)}$ and $\CA^{\mathrm (c)}$ may be studied in two different (but related) ways. First,  we may construct concrete \rep s of all \ca s $\CA_1^N$, $N<\infty$, as well as of $\CA_0^{\mathrm (c)}$ and $\CA_0^{\mathrm (q)}$ on a single \Hs; this approach leads to superselections rules in the traditional sense. This method will be taken up in the next  subsection. Second, we may  look at those families of states $(\om_1,\om_{1/2},\cdots,\om_{1/N},\cdots)$
(where $\om_{1/N}$ is a state on $\CA_1^N$) that admit limit states $\om_0^{\mathrm (c)}$ {\it and} $\om_0^{\mathrm (q)}$ on $\CA_0^{\mathrm (c)}$ and $\CA_0^{\mathrm (q)}$, respectively, such that the ensuing families of states $(\om_0^{\mathrm (c)},\om_1,\om_{1/2},\cdots)$ and $(\om_0^{\mathrm (q)},\om_1,\om_{1/2},\cdots)$ are  {\it continuous} fields of states on  $\CA^{\mathrm (c)}$ and on $\CA^{\mathrm (q)}$, respectively (cf.\ the end of Subsection \ref{CSR}).

Now, any state $\om_0^{\mathrm (q)}$ on $\CA^{\mathrm (q)}_0$ defines  a state
$\om_{0|1/N}^{\mathrm (q)}$ on $\CA_1^N$ by restriction, and the ensuing field of states on $\CA^{\mathrm (q)}$ is clearly continuous. Conversely, any continuous field 
$(\om_0^{\mathrm (q)},\om_1,\om_{1/2},\ldots, \om_{1/N},\ldots)$ of states on $\CA^{\mathrm (q)}$
becomes arbitrarily close to a field of the above type for $N$ large.\footnote{
For any fixed quasilocal sequence $(A_1,A_2,\cdots)$ 
and $\varepsilon>0$, there is an $N_{\varepsilon}$ such that $|\om_{1/N}(A_N)-
\om_{0|1/N}^{\mathrm (q)}(A_N)|<\varepsilon$ for all $N>N_{\varepsilon}$.} 
However, the restrictions $\om_{0|1/N}^{\mathrm (q)}$ of a given state $\om_0^{\mathrm (q)}$ on $\CA^{\mathrm (q)}_0$ to $\CA_1^N$ may not converge to a state $\om_0^{\mathrm (c)}$ on $\CA_0^{\mathrm (c)}$ for $N\raw\infty$.\footnote{See footnote  \ref{259} below for an example}. States $\om_{0}^{\mathrm (q)}$ on
$\ovl{\cup_{N\in\N} \CA_1^N}$ that do  have this property will here be called {\it classical}.
In other words, $\om_{0|1/N}^{\mathrm (q)}$ is classical when there exists a probability measure $\mu_0$ on $\CS(\CA_1)$ such that
\beq
\lim_{N\raw\infty} \int_{\CS(\CA_1)} d\mu_0(\rh)\, (\rh^N(A_N)-\om_{0|1/N}^{\mathrm (q)}(A_N))=0 \label{PMOQ}\eeq
for each (approximately) symmetric sequence $(A_1, A_2,\ldots)$. 
 To analyze this notion we need a brief intermezzo on general \ca s and their \rep s. 
\begin{itemize}
\item
A  {\it folium}  in the state space $\CS(\CB)$ of a \ca\ $\CB$ 
 is a convex, norm-closed subspace $\CF$ of $\CS(\CB)$ with the property that if $\om\in\CF$ and $B\in\CB$ such that $\om(B^*B)>0$, then the ``reduced" state $\om_B:A\mapsto \om(B^*AB)/\om(B^*B)$ must be in $\CF$ (Haag, Kadison, \&\ Kastler, 1970).\footnote{See also Haag (1992). The name `folium' is very badly chosen, since  $\CS(\CB)$ is by no means foliated by its folia; for example, a folium may contain subfolia. }  For example, 
if $\pi$ is a \rep\ of $\CB$ on a \Hs\ $\H$, then the set of all density matrices on $\H$ (i.e.\ the $\pi$-normal states on $\CB$)\footnote{A state $\om$ on $\CB$ is called  $\pi$-normal when it is of the form $\om(B)=\Tr \rh\pi(B)$ for some density matrix $\rh$. Hence the $\pi$-normal states are the  normal states on the von Neumann algebra $\pi(\CB)''$.} comprises a folium $\CF_{\pi}$. 
In particular, each state $\om$ on $\CB$ defines a folium $\CF_{\om}\equiv \CF_{\pi_{\om}}$ through its GNS-\rep\ $\pi_{\om}$.
\item 
Two \rep s $\pi$ and $\pi'$ are called {\it disjoint}, written $\pi\bot \pi'$, if no sub\rep\ of $\pi$ is (unitarily) equivalent to a sub\rep\ of $\pi'$ and vice versa. They are said to be {\it quasi-equivalent}, written $\pi\sim \pi'$, when $\pi$ has no sub\rep\ disjoint from $\pi'$, and vice versa.\footnote{Equivalently, two  \rep s $\pi$ and $\pi'$ are disjoint iff no $\pi$-normal state
is $\pi'$-normal and vice versa, and quasi-equivalent iff each $\pi$-normal state is $\pi'$-normal and vice versa.} Quasi-equivalence is an equivalence relation $\sim$ on the set of \rep s.  See Kadison \&\ Ringrose (1986), Ch.\ 10.
\item 
Similarly, two states $\rh,\sg$  are called either quasi-equivalent ($\rh\sim\sg$) or disjoint ($\rh\bot\sg$) when the corresponding GNS-\rep s have these properties.
\item  A state $\om$ is called {\it primary} when the corresponding von Neumann algebra $\pi_{\om}(\CB)''$ is a factor.\footnote{A 
von Neumann algebra $\CM$ acting on a \Hs\ is called a {\it factor} when its center $\CM\cap\CM'$ is trivial, i.e.\ consists of multiples of the identity.}
Equivalently, $\om$ is primary iff each sub\rep\ of $\pi_{\om}(\CB)$ is quasi-equivalent to  $\pi_{\om}(\CB)$, which is the case iff $\pi_{\om}(\CB)$ admits no (nontrivial) decomposition
as the direct sum of two disjoint sub\rep s.
\end{itemize}

 Now, there is a bijective correspondence between folia in $\CS(\CB)$ and quasi-equivalence classes of \rep s of $\CB$, in that $\CF_{\pi}=\CF_{\pi'}$ iff $\pi\sim\pi'$. Furthermore (as one sees from the GNS-construction), any folium $\CF\subset \CS(\CB)$ is of the form $\CF=\CF_{\pi}$ for some \rep\ $\pi(\CB)$. 
Note that if $\pi$ is injective (i.e.\ faithful),  then the corresponding folium  is dense in $\CS(\CB)$ in the weak$\mbox{}^*$-topology by Fell's Theorem. So in case that
$\CB$ is simple,\footnote{In the sense that it has no {\it closed} two-sided ideals.
For example, the matrix algebra $M_n(\C)$ is simple for any $n$, as is its infinite-dimensional analogue, the \ca\ of all compact operators on a \Hs. The \ca\ of quasilocal observables 
of an infinite quantum systems is typically simple as well.} 
  any folium is weak$\mbox{}^*$-dense in the state space. 

Two states need not be either disjoint or quasi-equivalent. This dichotomy does apply, however, within the class of primary states.
 Hence {\it two primary states are either disjoint or
quasi-equivalent}. If $\om$ is primary, then each state in the folium of
$\pi_{\om}$ is primary as well, and is quasi-equivalent to $\om$. 
If, on the other hand, $\rh$ and $\sg$ are primary and disjoint,  then $\CF_{\rh}\cap\CF_{\sg}=\emptyset$.
Pure states are, of course, primary.\footnote{Since the corresponding GNS-\rep\ $\pi_{\om}$ is irreducible,
$\pi_{\om}(\CB)''=\CB(\H_{\om})$ is a factor.} Furthermore, in thermodynamics
pure phases are described by primary KMS states (Emch \&\ Knops, 1970;
 Bratteli \&\  Robinson, 1981; Haag, 1992; Sewell, 2002). This apparent relationship between
 primary states and ``purity" of some sort is confirmed by 
  our description of macroscopic observables:\footnote{These claims easily follow from Sewell (2002), \S 2.6.5, which in turn relies on Hepp (1972).}  {\it
\begin{itemize}
\item  If  $\om_0^{\mathrm (q)}$ is a classical  primary state on $\CA^{\mathrm (q)}_0=\ovl{\cup_{N\in\N} \CA_1^N}$, then the corresponding limit state $\om_0^{\mathrm (c)}$ on $\CA_0^{\mathrm (c)}=C(\CS(\CA_1))$ is pure (and hence given by a point in $\CS(\CA_1)$).
\item  If  $\rh_0^{\mathrm (q)}$ and  $\sg_0^{\mathrm (q)}$ are classical  primary states  on $\CA^{\mathrm (q)}_0$, then 
\begin{eqnarray}
\rh_0^{\mathrm (c)}=\sg_0^{\mathrm (c)} &\Leftrightarrow&  \rh_0^{\mathrm (q)}\sim \sg_0^{\mathrm (q)}; \label{rssim} \\ 
\rh_0^{\mathrm (c)}\neq\sg_0^{\mathrm (c)} &\Leftrightarrow& 
\rh_0^{\mathrm (q)} \bot\, \sg_0^{\mathrm (q)}. \label{rsbot}
\end{eqnarray}
\end{itemize}}

 As in \er{PMOQ}, a general classical state $\om_0^{\mathrm (q)}$ with limit state $\om_0^{\mathrm (c)}$ on 
$C(\CS(\CA_1))$ defines a probability measure $\mu_0$ on $\CS(\CA_1)$ by
\beq
 \om_0^{\mathrm (c)}(f)=\int_{\CS(\CA_1)} d\mu_0\, f, \label{probm} \eeq
which describes the probability distribution of the macroscopic observables in that state.
As we have seen,  this distribution  is a delta function for primary states. In any case, it is insensitive to the microscopic details of $\om_0^{\mathrm (q)}$ in the sense that local modifications of $\om_0^{\mathrm (q)}$ do not affect the limit state $\om_0^{\mathrm (c)}$ (Sewell, 2002). Namely, it easily follows from \er{aprc}
and the fact that the GNS-\rep\ is cyclic that one can strengthen the second claim above:
\begin{quote} {\it Each state in the folium $\CF_{\om_0^{\mathrm (q)}}$ of a classical state $\om_0^{\mathrm (q)}$ is automatically classical and has the same limit state on $\CA_0^{\mathrm (c)}$ as $\om_0^{\mathrm (q)}$.}
\end{quote}

To make this discussion a bit more concrete, we now identify an important class of classical states on $\ovl{\cup_{N\in\N} \CA_1^N}$. 
We say that a state $\om$ on this \ca\  is {\it permutation-invariant} when each of its restrictions  to $\CA_1^N$ is invariant under the natural action of the symmetric group $\GS_N$ on  $\CA_1^N$ (i.e.\ $\sg\in\GS_N$ maps an elementary tensor $A_N=B_1\ot\cdots\ot B_N\in\CA_1^N$ to $B_{\sg(1)}\ot\cdots\ot B_{\sg(N)}$, cf.\ \er{landc}). The structure of the set $\CS^{\GS}$ of all permutation-invariant states in $\CS(\CA^{\mathrm (q)}_0)$ has been analyzed by St\o rmer (1969). Like any compact convex set, it is the (weak$\mbox{}^*$-closed)  convex hull of  its extreme boundary $\partial_e  \CS^{\GS}$.
The latter  consists of all infinite product states $\om= \rh^{\infty}$, where $\rh\in\CS(\CA_1)$. I.e.\ if $A_0\in \CA^{\mathrm (q)}_0$ is an equivalence class $[A_1,A_2,\cdots]$, then 
\beq \rh^{\infty}(A_0)=\lni \rh^N(A_N);\eeq
cf.\ \er{omN}.
 Equivalently, the restriction of $\om$ to any $\CA_1^N\subset \CA^{\mathrm (q)}_0$ is given by $\ot^N \rh$. Hence  $\partial_e  \CS^{\GS}$ is isomorphic
(as a compact convex set) to $\CS(\CA_1)$  in the obvious way, and
the primary states in $\CS^{\GS}$ are precisely the elements of $\partial_e  \CS^{\GS}$. 

 A general state $\om^{\mathrm (q)}_0$ in $\CS^{\GS}$  has a unique decomposition\footnote{This follows because $\CS^{\GS}$ is a so-called Bauer simplex (Alfsen, 1970). This is a compact convex set $K$ whose extreme boundary $\partial_e K$ is closed and for which 
every $\om\in K$ has a {\it unique} decomposition as a probability measure supported by $\partial_e K$, in the sense that $a(\om)=\int_{\partial_e K} d\mu(\rh)\, a(\rh)$ for any continuous affine function $a$ on $K$.
For a unital \ca\ $\CA$ the continuous affine functions on the state space $K=\CS(\CA)$ are precisely the elements $A$  of $\CA$, reinterpreted as
functions $\hat{A}$ on $\CS(\CA)$ by $\hat{A}(\om)=\om(A)$.
For example, the state space $\CS(\CA)$ of a commutative unital \ca\ $\CA$ is a 
 Bauer simplex, which consists of all (regular Borel) probability measures
 on the pre state space  $\CP(\CA)$.}
\beq \om^{\mathrm (q)}_0(A_0)=\int_{\CS(\CA_1)} d\mu(\rh)\, \rh^{\infty}(A_0), \label{Unn}
\eeq  where $\mu$ is a probability measure on $\CS(\CA_1)$ and $A_0\in \CA^{\mathrm (q)}_0$.\footnote{ This is a quantum analogue of De Finetti's \rep\ theorem
in classical probability theory (Heath \&\ Sudderth, 1976; van Fraassen, 1991); see also Hudson  \&\ Moody (1975/76) and  Caves et al.   (2002).}
The following beautiful illustration of the abstract theory 
(Unnerstall, 1990a,b) is then clear from \er{PMOQ} and \er{Unn}: 
\begin{quote}{\it 
 If $\om^{\mathrm (q)}_0$ is permutation-invariant, then it is classical.
The  associated limit state $\om_0^{\mathrm (c)}$
on $\CA_0^{\mathrm (c)}$ is characterized by the fact that
 the measure $\mu_0$ in \er{probm} coincides with the measure $\mu$ in \er{Unn}.}\footnote{In fact, each state in the folium $\CF^{\GS}$  in 
$\CS(\CA^{\mathrm (q)}_0)$ corresponding to the (quasi-equivalence class of)
the \rep\ $\oplus_{[\om\in \CS^{\GS}]}\pi_{\om}$ is classical.}
\end{quote}
\subsection{Superselection rules} \label{SE}
Infinite quantum systems are often associated with the notion of a superselection rule  (or sector), which was originally introduced by Wick, Wightman, \&\ Wigner (1952) in the setting of standard \qm\ on a \Hs\ $\H$. The basic idea may be illustrated in the example of the boson/fermion (or ``univalence") superselection rule.\footnote{See also Giulini (2003) for a  modern mathematical treatment.}  Here one has a {\it projective} unitary \rep\ $\CD$
of the rotation group $SO(3)$ on $\H$, for which $\CD(R_{2\pi})=\pm 1$ for  any rotation $R_{2\pi}$ of $2\pi$ around some axis. Specifically, on bosonic states $\Ps_B$ one has $\CD(R_{2\pi})\Ps_B=\Ps_B$, whereas on fermionic states $\Ps_F$ the rule is $\CD(R_{2\pi})\Ps_F=-\Ps_F$. Now the argument is that a rotation of $2\pi$ accomplishes nothing, so that it cannot change the physical state of the system. This requirement evidently holds on the subspace $\H_B\subset \H$ of bosonic states in $\H$, but it is equally well satisfied on the subspace $\H_F\subset \H$ of fermionic states, since $\Ps$ and $z\Ps$ with $|z|=1$ describe the same physical state. However, if $\Ps=c_B\Ps_B+c_F\Ps_F$ (with
$|c_B|^2+|c_F|^2=1$), then $\CD(R_{2\pi})\Ps=c_B\Ps_B-c_F\Ps_F$, which is not proportional to $\Ps$ and apparently describes a genuinely different  physical state from $\Ps$. 

The way out is to deny this conclusion by declaring that $\CD(R_{2\pi})\Ps$ and $\Ps$ {\it do} describe the same physical state, and this is achieved by postulating that no physical {\it observables} $A$
(in their usual mathematical guise as operators on $\H$) exist for which $(\Ps_B,A\Ps_F)\neq 0$. For in that case one has
\beq (c_B\Ps_B\pm c_F\Ps_F,A(c_B\Ps_B\pm c_F\Ps_F))=|c_B|^2(\Ps_B,A\Ps_B)+|c_F|^2(\Ps_F,A\Ps_F) \label{SSr}\eeq for any {\it observable} $A$, so that  $(\CD(R_{2\pi})\Ps, A\CD(R_{2\pi})\Ps)=(\Ps,A\Ps)$ for any $\Ps\in \H$. Since any quantum-mechanical prediction ultimately rests on expectation values $(\Ps,A\Ps)$ for physical observables $A$, the conclusion is that a rotation of $2\pi$ indeed does nothing to the system. This is codified by saying that superpositions of the type $c_B\Ps_B+c_F\Ps_F$ are {\it incoherent} (whereas  superpositions $c_1\Ps_1+c_2\Ps_2$ with $\Ps_1,\Ps_2$ both in either $\H_B$ or in $\H_F$ are {\it coherent}). Each of the subspaces $\H_B$ and $\H_F$ of $\H$ is said to be a {\it superselection sector}, and the statement that $(\Ps_B,A\Ps_F)=0$
for any observbale $A$ and $\Ps_B\in\H_B$ and $\Ps_F\in\H_F$ is called a {\it superselection rule}.\footnote{In an ordinary selection rule between $\Ps$ and $\Ph$ one merely has $(\Ps,H\Ph)=0$ for the Hamiltonian $H$.} 

The price one pays for this solution is that states of the form $c_B\Ps_B+ c_F\Ps_F$ with $c_B\neq 0$ and $c_F\neq 0$ are mixed, as one sees from \er{SSr}.
More generally, if $\H=\oplus_{\lm\in\Lm} \H_{\lm}$ with
$(\Ps,A\Ph)=0$ whenever $A$ is an observable, $\Ps\in \H_{\lm}$,
$\Ph\in \H_{\lm'}$, and $\lm\neq\lm'$, and if in addition for each $\lm$ and each pair $\Ps,\Ph\in\H_{\lm}$ there exists an observable $A$ for which $(\Ps,A\Ph)\neq 0$, then
the subspaces $\H_{\lm}$ are called superselection sectors in $\H$. Again a key consequence of the occurrence of superselection sectors is that unit vectors of the type $\Ps=\sum_{\lm} c_{\lm}\Ps_{\lm}$ with $\Ps\in\H_{\lm}$ (and $c_{\lm}\neq 0$ for at least two $\lm$'s)
define mixed states $$\ps(A)=(\Ps,A\Ps)=\sum_{\lm} |c_{\lm}|^2(\Ps_{\lm},A\Ps_{\lm})
=\sum_{\lm} |c_{\lm}|^2 \ps_{\lm}(A).$$

 This procedure is rather ad hoc.  A much deeper approach to superselection theory was developed by Haag and collaborators; see Roberts \&\ Roepstorff (1969) for an introduction.
Here the starting point is the abstract  \ca\ of  observables $\CA$  of a given quantum system, and superselection sectors are reinterpreted as equivalence classes (under unitary isomorphism) of \irrep s of $\CA$ (satisfying a certain selection criterion - see below).
 The connection between the concrete \Hs\ approach to superselection sectors discussed above and the abstract \ca ic approach is given by the following lemma (Hepp, 1972):\footnote{\label{fnhepp}Hepp proved a more general version of this lemma, in which  `Two pure states $\rh,\sg$ on a \ca\ $\CB$ define different sectors iff\ldots' is replaced by `Two states $\rh,\sg$ on a \ca\ $\CB$ are disjoint iff\ldots'}
\begin{quote}
{\it Two pure states $\rh,\sg$ on a \ca\ $\CA$ define different sectors iff for each \rep\ $\pi(\CA)$ on a \Hs\ $\H$ containing unit vectors $\Ps_{\rh},\Ps_{\sg}$ 
such that $\rh(A)=(\Ps_{\rh},\pi(A)\Ps_{\rh})$ and $\sg(A)=(\Ps_{\sg},\pi(A)\Ps_{\sg})$
for all $A\in\CA$, one has $(\Ps_{\rh},\pi(A)\Ps_{\sg})=0$
for all $A\in\CA$.}
\end{quote}

In practice, however, most \irrep s of a typical \ca\ $\CA$ used in physics are physically irrelevant mathematical artefacts. Such \rep s may be excluded from consideration by some {\it selection criterion}. 
 What this  means depends on the context. For example,  in quantum field theory this notion is made precise in the so-called DHR theory (reviewed by Roberts (1990), Haag (1992),  Araki (1999), and  Halvorson (2005)). In the class of theories discussed in the preceding two subsections, we take the algebra of observables $\CA$ to be $\CA^{\mathrm (q)}_0$ - essentially for reasons of human limitation - and for pedagogical reasons define (equivalence classes of) \irrep s of $\CA^{\mathrm (q)}_0$ as superselection sectors, henceforth often just called {\it sectors}, only when they are equivalent to the GNS-\rep\ given by a permutation-invariant pure state on  $\CA^{\mathrm (q)}_0$. In particular, such a state is classical. On this selection criterion, the results in the preceding subsection trivially imply that there is a bijective correspondence between pure states on $\CA_1$ and sectors of $\CA_0^{\mathrm (q)}$. The  sectors of the commutative \ca\  $\CA_0^{\mathrm (c)}$ are just the points of $\CS(\CA_1)$; note that a {\it mixed} state on $\CA_1$ defines a {\it pure} state on $\CA_0^{\mathrm (c)}$! The role of the sectors of $\CA_1$ in connection with those of $\CA_0^{\mathrm (c)}$ will be clarified in Subsection \ref{PSD}.

Whatever the model or the selection criterion, it is enlightening (and to some extent even in accordance with experimental practice) to consider superselection sectors entirely from the perspective of the pure states on the algebra of  observables $\CA$, removing $\CA$ itself and its \rep s from the scene. To do so, we equip the space $\CP(\CA)$ of pure states on $\CA$ with the structure of a  transition probability space (von Neumann, 1981; Mielnik, 1968).\footnote{See also  Beltrametti \&\ Cassinelli (1984) or Landsman (1998) for  concise reviews.}
 A {\it transition probability} on a set $\CP$ is a function 
\beq p:\CP\times\CP\raw[0,1] \label{tp1} \eeq
that satisfies \beq p(\rh,\sg)=1 \, \Longleftrightarrow \,\rh=\sg
\label{tp2} \eeq and \beq p(\rh,\sg)=0 \, \Longleftrightarrow \,
p(\sg,\rh)=0. \label{tp2half} \eeq A set with such a transition probability is called a {\it
transition probability space}.
 Now, the pure state space $\CP(\CA)$ of a \ca\ $\CA$ carries precisely this structure if we define\footnote{This definition applies to the case that $\CA$ is unital; see Landsman (1998) for the general case. An analogous formula defines a transition probability on the extreme boundary of any compact convex set.}
\beq p(\rh,\sg):=\inf\{\rh(A)\mid  A\in \CA, 0\leq A\leq 1, \sg(A)=1\}.\label{mtp} \eeq  
To give a more palatable formula, note that since pure states are primary, two pure states $\rh,\sg$ are either disjoint ($\rh\bot\sg$) or else (quasi, hence unitarily) equivalent ($\rh\sim\sg$). In the first case,  \er{mtp} yields
\beq p(\rh,\sg)=0\:\:\: (\rh\bot\sg).\eeq
Ine the second case it follows from Kadison's transitivity theorem  (cf.\ Thm.\ 10.2.6 in Kadison \&\ Ringrose (1986)) that the \Hs\ 
$\H_{\rh}$ from the GNS-\rep\ $\pi_{\rh}(\CA)$ defined by $\rh$ contains a unit vector $\Om_{\sg}$ (unique up to a phase) such that
\beq\sg(A)=(\Om_{\sg},\pi_{\rh}(A)\Om_{\sg}).\eeq
Eq.\ \er{mtp} then leads to the well-known expression
\beq
p(\rh,\sg)=|(\Om_{\rh},\Om_{\sg})|^2 \:\:\: (\rh\sim\sg). \label{tpsforca} \eeq
In particular, if $\CA$  is commutative, then 
\beq
p(\rh,\sg)=\dl_{\rh\sg}. \label{cltp}\eeq 
For $\CA=M_2(\C)$ one obtains 
\beq
p(\rh,\sg)=\half(1+\cos \theta_{\rh\sg}), \label{thetarhsg}\eeq
where $\theta_{\rh\sg}$ is the angular distance between $\rh$ and $\sg$ (seen as points on the two-sphere $S^2=\partial_e B^3$, cf.\ \er{gens2} etc.), measured along a great circle.

Superselection sectors may now be defined for any
transition probability spaces $\CP$.  A family of subsets of $\CP$ is called {\it orthogonal} if $p(\rh,\sg)=0$ whenever $\rh$ and $\sg$ do not lie in the same subset. The space $\CP$ is
called {\it reducible} if it is the union of two (nonempty) orthogonal subsets; if not, it is
said to be {\it irreducible}.  A {\it component} of $\CP$ is a
subset $\CC\subset \CP$ such that $\CC$ and $\CP\backslash \CC$ are
orthogonal. An irreducible component of  $\CP$ is called a {\it (superselection) sector}. Thus $\CP$ is the disjoint union of its sectors. 
For $\CP=\CP(\CA)$ this reproduces the algebraic definition of a superselection sector (modulo the selection criterion) via the correspondence between states and \rep s given by the GNS-constructions.
 For example, in the commutative case $\CA\cong C(X)$ each point in $X\cong \CP(\CA)$ is its own little sector. 
\subsection{A simple example: the infinite spin chain}\label{SC}
Let us illustrate the occurrence of superselection sectors  in a simple example, where the algebra of observables is $\CA^{\mathrm (q)}_0$ with
$\CA_1=M_2(\C)$. Let $\H_1=\C^2$, so that 
$\H_1^N=\ot^N\C^2$ is  the tensor product of $N$ copies of $\C^2$. It is clear that $\CA_1^N$ acts on $\H^N_1$ in a natural way (i.e.\ componentwise). This defines an \irrep\ $\pi_N$ of $\CA_1^N$, which is indeed its unique \irrep\ (up to unitary equivalence). In particular, for $N<\infty$ the quantum system whose algebra of observables is $\CA_1^N$ (such as a chain with $N$ two-level systems) has no superselection rules. We define
 the $N\raw\infty$  limit ``$(M_2(\C))^{\infty}$''  of the \ca s $(M_2(\C))^N$ as  the inductive limit $\CA^{\mathrm (q)}_0$ for $\CA_1=M_2(\C)$, as introduced in Subsection \ref{QLO}; see \er{ILCA}.   The definition of ``$\ot^{\infty} \C^2$'' is slightly more involved, as follows (von Neumann, 1938).

For any \Hs\ $\H_1$, let $\Ps$ be a sequence $(\Ps_1,\Ps_2,\ldots)$ with $\Ps_n\in\H_1$.
The space $\mathsf{H}_1$ of such sequences is a vector space in the obvious way. Now let $\Ps$ and
$\Ph$ be two such sequences, and write $(\Ps_n,\Ph_n)=\exp(i\al_n)|(\Ps_n,\Ph_n)|$.
If $\sum_n |\al_n|=\infty$, we define the (pre-) inner product $(\Ps,\Ph)$ to be zero. If $\sum_n |\al_n|<\infty$, we put
$(\Ps,\Ph)=\prod_n (\Ps_n,\Ph_n)$ (which, of course, may still be zero!). 
The (vector space) quotient of  $\mathsf{H}_1$ by the space of sequences $\Ps$ for which $(\Ps,\Ps)=0$ can be completed to  a \Hs\  $\H_1^{\infty}$ in the induced inner product, called the {\it complete} infinite tensor product of the \Hs\ $\H_1$ (over the index set $\N$).\footnote{Each fixed $\Ps\in\H_1$ defines an {\it incomplete} tensor product
$\H_{\Ps}^{\infty}$, defined as the closed subspace of $\H_1^{\infty}$
consisting of all $\Ph$ for which $\sum_n|(\Ps_n,\Ph_n)-1|<\infty$. If $\H_1$ is separable, then so is $\H_{\Ps}^{\infty}$ (in contrast to $\H_1^{\infty}$, which is an uncountable direct sum of the $\H_{\Ps}^{\infty}$).}  
We apply this construction with $\H_1=\C^2$. 
If $(e_i)$ is some basis of $\C^2$, an orthonormal basis of $\H_1^{\infty}$ then consists of all different infinite strings $e_{i_1}\ot \cdots e_{i_n}\ot \cdots$, where $e_{i_n}$ is $e_i$ regarded as a vector in $\C^2$.\footnote{The cardinality of the set of all such strings equals that of $\R$,  so that $\H^{\infty}_1$ is non-separable, as claimed.}  We denote the multi-index $(i_1,\ldots, i_n,\ldots)$ simply by $I$,
and the corresponding basis vector by $e_{I}$.  

 This  \Hs\ $\H^{\infty}_1$  carries a natural faithful \rep\ $\pi$ of $\CA^{\mathrm (q)}_0$:  if $A_0\in \CA^{\mathrm (q)}_0$ is an equivalence class $[A_1,A_2,\cdots]$, then $\pi(A_0)e_I=\lni A_Ne_i$, where $A_N$ acts on the first $N$ components of $e_I$ and leaves the remainder unchanged.\footnote{Indeed, this yields an alternative way of defining $\ovl{\cup_{N\in\N} \CA_1^N}$ as the norm closure of the union
of all $\CA_1^N$ acting on $\H^{\infty}_1$ in the stated way.} Now the point is that
although each $\CA_1^N$ acts irreducibly on $\H^N_1$, the \rep\ $\pi(\CA^{\mathrm (q)}_0)$
on $\H^{\infty}_1$ thus constructed is highly reducible. The reason for this is that by definition (quasi-) local elements of $\CA^{\mathrm (q)}_0$ leave the infinite tail of a vector in $\H^{\infty}_1$ (almost) unaffected, so that vectors with different tails lie in different superselection sectors. 
Without the quasi-locality condition on the elements of $\CA^{\mathrm (q)}_0$, no superselection rules would arise. 
 For example, in terms of the usual basis 
\beq \left\{\up=\left( \begin{array}{c} 1 \\ 0 \end{array} \right), \down=\left( \begin{array}{c} 0 \\ 1 \end{array} \right)\right\} \label{usualbasis}\eeq
 of $\C^2$, the vectors  $\Ps_{\up}=\up\ot\up\cdots \up\cdots$ (i.e.\ an infinite product of `up' vectors) and $\Ps_{\down}=\down\ot\down\cdots \down\cdots$ (i.e.\ an infinite product of `down' vectors) lie in different sectors.  The reason why the inner product
$(\Ps_{\up}, \pi(A)\Ps_{\down})$ vanishes  for any $A\in\CA^{\mathrm (q)}_0$ is that for local observables $A$ one has
$\pi(A)=A_M\ot 1\ot\cdots 1\cdots$ for some  $A_M\in\CB(\H_M)$;  
the inner product in question therefore involves infinitely many factors $(\up, 1\down)=(\up,\down)=0$. For quasilocal $A$ the operator $\pi(A)$ might have a small nontrivial tail, but the inner product  vanishes nonetheless by an approximation argument.

 More generally,
elementary analysis shows that $(\Ps_u, \pi(A)\Ps_v)=0$ whenever $\Ps_u=\ot^{\infty}u$ and 
$\Ps_v=\ot^{\infty}v$ for unit vectors $u,v\in\C^2$ with $u\neq v$. The corresponding vector states $\ps_u$ and $\ps_v$ on $\CA^{\mathrm (q)}_0$ (i.e.\ $\ps_u(A)=(\Ps_u, \pi(A)\Ps_u)$ etc.) are obviously permutation-invariant and hence classical. Identifying
$\CS(M_2(\C))$ with $B^3$, as in \er{gens2}, the corresponding limit state
$(\ps_u)_0$ on $\CA_0^{\mathrm (c)}$ defined by $\ps_u$ is  given by (evaluation at) the point $\til{u}=(x,y,z)$ of
$\partial_e  B^3=S^2$ (i.e.\ the two-sphere) for which the corresponding density matrix $\rh(\til{u})$ is the projection operator onto $u$. 
It follows that  $\ps_u$ and $\ps_v$ are disjoint; cf.\ \er{rsbot}.
We conclude that each unit vector $u\in\C^2$ determines a superselection sector $\pi_u$, namely the GNS-\rep\ of the corresponding state $\ps_u$, and that each such sector is realized as a subspace $\H_u$ of $\H^{\infty}_1$ (viz.\ $\H_u=\ovl{\pi(\CA^{\mathrm (q)}_0)\Ps_u}$). Moreover, since a permutation-invariant state on $\CA^{\mathrm (q)}_0$ 
is pure iff it is of the form $\ps_u$, we have found all superselection sectors of our system.
Thus in what follows we may concentrate our attention on the subspace 
(of $\H^{\infty}_1$) and sub\rep\ (of $\pi$)
\begin{eqnarray}
\H_{\GS}&=&\oplus_{\til{u}\in S^2} \H_u; \nn\\
\pi_{\GS}(\CA^{\mathrm (q)}_0)&=& \oplus_{\til{u}\in S^2} \pi_u(\CA^{\mathrm (q)}_0),\label{repsubrep}
\end{eqnarray}
where $\pi_u$ is simply the restriction of $\pi$ to $\H_u\subset \H^{\infty}_1$.

In the presence of superselection sectors one may construct operators that distinguish different sectors whilst being a multiple of the unit in each sector. In quantum field theory these are typically global charges, and in our example the macroscopic observables play this role. To see this, we return to Subsection \ref{MO}.
 It is not difficult to show that for any approximately symmetric sequence $(A_1,A_2,\cdots)$
the limit
\beq \ovl{A}=\lni\pi_{\GS}(A_N) \label{slim}\eeq
exists in the strong operator topology on $\CB(\H_{\GS})$ (Bona, 1988). Moreover, if $A_0\in \CA_0^{\mathrm (c)}=C(\CS(\CA_1))$ is the function defined by the given sequence,\footnote{Recall that $A_0(\om)=\lni \om^N(A_N)$.} then the map $A_0\mapsto \ovl{A}$ defines a faithful \rep\ of $\CA_0^{\mathrm (c)}$ on $\H_{\GS}$, which we call $\pi_{\GS}$ as well (by abuse of notation).  An easy calculation in fact shows that $\pi_{\GS}(A_0)\Ps= A_0(\til{u})\Ps$ for $\Ps\in \H_u$, or, in other words,
\beq
\pi_{\GS}(A_0)=\oplus_{\til{u}\in S^2} A_0(\til{u})1_{\H_u}. \eeq
Thus the $\pi_{\GS}(A_0)$ indeed serve as the operators in question.

To illustrate how delicate all this is, it may be interesting to note that even for symmetric sequences the limit $\lni\pi(A_N)$ does not exist on $\H^{\infty}_1$, not even in the strong topology.\footnote{\label{259} For example, let us take the sequence $A_N=j_{N1}(\mathrm{diag}(1,-1))$ and the vector
$\Ps=\up\down\down\up\up\up\up\down\down\down\down\down\down\down\down\up\up\up\up\up\up\up\up\up\up\up\up\up\up\up\up\cdots,$
where  a sequence of $2^N$ factors of $\up$ is followed by $2^{N+1}$ factors
of $\down$, etc. Then the sequence $\{\pi(A_N)\Ps\}_{N\in\N}$ in $\H^{\infty}_1$ diverges:
the subsequence where $N$ runs over all numbers $2^n$ with $n$ odd 
converges to $\third\Ps$, whereas the subsequence where $N$ runs over all $2^n$ with $n$ even converges
to $-\third\Ps$.} On the positive side, it can be shown  that 
 $\lni\pi(A_N)\Ps$ exists as an element of the von Neumann algebra $\pi(\CA^{\mathrm (q)}_0)''$ whenever the vector state $\ps$ defined by $\Ps$ lies in the folium $\CF^{\GS}$ generated by all permutation-invariant states (Bona, 1988; Unnerstall, 1990a).

This observation is part of a general theory of macroscopic observables in the setting of von Neumann algebras (Primas, 1983; Rieckers, 1984;  Amann, 1986, 1987; Morchio \&\ Strocchi, 1987; Bona, 1988, 1989; Unnerstall, 1990a, 1990b; Breuer, 1994; Atmanspacher, Amann, \&\ M\"{u}ller-Herold, 1999), which complements the purely \ca ic approach of Raggio \&\  Werner (1989, 1991),  Duffield \&\  Werner (1992a,b,c), and  Duffield, Roos, \&\  Werner (1992) explained so far.\footnote{Realistic models have been studied in the context of both the $C^*$-algebraic and the von Neumann algebraic approach by Rieckers and his associates. See, for example, Honegger \&\ Rieckers (1994), 
Gerisch,   M\"{u}nzner, \&\ Rieckers (1999), Gerisch, Honegger, \&\ Rieckers  (2003), and many other papers.  For altogether  different approaches to macroscopic observables see van Kampen (1954, 1988, 1993), Wan \&\ Fountain (1998), 
Harrison \&\ Wan (1997),   Wan et al. (1998),
Fr\"{o}hlich, Tsai,  \&\ Yau (2002), 
 and Poulin (2004).} In our opinion, the latter has the advantage that conceptually the passage to the limit $N\raw\infty$ (and thereby the idealization of a large system as an infinite one) is very satisfactory, especially in our reformulation in terms of continuous fields of \ca s. Here the commutative  \ca\ $\CA_0^{\mathrm (c)}$ of macroscopic observables of the infinite system is glued to the noncommutative algebras  $\CA_1^N$ of the corresponding finite systems in a continuous way, and the continuous sections of the ensuing continuous field of \ca s $\CA^{\mathrm (c)}$ exactly describe how {\it macroscopic} quantum observables of the finite systems converge to classical ones.  {\it Microscopic} quantum observables of the pertinent finite systems, on the other hand, converge to quantum observables of the infinite quantum system, and this convergence is described by the continuous sections  of the continuous field of \ca s $\CA^{\mathrm (q)}$. This entirely avoids the language of superselection rules,  which rather displays a shocking {\it dis}continuity between finite and infinite systems:  for superselection rules do not exist in finite systems!\footnote{We here refer to superselection rules in the traditional sense of inequivalent \irrep s of {\it simple} \ca s. For topological reasons certain finite-dimensional systems are described by (non-simple) \ca s that do admit inequivalent \irrep s (Landsman, 1990a,b).} 
\subsection{Poisson structure and dynamics}\label{PSD}
We now pass to the discussion of time-evolution in infinite systems of the type considered so far. We start with the observation  that the state space 
 $\CS(\CB)$ of a finite-dimensional  \ca\ $\CB$ (for simplicity assumed unital in what follows)  is a Poisson manifold (cf.\ Subsection \ref{DQsection})  in a natural way. A similar statement holds in the infinite-dimensional case, and we carry the reader through the necessary adaptations of the main argument by means of footnotes.\footnote{Of which this is the first. When  $\CB$ is infinite-dimensional, the state space $\CS(\CB)$ is no longer a manifold, let alone a Poisson manifold, but a {\it Poisson space} (Landsman, 1997, 1998).
 This  is a generalization of a Poisson manifold, which turns a crucial {\it property} of the latter into a {\it definition}. This property is the foliation of a Poisson manifold by its symplectic leaves (Weinstein, 1983), and the corresponding definition is as follows: {\it  A  Poisson space $P$ is a Hausdorff space of the form $P=\cup_{\al}S_{\al}$ (disjoint union), where each  $S_{\al}$ is a symplectic manifold (possibly infinite-dimensional)
and each injection $\iota_{\al}: S_{\al}\hookrightarrow P$ is continuous. Furthermore, one has a linear subspace $F\subset C(P,\R)$ that separates points and has the property that the restriction of each $f\in F$ to each $S_{\al}$ is smooth. Finally, if $f,g\in F$ then $\{f,g\}\in F$, where
the Poisson bracket is defined by $
\{f,g\}(\iota_{\al}(\sg))=\{\iota_{\al}^*f,\iota_{\al}^*g\}_{\al}(\sg)$.}
Clearly, a Poisson manifold $M$ defines a Poisson space  if one takes $P=M$, $F=\cin(M)$, and the $S_{\al}$ to be the symplectic leaves defined by the given Poisson bracket. Thus we refer to the manifolds $S_{\al}$ in the above definition as the {\it symplectic leaves} of $P$ as well.} We  write $K=\CS(\CB)$.

Firstly, an element $A\in\CB$ defines a linear function $\hat{A}$ on $\CB^*$ and hence on $K$ (namely by restriction) through $\hat{A}(\om)=\om(A)$.
 For such functions we define the Poisson bracket by
\beq \{\hat{A},\hat{B}\}=i\widehat{[A,B]}.\label{PBSC}\eeq 
Here the factor $i$ has been inserted in order to make the Poisson bracket of two real-valed functions real-valued again; for $\hat{A}$ is  real-valued on $K$ precisely when $A$ is self-adjoint, and if $A^*=A$ and
$B^*=B$, then $i[A,B]$ is self-adjoint (whereras $[A,B]$ is skew-adjoint). In general, for $f,g\in\cin(K)$ we put
\beq \{f,g\}(\om)=i\om([df_{\om},dg_{\om}]), \label{PBSS}\eeq
interpreted as follows.\footnote{In the infinite-dimensional case $\cin(K)$ is defined as the intersection of the smooth functions on $K$ with respect 
 to its Banach manifold structure and the space $C(K)$ of weak$\mbox{}^*$-continuous functions on $K$. The differential forms $df$ and $dg$ in \er{PBSS}
also require an appropriate definition;
see Duffield \&\ Werner (1992a), Bona (2000), and Odzijewicz \&\ Ratiu (2003)  for the technicalities.} 
 Let $\CB_{\R}$ be the self-adjoint part of $\CB$, and interpret $K$ as a subspace of $\CB_{\R}^*$; since a state $\om$ satisfies 
$\om(A^*)=\ovl{\om(A)}$ for all $A\in\CB$, it is determined by its values on self-adjoint elements. Subsequently, we identify the tangent space at $\om$ with \beq T_{\om}K = \{\rh\in \CB_{\R}^*\mid\rh(1)=0\}\subset\CB_{\R}^*
\label{TSpace}\eeq
 and the cotangent space at $\om$ with the quotient (of real Banach spaces)
 \beq T^*_{\om}K= \CB_{\R}^{**}/\R 1, \label{Cotspace} \eeq
 where the unit $1\in\CB$ is regarded
as an element of $\CB^{**}$ through the canonical embedding $\CB\subset\CB^{**}$. Consequently, the differential forms $df$ and $dg$
at $\om\in K$ define elements of $\CB_{\R}^{**}/\R 1$.
 The commutator in \er{PBSS} is then defined as follows: one lifts $df_{\om}\in \CB_{\R}^{**}/\R 1$ to  $\CB_{\R}^{**}$, and uses the natural isomorphism
 $\CB^{**}\cong \CB$ typical of finite-dimensional vector spaces.\footnote{In the infinite-dimensional case one uses the canonical identification between $\CB^{**}$ and the enveloping von Neumann algebra of $\CB$ to define the commutator.} The arbitrariness in this lift is a multiple of 1, which drops out of the commutator.  Hence $i[df_{\om},dg_{\om}]$ is an element of $\CB_{\R}^{**}\cong \CB_{\R}$,
 on which the value of the functional $\om$ is defined.\footnote{If  $\CB$ is infinite-dimensional, one here regards $\CB^*$ as the predual of the von Neumann algebra $\CB^{**}$.} This completes the definition of the Poisson bracket; one easily recovers \er{PBSC} as a special case of \er{PBSS}.

The symplectic leaves of the given Poisson structure on $K$ have been determined by  Duffield \&\ Werner (1992a).\footnote{See also Bona (2000) for the infinite-dimensional special case where $\CB$ is the \ca\ of compact operators.} Namely:
\begin{quote} {\it
  Two states $\rh$ and $\sg$ lie in the same symplectic leaf 
 of $\CS(\CB)$  iff 
$\rh(A)=\sg(UAU^*)$ for some unitary $U\in\CB$.}
\end{quote}
 When $\rh$ and $\sg$ are pure, this is the case iff  the corresponding GNS-representations $\pi_{\rh}(\CB)$ and $\pi_{\sg}(\CB)$ are unitarily equivalent,\footnote{Cf.\ Thm.\ 10.2.6 in Kadison \&\ Ringrose (1986).} but in general the implication holds only in one direction: if 
$\rh$ and $\sg$  lie in the same leaf, then they have unitarily equivalent GNS-\rep s.\footnote{An important step of the proof is the observation that the Hamiltonian vector field $\xi_f(\om)\in T_{\om}K\subset \CA^*_{\R}$ of $f\in\cin(K)$ is given by $\langle \xi_f(\om),B\rangle=i[df_{\om},B]$, where 
$B\in\CB_{\R}\subset \CB_{\R}^{**}$
and $df_{\om}\in \CB_{\R}^{**}/\R 1$. (For example, this gives $\xi_{\hat{A}}\hat{B}=i\widehat{[A,B]}=\{\hat{A},\hat{B}\}$ by \er{PBSC}, as it should be.)
If $\phv^h_t$ denotes the Hamiltonian flow of $h$ at time $t$, 
it follows (cf.\  Duffield, Roos,  \&\ Werner (1992), Prop.\ 6.1 or Duffield \&\ Werner (1992a), Prop.\ 3.1) that 
$\langle\phv_h^t(\om),B\rangle=\langle\om, U_t^h B (U_t^h)^*\rangle$ for some unitary $U_t^h\in\CB$. For example, if $h=\hat{A}$ then $U_t^h=\exp(itA)$.}

It follows from this characterization of the symplectic leaves of $K=\CS(\CB)$
that the pure state space $\partial_e K=\CP(\CB)$ inherits the Poisson bracket from $K$, and thereby becomes a Poisson manifold in its own right.\footnote{More generally, a Poisson space. The structure of $\CP(\CB)$
as a Poisson space was introduced by Landsman (1997, 1998) without recourse to the full state space or the work of Duffield \&\ Werner (1992a).}
This leads to an important connection between the superselection sectors of $\CB$ and the Poisson structure on $\CP(\CB)$ (Landsman, 1997, 1998):
\begin{quote} {\it The sectors of the pure state space $\CP(\CB)$ of a \ca\ $\CB$ as a transition probability space coincide with its symplectic leaves as a Poisson manifold.}
\end{quote}
 For example, when $\CB\cong C(X)$ is commutative, the space $\CS(C(X))$ of all (regular Borel) probability measures on $X$ acquires a Poisson bracket that is identically zero, as does its extreme boundary $X$.
It follows from \er{cltp} that the sectors in $X$ are its points, and so are its  symplectic leaves (in view of their definition and the vanishing Poisson bracket). The simplest noncommutative  case is $\CB=M_2(\C)$, for which  the symplectic leaves of the state space $K=\CS(M_2(\C))\cong B^3$ (cf.\ \er{gens2}) are the spheres with constant radius.\footnote{ Equipped with a multiple of the so-called  Fubini--Study symplectic structure; see Landsman (1998) or any decent book on differential geometry for this notion.
 This claim is immediate from footnote  \ref{SSlemma}. More generally, the pure 
 state space of $M_n(\C)$ is the projective space $\mathbb{P}\C^n$, which again becomes equipped with the Fubini--Study symplectic structure.  
 This is even true for $n=\infty$ if one defines $M_{\infty}(\C)$ as the \ca\ of compact operators on a separable \Hs\ $\H$:
   in that case one has $\CP(M_{\infty}(\C))\cong \mathbb{P}\H$.
   Cf.\ Cantoni (1977), Cirelli,  Lanzavecchia,  \&\  Mani\'{a} (1983),
Cirelli, Mani\'{a}, \&\ Pizzocchero (1990), 
Landsman (1998), Ashtekar \&\ Schilling (1999), Marmo et al. (2005), 
   etc.} The sphere with radius 1 consists of points in $B^3$ that correspond to pure states on $M_2(\C)$, all interior symplectic leaves of $K$  coming from mixed states on $M_2(\C)$.  
   
 The coincidence of sectors and symplectic leaves of  $\CP(\CB)$ is a compatibility condition between the transition probability structure and the Poisson structure. It is typical of the specific choices \er{mtp} and \er{PBSS}, respectively, and hence of quantum theory. In classical mechanics one has the freedom of equipping a manifold $M$ with an arbitrary Poisson structure, and yet use $C_0(M)$ as the commutative \ca\ of observables. The transition probability \er{cltp} (which follows from \er{mtp} in the commutative case) are clearly the correct ones in classical physics, but since the symplectic leaves of $M$ can be almost anything, the coincidence in question does not hold. 
 
However, there exists a compatibility condition between the transition probability structure and the Poisson structure, which is shared by classical and quantum theory. This is the property of  {\it unitarity} of a Hamiltonian flow, which in the present setting we formulate as follows.\footnote{All this can be boosted into an axiomatic structure into which both classical and quantum theory fit; see  Landsman (1997, 1998).} First, in quantum theory
with algebra of observables $\CB$ we define time-evolution 
(in the sense of an automorphic action of the abelian group $\R$ on $\CB$, i.e.\ a one-parameter group $\al$ of automorphisms on $\CB$) to be {\it Hamiltonian}
when $A(t)=\al_t(A)$ satisfies the Heisenberg equation $i\hbar dA/dt=[A,H]$
for some self-adjoint element $H\in\CB$. The corresponding flow on $\CP(\CB)$ - 
i.e.\ $\om_t(A)=\om(A(t))$ -  is equally well said to be Hamiltonian in that case.
In classical mechanics with Poisson manifold $M$ we similarly say that a flow on $M$ is Hamiltonian when it is the flow of a Hamiltonian vector field $\xi_h$ for some $h\in\cin(M)$. (Equivalently, the time-evolution of the observables $f\in\cin(M)$ is given by $df/dt=\{h,f\}$; cf.\ \er{ft} etc.) The point is that in either case the flow is unitary in the sense that
 \beq
p(\rh(t),\sg(t))=p(\rh,\sg)\label{unitarityeq} \eeq for all $t$ and all $\rh,\sg\in P$ with $P=\CP(\CB)$ (equipped with  the transition probabilities  \er{mtp} and the Poisson bracket \er{PBSS}) or $P=M$ (equipped with  the transition probabilities  \er{cltp} and any Poisson bracket).\footnote{In quantum theory the flow is defined for any $t$. In classical dynamics, \er{unitarityeq} holds for all $t$ for which $\rh(t)$ and $\sg(t)$ are defined, cf.\ footnote  \ref{crunch}.} 

In both cases $P=\CP(\CB)$ and $P=M$,  a Hamiltonian flow has the property (which is immediate from the definition of a symplectic leaf) that for all (finite) times $t$  a point  $\om(t)$ lies in the same symplectic leaf of $P$ as $\om=\om(0)$. In particular, in quantum theory  $\om(t)$ and $\om$ must lie in the same sector. In the quantum theory of infinite systems an automorphic time-evolution is rarely Hamiltonian, but one reaches a similar conclusion under a weaker assumption. Namely, if  a given one-parameter group of automorphisms $\al$ on $\CB$ is {\it implemented} in the GNS-\rep\ $\pi_{\om}(\CB)$ for some $\om\in \CP(\CB)$,\footnote{This  assumption means that there exists a unitary \rep\ $t\mapsto U_t$ of $\R$ on $\H_{\om}$ such that $\pi_{\om}(\al_t(A))=U_t \pi_{\om}(A)U_t^*$ for all $A\in\CB$ and all $t\in\R$.} then $\om(t)$ and $\om$  lie in the same sector and hence in the same symplectic leaf of $\CP(\CB)$.
    
 To illustrate these concepts, let us return to  our continuous field of \ca s $\CA^{\mathrm (c)}$; cf.\ \er{fibers}. It may not come as a great surprise that
 the canonical \ca ic  transition probabilities  \er{mtp}  on the pure state space of each fiber algebra $\CA^{\mathrm (c)}_{1/N}$  for  $N<\infty$ converge to 
the  classical  transition probabilities \er{cltp} on the commutative limit algebra
 $\CA^{\mathrm (c)}_0$. Similarly, the \ca ic  Poisson structure \er{PBSS} on each $\CP(\CA^{\mathrm (c)}_{1/N})$ converges to zero. However, we know from the  limit $\hbar\raw 0$ of \qm\ that in generating classical behaviour on the limit algebra of a continuous field of \ca s one should rescale the commutators; see Subsection \ref{DQsection} and Section \ref{S5}. Thus we replace the Poisson bracket \er{PBSS} for $\CA^{\mathrm (c)}_{1/N}$ by
 \beq \{f,g\}(\om)=iN\om([df_{\om},dg_{\om}]). \label{PBSSN}\eeq
 Thus rescaled, the Poisson brackets on the spaces $\CP(\CA^{\mathrm (c)}_{1/N})$ turn out to converge to the canonical Poisson bracket \er{PBSS}
 on $\CP(\CA^{\mathrm (c)}_0)=\CS(\CA_1)$, instead of the zero bracket expected from the commutative nature of the limit algebra $\CA^{\mathrm (c)}_0$. Consequently, the  symplectic leaves of the {\it full} state space $\CS(\CA_1)$ of  the fiber algebra $\CA^{\mathrm (c)}_1$ become the symplectic leaves of the {\it pure} state space $\CS(\CA_1)$  of the 
 fiber algebra $\CA^{\mathrm (c)}_0$. This is undoubtedly indicative of the origin of classical phase spaces and their Poisson structures in quantum theory.

More precisely, we have the following result (Duffield \&\ Werner, 1992a): 
\begin{quote}
{\it If $A=(A_0,A_1,A_2,\cdots)$ and $A'=(A_0',A_1',A_2',\cdots)$ are continuous sections 
of $\CA^{\mathrm (c)}$  defined by symmetric sequences,\footnote{\label{generalization}The result does not hold for all
continuous sections (i.e.\ for all approximately symmetric sequences), since, for example, the limiting functions $A_0$ and $A_0'$ may not be differentiable, so that their Poisson bracket does not exist. This problem occurs in all examples of deformation quantization. However, the class of sequences for which the claim is valid is larger than the symmetric ones alone. A sufficient condition on $A$ and $B$
for \er{choice} to make sense is that $A_N=\sum_{M\leq N} j_{NM}(A_M^{(N)})$
(with $A_M^{(N)}\in\CA_1^M$), such that $\lim_{N\raw\infty} A_M^{(N)}$ exists (in norm)
and $\sum_{M=1}^{\infty} M \sup_{N\geq M}\{ \| A_M^{(N)}\|\}<\infty$. See Duffield \&\ Werner (1992a).} then the sequence
\beq \left( \{A_0,A_0'\}, i[A_1,A_1'],\dots, iN [A_N,A_N'],\cdots\right) \label{choice}\eeq
defines a continuous section of $\CA^{\mathrm (c)}$.}
\end{quote}
This follows from an easy computation. In other words, although  the sequence of commutators $[A_N,A_N']$ converges to zero,
the rescaled commutators $iN [A_N,A_N']\in\CA_N$ converge to the macroscopic observable $\{A_0,A_0'\}\in \CA_0^{\mathrm (c)}=C(\CS(\CA_1))$.    Although it might seem perverse to reinterpret this result on the classical limit of a large quantum system in terms of quantization (which is the {\it opposite} of taking the classical limit), it is formally possible to do so (cf.\ Section \ref{DQsection}) if we put
\beq \hbar=\frac{1}{N}. \label{h1N}\eeq Using the axiom of choice if necessary, we devise a procedure that assigns a continuous section $A=(A_0,A_1,A_2,\cdots)$ of our field to a given function $A_0\in\CA_0^{\mathrm (c)}$. We write this as $A_N=\CQ_{\frac{1}{N}}(A_0)$, and similarly $A_N'=\CQ_{\frac{1}{N}}(A_0')$. This choice need not be such that the sequence \er{choice} is assigned to $\{A_0,A_0'\}$, but since the latter  is the unique limit of \er{choice}, it must be that
\beq  \lni \left\| iN \left[\CQ_{\frac{1}{N}}(A_0),\CQ_{\frac{1}{N}}(A_0')\right]-\CQ_{\frac{1}{N}}(
 \{A_0,A_0'\})\right\| =0.\eeq
Also note that \er{normcont} is just \er{normeq}. Consequently (cf.\ \er{Dirac} and surrounding text): 
\begin{quote}
{\it The continuous field of \ca s $\CA^{\mathrm (c)}$ defined by \er{fibers} and approximately symmetric sequences (and their limits) as continuous sections yields a deformation quantization of the phase space $\CS(\CA_1)$ (equipped with the Poisson bracket \er{PBSS}) for any quantization map $\CQ$.}
\end{quote}
For the dynamics this implies:
\begin{quote}
{\it Let $H=(H_0, H_1,H_2,\cdots)$ be a continuous section of $\CA^{\mathrm (c)}$ defined by a symmetric sequence,\footnote{Once again, the result in fact holds for a larger class of Hamiltonians, namely the ones satisfying the conditions specified in 
footnote  \ref{generalization} (Duffield \&\ Werner, 1992a). The assumption that each Hamiltonian $H_N$ lies in $\CA_1^N$ and hence is bounded is natural in lattice models, but is undesirable in general.} and let $A=(A_0,A_1,A_2,\cdots)$ be an arbitrary continuous section of $\CA^{\mathrm (c)}$ (i.e.\ an approximately symmetric sequence). Then the sequence
\beq 
\left(A_0(t), e^{iH_1 t}A_1e^{-iH_1 t},\cdots e^{iNH_N t}A_Ne^{-iNH_N t},\cdots\right), \label{DWEOM}\eeq
where $A_0(t)$ is the solution of the equations of motion with classical Hamiltonian $H_0$,\footnote{See \er{ft} and surrounding text.}
defines a continuous section of $\CA^{\mathrm (c)}$.}
\end{quote}
In other words, for bounded symmetric sequences of Hamiltonians $H_N$ the quantum dynamics restricted to macroscopic observables converges to the classical dynamics with Hamiltonian $H_0$. 
Compare the positions of $\hbar$ and $N$ in \er{HSEOM} and \er{DWEOM}, respectively, and rejoice in the reconfirmation of \er{h1N}. 

In contrast, the quasilocal observables are {\it not} well behaved as far as the $N\raw\infty$ limit of the dynamics defined by such Hamiltonians is concerned. 
Namely, if $(A_0,A_1,\cdots)$ is a section of the continuous field $\CA^{\mathrm (q)}$,
and $(H_1,H_2,\cdots)$ is any bounded symmetric  sequence of Hamiltonians, then the sequence $$\left(e^{iH_1 t}A_1e^{-iH_1 t},\cdots e^{iNH_N t}A_Ne^{-iNH_N t},\cdots\right)$$ has no limit for $N\raw\infty$, in that  it cannot be extended by some $A_0(t)$ to a continuous section of $\CA^{\mathrm (q)}$. Indeed, this was the very reason why macroscopic observables were originally introduced in this context (Rieckers, 1984; Morchio \&\ Strocchi, 1987; Bona, 1988; Unnerstall, 1990a;  Raggio \&\  Werner,  1989;  Duffield \&\  Werner, 1992a). Instead, the natural finite-$N$ Hamiltonians for which the limit $N\raw\infty$ of the time-evolution on $\CA_1^N$ exists as a one-parameter automorphism group on $\CA^{\mathrm (q)}$ satisfy an appropriate locality condition, which excludes the global averages defining  symmetric  sequences.
\subsection{Epilogue: Macroscopic observables and the measurement problem\label{hepps}}
In a renowned paper, Hepp (1972) suggested that macroscopic observables and superselection rules  should play a role in the solution of the measurement problem of \qm. He assumed that a macroscopic apparatus may be idealized as an infinite quantum system, whose algebra of observables $\CA_A$ has disjoint pure states.
Referring to our discussion in Subsection \ref{vNs} for context and notation, Hepp's basic idea (for which he claimed no originality) was that as a consequence of  the measurement process the initial state vector $\Om_I=\sum_n c_n\Ps_n\ot I$ of system plus apparatus 
evolves into a final state vector $\Om_F=\sum_n c_n\Ps_n\ot \Ph_n$,  in which each $\Ph_n$
lies in a different superselection sector of the \Hs\ of the apparatus (in other words, the corresponding states $\phv_n$ on $\CA_A$ are mutually disjoint). Consequently, 
although the initial state $\om_I$ is pure, the final state $\om_F$ is mixed. Moreover, 
because of the disjointness of the $\om_n$ the final state $\om_F$ has   a unique decomposition $\om_F=\sum_n |c_n|^2 \ps_n\ot\phv_n$ into pure states, and therefore admits a bona fide  ignorance interpretation. Hepp therefore claimed with some justification that the measurement ``reduces the wave packet", as desired in quantum measurement theory. 

Even apart from the usual conceptual problem of passing from the collective of all terms in the final mixture to one actual measurement outcome, Hepp himself indicated a serious mathematical problem with this program. Namely, if the initial state is pure it must lie in a certain superselection sector (or equivalence class of states); but then the final state must lie in the very same sector if the time-evolution is Hamiltonian, or, more generally, automorphic (as we have seen in the preceding subsection). Alternatively, it follows 
from a more general lemma Hepp (1972)  himself proved:
\begin{quote} {\it If two states $\rh,\sg$ on a \ca\ $\CB$ are disjoint and $\al:\CB\raw\CB$ is an automorphism of $\CB$, then $\rh\circ\al$ and $\sg\circ\al$ are disjoint, too.}
\end{quote}
To reach the negative conclusion above, one takes $\CB$ to be the algebra of observables of system and apparatus jointly, and computes back in time by choosing $\al=\al_{t_F-t_I}\inv$, where $\al_t$ is the one-parameter automorphism group on $\CB$ describing the joint time-evolution of system and apparatus (and $t_I$ and $t_F$ are the initial and final times of the measurement, respectively). However,  Hepp pointed out that this conclusion may be  circumvented  if one admits the possibility that a measurement takes infinitely long to complete. For the limit $A\mapsto \lim_{t\raw\infty} \al_t(A)$ (provided it exists in a suitable sense, e.g., weakly) does not necessarily yield an automorphism of $\CB$. Hence a state - evolving 
in the Schr\"{o}dinger picture by $\om_t(A)\equiv \om(\al_t(A))$ - may leave its sector in infinite time, a possibility Hepp actually demonstrated in a range of models; see also
Frigerio (1974), Whitten-Wolfe \&\  Emch (1976), Araki (1980), Bona (1980), Hannabuss (1984), Bub (1988), Landsman (1991), Frasca (2003, 2004), and many other papers. 

Despite the criticism that has been raised against the conclusion that a quantum-mechanical measurement requires an infinite apparatus and must take infinite time (Bell, 1975; Robinson, 1994; Landsman, 1995), and despite the fact that this procedure  is quite against the spirit of von Neumann (1932), in whose widely accepted description measurements are practically instantaneous, this conclusion resonates well with the modern idea that quantum theory is universally valid and the classical world has no absolute existence; cf.\ the Introduction. Furthermore, a quantum-mechanical measurement is nothing but a specific interaction,
comparable with a scattering process; and it is quite uncontroversial
that such a process  takes infinite time to complete.  Indeed, what would it mean for scattering to be over after some finite time? Which time? As we shall see in the next section, the theory of decoherence requires the limit $t\raw\infty$ as well, and largely for the same mathematical reasons. There as well as in Hepp's approach, the limiting behaviour
 actually tends to be approached very quickly (on the pertinent time scale), and one needs
to let $t\raw\infty$ merely to make terms $\sim\exp-\gm t$ (with $\gm>0$) zero rather than just very small. See also Primas (1997) for a less pragmatic point of view on the significance of this limit. 

A more serious problem with Hepp's approach lies in his assumption that the time-evolution on the quasilocal algebra of observables of the infinite measurement apparatus (which in our class of examples would be $\CA_0^{\mathrm (q)}$) is automorphic. This, however, is by no means always the case; cf.\ the references listed near the end of Subsection \ref{PSD}. As we have seen, for  certain natural Hamiltonian (and hence automorphic) time-evolutions at finite $N$ the dynamics  {\it has no limit $N\raw\infty$ on the algebra of quasilocal observables} - let alone an automorphic one. 

Nonetheless, Hepp's  conclusion remains valid if we use the algebra  $\CA_0^{\mathrm (c)}$ of macroscopic observables, on which (under suitable assumptions - see Subsection \ref{PSD}) Hamiltonian time-evolution on $\CA_1^N$  {\it does} have a limit as $N\raw\infty$. For, as pointed out in Subsection \ref{SE}, each superselection sector of $\CA_0^{\mathrm (q)}$ defines and is defined by a pure state on $\CA_1$, which in turn defines a sector of $\CA_0^{\mathrm (c)}$. Now the latter sector is simply a point in the pure state space $\CS(\CA_1)$ of the commutative \ca\ $\CA_0^{\mathrm (c)}$, so that  Hepp's lemma quoted above boils down to the claim that
if $\rh\neq \sg$, then $\rh\circ\al\neq\sg\circ\al$ for any automorphism $\al$.
This, of course, is a trivial property of any Hamiltonian time-evolution, and it follows once again that a transition from a pure pre-measurement state to a mixed post-measurement state on $\CA_0^{\mathrm (c)}$ is impossible in finite time. To avoid this conclusion, one should simply  avoid the limt $N\raw\infty$, which is the root of the $t\raw\infty$ limit; see Janssens (2004).
 
What, then, does all this formalism mean for Schr\"{o}dinger's cat?
In our opinion, it confirms the impression that the appearance of a 
 paradox rests upon an equivocation. Indeed, the problem arises because 
 one oscillates between two mutually exclusive interpretations.\footnote{Does {\it complementarity} re-enter through the back door?} 
 
 {\it Either} one is a bohemian theorist who,  in vacant or in pensive mood, puts off his or her  glasses and merely contemplates whether the cat is dead or alive. Such a person  studies the cat exclusively from the point of view of its macroscopic observables, so that he or she has to use a post-measurement state $\om_F^{\mathrm (c)}$ on the algebra $\CA_0^{\mathrm (c)}$.  If $\om_F^{\mathrm (c)}$ is pure, it lies  in $\CP(\CA_1)$ (unless the pre-measurement state was mixed). Such a state  corresponds to a single superselection sector $[\om_F^{\mathrm (q)}]$ of $\CA_0^{\mathrm (q)}$, so that  the cat is dead or alive. If, on the other hand,  $\om_F^{\mathrm (c)}$ is mixed (which is what occurs if  Schr\"{o}dinger has his way), there is no problem in the first place:  at the level of macroscopic observables one merely has a statistical description of the cat. 
 
 {\it Or} one is a hard-working experimental physicist of formidable power,   who investigates the detailed microscopic constitution of the cat. For him or her  the cat is always in a pure state on $\CA_1^N$ for some large $N$. This time the issue of life and death is not a matter of lazy observation and conclusion, but one of sheer endless experimentation and computation. From the point of view of such an observer, nothing is wrong with the cat being in a coherent superposition of two states that are actually quite close to each other microscopically - at least for the time being. 
 
Either way,  {\it the riddle does not exist} (Wittgenstein, TLP, \S 6.5).
\section{Why classical states and observables?}
\label{S7}\setcounter{equation}{0}
\begin{quote}
`We have found a strange footprint on the shores of the unknown. We have devised profound theories, one after another, to account for its origins. At last, we have succeeded in reconstructing the creature that made the footprint. And lo! It is our own.' (Eddington, 1920, pp.\ 200--201)
\end{quote} 
 
The conclusion of Sections \ref{S5} and \ref{S6} is that quantum theory may give rise to classical behaviour {\it in certain states} and {\it with respect to certain observables}.
For example, we have seen that in the limit $\hbar\raw 0$ coherent states and operators of the form $\qh(f)$, respectively, are appropriate, whereas in the limit $N\raw\infty$ one should use classical states ({\it nomen est omen}!) as defined in Subsection \ref{QLO} and macroscopic observables. If, instead, one uses superpositions of such states, or observables with the wrong limiting behaviour, no classical physics emerges. 
Thus the question remains why the world at large should happen to be in such states, and why we turn out to study this world with respect to the observables in question. This question found its original incarnation in the measurement problem (cf.\ Subsection \ref{vNs}), but this problem is really a figure-head for a much wider difficulty. 

Over the last 25 years,\footnote{Though some say the basic idea of decoherence goes back to Heisenberg and Ludwig.} two profound and original answers to this question have been proposed.
 \subsection{Decoherence}\label{DSS}
The first goes  under the name of {\it decoherence}. Pioneering papers include
van Kampen (1954),  Zeh (1970), Zurek (1981, 1982),\footnote{See also Zurek (1991) and the subsequent debate in {\it Physics Today} (Zurek, 1993), which  drew wide attention  to decoherence.} and Joos \&\ Zeh (1985), and some recent reviews are  Bub (1999), Auletta (2001), Joos et al. (2003), Zurek (2003),  Blanchard \&\  Olkiewicz (2003), Bacciagaluppi (2004) and Schlosshauer (2004).\footnote{The website \texttt{http://almaak.usc.edu/$\sim$tbrun/Data/decoherence$\mbox{}_{-}$list.html} contains an extensive list of references on decoherence.}  More references will be given in due course. 
The existence (and excellence) of these reviews obviates the need for a detailed treatment of decoherence in this article, 
all the more so since at the time of writing this approach appears to be in a transitional stage,  conceptually as well as  mathematically (as will be evident from what follows).
Thus we depart from the layout of our earlier chapters and restrict ourselves to a few  personal comments.
\begin{enumerate}
\item \label{DP1}  Mathematically, decoherence boils down to the idea of adding one more link to the von Neumann chain (see Subsection \ref{vNs}) beyond $S+A$ (i.e.\ the system and the apparatus). Conceptually, however, there is a major difference between decoherence and older approaches that took such a step: whereas previously (e.g., in the hands of von Neumann, London \&\ Bauer, Wigner, etc.)\footnote{See Wheeler \&\ Zurek (1983).} the chain {\it converged  towards the observer}, in decoherence it {\it diverges  away from the observer}. Namely, the third and final link is now taken to be the {\it environment} (taken in a fairly literal sense in agreement with the intuitive meaning of the word). In particular, in realistic models the environment is treated as an infinite system 
(necessitating the limit $N\raw\infty$), which has the consequence that 
(in simple models where the pointer has discrete spectrum) the 
post-measurement state $\sum_n c_n \Psi_n \ot\Phi_n\ot \ch_n$ (in which the
$\ch_n$ are mutually orthogonal) is only reached in the limit $t\raw\infty$. However, as already mentioned in Subsection \ref{hepps},  infinite time is 
only needed mathematically in order to make terms of the type $\sim\exp-\gm t$ (with $\gm>0$) zero rather than just very small: in many models
the inner products $(\ch_n,\ch_m)$ are actually negligible for $n\neq m$ 
within surprisingly short time scales.\footnote{Cf. Tables 3.1 and 3.2 on pp.\ 66--67 of Joos et al.\ (2003).}

If only in view of the need for limits of the type $N\raw\infty$ (for the environment) and $t\raw\infty$, in our opinion decoherence is best linked to stance 1 of the Introduction: its goal is to explain the approximate appearance of the classical world from quantum mechanics seen as a universally valid theory. However, decoherence has been claimed
to support almost any opinion on the foundations of \qm; cf.\  Bacciagaluppi (2004) and Schlosshauer (2004) for a critical overview and also see Point \ref{point6} below. 
\item Originally, decoherence  entered the scene
as a proposed solution to the measurement problem (in the precise form  stated at the end of Subsection \ref{vNs}). For the restriction of the state $\sum_n c_n \Psi_n \ot\Phi_n\ot \ch_n$ to $S+A$ (i.e.\ its  trace over the degrees of freedom of the environment) is mixed in the limit $t\raw\infty$, which means that the quantum-mechanical interference
between the states $ \Psi_n \ot\Phi_n$ for different values of $n$ has become `delocalized' to the environment, and accordingly is irrelevant if the latter is not observed (i.e.\ omitted from the description). Unfortunately, the application of the ignorance interpretation of the mixed post-measurement state of $S+A$ is illegal even from the point of view of  stance 1 of the Introduction. The ignorance interpretation is only valid if the environment is kept within the description  {\it and} is classical (in having a commutative \ca\ of observables). The latter assumption (Primas, 1983), however, makes the decoherence solution to the measurement problem circular.\footnote{On the other hand, treating the environment {\it as if} it were classical
might be an improvement on the Copenhagen ideology of treating the measurement apparatus {\it as if} it were classical (cf.\ Section \ref{S3}).}

 In fact, as quite rightly pointed out by Bacciagaluppi (2004), decoherence actually {\it aggravates} the measurement problem. Where previously this problem was  believed to be man-made and relevant only to rather unusual laboratory situations (important as these might be for the foundations of physics), it has now become clear that ``measurement" of a quantum system {\it by the environment} (instead of by an experimental physicist) happens everywhere and all the time: hence it remains even more  miraculous than before  that there is a single outcome after each such measurement. Thus decoherence as such does not provide a solution to the measurement problem (Leggett, 2002;\footnote{In fact, Leggett's argument
only applies to strawman 3 of the Introduction and loses its force against stance 1. For his argument is that decoherence just removes the {\it evidence} for a given state (of Schr\"{o}dinger's cat type) to be a superposition, and accuses those claiming that this solves the measurement problem of committing the logical fallacy that removal of the evidence for a crime would undo the crime. But according to stance 1 {\it the crime is only defined relative to the evidence!}  Leggett is quite right, however,  in insisting on the `from `` and" to ``or" problem' mentioned at the end of the Introduction.}
 Adler, 2003; Joos \&\ Zeh, 2003), but is in actual  fact parasitic on such a solution.
 \label{point5}
\item \label{point6} There have been various responses to this insight. The dominant one has been to combine decoherence with some interpretation of \qm:  decoherence then finds a home, while conversely the interpretation in question is usually enhanced by decoherence. In this context, the most popular of these has been the many-worlds interpretation, which,  after decades of  obscurity and derision, suddenly started to be greeted with a flourish of trumpets in the wake of the popularity of decoherence. See, for example, Saunders (1993, 1995), Joos et al.\ (2003) and Zurek (2003). In quantum cosmology circles, the consistent histories approach  has been a popular partner to decoherence, often in combination with many worlds;  see below.
 The importance of decoherence in the modal interpretation has been emphasized by  Dieks (1989b) and Bene \&\ Dieks (2002), and practically all authors on decoherence find the opportunity to pay some lip-service to Bohr in one way or another. See  Bacciagaluppi (2004) and Schlosshauer (2004) for a critical  assessment of all these combinations. 

In our opinion, none of the established  interpretations of \qm\ will do the job, leaving room for genuinely new ideas. One such idea is the  {\it return of the environment}: 
instead of ``tracing it out",  as in the original setting of decoherence theory,  the environment should {\it not} be ignored! The essence of measurement has now been recognized to be the {\it redundancy} of the  outcome (or ``record") of the measurement in the environment.  It is this very redundancy of  information about the underlying quantum object that ``objectifies" it, in that the information becomes accessible to a large number of observers without necessarily disturbing the object\footnote{Such objectification is claimed to yield an `operational definition of existence' (Zurek, 2003, p.\ 749.).} 
(Zurek, 2003;   Ollivier, Poulin, \&\ Zurek, 2004; Blume-Kohout \&\ Zurek, 2004, 2005). This insight (called ``Quantum Darwinism") 
has given rise to the  ``existential"  interpretation of \qm\ due to Zurek (2003). 
\item 
Another response to the failure of decoherence (and indeed all other approaches) to solve the measurement problem (in the sense of failing to win a general consensus)
has been of a somewhat more pessimistic (or, some would say, pragmatic) kind: all attempts to explain the quantum world  are given up, yielding to the point of view that `the appropriate aim of physics at the fundamental level then becomes the representation and manipulation 
of information' (Bub, 2004). Here `measuring instruments ultimately remain black boxes at some level', and one concludes that all efforts to {\it understand} measurement (or, for that matter, \epr-correlations) are futile and pointless.\footnote{
It is indeed in describing the transformation of quantum information (or entropy) to classical information during measurement that decoherence comes to its own and exhibits some of its greatest strength. Perhaps for this reason such thinking pervades also Zurek (2003).}
\item Night thoughts of a quantum physicist, then?\footnote{Kent, 2000. Pun on
 the title of McCormmach (1982).}  Not quite. Turning vice into virtue: rather than solving the measurement problem, the  true significance of the  decoherence program is that it gives
conditions under which there is no measurement problem!
Namely, foregoing an explanation of the transition from the state $\sum_n c_n \Psi_n \ot\Phi_n\ot \ch_n$ of $S+A+{\CE}$ to a single one of the states $\Psi_n \ot\Phi_n$ of $S+A$,  at the heart of decoherence is the claim  that  each of the latter states is {\it robust} against coupling to the environment 
(provided the Hamiltonian  is such that $\Psi_n \ot\Phi_n$ tensored with some initial state $I_{\CE}$ of the environment indeed evolves into $\Psi_n \ot\Phi_n\ot \ch_n$, as assumed so far). This implies that each state  $\Psi_n \ot\Phi_n$  remains pure after coupling to the environment and subsequent restriction to the original system plus apparatus, so that at the end of the day the environment has had no influence on it.  In other words, the real point of decoherence is the phenomenon of {\it einselection} (for {\it environment-induced superselection}), where a state is `einselected' precisely when (given some interaction Hamiltonian) it possesses the stability property just mentioned. The claim, then, is that einselected states are often classical, or at least that classical states (in the sense mentioned at the beginning of this section) are classical precisely because they are robust against coupling to the environment. Provided this scenario indeed gives rise to the classical world (which remains to be shown in detail), it  gives a dynamical explanation of it. But even short of having achieved this goal, the importance of the notion of  einselection cannot be overstated; in our opinion, it is the most important and powerful idea in quantum theory since entanglement (which einselection, of course, attempts to undo!).
 \item 
The measurement problem, and the associated distinction between system and apparatus on the one hand and environment on the other, can now be {\it omitted} from decoherence theory. 
Continuing the discussion in Subsection \ref{primas}, the goal of decoherence should simply be to find the robust or einselected states of a object $\CO$ coupled to an environment ${\CE}$, as well as the induced dynamics thereof (given the time-evolution of $\CO+\CE$). This search, however, must include the correct {\it identification} of the object $\CO$ within the total $\CS+\CE$, namely as a subsystem that actually {\it has} such robust states. Thus the Copenhagen idea that the Heisenberg cut between object and apparatus
be movable (cf.\ Subsection \ref{HC}) will not, in general, extend to the ``Primas--Zurek" cut between object and environment. In traditional physics terminology, the problem is to find the right ``dressing" of a quantum system so as to make at least some of its states robust 
against coupling to its environment (Amann \&\ Primas, 1997;
Brun \&\ Hartle, 1999;  Omn\`{e}s, 2002). In other words: {\it What is a system?}
To mark this change in perspective, we now change notation from $\CO$ (for ``object") to $\CS$ (for ``system"). 
Various tools for the solution of this problem within the decoherence program  have now  been developed - with increasing refinement and also increasing reliance on concepts from information theory (Zurek, 2003) - but the right setting for it seems the formalism of consistent histories, see below. 
\label{CHD}
\item Various dynamical regimes haven been unearthed, each of which leads to a different class of robust states (Joos et al., 2003; Zurek, 2003; 
 Schlosshauer, 2004). Here $H_{\CS}$ is the system Hamiltonian, $H_I$ is the interaction Hamiltonian between system and environment, and $H_{\CE}$ is the environment Hamiltonian. As stated, no reference to measurement, object or apparatus need be made here. 
\begin{itemize}  
\item In the regime $H_{\CS}<<H_I$, for suitable Hamiltonians  the robust states are the traditional pointer states of quantum measurement theory. This regime conforms to von Neumann's (1932) idea that
quantum measurements be almost instantaneous. If, moreover, $H_{\CE}<<H_I$ as well -  
with or without a measurement context - 
then the decoherence mechanism turns out to be universal in being independent of the details of $\CE$ and $H_{\CE}$ (Strunz,  Haake,  \&\ Braun, 2003).
\item If $H_{\CS}\approx H_I$, then (at least in models of quantum Brownian motion)
the robust states are coherent states (either of the traditional Schr\"{o}dinger type, or of a more general nature as defined in Subsection \ref{CSR}); see Zurek, Habib,  \&\  Paz  (1993)
and Zurek (2003). This case is, of course,  of supreme importance for the physical relevance of the results quoted in our Section \ref{S5} above, and  - if only for this reason - decoherence  theory would benefit from more interaction with mathematically rigorous results  on quantum stochastic analysis.\footnote{Cf.\ Davies (1976), 
 Accardi,  Frigerio, \&\  Lu (1990), Parthasarathy (1992), 
Streater (2000), K\"{u}mmerer (2002), Maassen (2003), etc. }
\item Finally, if  $H_{\CS}>> H_I$, then the robust states turn out to be eigenstates of the system Hamiltonian $H_{\CS}$ (Paz \&\ Zurek, 1999; Ollivier, Poulin \&\ Zurek, 2004). In view of our discussion of such states
in Subsections \ref{WKBS} and \ref{QC}, this shows that robust states are not necessarily classical. It should be mentioned that in this context decoherence theory largely coincides
with standard atomic physics, in which the atom is taken to be the system $\CS$ and the radiation field plays the role of the environment $\CE$; see  Gustafson \&\ Sigal (2003)
for a mathematically minded introductory treatment and Bach, Fr\"{o}hlich, \&\  Sigal (1998, 1999) for a full 
(mathematical) meal. 
\end{itemize}
\item Further to the above clarification of the role of energy eigenstates, decoherence 
also has had important things to say about quantum chaos (Zurek, 2003; Joos et al., 2003).
Referring to our discussion of wave packet revival in Subsection \ref{Ssection}, we have seen that in atomic physics wave packets do not behave classically on long time scales.
Perhaps surprisingly, this is even true for certain chaotic macroscopic systems: 
cf.\  the case of Hyperion mentioned in the Introduction and at the end of Subsection \ref{CEOM}.
 Decoherence now replaces the underlying 
superposition by a classical probability distribution, which reflects the chaotic nature of the limiting classical dynamics. Once again, the transition from the pertinent pure state of system plus environment to {\it a single}  observed system state remains clouded in mystery. But granted this transition, decoherence sheds new light on classical chaos and circumvents at least the most flagrant clashes with observation.\footnote{It should be mentioned, though, that any successful mechanism explaining the transition from quantum to classical should have this feature, so that at the end of the day decoherence might turn out to be a red herring here.}
\item Robustness and einselection form the state side or Schr\"{o}dinger picture of decoherence. Of course, there should also be a corresponding observable side or Heisenberg picture 
of decoherence. But the transition between the two pictures is more subtle than in the \qm\ of closed systems. In the Schr\"{o}dinger picture, the whole point of einselection is
that most pure states simply disappear from the scene. This may be beautifully  visualized on the example of a two-level system with \Hs\ $\H_{\CS}=\C^2$ (Zurek, 2003). If $\up$ and $\down$ (cf.\ \er{usualbasis}) happen to be the robust vector states of the system after coupling to an appropriate environment, and if we identify the corresponding density matrices with the north-pole $(0,0,1)\in B^3$ and the south-pole $(0,0,-1)\in B^3$, respectively (cf.\ \er{gens2}), then following decoherence all other states move towards the axis connecting the north- and south poles (i.e.\ the intersection of the $z$-axis with $B^3$) as $t\raw\infty$. In the  Heisenberg picture, this disappearance of all pure states except two
corresponds to the reduction of the full algebra of observables $M_2(\C)$ of the system
to its diagonal (and hence commutative) subalgebra $\C\oplus\C$ in the same limit. For it is only the latter algebra that contains enough elements to distinguish $\up$ and $\down$ without containing observables detecting interference terms between these pure states.
\item
To understand this in a more abstract and general way, we recall the mathematical relationship between pure states and observables (Landsman, 1998). The passage from a \ca\ $\CA$ of observables of a given system to its pure states is well known: as a set, the pure state space $\CP(\CA)$ is the extreme boundary of the total state space $\CS(\CA)$
(cf.\ footnote  \ref{EBfn}). In order to reconstruct $\CA$ from $\CP(\CA)$, the latter needs to be equipped with the structure of a transition probability space (see Subsection \ref{SE})
through \er{mtp}. Each element $A\in\CA$ defines a function $\hat{A}$ on $\CP(\CA)$ by $\hat{A}(\om)=\om(A)$. 
Now, in the simple case that $\CA$ is finite-dimensional (and hence a direct sum of matrix algebras), one can show that each function $\hat{A}$ is a finite linear combination of the form $\hat{A}=\sum_i p_{\om_i}$, where $\om_i\in \CP(\CA)$ and the elementary functions
$p_{\rh}$ on $\CP(\CA)$ are defined by $p_{\rh}(\sg)=p(\rh,\sg)$. Conversely, each such
 linear combination defines a function $\hat{A}$ for some $A\in\CA$. Thus  the elements of $\CA$ (seen as functions on the pure state space $\CP(\CA)$) are just the transition probabilities and linear combinations thereof. The algebraic structure of $\CA$ may then be reconstructed from the structure of $\CP(\CA)$ as a Poisson space with a transition probability (cf.\ Subsection \ref{PSD}). In this sense  $\CP(\CA)$ uniquely determines 
the algebra of observables of which it is the pure state space. 
For example, the space consisting of two points with classical transition probabilities \er{cltp} leads to the commutative algebra $\CA=\C\oplus\C$, whereas the unit two-sphere
in $\R^3$ with transition probabilities \er{thetarhsg} yields $\CA=M_2(\C)$. 

 This reconstruction procedure may be generalized to arbitrary \ca s (Landsman, 1998), and defines the precise connection 
between the Schr\"{o}dinger picture and the Heisenberg picture that is relevant to decoherence. These pictures are equivalent, but in practice the reconstruction procedure may be difficult to carry through.
\item For this reason it is of interest to have a direct description of decoherence in the Heisenberg picture. Such a description 
 has been developed by Blanchard \&\  Olkiewicz (2003), partly on the basis of earlier results by Olkiewicz (1999a,b, 2000).  Mathematically, their approach is more powerful than the Schr\"{o}dinger picture on which most of the literature on decoherence is based. Let $\CA_{\CS}=\CB(\H_{\CS})$ and
 $\CA_{\CE}=\CB(\H_{\CE})$, and assume one has a total Hamiltonian $H$ acting on $\H_{\CS}\ot \H_{\CE}$ as well as a fixed state
 of the environment, represented by a density matrix $\rh_{\CE}$ (often taken to be a thermal equilibrium state). 
 If $\rh_{\CS}$ is a density matrix on $\H_{\CS}$ (so that the total state is
 $\rh_{\CS}\ot \rh_{\CE}$), the Schr\"{o}dinger picture approach to decoherence (and more generally to the quantum theory of open systems) is based on the time-evolution
 \beq \rh_{\CS}(t)=\Tr_{\H_{\CE}}\left( e^{-\frac{it}{\hbar} H}\rh_{\CS}\ot \rh_{\CE}
 e^{\frac{it}{\hbar} H}\right).\eeq
 The Heisenberg picture, on the other hand, is based on the associated operator time-evolution for $A\in \CB(\H_{\CS})$ given by
 \beq A(t)=\Tr_{\H_{\CE}}\left(\rh_{\CE} e^{\frac{it}{\hbar} H}A \ot 1\,
 e^{-\frac{it}{\hbar} H}\right),\eeq
 since this yields the equivalence of the Schr\"{o}dinger and Heisenberg pictures expressed by
 \beq \Tr_{\H_{\CS}}\left( \rh_{\CS}(t)A\right)=  \Tr_{\H_{\CS}}\left( \rh_{\CS}A(t)\right).\eeq
 
 More generally, let $\CA_{\CS}$ and $\CA_{\CE}$ be unital \ca s with spatial tensor product $\CA_{\CS}\ot\CA_{\CE}$, equipped with a time-evolution $\al_t$ and a fixed state $\om_{\CE}$ on $\CA_{\CE}$. This defines a conditional expectation $P_{\CE}: \CA_{\CS}\ot\CA_{\CE}\raw \CA_{\CS}$ by linear and continuous extension of $P_{\CE}(A\ot B)=A\om_{\CE}(B)$, and consequently a reduced time-evolution $A\mapsto A(t)$ on $\CA_{\CS}$ via
 \beq A(t)=P_{\CE}(\al_t(A\ot 1)).\eeq
See, for example, Alicki \&\ Lendi (1987); in our context, this generality is crucial for the potential emergence of continuous classical phase spaces; see below.\footnote{For technical reasons
  Blanchard \&\  Olkiewicz (2003) assume $\CA_{\CS}$ to be a von Neumann algebra with trivial center.} Now the key point is that decoherence is described  
by a decomposition $\CA_{\CS}=\CA_{\CS}^{(1)}\oplus \CA_{\CS}^{(2)}$ {\it as a vector space} (not as a \ca), where $\CA_{\CS}^{(1)}$ is a \ca, 
 with the property that $\lim_{t\raw\infty} A(t)=0$
(weakly) for all $A\in \CA_{\CS}^{(2)}$, whereas  $A\mapsto A(t)$ is an automorphism on $\CA_{\CS}^{(1)}$  for each {\it finite} $t$ . Consequently, $\CA_{\CS}^{(1)}$ is the effective algebra of observables after decoherence, and it is precisely  the pure states
on $\CA_{\CS}^{(1)}$ that  are robust or einselected in the sense discussed before.
\item
For example, if $\CA_{\CS}=M_2(\C)$ and the states $\up$ and $\down$ are robust under decoherence, then $\CA_{\CS}^{(1)}=\C\oplus\C$ and $\CA_{\CS}^{(2)}$ consists of all $2\x 2$ matrices with zeros on the diagonal. In this example 
 $\CA_{\CS}^{(1)}$ is commutative hence classical, but this may not be the case in general. But if it is, the automorphic time-evolution on $\CA_{\CS}^{(1)}$
 induces a classical flow on its structure space, which should be shown to be Hamiltonian using the techniques of Section \ref{S6}.\footnote{Since on the assumption in the preceding footnote $\CA_{\CS}^{(1)}$ is a commutative von Neumann algebra one should define the structure space in an indirect way; see  Blanchard \&\  Olkiewicz (2003).} In any case, there will be some sort of classical behaviour of the decohered system whenever $\CA_{\CS}^{(1)}$ has a nontrivial center.\footnote{This is possible even when $\CA_{\CS}$ is a factor!} If this center is discrete, then the induced time-evolution on it is necessarily trivial, and one has the typical measurement situation where the center in question is generated by the projections on the eigenstates of a pointer observable with discrete spectrum. This is generic for the case where
 $\CA_{\CS}$ is a type {\sc i} factor. However, type {\sc ii} and {\sc iii} factors may give rise to continuous classical systems with nontrivial time-evolution; see  Lugiewicz \&\ Olkiewicz (2002, 2003). We cannot do justice here to the full technical details and complications involved here. But we would like to emphasize that further to  quantum field theory and the theory of the thermodynamic limit, the present context of decoherence should provide important motivation for specialists in the foundations of quantum theory  to learn the theory of operator algebras.\footnote{See the references in footnote  \ref{Cstarlit}.} 
\end{enumerate}
 \subsection{Consistent histories}\label{CHSS}
  Whilst doing so, one is well advised to work even harder and simultaneously familiarize  oneself with {\it consistent histories}.
This approach to quantum theory was pioneered by Griffiths (1984) and was subsequently taken up by Omn\`{e}s (1992) and others. Independently, Gell-Mann and Hartle (1990, 1993) arrived at analogous ideas. Like decoherence, the consistent histories method has been the subject of lengthy reviews (Hartle, 1995) and even books (Omn\`{e}s, 1994, 1999; Griffiths, 2002) by the founders.  
See also the reviews  by Kiefer (2003) and  Halliwell (2004), the critiques by
 Dowker \&\ Kent (1996), Kent (1998), Bub (1999), and  Bassi \&\  Ghirardi (2000),
as well as  the various mathematical reformulations and reinterpretations of the consistent histories program  (Isham, 1994, 1997; Isham \&\ Linden, 1994, 1995;  Isham, Linden \&\ Schreckenberg (1994); ÊIsham \&\ Butterfield, 2000; Rudolph, 1996a,b, 2000; Rudolph \&\ Wright, 1999).

The relationship between
consistent histories and decoherence is somewhat peculiar: on the one hand,
decoherence is a natural mechanism through which appropriate sets of histories become (approximately) consistent, but on the other hand these approaches appear to have quite different points of departure. Namely, where
decoherence starts from the idea that (quantum) systems are naturally coupled to their environments and therefore have to be treated as {\it open} systems, 
the aim of consistent histories is to deal with {\it closed} quantum systems such as the Universe, without a priori talking about measurements or observers. However, this distinction is merely historical: as we have seen in item \ref{CHD} in the previous subsection, the dividing line  between a system and its environment should be seen as a dynamical entity to be drawn according to  certain stability criteria, so that even in decoherence theory one should really study the system plus its environment as a whole from the outset.\footnote{This renders the distinction between ``open" and ``closed" systems a bit of a red herring, as even in decoherence theory the totality of the system plus its environment is treated as a closed system.}  And this is precisely what consistent historians do. 

As in the preceding subsection, and for exactly the same reasons, we format our treatment of consistent histories  as a list of items open to discussion. 
\begin{enumerate}
\item \label{HBC}
The starting point of the consistent histories formulation of quantum theory is conventional: one has a \Hs\ $\H$,  a state $\rh$, taken to be the initial state of the total system under consideration (realized as a density matrix on $\H$)
 and a Hamiltonian $H$ (defined as a self-adjoint operator on $\H$). What is unconventional is that this total system may well be the entire Universe. 
 Each property $\al$ of the total system is mathematically represented by a projection $P_{\al}$ on $\H$; for example, if $\al$ is the property that the energy takes some value $\epsilon$, then the operator  $P_{\al}$ is the projection onto the associated eigenspace (assuming $\epsilon$ belongs to the discrete spectrum of $H$). In the Heisenberg picture, $P_{\al}$ evolves in time as $P_{\al}(t)$ according to \er{HSEOM}; note that $P_{\al}(t)$ is once again a projection.
 
 A {\it history} $\mathbb{H}_A$ is a chain of properties (or propositions) $(\al_1(t_1),\ldots,\al_n(t_n))$  indexed by  $n$ different times $t_1<\ldots< t_n$; here  $A$ is a multi-label incorporating both the  properties $(\al_1,\ldots,\al_n)$ and the times $(t_1,\ldots,t_n)$. Such a history indicates that each  property $\al_i$ holds at time $t_i$, $i=1,\ldots,n$. Such  a history may be taken to be a collection 
 $\{\al(t)\}_{t\in\R}$ defined for all times, but for simplicity one usually assumes that $\al(t)\neq 1$ (where 1 is the trivial property that always holds) only for a finite set of times $t$; this set is  precisely  $\{t_1,\ldots,t_n\}$.   An example suggested by Heisenberg (1927) is to take $\al_i$ to be the property that a particle moving through a Wilson cloud chamber may be found in a cell $\Delta_i\subset \R^6$ of its phase space; the history $(\al_1(t_1),\ldots,\al_n(t_n))$ then denotes the state of affairs in which the particle is in cell $\Delta_1$ at time $t_1$, subsequently is in cell $\Delta_2$ at time $t_2$, etcetera. Nothing is stated about the particle's behaviour at intermediate times. Another example of a history is provided by the double slit experiment, where $\al_1$ is the particle's launch at the source at $t_1$ (which is usually omitted from the description), $\al_2$ is the particle passing through (e.g.) the upper slit at $t_2$, and $\al_3$ is the detection of the particle at some location $L$ at the screen at $t_3$. As we all know, there is a potential problem with this history, which will be clarified below in the present framework.
 
 The fundamental claim of the consistent historians seems to be  that quantum theory should do no more (or less) than making  predictions about the probabilities that histories occur.  What these probabilities actually mean remains obscure  (except perhaps when they are close to zero or one, or when reference is made to some measurement context; see Hartle (2005)), but let us first see when and how one can define them.  The  only potentially meaningful mathematical expression
 (within \qm)  for the probability of a history $\BBH_A$ with respect to a state $\rh$ is (Groenewold, 1952; Wigner, 1963)
 \beq p(\mathbb{H}_A) =\Tr(C_A\rh C_A^*), \label{Wig} \eeq
 where   
 \beq
 C_{A}=P_{\al_n}(t_n)\cdots P_{\al_1}(t_1).\label{defCA}\eeq
Note that $C_A$   is generally not a projection (and hence a property) itself (unless all $P_{\al_i}$ mutually commute). 
 In particular, when $\rh=[\Ps]$ is a pure state (defined by some unit vector $\Ps\in\H$), one simply has
   \beq p(\BBH_A) =\| C_A\Ps\|^2=\|P_{\al_n}(t_n)\cdots P_{\al_1}(t_1)\Ps\|^2.\eeq
 When $n=1$ this just yields the Born rule. Conversely, see  Isham (1994) for a derivation of \er{Wig} from the Born rule.\footnote{See also Zurek (2004) for a novel derivation of the Born rule, as well as the ensuing discussion in Schlosshauer (2004).} 
 \item
 Whatever one might think about the metaphysics of \qm, a probability makes no sense whatsoever when it is only attributed to a single history (except when it is exactly zero or one). The least one should have is something like  a sample space (or event space) of histories, each (measurable) subset  of which is assigned some probability such that the usual (Kolmogorov) rules are satisfied. This is a (well-known) problem even for a single time $t$ and  a single projection $P_{\al}$ (i.e.\ $n=1$).  In that case, the problem  is solved by finding a self-adjoint operator $A$ of which $P_{\al}$ is a spectral projection, so that 
 the sample space is taken to be the spectrum $\sg(A)$ of $A$, with  $\al\subset \sg(A)$. Given $P_{\al}$, the choice of $A$ is by no means unique, of course; different choices may lead to different and incompatible sample spaces. In practice, one usually starts from $A$ and derives the $P_{\al}$
 as its spectral projections $P_{\al}=\int_{\al} dP(\lm)$, given that the spectral resolution of $A$ is $A=\int_{\R} dP(\lm)\, \lm$.
  Subsequently,  one may then either {\it coarse-grain}  or {\it fine-grain} this sample space. The former is done by finding a partition  $\sg(A)=\coprod_i \al_i$ (disjoint union), and only admitting  elements of the $\sg$-algebra generated by the $\al_i$ as  events (along with the associated 
  spectral projection $P_{\al_i}$), instead of all (measurable) subsets of $\sg(A)$.  To perform fine-graining, one supplements $A$ by  operators that commute with $A$ as well as with each other, so that the new sample space is the joint spectrum of the ensuing family of mutually commuting operators. 
  
 In any case, in what follows it turns out to be convenient to work with the projections  $P_{\al}$ instead of the subsets $\al$ of the sample space; the above discussion then amounts to extending the  given projection on $\H$ to some Boolean sublattice of the lattice $\CP(\H)$ of all projections on $\H$.\footnote{This sublattice is supposed to the unit of  $\CP(\H)$, i.e.\ the unit operator on $\H$, as well as the zero projection. This comment also applies to
 the Boolean sublattice of  $\CP(\H^N)$ discussed below.}
 Any state $\rh$ then defines a probability measure on this sublattice in the usual way (Beltrametti \&\ Cassinelli, 1984).
  \item
 Generalizing this to the multi-time case is not a trivial task, 
somewhat facilitated by the following device (Isham, 1994).  Put $\H^N=\ot^N\H$, where $N$ is the cardinality  of the set of all times $t_i$ relevant to the histories in the given collection,\footnote{See the mathematical references above for the case $N=\infty$.} and, for a given history $\BBH_A$, define 
 \beq \mathbb{C}_A =P_{\al_n}(t_n)\otimes \cdots \otimes P_{\al_1}(t_1).\eeq
 Here $P_{\al_i}(t_i)$ acts on the copy of $\H$ in the tensor product $\H^N$
 labeled by $t_i$, so to speak. Note that $ \mathbb{C}_A$
 is a projection on $\H^N$ (whereas $C_A$ in \er{defCA} is generally {\it not} a projection on $\H$). Furthermore, given a density matrix  $\rh$ on $\H$ as above,  define the {\it decoherence functional} $d$ as a map from pairs of histories into $\C$ by
 \beq d(\BBH_A,\BBH_B)=\Tr(C_A\rh C_B^*).\eeq
  
 The main point of the consistent histories approach
may now be summarized as follows: a collection $\{\BBH_A\}_{A\in\mathbb{A}}$  of histories can be regarded as a sample space on which a state $\rh$ defines a probability measure via \er{Wig}, which of course amounts to \beq p(\BBH_A)=d(\BBH_A,\BBH_A), \label{CHP} \eeq
provided that:
 \begin{enumerate}
\item The operators $\{\mathbb{C}_A\}_{A\in\mathbb{A}}$  form a Boolean sublattice of the lattice $\CP(\H^N)$ of all projections on $\H^N$;
 \item The real part of $d(\BBH_A,\BBH_B)$ vanishes whenever $\BBH_A$ is disjoint from $\BBH_B$.\footnote{This means that $\mathbb{C}_A\mathbb{C}_B=0$; equivalently,  $P_{\al_i}(t_i)P_{\bt_i}(t_i)=0$
 for at least one time $t_i$. This condition guarantees that the probability \er{CHP} is additive on disjoint histories.} 
\end{enumerate}
In that case,  the set $\{\BBH_A\}_{A\in\mathbb{A}}$ is called  {\it consistent}.
It is important to realize that the possible consistency of a given set of histories depends (trivially) not only on this set, but in addition on the dynamics and on the initial state. 

Consistent sets of histories generalize families of commuting projections at a single time. There is no great loss in replacing the second condition by 
the vanishing of $d(\BBH_A,\BBH_B)$ itself, in which case the histories $\BBH_A$ and $\BBH_B$ are said to {\it decohere}.\footnote{Consistent historians use this terminology in a different way from decoherence theorists. By definition, any two histories involving only a single time are consistent (or, indeed, ``decohere") iff condition (a) above holds;
condition (b) is trivially satisfied in that case, and becomes relevant only when more than one time is considered. However, in decoherence theory the reduced density matrix at some given time does not trivially ``decohere" at all; the whole point of the (original) decoherence program was to provide models in which this happens (if only approximately)
because of the coupling of the system with its environment. Having said this, within the context of models there are close links between consistency (or decoherence) of multi-time histories and decoherence of  reduced density matrices, as the former is often (approximately) achieved by the same kind of dynamical mechanisms that lead to the latter.}
 For example,
in the double slit experiment the pair of histories $\{\BBH_A,\BBH_B\}$
where  $\al_1=\bt_1$ is the particle's launch at the source at $t_1$, $\al_2$ 
($\bt_2$) is the particle passing through  the upper (lower) slit at $t_2$, and $\al_3=\bt_3$ is the detection of the particle at some location $L$ at the screen,
is {\it not consistent}. It becomes consistent, however, when
the particle's passage through either one of the slits is recorded (or measured)
{\it without the recording device being included in the histories} (if it is, nothing would be  gained). This is reminiscent of the von Neumann chain in quantum measurement theory, which indeed provides an abstract setting for decoherence (cf.\ item \ref{DP1} in the preceding subsection). Alternatively, the set can be made consistent by omitting $\al_2$ and $\bt_2$. See Griffiths (2002) for a more extensive discussion of the double slit experiment in the language of consistent histories. 
 
More generally, coarse-graining by simply leaving out certain  properties is often a promising attempt  to make a given inconsistent set consistent; if the original history was already consistent, it can never become inconsistent by doing so. Fine-graining (by embedding into a larger set), on the other hand, is a dangerous act in that it may render a consistent set inconsistent.  
 \item What does it all mean? Each choice of a consistent set defines a ``universe of discourse" within which one can apply classical probability theory and  classical logic (Omn\`{e}s, 1992). In this sense the consistent historians are quite faithful to the Copenhagen spirit (as most of them acknowledge): {\it in order to understand it, the quantum world has to be looked at through classical glasses}. In our opinion, no convincing case has ever been made for the absolute necessity of this Bohrian stance (cf.\ Subsection \ref{Pcl}), but accepting it, the consistent histories approach is superior to Copenhagen in not relying on measurement as an a priori ingredient in the interpretation of \qm.\footnote{See Hartle (2005) for an analysis of the connection between consistent histories and the Copenhagen 
interpretation and others.}  It is also more powerful than the decoherence approach in turning the notion of a system 
 into a dynamical variable: different consistent sets describe different systems
 (and hence different environments, defined as the rest of the Universe); cf.\ item \ref{CHD} in the previous subsection.\footnote{Technically, as the commutant of the projections occurring in a given history.} In other words, the choice of a consistent set boils down to a choice of ``relevant variables" against ``irrelevant" ones omitted from the description. As indeed stressed in the literature, the act of identification of a certain consistent set as a universe of discourse is itself nothing but a coarse-graining of the Universe as a whole.
 \item \label{tag}
But these conceptual successes come with a price tag. Firstly, {\it consistent sets 
turn out not to exist in realistic models} (at least if the histories in the set  carry more than one time variable). 
 This has been recognized from the beginning of the program, the response being that one has to deal with approximately consistent sets for which  (the real part of) $d(\BBH_A,\BBH_B)$ is  merely very small. Furthermore, even the definition of a history often cannot be given  in terms of projections. For example, in Heisenberg's cloud chamber example (see item  \ref{HBC} above), because of his very own uncertainty principle it is impossible to write down the corresponding projections  $P_{\al_i}$. A natural candidate would be $P_{\al}=\qb(\ch_{\Delta})$, cf.\ \er{b2} and \er{qbttsrex}, but in view of \er{tpbt} this operator fails to satisfy $P_{\al}^2=P_{\al}$, so that it is not a projection (although it does satisfy the second defining property of a projection $P_{\al}^*=P_{\al}$). This merely reflects the usual property $\CQ(f)^2\neq \CQ(f^2)$ of any quantization method, and necessitates the use of  approximate projections (Omn\`{e}s, 1997). Indeed, this point calls for a reformulation of the entire consistent histories approach in terms of positive operators instead of projections (Rudolph, 1996a,b). 

These are probably not serious problems; indeed, the recognition that classicality emerges from quantum theory only in an approximate sense (conceptually as well as mathematically) is a profound  one (see the Introduction), and it rather should be counted among its blessings that the consistent histories program has so far confirmed it. 
\item
What is potentially  more troubling is that consistency by no means implies classicality  {\it beyond the ability (within a given consistent set) to assign classical probabilities and to use classical logic}. Quite to the contrary, neither Schr\"{o}dinger cat states nor histories that look classical at each time but follow  utterly unclassical trajectories in time are forbidden  by the consistency conditions alone (Dowker \&\ Kent, 1996). But is this a genuine problem, except to those who still believe that the earth is at the centre of the Universe and/or that humans are privileged observers? It just seems to be the case that - at least according to the consistent historians - the ontological landscape laid out by quantum theory is far more ``inhuman"  (or some would say ``obscure") than the one we inherited from Bohr, in the sense that most consistent sets bear no obvious relationship to the world that {\it we} observe. In attempting to make sense of these, no
appeal to ``complementarity" will do now: for one,  the complementary pictures of the quantum world called for by Bohr were classical in a much stronger sense than generic consistent sets are, and on top of that Bohr asked us to only think about two such pictures, as opposed to the innumerable  consistent sets  offered to us. Our conclusion is that, much as decoherence does not solve the measurement problem but rather aggravates it (see item \ref{point5} in the preceding subsection), also {\it consistent histories actually make the problem of interpreting \qm\ more difficult than  it was thought to be before}. 
In any case, it is beyond doubt that the consistent historians have significantly deepened our understanding of quantum theory - at the very least by providing a good bookkeeping device! 
\item 
Considerable progress has been made in the task of identifying at least {\it some} (approximately) consistent sets that display (approximate) classical behaviour in the full sense of the word (Gell-Mann \&\ Hartle, 1993; Omn\`{e}s, 1992, 1997; Halliwell, 1998, 2000, 2004;  Brun \&\ Hartle, 1999; Bosse \&\ Hartle, 2005). Indeed, in our opinion studies of this type form the main concrete outcome of the consistent histories program.
The idea is to find a consistent set $\{\BBH_A\}_{A\in\mathbb{A}}$ 
with three decisive properties:\begin{enumerate}
\item Its elements (i.e.\ histories) are strings of  propositions with a classical interpretation;
\item Any history in the set that delineates a classical trajectory (i.e.\ a solution of appropriate classical equations of motion) has probability \er{CHP} close to unity, and any history following a classically impossible trajectory has probability close to zero;
\item The description is sufficiently coarse-grained to achieve consistency, but is sufficiently fine-grained to turn the deterministic equations of motion following from (b) into a closed system. 
\end{enumerate}
When these goals are met, it is in this sense (no more, no less) that the consistent histories program can claim with some justification that it has indicated (or even explained) `How the quantum Universe becomes classical' (Halliwell, 2005).

Examples of propositions with a classical interpretation are quantized classical observables with a recognizable interpretation (such as the operators $\qb(\ch_{\Delta})$ mentioned in item \ref{tag}), macroscopic observables of the kind studied in Subsection \ref{MO}, and hydrodynamic variables (i.e.\ spatial integrals over conserved currents). 
These represent three different levels of classicality, which in principle are connected
through mutual fine- or coarse-grainings.\footnote{The study of these connections is 
relevant to the program laid out in this paper, but really belongs to classical physics 
{\it per se}; think of the derivation of the Navier--Stokes equations from Newton's equations.} 
The first are sufficiently coarse-grained to achieve consistency only in the limit $\hbar\raw 0$ (cf.\ Section \ref{S5}), whereas the latter two are already  coarse-grained by their very nature. Even so, also the initial state will have to be ``classical'' in some sense in order te achieve the three targets (a) - (c).  
\end{enumerate}

All this is quite impressive, but we would like  to state our opinion that neither decoherence nor consistent histories can stand on their  own in explaining the appearance of the classical world. Promising as these approaches are, they have to be combined at least with limiting techniques of the type described in Sections \ref{S5} and \ref{S6} - not to speak of the need for a new metaphysics! For even if it is granted that decoherence yields the disappearance of superpositions of Schr\"{o}dinger cat type, or that consistent historians give us consistent sets none of whose elements contain  such superpositions among their properties,  this by no means suffices to explain the emergence of classical phase spaces and  flows thereon determined by classical equations of motion. Since so far  the approaches cited in  Sections \ref{S5} and \ref{S6} have hardly been combined with the decoherence and/or the consistent histories program,  a full explanation of the classical world from quantum theory is still in its infancy. This is not merely true at the technical level, but also conceptually; what has been done so far only represents a modest beginning.  On the positive side, here lies an attractive challenge for mathematically minded researchers in the foundations of physics!
\section{Epilogue}\label{S8}
As a sobering closing note, one should not forget that whatever one's achievements in identifying a ``classical realm" in \qm, the theory continues to incorporate another realm, the pure quantum world, that  the young Heisenberg first gained access to, if not through his mathematics, then perhaps through
the music of his favourite composer, Beethoven. This world beyond ken has never been better described than by  Hoffmann (1810) in his  essay on Beethoven's instrumental music, and we find it appropriate to end this paper by quoting at some length from it:\footnote{Translation copyright: Ingrid  Schwaegermann (2001).}
\begin{quote}
Should one, whenever music is discussed as an independent art, not always be referred to instrumental music which, refusing the help of any other art (of poetry), expresses the unique essence of art that can only be recognized in it? It is the most romantic of all arts, one would almost want to say, the only truly romantic one, for only the infinite is its source. Orpheus' lyre opened the gates of the underworld.Ê Music opens to man an unknown realm, a world that has nothing in common with the outer sensual world that surrounds him, a realm in which he leaves behind all of his feelings of certainty, in order to abandon himself to an unspeakable longing.  (\ldots)

Beethoven's instrumental music opens to us the realm of the gigantic and unfathomable.Ê Glowing rays of light shoot through the dark night of this realm, and we see gigantic shadows swaying back and forth, encircling us closer and closer, destroying us (\ldots)  Beethoven's music moves the levers of fear, of shudder, of horror, of pain and thus awakens that infinite longing that is the essence of romanticism.Ê Therefore, he is a purely romantic composer, and may it not be because of it, that to him, vocal music that does not allow for the character of infinite longing - but, through words, achieves certain effects, as they are not present in the realm of the infinite -  is harder?Ê(\ldots)

What instrumental work of Beethoven confirms this to a higher degree than his magnificent and profound Symphony in c-Minor.ÊÊ 
Irresistibly, this wonderful composition leads its listeners in an increasing climax towards the realm of the spirits and the infinite.ÊÊÊÊ(\ldots)
Only that composer truly penetrates into the secrets of harmony who is able to have an effect on human emotions through them; to him, relationships of numbers, which, to the Grammarian, must remain dead and stiff mathematical examples without genius, are magic potions from which he lets a miraculous world emerge. (\ldots)

Instrumental music, wherever it wants to only work through itself and not perhaps for a certain dramatic purpose, has to avoid all unimportant punning, all dallying.ÊÊÊ It seeks out the deep mind for premonitions of joy that, more beautiful and wonderful than those of this limited world, have come to us from an unknown country, and spark an inner, wonderful flame in our chests, a higher expression than mere words - that are only of this earth - can spark.Ê 
\end{quote}
\newpage
\section{References} 
%\addcontentsline{toc}{section}{References}
\begin{trivlist}
\item Abraham, R. \&\
 Marsden, J.E. (1985). \textit{Foundations of Mechanics}, 2nd ed. Addison
Wesley, Redwood City.
\item  Accardi, L.,  Frigerio, A., \&\  Lu, Y.  (1990).  The weak coupling limit as a quantum
	functional central limit.   {\it  Communications in Mathematical Physics} 131, 537--570.
\item Adler, S.L. (2003). Why decoherence has not solved the measurement problem: A response to P.W. Anderson.  {\it Studies in History and Philosophy of Modern Physics}  34B, 135--142.
 \item Agmon, S. (1982). {\it Lectures on Exponential Decay of Solutions of Second-Order Elliptic Equations}.  Princeton: Princeton University Press.
\item Albeverio, S.A. \&\ H\o egh-Krohn, R.J. (1976).
 {\it  Mathematical Theory of Feynman Path Integrals}.
  Berlin: Springer-Verlag.
\item  Alfsen, E.M. (1970). {\it Compact
Convex Sets and Boundary Integrals}. Berlin: Springer. 
\item   Ali, S.T.,  Antoine,  J.-P.,  Gazeau, J.-P. \&\ 
Mueller, U.A. (1995). Coherent states and their generalizations: 
a mathematical overview. {\it Reviews in Mathematical
Physics}  7, 1013--1104.
\item   Ali, S.T.,  Antoine,  J.-P.,  \&\ Gazeau, J.-P. (2000).
{\it Coherent States, Wavelets and their Generalizations}. New York: Springer-Verlag.
\item Ali, S.T. \&\ Emch, G.G. (1986). Geometric quantization: modular reduction theory and coherent  states.  {\it Journal of Mathematical Physics}  27, 2936--2943. 
\item Ali, S.T \&\ Englis, M. (2004).  Quantization methods: a guide for physicists and analysts. \\ \texttt{arXiv:math-ph/0405065}.
\item Alicki, A. \&\ Fannes, M. (2001). {\it Quantum Dynamical Systems}.
Oxford: Oxford University Press. 
\item Alicki, A. \&\ Lendi, K. (1987). {\it Quantum Dynamical Semigroups and Applications}. Berlin: Springer. 
\item Amann, A. (1986). Observables in $W^*$-algebraic \qm. {\it Fortschritte der Physik} 34, 167--215.
\item Amann, A. (1987). Broken symmetry and the generation of classical observables in large systems. {\it Helvetica Physica Acta} 60, 384--393. 
\item Amann, A. \&\ Primas, H. (1997). What is the referent of a non-pure quantum state?
{\it Experimental Metaphysics: Quantum Mechanical Studies in Honor of Abner Shimony},  S. Cohen, R.S., Horne, M.A., \&\  Stachel, J. (Eds.). Dordrecht: Kluwer Academic Publishers.
\item Arai, T. (1995). Some extensions
of the semiclassical limit $\hbar\raw 0$ for Wigner functions on phase
space. {\em Journal of Mathematical\ Physics}  36, 622--630.
 \item Araki, H. (1980). A remark on the Machida-Namiki theory of measurement.
{\it Progress in  Theoretical  Physics } 64,  719--730.
\item Araki, H. (1999). {\it Mathematical Theory of Quantum Fields}. New York: Oxford University Press.
\item ÊArnold, V.I. (1989). {\it Mathematical Methods of Classical Mechanics.} Second edition.   New York: Springer-Verlag.
\item Ashtekar, A. \&\  Schilling, T.A. (1999). Geometrical formulation of quantum mechanics. 
{\it On Einstein's Path (New York, 1996)}, pp.\  23--65. 
 New York:  Springer.
\item Atmanspacher, H.,  Amann, A., \&\   M\"{u}ller-Herold, U. (Eds.). (1999).
{\it On Quanta, Mind and Matter: Hans Primas in Context.} Dordrecht: Kluwer Academic Publishers.
\item Auletta, G. (2001). {\it Foundations and Interpretation of Quantum Mechanics}. Singapore: World Scientific. 
\item Bacciagaluppi, G. (1993). Separation theorems and Bell inequalities in algebraic quantum mechanics. {\it Proceedings of the Symposium on the Foundations of Modern Physics
(Cologne, 1993)}, pp.\ 29--37. Busch, P., Lahti, P.J., \&\ Mittelstaedt, P. (Eds.).
Singapore: World Scientific.
\item Bacciagaluppi, G. (2004). The Role of Decoherence in Quantum Theory.
{\it Stanford Encyclopedia of Philosophy},  (Winter 2004 Edition),  Zalta, E.N. (Ed.).
Online only at \\  {\texttt http://plato.stanford.edu/archives/win2004/entries/qm-decoherence/}.         
\item Bach, V., Fr\"{o}hlich, J., \&\  Sigal, I.M. (1998).  Quantum electrodynamics of confined nonrelativistic particles.  {\it Advances in  Mathematics}  137, 299--395.
\item Bach, V., Fr\"{o}hlich, J., \&\  Sigal, I.M. (1999). 
Spectral analysis for systems of atoms and molecules coupled to the  quantized radiation field. {\it  Communications in Mathematical Physics}  207, 249--290.
\item Baez, J. (1987).  Bell's inequality for $C\sp *$-algebras.  
  {\it Letters in Mathematical Physics} 13,  135--136.
  \item Bagarello, F. \&\ Morchio, G. (1992). Dynamics of mean-field spin models from basic results in abstract differential equations. {\it Journal of Statistical Physics} 66, 849--866.
\item  Ballentine, L.E.  (1970). The statistical interpretation of quantum mechanics.
{\it Reviews of Modern Physics} 42, 358--381. 
\item  Ballentine, L.E. (1986). Probability theory in quantum mechanics. {\it American Journal of Physics} 54, 883--889. 
\item  Ballentine, L.E. (2002). Dynamics of quantum-classical differences for chaotic systems.
{\it Physical Review} A65, 062110-1--6.
\item  Ballentine, L.E. (2003). The classical limit of \qm\ and its implications for the foundations of \qm. {\it Quantum Theory: Reconsideration of Foundations -- 2}, pp.\ 71--82.
Khrennikov, A. (Ed.). V\"{a}xj\"{o}:  V\"{a}xj\"{o} University Press. 
\item  Ballentine, L.E., Yang, Y. \&\ Zibin, J.P. (1994). Inadequacy of Ehrenfest's theorem to characterize the classical regime. {\it Physical Review} A50, 2854--2859.
  \item Balian, R. \&\ Bloch, C. (1972). Distribution of eigenfrequencies for the wave equation in a finite domain. III. Eigenfrequency density oscillations. {\it Annals of  Physics  } 69, 76--160.
   \item Balian, R. \&\ Bloch, C. (1974). Solution of the Schr\"odinger equation in terms of classical paths. {\it Annals of  Physics  } 85, 514--545.
\item Bambusi, D.,  Graffi, S., \&\  Paul, T. (1999). 
  Long time semiclassical approximation of quantum flows: a proof of the  Ehrenfest time.  
 {\it  Asymptotic Analysis}  21, 149--160.
\item Barrow-Green, J. (1997).  {\it Poincar\'e and the Three Body Problem}. 
 Providence, RI: (American Mathematical Society.
\item Barut,
A.O. \&\  Ra\c{c}ka, R. (1977). \textit{Theory of Group Representations and
Applications}. Warszawa: PWN.
\item  Bassi, A. \&\  Ghirardi, G.C. (2000).
  Decoherent histories and realism.  
 {\it Journal of Statistical Physics }  98, 457--494. Reply by Griffiths, R.B.  (2000). 
 {\it ibid.} 99,  1409--1425. Reply to this reply by
  Bassi, A. \&\  Ghirardi, G.C.   (2000). {\it ibid.} 99, 1427.
\item Bates, S. \&\ Weinstein, A. (1995). 
{\it Lectures on the Geometry of Quantization}. {\it Berkeley
Mathematics Lecture Notes}  8. University of California,
Berkeley. Re-issued by the American Mathematical Society. 
\item Batterman, R.W. (2002). {\it The Devil in the Details: Asymptotic Reasoning in Explanation, Reduction, and Emergence}. Oxford: Oxford University Press.
\item Batterman, R.W. (2005). Critical phenomena and breaking drops: Infinite idealizations in physics. 
 {\it Studies in History and Philosophy of Modern Physics} 36, 225--244.  
\item  Baum, P.,   Connes, A. \&\  Higson, N. (1994).
 Classifying space for proper actions
and K-theory of group $C^*$-algebras.   {\it Contemporary\ Mathematics}
167, 241--291.  
  \item Bayen, F., Flato,  M.,  Fronsdal, C.,  Lichnerowicz, A. \&\ 
 Sternheimer, D.  (1978). Deformation theory and quantization I, II.  {\it
Annals of  Physics \ } 110, 61--110, 111--151.
\item  Bell, J.S. (1975). On wave packet reduction in the Coleman--Hepp model.
{\it Helvetica Physica Acta} 48, 93--98.
\item  Bell, J.S. (1987). {\it Speakable and Unspeakable in Quantum Mechanics}. 
Cambridge: Cambridge University Press. 
\item  Bell, J.S. (2001).  {\it John S. Bell on the Foundations of Quantum Mechanics}.
Singapore: World Scientific.
\item  Beller, M. (1999). {\it Quantum Dialogue}.  Chicago: University of Chicago Press.
\item Bellissard, J. \&\  Vittot, M. (1990).
  Heisenberg's picture and noncommutative geometry of the semiclassical  limit in quantum mechanics. {\it
  Annales de l' Institut Henri Poincar\'{e} -  Physique Th\'{e}orique}  52, 175--235.
\item Belot, G. (2005). Mechanics: geometrical. This volume. 
\item Belot, G. \&\ Earman, J. (1997). Chaos out of order: \qm, the correspondence principle and chaos. {\it Studies in History and Philosophy of Modern Physics} 28B, 147--182.
\item Beltrametti, E.G. \&\ Cassinelli, G. (1984). \textit{The
Logic of Quantum Mechanics}. Cambridge: Cambridge University Press.
\item Benatti, F. (1993).
 {\it Deterministic Chaos in Infinite Quantum Systems}. Berlin:
 Springer-Verlag. 
\item Bene, G. \& Dieks, D. (2002). A Perspectival Version of the Modal Interpretation of Quantum Mechanics and the Origin of Macroscopic Behavior. {\it Foundations of  Physics } 
32, 645-671.
\item Berezin, F.A. (1974). Quantization. {\it Mathematical\ USSR Izvestia} 8,
1109--1163.
\item Berezin, F.A. (1975a). Quantization in complex
symmetric spaces. {\it Mathematical\ USSR Izvestia}  9, 341--379.
\item
Berezin, F.A. (1975b). General concept of quantization.  {\it Communications in \
Mathematical\ Physics  }  40, 153--174.
\item Berry, M.V. (1977a). Semi-classical mechanics in phase space: a study of Wigner's function. {\it Philosophical Transactions of the  Royal Society} 287, 237--271.
\item Berry, M.V. (1977b). Regular and irregular semi-classical wavefunctions.
{\it Journal of Physics } A10, 2083--2091.
\item Berry, M.V.,   Balazs, N.L., Tabor, M., \&\  Voros, A. (1979).
Quantum maps. {\it Annals of Physics } 122, 26--63.
\item Berry, M.V. \& Tabor, M. (1977). Level clustering in the regular spectrum. {\it Proceedings of the Royal Society} A356, 375--394.
\item Berry, M.V. \&\ Keating, J.P.  (1999). The Riemann zeros and eigenvalue asymptotics.  {\it SIAM Review}  41, 236--266.
\item  Binz,
E., J.,  \'{S}niatycki, J. \&\  Fischer, H.  (1988).
{\it  The Geometry of Classical Fields}. Amsterdam: North--Holland.
\item Birkhoff, G. \&\ von Neumann, J.  (1936). The logic of quantum mechanics.  
{\it Annals of Mathematics} (2)  37, 823--843.
\item Bitbol, M. (1996). {\it Schr\"{o}dinger's Philosophy of Quantum Mechanics}.
Dordrecht: Kluwer Academic Publishers.
\item Bitbol, M. \&\ Darrigol, O. (Eds.) (1992).  {\it Erwin Schr\"{o}dinger:  Philosophy and the Birth of Quantum Mechanics}.
Dordrecht: Kluwer Academic Publishers.
\item Blackadar, B. (1998). {\it $K$-Theory for Operator Algebras}. Second edition.  Cambridge: Cambridge University Press.
\item Blair Bolles, E. (2004). {\it Einstein Defiant: Genius versus Genius in the Quantum Revolution}. Washington: Joseph Henry Press.
\item Blanchard,
E. (1996).  Deformations de $C^*$-algebras de Hopf. \textit{Bulletin de la  Soci\'{e}t\'{e}
math\'{e}matique de  France} 124, 141--215.
\item Blanchard, Ph. \&\  Olkiewicz, R. (2003). Decoherence induced transition from quantum to classical dynamics.  {\it Reviews in Mathematical  Physics}  15, 217--243.
\item Bohigas, O., Giannoni, M.-J., \&\ Schmit, C. (1984). Characterization of chaotic quantum spectra and universality of level fluctuation laws. {\it Physical Review Letters} 52, 1--4.
\item Blume-Kohout, R. \&\   Zurek, W.H. (2004). 
A simple example of "Quantum Darwinism": Redundant information storage   in many-spin environments. {\it Foundations of Physics}, to appear.\\  \texttt{arXiv:quant-ph/0408147}.
\item Blume-Kohout, R. \&\   Zurek, W.H. (2005). Quantum Darwinism: Entanglement, branches, and the emergent classicality   of redundantly stored quantum information.
{\it Physical Review} A, to appear.  \texttt{arXiv:quant-ph/0505031}.
\item Bohr, N. (1927) The quantum postulate and the recent development of atomic theory. {\it Atti del Congress Internazionale dei Fisici (Como, 1927)}.
Reprinted in Bohr (1934), pp.\ 52--91.
\item Bohr, N. (1934). {\it Atomic Theory and the Description of Nature}.
Cambridge: Cambridge University Press.
\item Bohr, N. (1935). Can quantum-mechanical description of physical reality be considered complete? {\it Physical Review} 48, 696--702.
\item Bohr, N. (1937). Causality and complementarity. {\it Philosophy of Science} 4, 289--298.
\item Bohr, N. (1949). Discussion with Einstein on epistemological problems in atomic physics. {\it Albert Einstein: Philosopher-Scientist}, pp.\ 201--241. P.A. Schlipp (Ed.). La Salle: Open Court. 
\item Bohr, N. (1958). {\it Atomic Physics and Human Knowlegde}. New York: Wiley. 
\item Bohr, N. (1985). {\it Collected Works. Vol.\ 6: Foundations of Quantum Physics {\sc i} (1926--1932)}. Kalckar, J. (Ed.). Amsterdam: North-Holland.
\item Bohr, N. (1996). {\it Collected Works. Vol.\ 7: Foundations of Quantum Physics {\sc ii} (1933--1958)}. Kalckar, J. (Ed.). Amsterdam: North-Holland.
\item Bogoliubov, N.N. (1958). On a new method in the theory of superconductivity. {\it Nuovo Cimento} 7, 794--805.
\item Bona, P. (1980). A solvable model of particle detection in quantum theory. {\it Acta Facultatis Rerum Naturalium Universitatis Comenianae Physica} XX, 65--94.
\item Bona, P. (1988). The dynamics of a class of mean-field theories. {\it Journal of Mathematical Physics} 29, 2223--2235.
\item Bona, P. (1989). Equilibrium states of a class of  mean-field theories. {\it Journal of Mathematical Physics} 30, 2994--3007.
\item Bona, P. (2000). Extended quantum mechanics. {\it Acta Physica Slovaca} 50, 1--198. 
 \item  Bonechi, F. \&\  De Bi\`{e}vre, S. (2000).  Exponential mixing and $\ln
\hbar$ time scales in quantized hyperbolic maps on the torus.
{\it Communications in Mathematical Physics} 211, 659--686.
\item Bosse, A.W. \&\ Hartle, J.B. (2005). Representations of spacetime alternatives and their classical limits. \texttt{arXiv:quant-ph/0503182}. 
\item Brack, M. \&\ Bhaduri, R.K. {\it Semiclassical Physics}. Boulder: Westview Press.
\item Bratteli, O. \&\  Robinson, D.W. (1987). {\it Operator
Algebras and Quantum Statistical Mechanics. Vol.\ I: $C^*$- and
$W^*$-Algebras, Symmetry Groups, Decomposition of States}. 2nd
Ed. Berlin: Springer.
\item  Bratteli, O. \&\  Robinson, D.W. (1981). {\it
Operator Algebras and Quantum Statistical Mechanics. Vol.\ II:
Equilibrium States, Models in Statistical Mechanics}. Berlin: Springer.
\item Brezger, B., Hackerm\"{u}ller, L., Uttenthaler, S., Petschinka, J., Arndt, M., \&\ Zeilinger, A. (2002). Matter-Wave Interferometer for Large Molecules. {\it Physical Review Letters} 88, 100404.
\item Breuer T.  (1994). {\it Classical Observables, Measurement, and Quantum Mechanics}. Ph.D.\ Thesis,  University of Cambridge. 
\item Br\"ocker, T. \&\ Werner, R.F. (1995).
  Mixed states with positive Wigner functions.  
 {\it Journal of Mathematical Physics}  36, 62--75.
 \item Brun, T.A. \&\ Hartle, J.B.  (1999). Classical dynamics of the quantum harmonic chain. {\it  Physical Review} D60,  123503-1--20.
 \item Brush, S.G. (2002). Cautious revolutionaries: Maxwell, Planck, Hubble.
{\it American Journal of Physics} 70, 119-Ð127.
\item   Bub, J. (1988). How to Solve the Measurement Problem of Quantum Mechanics.
{\it Foundations of Physics} 18,Ê 701--722. 
\item Bub, J. (1999). {\it Interpreting the Quantum World}.  Cambridge: Cambridge University Press.
\item Bub, J. (2004). Why the quantum? {\it Studies in History and Philosophy of Modern Physics } 35B, 241--266.
\item Busch, P.,  Grabowski, M. \&\ Lahti,  P.J. (1998).
{\it Operational Quantum Physics}, 2nd corrected ed.
 Berlin: Springer. 
 \item Busch, P.,   Lahti,  P.J., \&\ Mittelstaedt, P. (1991). {\it The Quantum Theory of Measurement}. Berlin: Springer. 
\item  Butterfield, J. (2002). Some Worlds of Quantum Theory. R.Russell, J. Polkinghorne et al (Ed.). {\it Quantum Mechanics} (Scientific Perspectives on Divine Action vol 5), pp.\ 111-140. Rome: Vatican Observatory Publications, 2. \texttt{arXiv:quant-ph/0105052}; PITT-PHIL-SCI00000204.
\item  Butterfield, J. (2005). On symmetry, conserved quantities and symplectic reduction in classical mechanics. {\it This volume}. 
\item B\"{u}ttner, L.,  Renn, J., \&\ Schemmel, M. (2003). Exploring the limits of classical physics: Planck, Einstein, and the structure of a scientific revolution.
{\it Studies in History and Philosophy of Modern Physics} 34B, 37--60.
\item Camilleri, K. (2005). {\it Heisenberg and Quantum Mechanics: The Evolution of a Philosophy of Nature}. Ph.D.\  Thesis, University of Melbourne. 
 \item Cantoni, V. (1975). 
Generalized ``transition probability''. \textit{Communications in Mathematical\ Physics }
 44, 125--128. 
 \item Cantoni, V. (1977). The Riemannian structure on the states of quantum-like systems.
 \textit{Communications in Mathematical\ Physics } 56, 189--193.
\item Carson, C. (2000). Continuities and discontinuities in Planck's {\it Akt der Verzweiflung}. {\it Annalen der Physik} 9, 851--960.
\item  Cassidy, D.C. (1992). {\it Uncertainty: the Life and Science of Werner Heisenberg}. New York: Freeman.
\item Castrigiano, D.P.L. \&\  Henrichs, R.W. (1980).
Systems of covariance and
subrepresentations of induced representations. {\it Letters in Mathematical
 Physics} 4, 169-175.
\item Cattaneo, U. (1979). On Mackey's imprimitivity theorem. 
 {\it Commentari 
Mathematici Helvetici}  54, 629-641.
\item Caves, C.M., Fuchs, C.A., \&\ Schack, R.  (2002). Unknown quantum states: the quantum de Finetti representation. Quantum information theory.  {\it Journal of Mathematical Physics}  43,   4537--4559.
\item Charbonnel, A.M. (1986). Localisation et d\'{e}veloppement asymptotique des \'{e}l\'{e}ments du spectre conjoint d' op\'{e}rateurs psuedodiff\'{e}rentiels qui commutent. 
{\it Integral Equations Operator Theory} 9, 502--536.
\item Charbonnel, A.M. (1988).  Comportement semi-classiques du spectre conjoint d' op\'{e}rateurs psuedodiff\'{e}rentiels qui commutent. {\it Asymptotic Analysis} 1, 227--261.
\item Charbonnel, A.M. (1992).   Comportement semi-classiques des syst\`{e}mes ergodiques.
{\it  Annales de l' Institut Henri Poincar\'{e} -  Physique Th\'{e}orique} 56, 187--214.
\item Chernoff, P.R. (1973). Essential
self--adjointness of powers of generators of hyperbolic equations.
\textit{Journal of  Functional Analysis}  12, 401--414.
\item Chernoff, P.R. (1995). 
Irreducible representations of infinite dimensional transformation
groups and Lie algebras I. {\it Journal of  Functional \ Analysis} 130,
255--282.
\item Chevalley, C. (1991). Introduction: Le dessin et la couleur. {\it Niels Bohr: Physique Atomique et Connaissance Humaine}. (French translation of Bohr, 1958). Bauer, E. \&\ Omn\`{e}s, R. (Eds.), pp.\ 17--140. Paris: Gallimard.
\item Chevalley, C. (1999). Why do we find Bohr obscure?  \textit{Epistemological and Experimental Perspectives on Quantum Physics}, pp.\ 59--74.
 Greenberger, D.,  Reiter, W.L., \&\  Zeilinger, A. (Eds.).
  Dordrecht: Kluwer Academic Publishers.
\item  Chiorescu, I.,  Nakamura, Y.,  Harmans, C.J.P.M., \&\ Mooij, J.E. (2003).
Coherent Quantum Dynamics of a Superconducting Flux Qubit. {\it Science} 299, Issue 5614, 1869--1871.      
\item  
Cirelli, R., Lanzavecchia, P., \&\  Mania, A. (1983).  Normal pure states
of the von Neumann algebra of bounded operators as a K\"{a}hler
manifold.  {\it Journal of  Physics } A16, 3829--3835.
\item Cirelli, R.,
 Mani\'{a}, A., \&\  Pizzocchero, L. (1990). Quantum mechanics as an
infinite-dimensional Hamiltonian system with uncertainty
structure. I, II. \textit{Journal of  Mathematical\ Physics }  31, 2891--2897, 2898--2903.
\item Colin de Verdi\`{e}re, Y.  (1973). Spectre du laplacien et longueurs des g\'{e}od\'{e}siques p\'{e}riodiques. I, II. {\it  Compositio Mathematica} 27, 83--106, 159--184. 
\item Colin de Verdi\`ere, Y. (1977). Quasi-modes sur les vari\'et\'es Riemanniennes. {\it Inventiones Mathematicae}  43, 15--52.
\item Colin de Verdi\`ere, Y. (1985). Ergodicit\'e et fonctions propres du Laplacien. {\it Communications in  Mathematical Physics} 102, 497--502.
\item Colin de Verdi\`ere, Y. (1998). Une introduction ˆ la mŽcanique semi-classique. {\it  l'Enseignement  Mathematique (2)}  44, 23--51.
\item Combescure, M. (1992). The squeezed state approach of the
semiclassical limit of the time-dependent Schr\"{o}dinger equation.
{\it Journal of  Mathematical\ Physics }  33, 3870--3880.
\itemÊCombescure, M., Ralston, J.,  \&\ Robert, D. (1999). A proof of the Gutzwiller semiclassical trace formula using coherent states decomposition.
{\it Communications in Mathematical\ Physics } 202, 463--480.
\itemÊCombescure, M. \&\ Robert, D. (1997). Semiclassical spreading of quantum wave packets and applications near  unstable fixed points of the classical flow.  {\it Asymptotic Analysis}  14,  377--404.
\item Corwin, L. \&\ Greenleaf, F.P. (1989). \textit{Representations of Nilpotent Lie Groups and
Their Applications}, Part I. Cambridge: Cambridge University Press.
\item Cucchietti, F.M. (2004). {\it The Loschmidt Echo in Classically Chaotic Systems: Quantum Chaos,  Irreversibility and Decoherence}. Ph.D. Thesis, Universidad Nacional de C\'{o}rdobo. \texttt{arXiv:quant-ph/0410121}.
\item Cushing, J.T. (1994). {\it Quantum Mechanics: Historical Contingency and the Copenhagen Hegemony}.   Chicago: University of Chicago Press.
\item Cvitanovic, P.  et al. (2005). {\it Classical and Quantum Chaos}.
\texttt{http://ChaosBook.org}. 
\item ÊCycon, H. L., Froese, R. G., Kirsch, W., \&\ Simon, B. (1987).
{\it Schr\"{o}dinger Operators with Application to Quantum Mechanics and  Global Geometry}. Berlin: Springer-Verlag.  
\item Darrigol, O. (1992). {\it From c-Numbers to q-Numbers}.  Berkeley: University of California Press. 
\item Darrigol, O. (2001). The Historians' Disagreements over the Meaning of Planck's Quantum. {\it  Centaurus}  43, 219--239.  
\item Davidson, D. (2001). {\it Subjective, Intersubjective, Objective}. Oxford: Clarendon Press
\item
 Davies, E.B. (1976). {\it Quantum Theory of Open Systems}.
London: Academic Press.
\item  De Bi\`evre, S.  (1992).
 Oscillator eigenstates concentrated on classical trajectories. {\it
Journal of Physics}    A25, 3399-3418.
\item  De Bi\`evre, S.  (2001). Quantum chaos: a brief first visit.  {\it Contemporary Mathematics}  289,  
  161--218.
\item De Bi\`{e}vre, S. (2003). Local states of  free bose fields. Lectures given at the Summer School  on Large Coulomb Systems, Nordfjordeid. 
\item  De Bi\`evre, S.,   Irac-Astaud, M, \&\  Houard, J.C. (1993). Wave packets localised on closed classical trajectories. {\it  Differential Equations and
Applications in Mathematical Physics}, pp.\ 
 25--33.   Ames, W.F. \&\  Harrell, E.M., \&\ Herod, J.V.  (Eds.). New York:
 Academic Press.
\item De Muynck, W.M.  (2002) {\it Foundations of Quantum Mechanics: an Empiricist Approach}. Dordrecht:  Kluwer Academic Publishers. 
\item  Devoret, M.H., Wallraff,  A., \&\ Martinis,  J.M.  (2004).
Superconducting Qubits: A Short Review. \texttt{arXiv:cond-mat/0411174}.
\item Diacu, F. \&\ Holmes, P.  (1996). {\it Celestial Encounters. The Origins of Chaos and Stability}. Princeton: Princeton University Press.
\item Dickson, M. (2005). Non-relativistic \qm. {\it This Volume}.  
\item  Dieks, D. (1989a). Quantum mechanics without the projection postulate and its realistic interpretation. {\it Foundations of Physics} 19, 1397--1423.
\item  Dieks, D. (1989b). Resolution of the measurement problem through
decoherence of the quantum state. {\it Physics Letters} 142A,  439--446.
\item  Dimassi, M. \&\  Sj\"ostrand, J. (1999). {\it Spectral Asymptotics in the Semi-Classical Limit}.  Cambridge: Cambridge University Press.
\item Dirac, P.A.M. (1926). The fundamental equations of \qm. {\it Proceedings of the Royal Society} A109, 642--653. 
\item Dirac, P.A.M. (1930). {\it The Principles of Quantum
Mechanics}. Oxford: Clarendon Press.
\item Dirac, P.A.M. (1964).  \textit{Lectures on Quantum
Mechanics}.  New York: Belfer School of Science, Yeshiva University.
\item Dixmier, J. (1977). {\it
$C^*$-Algebras}. Amsterdam: North--Holland.    
\item Doebner, H.D. \&\ J. Tolar (1975). Quantum mechanics
on homogeneous spaces. {\em Journal of Mathematical Physics}  16, 975--984.                                                                         
\item Dowker, F. \&\ Kent, A. (1996). On the Consistent Histories Approach to Quantum Mechanics. {\it Journal of Statistical Physics } 82, 1575--1646.
\item Dubin, D.A., Hennings, M.A., \&\ Smith, T.B.  (2000).
{\it Mathematical Aspects of Weyl Quantization and Phase}. Singapore: World Scientific.
\item ÊDuclos, P. \&\ Hogreve, H. (1993). On the semiclassical localization of the quantum probability. {\it Journal of Mathematical Physics}  34, 1681--1691.
\item Duffield, N.G. (1990). Classical and
thermodynamic limits for generalized quantum spin systems.  {\em
Communications in Mathematical\ Physics } 127, 27--39.
\item Duffield, N.G. \&\ Werner, R.F. (1992a). Classical Hamiltonian dynamics for quantum Hamiltonian mean-field limits. {\it Stochastics and Quantum Mechanics: Swansea, Summer 1990}, pp.\ 115--129. Truman, A. \&\ Davies, I.M. (Eds.). Singapore: World Scientific. 
\item Duffield, N.G. \&\ Werner, R.F. (1992b). On mean-field dynamical semigroups on \ca s.
{\it Reviews in Mathematical Physics} 4, 383--424.
\item Duffield, N.G. \&\ Werner, R.F. (1992c). Local dynamics of mean-field quantum systems. {\it Helvetica Physica Acta} 65, 1016--1054. 
\item Duffield, N.G., Roos, H.,  \&\ Werner, R.F. (1992).
Macroscopic limiting dynamics of a class of inhomogeneous mean field quantum systems. {\it  Annales de l' Institut Henri Poincar\'{e} -  Physique Th\'{e}orique}  56, 143--186. 
\item ÊDuffner, E. \&\ Rieckers, A. (1988). On the global quantum dynamics of multilattice systems with nonlinear  classical effects.  {\it Zeitschrift f\"{u}r Naturforschung}  A43, 521--532.
\item Duistermaat, J.J. (1974). Oscillatory integrals, Lagrange immersions and unfolding of singularities. {\it Communications in  Pure and Applied Mathematics} 27, 207--281.
\item Duistermaat, J.J. \&\ Guillemin, V. (1975). The spectrum of positive elliptic operators and periodic bicharacteristics. {\it  Inventiones Mathematicae} 29, 39--79.
\item Duistermaat, J.J. (1996). {\it Fourier Integral Operators}.
Original Lecture Notes from 1973. Basel: Birkh\"auser.
\item ÊDuval, C., Elhadad, J., Gotay, M.J., \'Sniatycki, J., \&\  Tuynman, G.M. (1991).
Quantization and bosonic BRST theory.  {\it Annals of  Physics}  206, 1--26.
\item Earman, J. (1986). {\it A Primer on Determinism}.  Dordrecht: Reidel.
\item Earman, J. (2005). Aspects of determinism in modern physics. {\it This volume}. 
\item Earman, J. (2006). Essential self-adjointness: implications for determinism and the classical-quantum correspondence. {\it Synthese}, to appear. 
\item   Echterhoff, S.,  Kaliszewski, S., Quigg, J., \&\ Raeburn, I. (2002).
A categorical approach to imprimitivity theorems for C*-dynamical  systems.    \texttt{arXiv:math.OA/0205322}.
\item Eddington, A.S. (1920). {\it Space, Time, and Gravitation: An Outline of the General Relativity Theory}. Cambridge:  Cambridge University Press.
\item  Effros, E.G. \&\ Hahn, F. (1967). Locally compact transformation groups and $C\sp{*} $-  algebras. {\it Memoirs of the American Mathematical Society} 75.
\item Ehrenfest, P. (1927). Bemerkung \"{u}ber die angen\"{a}herte Gultigkeit der klassischen Mechanik innerhalb der Quantenmechanik. {\it Zeitschrift f\"{u}r Physik} 45, 455--457.
\item Einstein, A. (1905). \"{U}ber einen die Erzeugung und Verwandlung des Lichtes betreffenden heuristischen Gesichtpunkt. {\it Annalen der Physik} 17, 132--178.   
\item Einstein, A. (1917). Zum Quantensatz von Sommerfeld und Epstein. {\it Verhandlungen der deutschen Physikalischen Geselschaft (2)} 19, 82--92.
\item Einstein, A. (1949). Remarks to the essays appearing in this collective volume.
(Reply to criticisms). 
 {\it Albert Einstein: Philosopher-Scientist}, pp.\ 663--688. Schilpp, P.A. (Ed.). La Salle: Open Court. 
\item ÊEmch, G.G. \&\ Knops, H.J.F. (1970). Pure thermodynamical phases as extremal KMS states. {\it Journal of Mathematical Physics} 11, 3008--3018.
\item ÊEmch, G.G. (1984) {\it Mathematical and conceptual foundations of 20th-century physics}. Amsterdam: North-Holland. 
\item Emch, G.G. \&\ Liu, C. (2002). {\it The Logic of Thermostatistical Physics}.
Berlin: Springer-Verlag. 
\item Emch, G.G., Narnhofer, H., Thirring, W., \&\ Sewell, G. (1994).
  Anosov actions on noncommutative algebras.
 {\it  Journal of Mathematical Physics}  35, 5582--5599.
\item d'Espagnat, B. (1995). {\it Veiled Reality: An Analysis of Present-Day Quantum Mechanical Concepts}.  Reading (MA): Addison-Wesley.
\item Esposito, G., Marmo, G., \&\  Sudarshan, G. (2004). {\it From Classical to Quantum Mechanics:  An Introduction to the Formalism, Foundations and Applications}. Cambridge: Cambridge University Press.
\item Enz, C.P. (2002). {\it No Time to be Brief: A Scientific Biography of Wolfgang Pauli}.
Oxford: Oxford University Press.
\item Everett, H. {\sc iii} (1957). ``Relative state" formulation of \qm. {\it Reviews in Modern Physics} 29, 454--462.
\item   Faye, J. (1991). {\it Niels Bohr: His Heritage and Legacy. An Anti-Realist View of Quantum Mechanics}. Dordrecht: Kluwer Academic Publishers.  
\item   Faye, J. (2002). Copenhagen Interpretation of Quantum Mechanics.
{\it The Stanford Encyclopedia of Philosophy (Summer 2002 Edition)}.  Zalta, E.N.Ê(Ed.). \\ \texttt{http://plato.stanford.edu/archives/sum2002/entries/qm-copenhagen/}.         
\item   Faye, J. \&\ Folse, H. (Eds.) (1994). {\it Niels Bohr and Contemporary Philosophy}.  Dordrecht: Kluwer Academic Publishers.
\item  Fell, J.M.G. \&\  Doran, R.S. (1988). {\it
Representations of $\mbox{}^*$-Algebras, Locally Compact Groups and
Banach $\mbox{}^*$-Algebraic Bundles, Vol.\ 2}. Boston: Academic Press.
\item Feyerabend, P. (1981). Niels Bohr's world view. {\it Realism, Rationalism \&\ Scientific Method: Philosophical Papers Vol.\ 1}, pp.\ 247--297.
Cambridge: Cambridge University Press. 
\item Fleming, G. \&\  Butterfield, J. (2000). Strange positions. {\it From Physics to Philosophy}, pp.\ 108--165. Butterfield, J. \&\  Pagonis, C. (Eds.). 
Cambridge: Cambridge University Press.
\item
Folse, H.J. (1985). {\it  The Philosophy of Niels Bohr}. Amsterdam: North-Holland.
\item Ford, J. (1988). Quantum chaos. Is there any? {\it Directions in Chaos, Vol. 2}, pp.\ 128--147. Bai-Lin, H. (Ed.). Singapore: World Scientific. 
\item Frasca, M. (2003). General theorems on decoherence in the thermodynamic limit. {\it Physics Letters} A308, 135--139.
\item Frasca, M. (2004). Fully polarized states and decoherence. \texttt{arXiv:cond-mat/0403678}. 
\item Frigerio, A. (1974). Quasi-local observables and the problem of
measurement in quantum mechanics. {\it  Annales de l' Institut Henri Poincar\'{e}} A3,  259--270. 
\item Fr\"{o}hlich, J., Tsai, T.-P., \&\ Yau, H.-T. (2002).  On the point-particle (Newtonian) limit of the non-linear Hartree  equation.  {\it Communications in Mathematical Physics}  225, 223--274.
\item Gallavotti, G. (1983).  {\it The Elements of Mechanics}. Berlin: Springer-Verlag.                                                 
\item Gallavotti, G., Bonetto, F., \&\  Gentile, G. (2004). {\it Aspects of Ergodic, Qualitative and Statistical Theory of Motion}. New York: Springer.   
\item Gell-Mann, M. \&\ Hartle, J.B. (1990). Quantum mechanics in the light of
quantum cosmology. {\it Complexity, Entropy, and the Physics of Information},  pp.\ 425--458.
Zurek, W.H. (Ed.). Reading, Addison-Wesley.
\item  Gell-Mann, M. \&\ Hartle, J.B. (1993). Classical equations for quantum systems.
 {\it Physical Review} D47, 3345--3382. 
\item G\'erard, P. \&\ Leichtnam, E. (1993). Ergodic properties of eigenfunctions for the Dirichlet problem.  {\it Duke Mathematical Journal}  71, 559--607.   
\item Gerisch, T.,  M\"{u}nzner, R., \&\ Rieckers, A.  (1999).
 Global $C^*$-dynamics and its KMS states of weakly inhomogeneous  bipolaronic superconductors.  {\it Journal of Statistical Physics }  97, 751--779.   
\item Gerisch, T.,  Honegger, R., \&\ Rieckers, A.  (2003). Algebraic quantum theory of the Josephson microwave radiator.  {\it  Annales Henri Poincar\'{e}}   4, 1051--1082.  
\item Geyer, B., Herwig, H.,  \&\ Rechenberg, H. (Eds.) (1993). {\it Werner Heisenberg: Physiker und Philosoph}. Leipzig: Spektrum. 
\item Giulini, D. (2003).  Superselection rules and symmetries. {\it Decoherence and the Appearance of a Classical World in Quantum Theory}, pp.\ 259--316. Joos, E. et al. (Eds.).
Berlin: Springer.                                                                                                                                                                  
\item Glimm, J. \&\ Jaffe, A. (1987). {\it Quantum Physics. A Functional Integral Point of View}. New York: Springer-Verlag.
\item Gotay, M.J. (1986). Constraints, reduction, and
quantization.  {\it Journal of  Mathematical\ Physics }  27, 2051--2066.
\item Gotay, M.J. (1999). On the Groenewold-Van Hove problem for $\mathbf{R}\sp {2n}$.  {\it Journal of Mathematical Physics}  40, 2107--2116.
\item Gotay, M.J., Grundling, H.B., \&\ Tuynman, G.M. (1996). Obstruction results in quantization theory.  {\it Journal of Nonlinear Science}  6, 469--498. 
\item Gotay, M.J.,
 Nester, J.M., \&\  Hinds, G. (1978). Presymplectic manifolds and the
Dirac--Bergmann theory of constraints. {\it Journal of  Mathematical\ Physics } 19, 2388--2399.
\item G\"{o}tsch, J. (1992). {\it Erwin Schr\"{o}dinger's World View: The Dynamics of Knowlegde and Reality}. Dordrecht: Kluwer Academic Publishers. 
 \item
Govaerts, J. (1991). {\it Hamiltonian Quantization and Constrained
Dynamics}. Leuven: Leuven University Press.
 \item     Gracia-Bond\'{\i}a, J.M.,   V\'{a}rilly, J.C., \&\ Figueroa, H. (2001).
\textit{Elements of Noncommutative Geometry}. Boston:
Birkh\"{a}user.
\item Griesemer, M., Lieb, E.H., \&\ Loss, M. (2001). Ground states in non-relativistic quantum electrodynamics.  {\it Inventiones Mathematicae}  145, 557--595.
 \item  Griffiths, R.B. (1984). Consistent histories and the interpretation of \qm. {\it Journal of Statistical Physics} 36, 219--272.
\item  Griffiths, R.B. (2002).  {\it   Consistent Quantum Theory}.
Cambridge: Cambridge University Press. 
\item Grigis, A. \&\ Sj\"ostrand, J. (1994). {\it Microlocal Analysis for Differential Operators.} Cambridge: Cambridge University Press.
\item Groenewold, H.J. (1946). On the principles of elementary \qm. {\it Physica} 12, 405--460.
\item Groenewold, H.J. (1952). Information in quantum measurements. 
{\it  Proceedings Koninklijke Nederlandse Akademie van Wetenschappen}
B55, 219--227. 
\item Guhr, T., M\"uller-Groeling, H., \&\ Weidenm\"uller, H. (1998). Random matrix theories in quantum physics: common concepts. {\it Physics Reports} 299, 189--425. 
\item  Guillemin, V.,  Ginzburg, V. \&  Karshon, Y. (2002). 
{\it Moment Maps, Cobordisms, and Hamiltonian Group Actions}. 
Providence (RI): American Mathematical Society.
\item Guillemin, V. \&\ Sternberg, S. (1977). {\it Geometric
Asymptotics}.  Providence (RI): American Mathematical
Society.
\item Guillemin, V. \&\ Sternberg, S. (1990). {\it  Variations on a Theme by Kepler}.  Providence (RI): American Mathematical
Society.
\item  Guillemin, V. \&\ Uribe, A. (1989). Circular symmetry and the trace formula. {\it Inventiones Mathematicae} 96, 385--423.
\item  Gustafson, S.J. \&\ Sigal, I.M. (2003). {\it Mathematical concepts of quantum mechanics}. Berlin: Springer.
\item  Gutzwiller, M.C. (1971). Periodic orbits and classical quantization conditions. {\it Journal of Mathematical Physics} 12, 343--358.
\item 
Gutzwiller, M.C. (1990). {\it Chaos in Classical and Quantum
Mechanics}. New York: Springer-Verlag.
\item Gutzwiller, M.C. (1992). Quantum chaos. {\it Scientific American} 266,  78--84.
\item Gutzwiller, M.C. (1998). Resource letter ICQM-1: The interplay between classical and quantum mechanics. {\it American  Journal of Physics } 66, 304--324.
\item Haag, R. (1962). The mathematical structure of the Bardeen--Cooper--Schrieffer model. {\it Nuovo Cimento} 25, 287--298. 
\item Haag, R., Kadison, R., \&\ Kastler, D. (1970). Nets of \ca s and classification of states. {\it Communications in Mathematical Physics} 16, 81--104.
\item   Haag, R. (1992).   {\it Local Quantum Physics: Fields, Particles, Algebras}. 
Heidelberg: Springer-Verlag.  
\item Haake, F. (2001). {\it Quantum Signatures of Chaos}. Second Edition. New York: Springer-Verlag. 
\item Hagedorn, G.A. (1998). Raising and lowering operators for semiclassical wave packets.  {\it Annals of  Physics}  269,   77--104.
\item Hagedorn, G.A. \&\ Joye, A. (1999). Semiclassical dynamics with exponentially small error estimates. {\it Communications in Mathematical Physics}  207, 439--465.
\item Hagedorn, G.A. \&\ Joye, A. (2000). Exponentially accurate semiclassical dynamics: propagation,  localization, Ehrenfest times, scattering, and more general states.  {\it Annales Henri Poincar\'e}  1, 837--883.
\item  Halliwell, J.J.  (1998). Decoherent histories and hydrodynamic equations. {\it   Physical Review} D58, 105015-1--12.
\item  ÊHalliwell, J.J. (2000). The emergence of hydrodynamic equations from quantum theory: a  decoherent histories analysis. {\it   International Journal of Theoretical Physics }  39, 1767--1777.
\item  Halliwell, J.J.   (2004).Some recent developments in the decoherent histories approach to  quantum theory. {\it   Lecture Notes in Physics}  633, 63--83.
  \item  ÊHalliwell, J.J. (2005). How the quantum Universe becomes classical.
   \texttt{arXiv:quant-ph/0501119}.
\item ÊHalvorson, H. (2004). Complementarity of representations in quantum mechanics.  {\it Studies in History and Philosophy of Modern Physics }   B35, 45--56.  
\item ÊHalvorson, H. (2005). Algebraic quantum field theory. This volume.
\item Halvorson, H. \&\ Clifton, R.  (1999). Maximal beable subalgebras of quantum-mechanical observables. {\it International Journal of Theoretical Physics } 38, 2441--2484. 
\item Halvorson, H. \&\ Clifton, R.  (2002). Reconsidering Bohr's reply to \epr. {\it Non-locality and Modality}, pp.\ 3--18. Placek, T. \&\ Butterfield, J. (Eds.).
Dordrecht: Kluwer Academic Publishers.       
\item Hannabuss, K.C. (1984). Dilations of a quantum measurement.
{\it Helvetica Physica Acta} 57,  610--620.   
\item Harrison, F.E. \&\ Wan, K.K. (1997). Macroscopic quantum systems as measuring devices: dc SQUIDs and  superselection rules. {\it  Journal of Physics}  A30,  4731--4755.     
\item Hartle, J.B.  (1995). Spacetime quantum mechanics and the quantum mechanics of  spacetime. {\it   Gravitation et Quantifications (Les Houches, 1992)}, pp.\   285--480. Amsterdam: North-Holland. 
\item Hartle, J.B.  (2005).
 What connects different interpretations of quantum
mechanics? {\it  Quo Vadis Quantum Mechanics}, pp.\ 73-82.
Elitzur, A.,  Dolev, S., \&\ Kolenda, N. (Eds.).  Heidelberg: Springer-Verlag.
\texttt{arXiv:quant-ph/0305089}.
\item Heath, D. \&\ Sudderth, W.  (1976).  De Finetti's theorem on exchangeable variables.  {\it American Statistics}  30, 188--189.        
\item Heelan, P. (1965). {\it Quantum Mechanics and Objectivity: A Study of the Physical Philosophy of Werner Heisenberg}. Den Haag: Martinus Nijhoff.                  
\item Heilbron, J. (2000). {\it The Dilemmas of an Upright Man: Max Planck as a Spokesman for German Science}.  Second Edition. Los Angeles: University of California Press.  
\item  Heisenberg, W. (1925). 
\"{U}ber die quantentheoretische Umdeutung kinematischer und mechanischer Beziehungen. {\it  Zeitschrift f\"{u}r Physik} 33,  879-893.
\item  Heisenberg, W. (1927). 
\"{U}ber den anschaulichen Inhalt der  quantentheoretischen Kinematik und Mechanik.
{\it  Zeitschrift f\"{u}r Physik} 43, 172--198.
\item  Heisenberg, W. (1930). {\it The Physical Principles of the Quantum Theory}. Chicago: University of Chicago Press.
\item Heisenberg, W. (1942). Ordnung der Wirklichkeit. 
In Heisenberg (1984a), pp.\ 217--306. 
 Also available at 
\texttt{http://werner-heisenberg.unh.edu/Ordnung.htm}. 
\item Heisenberg, W. (1958). {\it Physics and Philosophy: The Revolution in Modern Science}. London: Allen \&\ Unwin.
\item  Heisenberg, W. (1969). {\it Der Teil und das Ganze: Gespr\"{a}che im Umkreis der Atomphysik}.  M\"{u}nchen: Piper. English translation as
Heisenberg (1971).
\item  Heisenberg, W. (1971). 
{\it Physics and Beyond}. New York: Harper and Row. Translation of Heisenberg (1969). 
\item  Heisenberg, W. (1984a). {\it Gesammelte Werke. Series C: Philosophical and Popular Writings, Vol {\sc i}: Physik und Erkenntnis 1927--1955}. Blum, W., D\"{u}rr, H.-P., \&\ Rechenberg, H. (Eds.). M\"{u}nchen: Piper.
\item  Heisenberg, W. (1984b). {\it Gesammelte Werke. Series C: Philosophical and Popular Writings, Vol {\sc ii}: Physik und Erkenntnis 1956--1968}. Blum, W., D\"{u}rr, H.-P., \&\ Rechenberg, H. (Eds.). M\"{u}nchen: Piper.
\item  Heisenberg, W. (1985). {\it Gesammelte Werke. Series C: Philosophical and Popular Writings, Vol {\sc iii}: Physik und Erkenntnis 1969--1976}. Blum, W., D\"{u}rr, H.-P., \&\ Rechenberg, H. (Eds.). M\"{u}nchen: Piper.
\item Held, C. (1994). The Meaning of Complementarity. {\it Studies in History and Philosophy of Science} 25, 871--893.
\item Helffer, B. (1988) {\it
Semi-classical Analysis for the
Schr\"{o}dinger Operator and Applications}.  {\it Lecture Notes in
Mathematics} 1336. Berlin: Springer-Verlag.
\item Heller, E.J. \&\ Tomsovic, S. (1993). Postmodern quantum mechanics.
{\it Physics Today} July, 38--46.
\item Hendry, J. (1984). {\it The Creation of Quantum Mechanics and the Bohr-Pauli Dialogue}.  Dordrecht: D. Reidel.
\item ÊHenneaux, M. \&\ Teitelboim, C. (1992). {\it Quantization of Gauge Systems}. Princeton: Princeton University Press. 
\item Hepp, K. (1972). Quantum theory of measurement and macroscopic observables.
{\it Helvetica Physica Acta} 45, 237--248. 
 \item Hepp, K. (1974). The classical limit of quantum mechanical correlation
functions.  {\it Communications in Mathematical\ Physics }  35, 265--277.
  \item Hepp, K. \&\ Lieb, E. (1974). Phase transitions in reservoir driven open systems with applications to lasers and superconductors. {\it Helvetica Physica Acta} 46, 573--602.
\item Higson, N. (1990).  
  A primer on $KK$-theory.  {\it Operator Theory: Operator Algebras and Applications.  Proceedings Symposia in Pure Mathematical, 51, Part 1}, pp.\   239--283.   Providence, RI:
 American  Mathematical Society
\item Hillery, M.,
 O'Connel,  R.F.,  Scully, M.O., \&\  Wigner, E.P. (1984). Distribution
 functions in physics -- Fundamentals. \textit{Physics \ Reports}  106,
 121--167.
\item Hislop, P. D. \&  Sigal, I. M. (1996). {\it Introduction to Spectral Theory. With Applications to Schr\"{o}dinger Operators}. New York: Springer-Verlag.
\item Hoffmann, E.T.A. (1810). 
{\it Musikalische Novellen und Aufs\"{a}tze}. Leipzig: Insel-B\"{u}cherei.
\item Holevo, A.S. (1982). {\it Probabilistic and Statistical Aspects of Quantum Theory}.  Amsterdam: North-Holland Publishing Co.  
\item Hogreve, H.,  Potthoff, J., \&\  Schrader, R. (1983).
 Classical limits for quantum particles in external Yang--Mills
 potentials. {\it Communications in Mathematical\ Physics }  91, 573--598.
\item Honegger, R. \&\ Rieckers, A. (1994).  Quantized radiation states from the infinite Dicke model.  {\it Publications of the  Research Institute for Mathematical Sciences (Kyoto)}  30, 111--138.
\item Honner, J. (1987). {\it The Description of Nature: Niels Bohr and the Philosophy of Quantum Physics}. Oxford: Oxford University Press.
\item Hooker, C.A. (1972). The nature of quantum mechanical reality: Einstein versus Bohr.
{\it Paradigms \&\ Paradoxes: The Philosophical Challenges of the Quantum Domain}, pp.\ 67--302. Colodny, J. (Ed.). Pittsburgh: University of Pittsburgh Press.
\item H\"{o}rmander, L. (1965).   Pseudo-differential operators. {\it Communications in  Pure Applied Mathematical} 18, 501--517
\item H\"{o}rmander, L. (1979). The Weyl calculus of
 pseudo-differential operators. {\it Communications in Pure Applied\ Mathematical} 
 32, 359--443.
\item H\"{o}rmander, L. (1985a). {\it The Analysis of Linear
 Partial Differential Operators, Vol.\ III}.  Berlin: Springer-Verlag.
\item H\"{o}rmander, L. (1985b). {\it The Analysis of Linear
 Partial Differential Operators, Vol.\ IV}.  Berlin: Springer-Verlag.
\item  Horowitz,
 G.T. \&\ Marolf, D. (1995).  Quantum probes of spacetime
 singularities. \textit{Physical Review}   D52, 5670--5675.
\item H\"{o}rz, H. (1968). {\it Werner Heisenberg und die Philosophie}. Berin: VEB Deutscher Verlag der Wissenschaften.
\item Howard, D. (1990). `Nicht sein kann was nicht sein darf', or the Prehistory of \epr, 1909-1935: Einstein's early worries about the \qm\ of composite systems. {\it 
 Sixty-Two Years of Uncertainty}, pp.\ 61--11. Miller, A.I. (Ed.). New
 York: Plenum.
\item Howard, D. (1994). What makes a classical concept classical? Towards a reconstruction of Niels Bohr's philosophy of physics. {\it Niels Bohr and Contemporary Philosophy}, 
pp.\ 201--229. Faye, J. \&\ Folse, H. (Eds.).  Dordrecht: Kluwer Academic Publishers.
\item Howard, D. (2004). Who Invented the Copenhagen Interpretation? 
{\it Philosophy of Science} 71, 669-682.
\item Howe, R. (1980).
Quantum mechanics and partial differential equations. 
{\it Journal of Functional  Analysis} 38, 188--254
\item Hudson, R.L. (1974).
  When is the Wigner quasi-probability density non-negative?  
  {\it Reports of Mathematical Physics}  6, 249--252.
\item Hudson, R.L. \&\ Moody, G.R. (1975/76). 
  Locally normal symmetric states and an analogue of de Finetti's  theorem.  
{\it  Z. Wahrscheinlichkeitstheorie und Verw. Gebiete}  33,   343--351
\item Hunziker, W. \&\ Sigal, I. M. (2000). The quantum $N$-body problem.  {\it Journal of Mathematical Physics}  41, 3448--3510.
\item Husimi, K. (1940). Some formal
 properties of the density matrix.  \textit{Progress of the Physical and  Mathematical Society of Japan} 22, 264--314.
 \item ÊIsham, C.J. (1984). Topological and global aspects of quantum theory.  {\it  Relativity, Groups and Topology, II (Les Houches, 1983)},   1059--1290.
Amsterdam: North-Holland.  
 \item Isham, C.J. (1994). Quantum logic and the histories approach to quantum theory.  {\it  Journal of Mathematical Physics}  35, 2157--2185.
 \item Isham, C.J. (1997). Topos theory and consistent histories: the internal logic of the set of  all consistent sets. {\it  International Journal of Theoretical Physics }  36, 785--814.
 Ê\item Isham, C.J. \&\ Butterfield, J. (2000). Some possible roles for topos theory in quantum theory and quantum  gravity. {\it  Foundations of  Physics }  30, 1707--1735.
 \item Isham, C.J. \&\  Linden, N. (1994). Quantum temporal logic and decoherence functionals in the histories  approach to generalized quantum theory. {\it    Journal of Mathematical Physics}  35, 5452--5476.
 \item ÊIsham, C.J. \&\ Linden, N. (1995). Continuous histories and the history group in generalized quantum  theory.  {\it  Journal of Mathematical Physics}  36, 5392--5408.
 \item ÊIsham, C.J., Linden, N., \&\  Schreckenberg, S.  (1994). The classification of decoherence functionals: an analog of Gleason's  theorem.  {\it  Journal of Mathematical Physics}  35, 6360--6370.
\item Ivrii, V. (1998). {\it Microlocal Analysis and Precise Spectral Asymptotics}. New York: Springer-Verlag.
\item  Jalabert, R.O. \&\   Pastawski, H.M. (2001). Environment-Independent Decoherence Rate in Classically Chaotic Systems. {\it Physical Review Letters} 86, 2490--2493.
\item Jammer, M. (1966). {\it The Conceptual Development of Quantum Mechanics}. 
New York: McGraw-Hill.
\item Jammer, M. (1974). {\it The Philosophy of Quantum Mechanics}.  New York: Wiley.
\item Janssens, B. (2004). {\it Quantum Measurement: A Coherent Description}.
M.Sc.\ Thesis, Radboud Universiteit Nijmegen. 
\item Jauch, J.M. (1968). {\it Foundations of Quantum Mechanics}. Reading (MA): Addison-Wesley.
\item Joos, E. \&\ Zeh, H.D. (1985). The emergence of classical properties through interaction with the environment. {\it Zeitschrift f\"{u}r Physik} B59, 223--243. 
\item Joos, E., Zeh, H.D., Kiefer, C., Giulini, D., Kupsch, J., \&\ Stamatescu, I.-O. (2003).
{\it Decoherence and the Appearance of a Classical World in Quantum Theory}. Berlin: Springer-Verlag. 
\item J\o rgensen, P.E.T. \&\ Moore, R.T. (1984). {\it Operator Commutation Relations}.
Dordrecht: Reidel. 
\item ÊKaplan, L. \&\ Heller, E.J.  (1998a). Linear and nonlinear theory of eigenfunction scars.  {\it Annals of  Physics}  264, 171--206. 
\item Kaplan, L. \&\ Heller, E.J. (1998b). Weak quantum ergodicity. {\it Physica D}  121, 1--18. 
\item Kaplan, L. (1999). Scars in quantum chaotic wavefunctions.  {\it Nonlinearity}  12, R1--R40.
\item Katok, A. \&\ Hasselblatt, B. (1995). {\it Introduction to the Modern Theory of Dynamical Systems}. Cambridge: Cambridge University Press.
\item Kadison, R.V. \&\ Ringrose, J.R. (1983). {\it Fundamentals of the theory of operator algebras. Vol. 1: Elementary Theory}.   New York: Academic Press.
\item Kadison, R.V. \&\ Ringrose, J.R. (1986). {\it Fundamentals of the theory of operator algebras. Vol. 2: Advanced  Theory}.   New York: Academic Press.
\item  Karasev, M.V. (1989). The Maslov quantization conditions in higher cohomology and analogs of  notions developed in Lie theory for canonical fibre bundles of symplectic  manifolds. I, II.  {\it Selecta Mathematica Formerly Sovietica 
}  8, 213--234, 235--258. 
\item Kent, A. (1990). Against Many-Worlds Interpretations.
{\it  International Journal of Modern Physics }  A5 (1990) 1745.
\item Kent, A. (1997). Consistent sets yield contrary inferences in quantum  theory. {\it Physical Review Letters}  78, 2874--2877. Reply by
Griffiths, R.B. \&\ Hartle, J.B.  (1998). {\it ibid.} 81, 1981. Reply to this reply by
Kent, A.  (1998).  {\it ibid.} 81, 1982. 
\item Kent, A. (1998). Quantum histories. {\it Physica Scripta} T76, 78--84. 
\item Kent, A. (2000). Night thoughts of a quantum physicist.
{\it Philosophical Transactions of the  Royal Society of London} A 358, 75--88.
\item Kiefer, C. (2003). Consistent histories and decoherence. {\it Decoherence and the Appearance of a Classical World in Quantum Theory}, pp.\ 227--258. Joos, E. et al. (Eds.).
Berlin: Springer-Verlag.
\item Kirchberg, E. \&\
  Wassermann, S. (1995). Operations on continuous bundles of
 $C^*$-algebras. {\it Mathematische  Annalen}  303, 677--697.
\item
 Kirillov, A.A. (1990). Geometric Quantization.  {\it Dynamical Systems IV}, pp.\ 137--172. Arnold, V.I. \&\ S.P. Novikov (Eds.).   Berlin: Springer-Verlag.
\item  Kirillov,  A.A. (2004). {\it Lectures on the Orbit Method}.  Providence, RI: American Mathematical Society.
\item Klauder, J.R. \&\ B.-S. Skagerstam (Eds.). (1985).  {\it Coherent
 States}. Singapore: World Scientific.
\item
 Klingenberg, W. (1982). \textit{Riemannian Geometry}. de Gruyter,
 Berlin.
\item Kohn, J.J. \&\  Nirenberg, L. (1965).
An algebra of pseudo-differential operators. 
{\it Communications in  Pure and Applied Mathematics} 18  269--305.
\item Koopman, B.O. (1931). Hamiltonian systems and transformations in \Hs. {\it Proceedings of the National Academy of  Sciences} 18, 315--318.
\item Kostant, B. (1970). Quantization and unitary
 representations.  {\it Lecture Notes in Mathematics} 170,
 87--208.
\item  Krishnaprasad, P. S. \&\ Marsden, J. E. (1987). Hamiltonian structures and stability for rigid bodies with flexible  attachments.  {\it Archive of Rational Mechanics and  Analysis}  98, 71--93. 
\item Kuhn,  T. S.  (1978). {\it Black-body Theory and the Quantum Discontinuity: 1894Ð1912}.  New York:  Oxford University Press. 
\item K\"{u}mmerer, B. (2002).  Quantum Markov processes. {\it Coherent Evolution in Noisy Environment (Lecture Notes in Physics Vol. 611)}, Buchleitner, A. \&\  Hornberger, K. (Eds.), pp.\ 139-198. Berlin: Springer-Verlag.
\item Lahti, P. \&\ Mittelstaedt, P. (Eds.) (1987). {\it The Copenhagen Interpretation 60 Years After the Como Lecture}. Singapore: World Scientific.
\item Landau, L.D. \&\ Lifshitz, E.M. (1977). {\it Quantum Mechanics: Non-relativistic Theory}.
3d Ed. Oxford: Pergamon Press.
\item Landsman, N.P. (1990a). 
 Quantization and superselection sectors I.  Transformation group
 $C^*$-algebras. \textit{Reviews in Mathematical Physics }  2, 45--72.
\item
 Landsman, N.P. (1990b). Quantization and superselection sectors
 II. Dirac Monopole and Aharonov--Bohm effect. \textit{Reviews in Mathematical 
 Physics } 2, 73--104.
\item Landsman, N.P. (1991). Algebraic theory of superselection sectors and the measurement problem in quantum mechanics. {\it International Journal of Modern Physics} A30, 5349--5371. 
\item Landsman, N.P. (1992) Induced representations, gauge fields, and
 quantization on homogeneous spaces.  \textit{Reviews in Mathematical Physics } 
 4, 503--528.
\item  Landsman, N.P.  (1993). Deformations of algebras
 of observables and the classical limit of quantum mechanics. {\it
 Reviews in Mathematical Physics }  5, 775--806.
\item Landsman, N.P. (1995). Observation and superselection in quantum mechanics.
{\it Studies in History and Philosophy of Modern Physics } 26B, 45--73.
\item Landsman, N.P. (1997). Poisson spaces with a
 transition probability. \textit{Reviews in Mathematical Physics }  9,
 29--57.
\item Landsman, N.P. (1998). {\it Mathematical Topics Between Classical and Quantum Mechanics}. New York: Springer-Verlag. 
\item Landsman, N.P. (1999a).
Quantum Mechanics on phase Space.  {\it Studies in History and Philosophy of Modern Physics } 30B, 287--305.
\item  Landsman, N.P.  (1999b). Lie groupoid $C\sp *$-algebras and Weyl quantization.  {\it Communications in Mathematical Physics}  206, 367--381.
\item  Landsman, N.P. (2001).  Quantized reduction as a tensor product.  {\it Quantization of Singular Symplectic Quotients}, pp.\ 137--180. Landsman, N.P., Pflaum, M.J., \&\ Schlichenmaier, M. (Eds.).
Basel: Birkh\"{a}user.
\item Landsman, N.P.  (2002). Quantization as a functor. {\it Contemporary Mathematics}
315, 9--24. 
\item Landsman, N.P.  (2005a). Functorial quantization and the Guillemin--Sternberg conjecture. {\it Twenty Years of Bialowieza: A Mathematical Anthology}, pp.\ 23--45.
Ali, S.T., Emch, G.G., Odzijewicz, A., Schlichenmaier, M., \&\ Woronowicz, S.L.  (Eds). Singapore: World Scientific. 
\texttt{arXiv:math-ph/0307059}.
\item Landsman, N.P. (2005b). Lie Groupoids and Lie algebroids in physics and noncommutative geometry. {\it Journal of Geom. Physics }, to appear.
\item Landsman, N.P. (2006). 
When champions meet:  Rethinking the Bohr--Einstein debate.
{\it Studies in History and Philosophy of Modern Physics}, to appear.
\texttt{arXiv:quant-ph/0507220}.
\itemÊLandsman, N.P. \&\ Ramazan, B.  (2001). Quantization of Poisson algebras associated to Lie algebroids. {\it  Contemporary Mathematics} 282, 159--192.
\item Laurikainen, K.V. (1988). {\it Beyond the Atom: The Philosophical Thought of Wolfgang Pauli}. Berlin: Springer-Verlag.
\item Lazutkin, V.F. (1993). {\it KAM Theory and Semiclassical Approximations to Eigenfunctions}. Berlin: Springer-Verlag. 
\item Leggett, A.J. (2002). Testing the limits of quantum mechanics: motivation, state of play, prospects.  {\it Journal of Physics: Condensed Matter} 14, R415--R451.
\item Liboff, R.L. (1984). The correspondence principle revisited. {\it Physics Today} February, 50--55. 
\item Lieb, E.H. (1973).  The classical limit of quantum spin systems.  
 {\it Communications in Mathematical Physics}  31, 327--340.
\item Littlejohn, R.G. (1986). The semiclassical evolution of wave packets.  {\it Physics  Reports}  138, 193--291. 
\item  Ludwig, G. (1985).
{\it   An Axiomatic Basis for Quantum Mechanics. Volume 1:
                 Derivation of Hilbert Space Structure}. Berlin: Springer-Verlag.
\item ÊLugiewicz, P. \&\ Olkiewicz, R. (2002). Decoherence in infinite quantum systems.  {\it Journal of Physics }  A35, 6695--6712.
\item ÊLugiewicz, P. \&\ Olkiewicz, R. (2003).  Classical properties of infinite quantum open systems.  Communications in Mathematical Physics  239, 241--259. 
\item Maassen, H.  (2003). Quantum probability applied to the damped harmonic oscillator.  {\it Quantum Probability Communications, Vol. XII (Grenoble, 1998)},  pp.\  23--58. 
 River Edge, NJ: World Scientific Publishing. 
\item Mackey, G.W. (1962). {\it   The Mathematical Foundations of Quantum Mechanics}.
New York: Benjamin.
\itemÊMackey, G.W.  (1968). {\it Induced Representations of Groups and Quantum Mechanics}. New York: W. A. Benjamin; Turin: Editore Boringhieri. 
\itemÊMackey, G.W.  (1978). {\it Unitary Group Representations in Physics, Probability, and Number  Theory}. Reading, Mass.:  Benjamin/Cummings Publishing Co.
\itemÊMackey, G.W.  (1992). {\it  The Scope and History of Commutative and Noncommutative Harmonic  Analysis}.   Providence, RI: American Mathematical Society.
\item Majid, S. (1988). Hopf algebras for physics at the Planck scale.  {\it Classical \&\ Quantum Gravity}  5, 1587--1606.
\item Majid, S. (1990). Physics for algebraists: noncommutative and noncocommutative Hopf  algebras by a bicrossproduct construction. {\it  Journal of Algebra}  130, 17--64.
\item Marmo, G., Scolarici, G., Simoni, A., \&\ Ventriglia, F. (2005).
The quantum-classical transition: the fate of the complex structure.
{\it International Journal of Geometric Methods in Physics} 2, 127--145.
\item Marsden, J.E. (1992). {\it Lectures on Mechanics}.  Cambridge: Cambridge University Press.
 \item Marsden, J.E. \&\ T.S. Ratiu (1994).
{\it Introduction to Mechanics and Symmetry}. New York: Springer-Verlag.
 \item Marsden, J.E., Ra\c tiu, T., \&\  Weinstein, A. (1984). Semidirect products and reduction in mechanics.  {\it Transactions of the  American  Mathematical Society}  281, 147--177.
\item Marshall, W.,  Simon, C., Penrose,  R., \&\ Bouwmeester, D. (2003).
Towards quantum superpositions of a mirror. {\it Physical Review Letters} 91, 130401-1--4.
 \item Martinez, A. (2002). {\it An Introduction to Semiclassical and Microlocal Analysis}. New York: Springer-Verlag. 
 \item Maslov, V.P. \&\ Fedoriuk, M.V. (1981). {\it Semi-Classical Approximation in Quantum Mechanics}. Dordrecht: Reidel.
 \item McCormmach, R. (1982). {\it Night Thoughts of a Classical Physicist}.
 Cambridge (MA): Harvard University Press.                  
\item  Mehra, J. \&\ and Rechenberg, H. (1982a). {\it The Historical Development of Quantum Theory. Vol.\ 1: The Quantum Theory of Planck, Einstein, Bohr, and Sommerfeld: Its Foundation and the Rise of Its Difficulties}. New York: Springer-Verlag.
\item  Mehra, J. \&\ and Rechenberg, H. (1982b). {\it The Historical Development of Quantum Theory. Vol.\ 2: The Discovery of Quantum Mechanics }. New York: Springer-Verlag.
\item  Mehra, J. \&\ and Rechenberg, H. (1982c). {\it The Historical Development of Quantum Theory. Vol.\ 3: The  formulation of matrix mechanics and its modifications, 1925-1926.}  New York: Springer-Verlag.
\item  Mehra, J. \&\ and Rechenberg, H. (1982d). {\it The Historical Development of Quantum Theory. Vol.\ 4: The  fundamental equations of quantum mechanics 1925-1926. The reception of the new quantum mechanics.}  New York: Springer-Verlag.
\item  Mehra, J. \&\ and Rechenberg, H. (1987). {\it The Historical Development of Quantum Theory. Vol.\ 5:  Erwin Schr\"{o}dinger and the Rise of Wave Mechanics.} New York: Springer-Verlag.
\item  Mehra, J. \&\ and Rechenberg, H. (2000). {\it The Historical Development of Quantum Theory. Vol.\ 6: The Completion of Quantum Mechanics 1926--1941.    Part 1: The probabilistic Interpretation and the Empirical and Mathematical Foundation of Quantum Mechanics, 1926-1936.} New York: Springer-Verlag.
\item  Mehra, J. \&\ and Rechenberg, H. (2001). {\it The Historical Development of Quantum Theory. Vol.\ 6: The Completion of Quantum Mechanics 1926--1941.    Part 2: The Conceptual Completion of Quantum Mechanics.}  New York: Springer-Verlag.
\item Meinrenken, E. (1998).
Symplectic surgery and the $\spinc$-Dirac operator. {\it Adv.\ Mathematical}
 134, 240--277.
\item Meinrenken, E. \&\ Sjamaar, R.  (1999).
Singular reduction and quantization. {\it Topology} 38,  699--762.
\item  Mermin, N.D. (2004).  What's wrong with this quantum world? {\it Physics Today}
57 (2), 10.
\item Mielnik, B. (1968). Geometry of quantum states. {\it
Communications in Mathematical\ Physics }  9, 55--80.
\item Miller, A.I. (1984). {\it
 Imagery in Scientific Thought: Creating 20th-Century Physics}.  Boston: Birkh\"{a}user. 
\item Mirlin, A.D. (2000). Statistics of energy levels and eigenfunctions in disordered systems.
{\it Physics Reports} 326, 259--382.
\item Mittelstaedt, P. (2004). {\it The Interpretation of Quantum Mechanics and the Measurement Process}. Cambridge: Cambridge University Press.
\item Moore, G.E. (1939). Proof of an external world. {\it Proceedings of the British Academy} 25, 273--300. Reprinted in {\it Philosophical Papers} (George, Allen and Unwin, London, 1959) and in {\it Selected Writings} (Routledge, London, 1993). 
\item Moore, W. (1989). {\it Schr\"{o}dinger: Life and Thought}. Cambridge: Cambridge University Press.
\item Morchio, G. \&\ Strocchi, F. (1987). Mathematical structures for long-range dynamics and symmetry breaking. {\it Journal of Mathematical Physics} 28, 622--635.
\item Muller, F.A. (1997). The equivalence myth of quantum mechanics I, II.
{\it Studies in History and Philosophy of Modern Physics } 28  35--61, 219--247.
\item Murdoch, D. (1987). {\it Niels BohrÕs Philosophy of Physics}. Cambridge: Cambridge University Press. 
\item Nadirashvili, N., Toth, J., \&\ Yakobson, D. (2001).
  Geometric properties of eigenfunctions. {\it Russian Mathematical Surveys}  56,  1085--1105.
\item ÊNagy, G. (2000). A deformation quantization procedure for $C^*$-algebras. {\it Journal of Operator Theory}  44, 369--411.
\item Narnhofer, H. (2001). Quantum K-systems and their abelian models.  {\it Foundations of Probability and Physics},  pp.\ 274--302. River Edge, NJ: World Scientific. 
\item ÊNatsume, T. \&\  Nest, R.  (1999). Topological approach to quantum surfaces. 
{\it Communications in Mathematical Physics}  202, 65--87.
\item ÊNatsume, T.,  Nest, R., \&\  Ingo, P. (2003).  Strict quantizations of symplectic manifolds.  {\it Letters  in Mathematical Physics}   66, 73--89. 
 \item Nauenberg, M. (1989). Quantum wave packets on Kepler elliptic orbits. {\it Physical Review} A40, 1133--1136.
\item Nauenberg, M., Stroud, C., \&\ Yeazell, J. (1994). The classical limit of an atom. {\it Scientific American } June, 24--29. 
\item Neumann, J. von (1931). Die Eindeutigkeit der Schr\"{o}dingerschen
Operatoren. {\it Mathematische Annalen} 104, 570--578.
\item Neumann, J. von (1932). {\it Mathematische Grundlagen der Quantenmechanik.}
Berlin: Springer--Verlag. English translation (1955): {\it Mathematical Foundations of Quantum Mechanics}. Princeton: Princeton University Press.
\item Neumann, J. von (1938). On infinite direct products. {\it Compositio Mathematica} 6, 1--77.
\item Neumann, J.  von (1981).  
Continuous geometries with a transition probability. \textit{Memoirs of the American 
Mathematical Society}  252, 1--210. (Edited by I.S. Halperin. MS from
1937).
\item Neumann, H. (1972). Transformation properties of observables.
{\it Helvetica Physica Acta}  25, 811-819.
\item Nourrigat, J. \&\ Royer, C.  (2004). Thermodynamic limits for Hamiltonians defined as pseudodifferential  operators.  {\it Communications in  Partial Differential Equations}  29,   383--417.
\item O'Connor, P.W.,  Tomsovic, S., \&\  Heller, E.J. (1992).
Semiclassical dynamics in the strongly chaotic regime: breaking the log time barrier.
{\it Physica } D55, 340--357.
\item Odzijewicz, A. (1992). Coherent states and geometric quantization. {\it Communications in Mathematical Physics} 150, 385--413.
\item Odzijewicz, A. \&\ Ratiu, T.S.  (2003).  Banach Lie-Poisson spaces and reduction.  {\it Communications   in Mathematical Physics}  243, 1--54.
\item ÊOlkiewicz, R. (1999a). Dynamical semigroups for interacting quantum and classical  systems.  {\it Journal of Mathematical Physics}  40, 1300--1316.
\item Olkiewicz, R.  (1999b). Environment-induced superselection rules in Markovian regime.  {\it Communications in  in Mathematical Physics}  208, 245--265. 
\item Olkiewicz, R. (2000). Structure of the algebra of effective observables in quantum  mechanics. {\it  Annals of  Physics}  286, 10--22. 
\item Ollivier, H., Poulin, D., \&\ Zurek, W.H. (2004). Environment as witness: selective proliferation of information and emergence of objectivity. \texttt{arXiv: quant-ph/0408125}. 
\item Olshanetsky, M.A. \&\ Perelomov, A.M.  (1981). Classical integrable finite-dimensional systems related to Lie  algebras.  {\it Physics  Reports}  71,  313--400. 
\item Olshanetsky, M.A. \&\ Perelomov, A.M. (1983).
Quantum integrable systems related to Lie algebras.  {\it Physics  Reports}  94, 313--404. 
  \item Ê   Omn\`{e}s, R.  (1992). Consistent interpretations of quantum mechanics. {\it  Reviews of Modern Physics }  64, 339--382.
  \item   Omn\`{e}s, R. (1994). {\it  The Interpretation of Quantum Mechanics}. 
Princeton: Princeton University Press.
\item   Omn\`{e}s, R. (1997). Quantum-classical correspondence using projection operators. {\it   Journal of Mathematical Physics}  38, 697--707. 
\item  Omn\`{e}s, R. (1999). {\it  Understanding Quantum Mechanics}.
Princeton: Princeton University Press. 
\item \O rsted, B. (1979). Induced representations and a new proof of the
imprimitivity theorem. {\it Journal of  Functional \ Analysis} 31,
355--359.
\item Ozorio de Almeida, A.M.  (1988). {\it Hamiltonian Systems: Chaos and Quantization}. Cambridge: Cambridge University Press.
\item Pais, A. (1982). {\it Subtle is the Lord: The Science and Life of Albert Einstein}. Oxford: Oxford University Press.
\item Pais, A. (1991). {\it
Niels BohrÕs Times: In Physics, Philosophy, and Polity}.  Oxford: Oxford University Press. 
\item Pais, A. (1997). {\it A Tale of Two Continents: A Physicist's Life in a Turbulent World}.
 Princeton: Princeton University Press.
\item Pais, A. (2000). {\it The Genius of Science}.  Oxford: Oxford University Press. 
\item  Parthasarathy, K.R. (1992). {\it An Introduction to Quantum Stochastic Calculus}.
Basel: Birkh\"{a}user.
\item Paul, T. \&\ Uribe, A.  (1995). The semi-classical trace formula and propagation of wave packets.  {\it Journal of Functional  Analysis}  132, 192--249.
\item Paul, T. \&\ Uribe, A. (1996). On the pointwise behavior of semi-classical measures.  {\it Communications in Mathematical Physics}  175, 229--258. 
\item Paul, T. \&\ Uribe, A.  (1998). A. Weighted trace formula near a hyperbolic trajectory and complex  orbits. {\it Journal of Mathematical Physics}  39, 4009--4015. 
\item Pauli, W. (1925).  \"{U}ber den Einflu\ss\ der Geschwindigkeitsabh\"{a}ngigkeit der Elektronenmasse auf den Zeemaneffekt.  {\it Zeitschrift f\"{u}r Physik} 31, 373--385.
\item Pauli, W. (1933). {\it Die allgemeinen Prinzipien der Wellenmechanik}. Fl\"{u}gge, S. (Ed.). {\it Handbuch der Physik}, Vol. V, Part I. Translated as Pauli, W. (1980). {\it General Principles of Quantum Mechanics}. Berlin: Springer-Verlag.
\item Pauli, W. (1949). Die philosophische Bedeutung der Idee der Komplementarit\"{a}t.
Reprinted in von Meyenn, K. (Ed.) (1984). {\it Wolfang Pauli:  Physik und Erkenntnistheorie}, pp.\ 10--23.  Braunschweig: Vieweg Verlag. English translation in Pauli (1994).
\item Pauli, W. (1979). {\it Wissenschaftlicher Briefwechsel mit Bohr, Einstein, Heisenberg. Vol 1: 1919--1929}.  Hermann, A., von Meyenn, K., \&\ Weisskopf, V. (Eds.).  New York: Springer-Verlag.
\item Pauli, W. (1985). {\it Wissenschaftlicher Briefwechsel mit Bohr, Einstein, Heisenberg. Vol 2: 1930--1939}. von Meyenn, K. (Ed.).  New York: Springer-Verlag. 
\item Pauli, W. (1994). {\it  Writings on Physics and Philosophy}.  Enz, C.P. \&\ von Meyenn, K. (Eds.). Berlin: Springer-Verlag.
\item Paz, J.P. \&\ Zurek, W.H. (1999). Quantum limit of decoherence: environment induced superselection of energy eigenstates. {\it  Physical Review Letters} 82, 5181--5185.
\item Pedersen, G.K.  (1979.) {\it \ca s and their Automorphism Groups}.
 London: Academic Press. 
\item Pedersen, G.K.  (1989). {\it Analysis Now}. New York: Springer-Verlag.
 \item Perelomov, A. (1986.) {\it Generalized Coherent States and their Applications}.  Berlin: Springer-Verlag.
\item Peres, A. (1984). Stability of quantum motion in chaotic and regular systems.
{\it Physical Review} A30, 1610--1615.
\item Peres, A. (1995). {\it Quantum Theory: Concepts and Methods}.  Dordrecht: Kluwer Academic Publishers.
\item  Petersen, A.  (1963). The Philosophy of Niels Bohr.  {\it Bulletin of the  Atomic Scientists} 19, 8--14.  
\item Pitowsky, I. (1989). {\it Quantum Probability - Quantum Logic}. Berlin: Springer-Verlag.
\item Planck, M. (1906). {\it Vorlesungen \"{U}ber die Theorie der W\"armestrahlung} Leipzig: J.A. Barth.
\item Poincar\'e, H. (1892--1899). {\it Les M\'ethodes Nouvelles de la M\'echanique C\'eleste}.
Paris: Gauthier-Villars.
\item Popov, G. (2000). Invariant tori, effective stability, and quasimodes with exponentially  small error terms. I \&\ II. {\it  Annales  Henri Poincar\'e}  1, 223--248 \&\ 249--279.
\item Poulin, D. (2004). Macroscopic observables. \texttt{arXiv:quant-ph/0403212}.
\item Poulsen, N.S. (1970). {\it Regularity Aspects of the Theory of
Infinite-Dimensional Representations of Lie Groups}. Ph.D Thesis, MIT.
\item Primas, H. (1983). {\it Chemistry, Quantum Mechanics and Reductionism}. Second Edition. Berlin: Springer-Verlag. 
\item  Primas, H. (1997). The representation of facts in physical theories. {\it  Time, Temporality, Now}, pp.\ 241-263.  Atmanspacher, H. \&\ Ruhnau, E. (Eds.). Berlin: Springer-Verlag.  
\item Prugovecki, E. (1971). {\it Quantum Mechanics in Hilbert Space}. New York: Academic Press. 
\item Puta, M. (1993). {\it Hamiltonian Dynamical Systems and Geometric Quantization}. Dordrecht: D. Reidel.
\item Raggio, G.A. (1981). {\it States and Composite Systems in $W^*$-algebras Quantum Mechanics}. Ph.D Thesis, ETH Z\"{u}rich. 
\item Raggio, G.A. (1988). 
  A remark on Bell's inequality and decomposable normal states.  
  {\it Letters in Mathematical Physics}  15, 27--29.
   \item Raggio, G.A. \&\ Werner, R.F. (1989). Quantum statistical mechanics
  of general mean field systems. {\it Helvetica Physica Acta} 62, 980--1003.
 \item Raggio, G.A. \&\ Werner, R.F. (1991). The Gibbs variational principle for inhomogeneous mean field systems.  {\it Helvetica Physica Acta} 64, 633--667.
\item Raimond, R.M., Brune, M., \&\ Haroche, S. (2001). Manipulating quantum entanglement with atoms and photons in a cavity. {\it Reviews of Modern Physics} 73, 565--582.
\item R\'{e}dei, M. (1998). {\it Quantum logic in algebraic approach}. Dordrecht: Kluwer Academic Publishers.
\item  R\'{e}dei, M.  \&\ St\"{o}ltzner, M. (Eds.). (2001). {\it  John von Neumann and the Foundations of Modern Physics}. Dordrecht: Kluwer Academic Publishers.
\item Reed, M. \&\ Simon, B. (1972).  {\it Methods of Modern Mathematical Physics. Vol I. Functional Analysis.}  New York: Academic Press.
\item Reed, M. \&\ Simon, B. (1975).  {\it  Methods of Modern Mathematical Physics. Vol II. Fourier Analysis,  Self-adjointness.} New York: Academic Press.
\item Reed, M. \&\ Simon, B. (1979).  {\it Methods of Modern Mathematical Physics. Vol III.
Scattering Theory}.  New York: Academic Press.
\item Reed, M. \&\ Simon, B. (1978).  {\it Methods of Modern Mathematical Physics. Vol IV. Analysis of Operators.} New York: Academic Press.
\item Reichl, L.E. (2004). {\it The Transition to Chaos in Conservative Classical Systems: Quantum Manifestations}. Second Edition. New York: Springer-Verlag. 
\item Rieckers, A. (1984). On the classical part of the mean field dynamics for quantum lattice systems in grand canonical representations. {\it Journal of Mathematical Physics} 25, 2593--2601.
\item Rieffel, M.A. (1989a). Deformation quantization of Heisenberg manifolds.  {\it Communications in Mathematical Physics}  122, 531--562. 
\item Rieffel, M.A. (1989b).  Continuous
fields of $C^*$-algebras coming from group cocycles and actions. {\it
Mathematical Annals}  283, 631--643.
\item Rieffel, M.A. (1994). 
  Quantization and $C\sp *$-algebras.   {\it Contemporary Mathematics} 167,  66--97.
\item Riesz, F. \&\  Sz.-Nagy, B. (1990). 
{\em Functional Analysis}. New York: Dover.
\item  Robert,  D. (1987) {\it Autour de l'Approximation
 Semi-Classique}.  Basel: Birkh\"{a}user. 
\item Robert, D.  (Ed.). (1992) {\it  M\'{e}thodes Semi-Classiques}.
{\it Ast\'{e}risque}  207, 1--212, {\it ibid.}\ 210, 1--384.
\item Robert, D.   (1998).  Semi-classical approximation in quantum
mechanics.  A survey of old and recent mathematical results. {\it
Helvetica Physica Acta} 71, 44--116.
\item Roberts, J.E. (1990). Lectures on algebraic quantum field theory. {\it The Algebraic Theory of Superselection Sectors. Introduction and Recent Results}, pp.\ 1--112.  Kastler, D. (Ed.).  River Edge, NJ: World Scientific Publishing Co.  
\item Roberts, J.E. \&\ Roepstorff, G. (1969).
 Some basic concepts of algebraic quantum theory.  
{\it  Communications in Mathematical Physics } 11, 321--338.
\item Robinett, R.W. (2004). Quantum wave packet revival. {\it Physics  Reports} 392, 1-119.
\item Robinson, D. (1994). Can Superselection Rules Solve the Measurement Problem?
{\it British Journal for the Philosophy of Science}  45, 79-93.
\item Robinson, S.L. (1988a). The semiclassical limit of
quantum mechanics. I. Time evolution.  {\it Journal of  Mathematical\ Physics } 
29, 412--419.
\item Robinson, S.L. (1988b). The semiclassical limit of
quantum mechanics. II. Scattering theory.  {\it
 Annales de l' Institut Henri Poincar\'{e}} 
  A48, 281--296.
\item Robinson, S.L. (1993).
Semiclassical mechanics for time-dependent Wigner functions.  \textit{Journal of 
Mathematical\ Physics } 34, 2185--2205.
\item Robson, M.A. (1996). Geometric quantization of reduced cotangent bundles.  {\it Journal of Geometry and  Physics}  19, 207--245.
\item Rosenfeld, L. (1967). Niels Bohr in the Thirties. Consolidation and extension of the conception of complementarity. {\it Niels Bohr: His Life and Work as Seen by His Friends and Colleagues}, pp.\ 114--136. Rozental, S. (Ed.). Amsterdam: North-Holland. 
Ê\item Rudolph, O. (1996a). Consistent histories and operational quantum theory.  {\it  International Journal of Theoretical Physics }  35, 1581--1636.
\item Rudolph, O. (1996b). On the consistent effect histories approach to quantum mechanics. {\it   Journal of Mathematical Physics}  37, 5368--5379. 
\item Rudolph, O. (2000). The representation theory of decoherence functionals in history quantum  theories. {\it    International Journal of Theoretical Physics }  39, 871--884.
\item Rudolph, O. \&\  Wright, J.D. M.  (1999). Homogeneous decoherence functionals in standard and history quantum  mechanics. {\it  Communications in Mathematical Physics}  204, 249--267.
\item Sarnak, P. (1999).  Quantum chaos, symmetry and zeta functions.  I. \&\ II. 
{\it  Quantum  Chaos.  Current Developments in Mathematics, 1997 (Cambridge, MA)}, pp.\  127--144 \&\ 145--159. Boston: International Press. 
\item Saunders, S. (1993). Decoherence, relative states, and evolutionary adaptation. {\it Foundations of Physics} 23, 1553--1585.
 \item Saunders, S. (1995) Time, quantum mechanics, and decoherence.
{\it Synthese} 102, 235--266.
 \item Saunders, S. (2004). Complementarity and Scientific Rationality.
\texttt{arXiv:quant-ph/0412195}.
\item Scheibe, E. (1973). {\it The Logical Analysis of Quantum Mechanics}. Oxford: Pergamon Press.
\item Scheibe, E. (1991). J. v. Neumanns und J. S. Bells Theorem. Ein Vergleich.
{\it Philosophia Naturalis} 28, 35--53. English translation in Scheibe, E. (2001).
{\it Between Rationalism and Empiricism: Selected Papers in the Philosophy of Physics}.
New York: Springer-Verlag. 
\item    Scheibe, E. (1999).  {\it Die Reduktion Physikalischer Theorien.
  Ein Beitrag zur Einheit der Physik. Teil II: Inkommensurabilit\"{a}t und Grenzfallreduktion.}
Berlin: Springer-Verlag.
\item Schlosshauer, M. (2004). Decoherence, the measurement problem, and interpretations of quantum mechanics. {\it Reviews of Modern Physics } 76, 1267--1306.
\item Schm\"{u}dgen, K. (1990). {\it Unbounded Operator Algebras and Representation Theory}. Basel: Birkh\"{a}user Verlag. 
\item  Schr\"{o}dinger, E. (1926a). Quantisierung als Eigenwertproblem. I.-IV.
{\it  Annalen der Physik } 79, 361--76, 489--527, {\it ibid.}  80, 437--90, {\it ibid.} 81, 109--39.
English translation in Schr\"{o}dinger  (1928).
\item  Schr\"{o}dinger, E. (1926b). Der stetige \"{U}bergang von der Mikro-zur Makromekanik. 
{\it Die Naturwissenschaften}  14, 664--668. English translation in Schr\"{o}dinger  (1928).
\item  Schr\"{o}dinger, E. (1926c). \"{U}ber das Verhaltnis der Heisenberg--Born--Jordanschen Quantenmechanik zu der meinen. 
{\it  Annalen der Physik} 79, 734--56. English translation in Schr\"{o}dinger  (1928).
\item  Schr\"{o}dinger, E. (1928). {\it Collected Papers on Wave Mechanics}. London: Blackie and Son.
\item Schroeck, F.E., Jr. (1996). {\it Quantum Mechanics on Phase Space}. Dordrecht: Kluwer Academic Publishers.
\item Scutaru, H. (1977). Coherent states and induced representations.
{\it Letters in Mathematical
 Physics}   2, 101-107.
\item Segal, I.E. (1960). Quantization of nonlinear systems. {\it  Journal of Mathematical Physics}  1,   468--488.
\item Sewell, G. L. (1986). {\it Quantum Theory of Collective Phenomena}. New York:
 Oxford University Press.
\item Sewell, G. L. (2002). {\it Quantum Mechanics and its Emergent Macrophysics}. Princeton: Princeton University Press.
\item  Simon, B. (1976). Quantum dynamics: from automorphism to Hamiltonian. {\it Studies in Mathematical Phyiscs: Essays in Honour of Valentine Bargmann}, pp.\ 327--350. Lieb, E.H., Simon, B. \&\ Wightman, A.S. (Eds.).  Princeton: Princeton University Press. 
\item  Simon, B. (1980).  The classical limit of quantum partition functions.  
 {\it Communications in Mathematical Physics}  71, 247--276.
\item ÊSimon, B. (2000) Schr\"{o}dinger operators in the twentieth century.  {\it Journal of Mathematical Physics}  41, 3523--3555.
\item \'{S}niatycki, J. (1980). {\it Geometric Quantization
and Quantum Mechanics}. Berlin: Springer-Verlag.
\item Snirelman, A.I. (1974). Ergodic properties of eigenfunctions. {\it Uspekhi Mathematical Nauk} 29, 181--182. 
\item Souriau, J.-M. (1969). {\it Structure des syst\`{e}mes dynamiques}. Paris:  Dunod.
Translated as Souriau, J.-M. (1997).
\item Souriau, J.-M. (1997). {\it Structure of Dynamical Systems. A Symplectic View of Physics}. Boston: Birkh\"{a}user.
\item  Stapp, H.P. (1972). The Copenhagen Interpretation. 
{\it  American Journal of Physics} 40, 1098--1116.
\item Steiner, M. (1998). {\it The Applicability of Mathematics as a Philosophical Problem}.
Cambridge (MA): Harvard University Press.
\item
Stinespring, W. (1955). Positive functions on $C^*$-algebras.
{\it Proceedings of the American Mathematical Society} 
6, 211--216.
\item St\o rmer, E. (1969). Symmetric states of infinite tensor products of \ca s. {\it Jornal of Functional Analysis} 3, 48--68.
\item Strawson, P.F. (1959). {\it Individuals: An Essay in Descriptive Metaphysics}. London: Methuen.
\item Streater, R.F. (2000). Classical and quantum probability. {\it Journal of Mathematical Physics} 41, 3556--3603. 
\item Strichartz,
R.S. (1983). Analysis of the Laplacian on a complete Riemannian
manifold.  \textit{Journal of  Functional \ Analysis}  52, 48--79.
\item ÊStrocchi, F. (1985). {\it Elements of Quantum mechanics of Infinite Systems}.  Singapore: World Scientific.
\item Strunz, W.T., Haake, F., \&\ Braun, D. (2003). Universality of decoherence for macroscopic quantum superpositions. {\it Physical Review}  A 67, 022101--022114. 
\item  Summers, S.J. \&\  Werner, R. (1987). BellÕs inequalities and quantum field theory, I,  II.  {\it Journal of Mathematical Physics} 28, 2440--2447, 2448--2456.
\item Sundermeyer, K. (1982). {\it Constrained
Dynamics}.  Berlin: Springer-Verlag.
\item Takesaki, M. (2003). {\it Theory of Operator Algebras. Vols.\ I-III}.  New York: Springer-Verlag.
\item Thirring, W. \&\ Wehrl, A. (1967). On the mathematical structure of the BCS model. I.  {\it Communications in Mathematical Physics} 4, 303--314. 
\item Thirring, W. (1968). On the mathematical structure of the BCS model. II.  {\it Communications in Mathematical Physics} 7, 181--199. 
\item Thirring, W. (1981).  {\it A Course in Mathematical Physics. Vol.\ 3: Quantum Mechanics of Atoms and Molecules}. New York: Springer-Verlag.
\item Thirring, W. (1983).  {\it A Course in Mathematical Physics. Vol.\ 4:
Quantum Mechanics of Large Systems}. New York: Springer-Verlag.
\item ÊTomsovic, S. \&\ Heller, E.J. (1993). Long-time semiclassical dynamics of chaos: The stadium billiard. {\it Physical Review} E47, 282--299.
\item ÊTomsovic, S. \&\ Heller, E.J. (2002). Comment on" Ehrenfest times for classically chaotic systems".  {\it Physical Review}  E65, 035208-1--2.
\item Toth, J.A. (1996). Eigenfunction localization in the quantized rigid body.  
 {\it Journal of Differential Geometry}  43, 844--858.
\item Toth, J.A. (1999). On the small-scale mass concentration of modes. {\it  Communications in Mathematical Physics} 206, 409--428.
\item Toth, J.A. \&\ Zelditch, S. (2002). Riemannian manifolds with uniformly bounded eigenfunctions. {\it  Duke Mathematical Journal}  111, 97--132.
\item Toth, J.A. \&\ Zelditch, S. (2003a). $L^p$ norms of eigenfunctions in the completely integrable  case.  {\it Annales Henri Poincar\'e}  4, 343--368. 
\item Toth, J.A. \&\ Zelditch, S. (2003b). Norms of modes and quasi-modes revisited.  {\it Contemporary Mathematics} 320, 435--458.
\item ÊTuynman, G.M. (1987).  Quantization: towards a comparison between methods.  
 {\it Journal of Mathematical Physics}  28, 2829--2840.
\item ÊTuynman, G.M. (1998). Prequantization is irreducible. {\it Indagationes Mathematicae (New Series)}  9,  607--618.
\item Unnerstall, T. (1990a). Phase-spaces and dynamical descriptions of infinite mean-field quantum systems. {\it Journal of Mathematical Physics} 31, 680--688.
\item Unnerstall, T. (1990b). Schr\"{o}dinger dynamics and physical folia of infinite mean-field quantum systems.   {\it Communications in Mathematical\ Physics  } 130, 237--255.
\item Vaisman, I.  (1991). On the geometric quantization of Poisson manifolds.  {\it Journal of Mathematical Physics}  32,   3339--3345.
\item  van Fraassen, B.C.  (1991). {\it  Quantum Mechanics: An Empiricist View}. Oxford: Oxford University Press.
\item van Hove, L. (1951). Sur certaines repr\'{e}sentations unitaires d'un
 groupe infini de transformations. \textit{Memoires de l'Acad\'{e}mie Royale de
 Belgique, Classe des Sciences}  26, 61--102.
\item Van Vleck, J.H. (1928). The Correspondence Principle in the Statistical Interpretation of Quantum Mechanics. {\it Proceedings of the National Academy of Sciences} 14, 178--188.
\item  van der Waerden, B.L. (Ed.).  (1967).  {\it Sources of Quantum Mechanics}.
Amsterdam: North-Holland. 
\item van Kampen, N. (1954). Quantum statistics of irreversible processes.
{\it Physica} 20, 603--622. 
\item van Kampen, N. (1988). Ten theorems about quantum mechanical measurements.
{\it Physica} A153, 97--113. 
\item van Kampen, N. (1993). Macroscopic systems in \qm. {\it Physica} A194, 542--550.
\item  Vanicek, J. \&\ Heller, E.J. (2003). Semiclassical evaluation of quantum fidelity
{\it Physical Review} E68, 056208-1--5. 
\item Vergne, M. (1994). Geometric quantization and equivariant cohomology. {\it First European Congress in Mathematics, Vol.\ 1}, pp.\ 249--295.
Boston: Birkh\"{a}user.
\item Vermaas, P. (2000). {\it A Philosopher's Understanding of Quantum Mechanics:
Possibilities and Impossibilities of a Modal Interpretation}. Cambridge: Cambridge University Press.
\item Vey, J. (1975).
D\'{e}formation du crochet de Poisson sur une vari\'{e}t\'{e}
symplectique. {\it Commentarii Mathematici Helvetici}
50, 421--454.
\item Voros, A. (1979). Semi-classical ergodicity of quantum eigenstates in the Wigner representation. {\it Stochastic Behaviour in Classical and Quantum Hamiltonian Systems}. {\it Lecture Notes in Physics} 93, 326--333.
\item Wallace, D. (2002). Worlds in the Everett interpretation. {\it Studies in History and Philosophy of Modern Physics  } 33B, 637--661.
\item Wallace, D. (2003). Everett and structure. {\it Studies in History and Philosophy of Modern Physics} 34B, 87--105.
\item ÊWan, K.K. \&\ Fountain, R.H. (1998). Quantization by parts, maximal symmetric operators, and quantum  circuits. {\it  International Journal of Theoretical Physics }  37, 2153--2186. 
\item Wan, K.K.,  Bradshaw, J.,  Trueman, C., \&\ Harrison, F.E. (1998). Classical systems, standard quantum systems, and mixed quantum systems  in Hilbert space.  {\it Foundations of  Physics }  28, 1739--1783. 
\item Wang, X.-P. (1986). Approximation semi-classique
de l'equation de Heisenberg.  {\it Communications in Mathematical\ Physics  } 104,
77--86. 
\item  Wegge-Olsen, N.E. (1993). \textit{K-theory and \ca s}. Oxford: Oxford University Press. 
\item Weinstein, A. (1983). The local structure of Poisson manifolds.  {\it Journal of Differential Geometry}  18, 523--557.
\item  Werner, R.F. (1995). The classical limit of quantum theory.
\texttt{arXiv:quant-ph/9504016}. 
\item Weyl, H. (1931).  {\it The Theory of Groups and Quantum Mechanics}.
 New York: Dover.
\item  Wheeler, J.A. \&\  and Zurek, W.H.  (Eds.)  (1983). {\it Quantum Theory and Measurement}. Princeton: Princeton University Press.
\item  Whitten-Wolfe, B. \&\   Emch,  G.G. (1976). A mechanical quantum
measuring process. {\it Helvetica Physica Acta} 49, 45-55.
\item Wick, C.G., Wightman, A.S., \&\ Wigner, E.P.  (1952). The intrinsic parity of elementary
particles. {\it Physical Review}  88, 101--105.
\item Wightman, A.S. (1962) On the localizability of quantum mechanical
systems. {\it Reviews of Modern Physics} 34, 845-872.
\item Wigner, E.P. (1932). On the quantum correction for
 thermodynamic equilibrium. \textit{Physical Review} 40, 749--759.
\item Wigner, E.P. (1939) Unitary representations of the inhomogeneous
Lorentz group. {\it Annals of Mathematics} 40, 149-204.
\item Wigner, E.P. (1963). The problem of measurement.
 {\it American Journal of Physics} 31, 6--15.
\item Williams, F.L. (2003). {\it Topics in Quantum Mechanics}. Boston: Birkh\"{a}user.
\item ÊWoodhouse, N. M. J. (1992). {\it Geometric Quantization}. Second edition. Oxford: The Clarendon Press.
\item Yajima, K. (1979). The quasi-classical limit of quantum scattering theory.
 {\it Communications in Mathematical\ Physics  }  69, 101--129.
\item Zaslavsky, G.M. (1981). Stochasticity in quantum systems. {\it Physics Reports}
80, 157--250.
\item Zeh, H.D. (1970). On the interpretation of measurement in quantum theory.
{\it Foundations of Physics} 1, 69--76. 
\item Zelditch, S. (1987). Uniform distribution of eigenfunctions on compact hyperbolic surfaces. {\it Duke Mathematical J.} 55, 919--941.
\item Zelditch, S. (1990). Quantum transition amplitudes for ergodic and for completely integrable systems. {\it Journal of Functional  Analysis} 94, 415--436.
\item Zelditch, S.  (1991). Mean Lindel\"of hypothesis and equidistribution of cusp forms and  Eisenstein series.  {\it Journal of Functional  Analysis}  97, 1--49.
\item Zelditch, S. (1992a). On a ``quantum chaos'' theorem of R. Schrader and M. Taylor. {\it Journal of Functional  Analysis} 109, 1--21. 
\item Zelditch, S. (1992b). Quantum ergodicity on the sphere.  {\it Communications in Mathematical\ Physics  } 146, 61--71.
\item Zelditch, S. (1996a). Quantum dynamics from the semiclassical viewpoint. Lectures at the Centre E. Borel. Available at \texttt{http://mathnt.mat.jhu.edu/zelditch/Preprints/preprints.html}.
\item  Zelditch, S. (1996b).  Quantum mixing.  {\it Journal of Functional  Analysis}  140, 68--86.
\item  Zelditch, S. (1996c). Quantum ergodicity of $C\sp *$ dynamical systems. {\it Communications in Mathematical Physics}  177, 507--528.
\item Zelditch, S. \&\  Zworski, M. (1996). Ergodicity of eigenfunctions for ergodic billiards.  {\it Communications in Mathematical Physics}  175, 673--682. 
\item  Zurek, W.H. (1981). Pointer basis of quantum apparatus: into what mixture does the wave packet collapse? {\it Physical Review} D24, 1516--1525.
\item  Zurek, W.H. (1982) Environment-induced superselections rules. {\it Physical Review} D26, 1862--1880.
\item  Zurek, W.H. (1991). Decoherence and the transition from quantum to classical. {\it Physics Today} 44 (10), 36--44. 
\item  Zurek, W.H. (1993). Negotiating the tricky border between quantum and classical.  {\it Physics Today} 46 (4), 13--15, 81--90. 
\item Zurek, W.H. (2003). Decoherence, einselection, and the quantum origins of the classical.
{\it Reviews of Modern Physics} 75, 715--775.
\item Zurek, W.H. (2004). Probabilities from entanglement, Born's rule from envariance.\\
\texttt{arXiv:quant-ph/0405161}.
\item Zurek, W.H., Habib, S., \&\  Paz, J.P. (1993). Coherent states via decoherence.
{\it Physical Review Letters} 70, 1187--1190.
\item Zurek, W.H. \&\  Paz, J.P. (1995).  Why We Don't Need Quantum Planetary Dynamics: Decoherence and the   Correspondence Principle for Chaotic Systems.
{\it Proceedings of the Fourth Drexel Meeting}.  Feng, D.H. et al. (Eds.). New York: Plenum.   
\texttt{arXiv:quant-ph/9612037}.

\end{trivlist}
\end{document}